\documentclass{aa}

\usepackage{graphicx,txfonts,color}
\usepackage{longtable,lscape}
\usepackage{multirow,multicol}
\usepackage{natbib}
\usepackage{xspace}
\usepackage{amsmath}

\bibpunct{(}{)}{;}{a}{}{,}

\newcommand{\kms}{km\,s$^{-1}$\xspace}

\newcommand{\myrkmsfrac}{$\frac{\mathrm{M}_\odot\ \mathrm{yr}^{-1}}{\mathrm{km}\ \mathrm{s}^{-1}}$\xspace}
\newcommand{\myrkms}{$\mathrm{M}_\odot\ \mathrm{yr}^{-1}\ \mathrm{km}^{-1}\ \mathrm{s}$\xspace}
\newcommand{\mic}{$\mu$m\xspace}

\newcommand{\gastronoom}{\emph{GASTRoNOoM}\xspace}
\newcommand{\mcmax}{\emph{MCMax}\xspace}

\newcommand{\water}{H$_2$O\xspace}
\newcommand{\waterih}{H$_2^{18}$O\xspace}
\newcommand{\wateril}{H$_2^{17}$O\xspace}

\newcommand{\coabun}{$n_{\mathrm{CO}}/n_{\mathrm{H}_2}$\xspace}
\newcommand{\waterabun}{$n_{\mathrm{H}_2\mathrm{O}}/n_{\mathrm{H}_2}$\xspace}

\newcommand{\rstar}{R$_\star$\xspace}
\newcommand{\mg}{\dot{M}_\mathrm{g}}
\newcommand{\shellmass}{\dot{M}_\mathrm{g}/\varv_{\infty\mathrm{,g}}}

\newcommand{\vs}{\varv_\mathrm{stoch}}
\newcommand{\vg}{\varv_{\infty\mathrm{,g}}}

\newcommand{\msunyr}{\mathrm{M}_\odot\ \mathrm{yr}^{-1}}

\begin{document}

%\thesaurus{06(08.01.1; 08.03.1; 08.05.3; 08.16.4)}

\title{Constraints on the \water formation mechanism in the wind of carbon-rich AGB stars\thanks{\emph{Herschel} is an ESA space observatory with science instruments provided by European-led Principal Investigator consortia and with important participation from NASA.
}}
%\subtitle{}

%\author{R.~Lombaert \inst{1} et al.}
\author{R.~Lombaert \inst{1,2} \and L.~Decin \inst{2,3} \and P.~Royer \inst{2} \and A.~de Koter \inst{2,3} \and N.L.J.~Cox \inst{2} \and E.~Gonz\'alez-Alfonso \inst{4} \and D.~Neufeld \inst{5} \and J.~De Ridder \inst{2} \and M.~Ag\'undez \inst{6} \and J.A.D.L.~Blommaert \inst{2,7} \and T.~Khouri \inst{1,3} \and M.A.T.~Groenewegen \inst{8} \and F.~Kerschbaum \inst{9} \and J.~Cernicharo \inst{6} \and B.~Vandenbussche \inst{2} \and C.~Waelkens \inst{2}}

%\and G.J.~Melnick \inst{10}
%\and A.~Jorissen \inst{8}

\offprints{R.~Lombaert, lombaert@chalmers.se}

\institute{Department of Earth and Space Sciences, Chalmers University of Technology, Onsala Space Observatory, 439 92 Onsala, Sweden
\and KU~Leuven, Instituut voor Sterrenkunde, Celestijnenlaan 200D B-2401, 3001 Leuven, Belgium 
 \and University of Amsterdam, Astronomical Institute ``Anton Pannekoek'', P.O.~Box 94249, 1090 GE Amsterdam, The Netherlands
 \and Universidad de Alcal\'a de Henares, Departamento de F\'isica y Matem\'aticas, Campus Universitario, 28871 Alcal\'a de Henares, Madrid, Spain
 \and Johns Hopkins University, Department of Physics and Astronomy, 3400 North Charles Street, Baltimore, MD 21218, USA
 %\and Universit\'e Bordeaux, LAB, UMR 5804, 33270 Floirac, France
 \and Group of Molecular Astrophysics, Instituto de Ciencia de Materiales de Madrid, CSIC, C/Sor Juana In\'es de La Cruz N3, E-28049 Cantoblanco, Madrid, Spain
 \and Vrije Universiteit Brussel, Department of Physics and Astrophysics, Pleinlaan 2, 1050 Brussels, Belgium
 \and Koninklijke Sterrenwacht van Belgi\"e, Ringlaan 3, 1180 Brussels, Belgium
 %\and Departamento de Astrof\'isica, Centro de Astrobiolog\'ia, CSIC-INTA, Ctra. de Torrejon a Ajalvir km 4, Torrej\'on de Ardoz, 28850 Madrid, Spain
 %\and Institut d’Astronomie et d’Astrophysique, Universit\'e Libre de Bruxelles, CP 226, Boulevard du Triomphe, 1050 Brussels, Belgium
 \and University of Vienna, Department of Astrophysics, T\"urkenschanzstra\ss e 17, 1180 Wien, Austria
 %\and Harvard-Smithsonian Center for Astrophysics, 60 Garden Street, MS 66, Cambridge, MA 02138, USA
 %\and Astronomical Institute Utrecht, University Utrecht, P.O. Box 80000, 3508 TA Utrecht, The Netherlands
 %\and Max Planck Institut f\"ur Radioastronomie, Auf dem H\"ugel 69, D-53121 Bonn, Germany}
 %\and Netherlands Institute for Space Research, Sorbonnelaan 2, 3584 CA Utrecht, The Netherlands}
}

\date{Received / Accepted}

\authorrunning{R. Lombaert et al.}
\titlerunning{Constraints on the \water formation mechanism in the wind of carbon-rich AGB stars}

%*****************************************************************************
%      ABSTRACT
%*****************************************************************************
\abstract
%context
{The recent detection of warm \water vapor emission from the outflows of carbon-rich asymptotic giant branch (AGB) stars challenges the current understanding of circumstellar chemistry. Two mechanisms have been invoked to explain warm \water vapor formation. In the first, periodic shocks passing through the medium immediately above the stellar surface lead to \water formation. In the second, penetration of ultraviolet interstellar radiation through a clumpy circumstellar medium leads to the formation of \water molecules in the intermediate wind. 
%The Fischer-Tropsch mechanism invokes iron grains as catalysts and contributes to cold \water formation in the intermediate wind.
}
%aims
{We aim to determine the properties of \water emission for a sample of 18 carbon-rich AGB stars and subsequently constrain which of the above mechanisms provides the most likely warm \water formation pathway.}
%Methods 
{Using far-infrared spectra taken with the PACS instrument onboard the \emph{Herschel} telescope, we combined two methods to identify \water emission trends and interpreted these in terms of theoretically expected patterns in the \water abundance. Through the use of line-strength ratios, we analyzed the correlation between the strength of \water emission and the mass-loss rate of the objects, as well as the radial dependence of the \water abundance in the circumstellar outflow per individual source. We computed a model grid to account for radiative-transfer effects in the line strengths.}
%Results
{We detect warm \water emission close to or inside the wind acceleration zone of {all} sample stars, irrespective of their stellar or circumstellar properties. The predicted \water abundances in carbon-rich environments are in the range of $10^{-6}$ up to $10^{-4}$ for Miras and semiregular-a objects, and cluster around $10^{-6}$ for semiregular-b objects. These predictions are up to three orders of magnitude greater than what is predicted by state-of-the-art chemical models. We find a negative correlation between the \water/CO line-strength ratio and gas mass-loss rate for $\mg > 5\times10^{-7}$ $\msunyr$, regardless of the upper-level energy of the relevant transitions. This implies that the \water formation mechanism becomes less efficient with increasing wind density. The negative correlation breaks down for the sources of lowest mass-loss rate, the semiregular-b objects.
}
%Conclusion
{Observational constraints suggest that pulsationally induced shocks play an important role in warm \water formation in carbon-rich AGB stars, although photodissociation by interstellar UV photons may still contribute. Both mechanisms fail in predicting the high \water abundances we infer in Miras and semiregular-a sources{, while our results for the semiregular-b objects are inconclusive.}
}

\maketitle

%******************************************************************
%     INTRODUCTION
%******************************************************************

\section{Introduction}\label{sect:intro}
It has long been assumed that the chemistry in asymptotic giant branch (AGB) photospheres, and consequently in AGB circumstellar envelopes, occurs in thermodynamic equilibrium (TE). The formation of carbon monoxide (CO) drives TE chemistry, followed by the formation of oxygen-based molecules for a carbon-to-oxygen ratio C/O $< 1$, or carbon-based molecules for C/O $> 1$ \citep{hab2003}. However, during the past two decades observations of both oxygen-rich and carbon-rich winds have revealed anomalous molecular abundances indicating that nonequilibrium effects play an important role in AGB circumstellar chemistry {(e.g.,~\citeauthor{mil2003}~\citeyear{mil2003}, \citeyear{mil2015}, \citeauthor{che2006}~\citeyear{che2006}, \citeauthor{dec2012}~\citeyear{dec2012})}. A prime example was the unexpected detection of {cold} \water vapor emission in CW~Leo, the carbon-rich AGB star closest to the solar system, by \citet{mel2001} with the \emph{Submillimeter Wave Astronomy Satellite} (SWAS; \citeauthor{mel2000}~\citeyear{mel2000}). Follow-up observations of \water emission with the ODIN satellite \citep{nor2003,has2006} and the detection of the 1665 MHz and 1667 MHz maser lines of OH \citep{for2003}, of which \water is the parent molecule, confirmed the presence of \water vapor in the carbon-rich environment of this star. The launch of the \emph{Herschel} space observatory \citep{pil2010} provided an opportunity to perform an unbiased \water survey in a much broader sample of carbon-rich AGB stars. Quickly after the launch, all three instruments onboard \emph{Herschel} revealed the widespread occurrence of not only cold, but also warm \water vapor in all these carbon-rich winds \citep{dec2010c,neu2010,neu2011b,neu2011a}, challenging our understanding of circumstellar chemistry in these environments. 

Several chemical processes have been suggested to be responsible for the production of {cold} \water vapor in carbon-rich environments. Firstly, evaporation of icy bodies was invoked as an explanation when \water vapor was first discovered in CW~Leo \citep{mel2001,for2001}. However, spectroscopically resolved \emph{Herschel} observations of \water emission in several carbon-rich AGB stars ruled this out as a dominant \water formation mechanism \citep{neu2011b,neu2011a}. Secondly, in nonlocal thermodynamic equilibrium (NLTE) conditions, gas-phase radiative association of H$_2$ with atomic O can also form \water vapor in a cold environment \citep{agu2006}, although recent results indicate that the expected rate constant for this reaction is too low to explain the observed amounts \citep{tal2010}. {Thirdly, \citet{wil2004} proposed that Fischer-Tropsch catalysis on the surfaces of small metallic Fe grains {at intermediate distances from the star} contributes to \water formation.}

{To explain the recently discovered {warm} \water emission, two mechanisms have been proposed. \citet{dec2010c} and \citet{agu2010} proposed the photodissociation of $^{13}$CO and SiO in the inner wind by interstellar ultraviolet (UV) radiation that can penetrate deeply into the wind if the medium is clumpy. As a result, atomic O is available to form \water vapor through two subsequent reactions with molecular hydrogen, for which the rate constant is high enough at temperatures above $\sim 300$K. Alternatively, \citet{che2011} has suggested the dynamically unstable environment close to the stellar surface as a means to produce free atomic O through collisional destruction of CO in shocked gas. Originally, \citet{che2006} predicted that such a shock-induced mechanism could not account for a large \water vapor abundance, as observed with \emph{Herschel}. However, by modifying the poorly constrained reaction rates of some reactions occurring in the shocked gas, the expected \water abundance can be boosted by several orders of magnitude, bringing them in agreement with the measured \water line strengths. Moreover, \citet{che2011} predicted \water emission to be variable in time, depending on the pulsational phase in which the observations were taken.}

In this paper, we present \water vapor emission measurements of a sample of 18 carbon-rich AGB stars observed with the \emph{Photodetecting Array Camera and Spectrometer} (PACS; \citeauthor{pog2010}~\citeyear{pog2010}) onboard \emph{Herschel}. {We constrain \water abundances and search for correlations between physical, chemical, and dynamical conditions that are implied and/or suggested by the different \water formation mechanisms in carbon-rich environments with the aim to discriminate between the proposed mechanisms.} 

In Sect.~\ref{sect:data}, we describe the selected sample and the data reduction. We analyze the sample-wide trends in the observed \water emission in Sect.~\ref{sect:corr}. In Sect.~\ref{sect:comp}, we compare the measured line strengths with a set of theoretical models, and investigate the possibility of a radial dependence of the \water abundance in individual sources in Sect.~\ref{sect:grad}. We follow up these results with a discussion in Sect.~\ref{sect:disc} and end this study with conclusions in Sect.~\ref{sect:conc}.

\section{Data} \label{sect:data}
\subsection{Target selection and observation strategy}\label{sect:targetsel}
The sample presented in Table~\ref{table:obs} consists of 19 carbon-rich AGB stars observed with \emph{Herschel}, and includes both Mira-type variables and semiregular (SR) pulsators covering a broad range of mass-loss rates and outflow velocities. Full PACS spectra were taken for six targets in the framework of the \emph{Mass loss of Evolved StarS} (MESS) guaranteed-time key project \citep{gro2011}. However, because MESS was biased toward sources with high mass-loss rates, additional deep line scans were gathered for 14 stars in the framework of a \emph{Herschel} open time 2 (OT2) program (P.I.:~L.~Decin) to complement the MESS~program with targets with lower mass-loss rates as well as different outflow velocities and variability types. LL~Peg was observed in both SED-scan mode and line-scan mode allowing for a consistency check between both observing modes. The observation settings are listed in Table~\ref{table:obs} for all spectra of carbon stars observed in the MESS~program and for all line scans taken in the OT2~program.
\begin{table*}[!t]
  {
  \setlength{\tabcolsep}{4pt}
  \caption{Observation settings of carbon-rich AGB stars observed with the PACS instrument onboard \emph{Herschel} in the MESS and OT2~programs. Given are the right ascension (R.A.) and declination (Dec.), observation identifier (Obsid), day of observation from the start of operations (OD), date of observation, total observation time including overhead ($t_\mathrm{obs}$), observation mode (SED for full spectral-range scan from the MESS~program, or LINE for line scan from the OT2~program), and bands in which spectra were taken. All observations were single pointings and were performed in chop-nodded mode. Line scans denoted as LINE$^\star$ were observed with the \emph{range-scan} observing template and were treated as line scans in the data reduction. LL~Peg is listed twice, as it was observed in both the MESS and OT2~programs. The OT2-program target R~Scl is included for completeness, but is not used in the remainder of this study.}\label{table:obs}
  \vspace{-0.5cm}
  \begin{center}
  \small
  \begin{tabular}[c]{lllllllll}\hline\hline\rule[0mm]{0mm}{3mm}Target &R.A. & Dec. & Obsid & OD & Date of obs. (UTC) & $t_\mathrm{obs}$ (s)& Mode & Bands \\\hline
\rule{0pt}{10pt}RW~LMi & 10:16:02.27 & 30:34:18.60 & 1342197799 & 387  & Jun 05 14:38:11 2010 & 2373 & SED & B2B-R1B \\
  &    &    & 1342197800 & 387 & Jun 05 15:08:40 2010 & 1125 & SED & B2A-R1A \\
V~Hya & 10:51:37.25 & -21:15:00.30 & 1342197790 & 387 & Jun 05 07:56:09 2010 & 4605 & SED & B2B-R1B \\ 
  &    &    & 1342197791 & 387 & Jun 05 08:54:29 2010 & 2124 & SED & B2A-R1A \\
II~Lup & 15:23:04.91 & -51:25:59.00 & 1342215685 & 665 & Mar 10 07:59:32 2011 & 2373 & SED & B2B-R1B \\
  &    &    & 1342215686 & 665 & Mar 10 08:30:02 2011 & 1125 & SED & B2A-R1A \\
V~Cyg & 20:41:18.27 & 48:08:28.80 & 1342208939 & 550 & Nov 15 02:00:33 2010 & 1125 & SED & B2A-R1A \\
  &    &    & 1342208940 & 550 & Nov 15 02:31:02 2010 & 2373 & SED & B2B-R1B \\
LL~Peg & 23:19:12.39 & 17:11:35.40 & 1342199417 & 412 & Jun 30 09:29:02 2010 & 1125 & SED & B2A-R1A \\
  &    &    & 1342199418 & 412 & Jun 30 09:59:30 2010 & 2373 & SED & B2B-R1B \\
LP~And & 23:34:27.66 & 43:33:02.40 & 1342212512 & 607 & Jan 11 00:55:51 2011 & 2124 & SED & B2A-R1A \\
  &    &    & 1342212513 & 607 & Jan 11 01:54:11 2011 & 4605 & SED & B2B-R1B \\\hline
\rule{0pt}{10pt}R~Scl & 01:26:58.09 & -32:32:35.40 & 1342247730 & 1149 & Jul 06 01:16:10 2012 & 2863 & LINE & B2B-R1A-R1B \\
  &    &    & 1342247731 & 1149 & Jul 06 01:44:54 2012 & 547 & LINE & B3A \\
V384~Per& 03:26:29.51 & 47:31:48.60 & 1342250571 & 1209 & Sep 04 02:22:04 2012 & 1261 & LINE & R1A \\
  &    &    & 1342250572 & 1209 & Sep 04 02:59:51 2012 & 3231 & LINE$^\star$ & B2B-R1B\\
  &    &    & 1342250573 & 1209 & Sep 04 03:31:41 2012 & 547 & LINE & B3A \\
R~Lep & 04:59:36.35 & -14:48:22.50 & 1342249508 & 1188 & Aug 14 11:43:13 2012 & 895 & LINE & R1A \\
  &    &    & 1342249509 & 1188 & Aug 14 12:11:34 2012 & 2465 & LINE$^\star$ & B2B-R1B\\
  &    &    & 1342249510 & 1188 & Aug 14 12:37:01 2012 & 547 & LINE & B3A\\
W~Ori & 05:05:23.72 & 01:10:39.50 & 1342249502 & 1188 & Aug 14 09:40:23 2012 & 1993 & LINE & R1A\\
  &    &    & 1342249503 & 1188 & Aug 14 10:24:16 2012 & 3231 & LINE$^\star$ & B2B-R1B\\
  &    &    & 1342249504 & 1188 & Aug 14 10:56:06 2012 & 547 & LINE & B3A\\
S~Aur & 05:27:07.45 & 34:08:58.60 & 1342250895 & 1216 & Sep 11 13:24:59 2012 & 1627 & LINE & R1A\\
  &    &    & 1342250896 & 1216 & Sep 11 14:18:34 2012 & 4761 & LINE$^\star$ & B2B-R1B\\
  &    &    & 1342250897 & 1216 & Sep 11 15:09:57 2012 & 1363 & LINE & B3A\\
U~Hya & 10:37:33.27 & -13:23:04.40 & 1342256946 & 1307 & Dec 11 11:07:39 2012 & 1627 & LINE & R1A\\ 
  &    &    & 1342256947 & 1307 & Dec 11 11:48:29 2012 & 3231 & LINE$^\star$ & B2B-R1B\\
  &    &    & 1342256948 & 1307 & Dec 11 12:20:19 2012 & 547 & LINE & B3A\\
QZ~Mus & 11:33:57.91 & -73:13:16.30 & 1342247718 & 1148 & Jul 05 06:26:18 2012 & 3609 & LINE & B2B-R1A-R1B\\
  &    &    & 1342247719 & 1148 & Jul 05 07:04:35 2012 & 547 & LINE & B3A \\
Y~CVn & 12:45:07.83 & 45:26:24.90 & 1342254304 & 1269 & Nov 02 17:45:29 2012 & 1261 & LINE & R1A\\ 
  &    &    & 1342254305 & 1269 & Nov 02 18:16:53 2012 & 2465 & LINE$^\star$ & B2B-R1B\\
  &    &    & 1342254306 & 1269 & Nov 02 18:45:44 2012 & 955 & LINE & B3A\\
AFGL~4202& 14:52:24.29 & -62:04:19.90 & 1342250003 & 1195 & Aug 21 10:00:56 2012 & 895 & LINE & R1A\\
  &    &    & 1342250004 & 1195 & Aug 21 10:29:17 2012 & 2465 & LINE$^\star$ & B2B-R1B\\
  &    &    & 1342250005 & 1195 & Aug 21 10:54:44 2012 & 547 & LINE & B3A\\
V821~Her& 18:41:54.39 & 17:41:08.50 & 1342244456 & 1068 & Apr 16 12:01:59 2012 & 2863 & LINE & B2B-R1A-R1B\\
  &    &    & 1342244457 & 1068 & Apr 16 12:30:43 2012 & 547 & LINE & B3A\\
V1417~Aql& 18:42:24.68 & -02:17:25.20 & 1342244470 & 1068 & Apr 16 20:05:15 2012 & 2863 & LINE & B2B-R1A-R1B\\
  &    &    & 1342244471 & 1068 & Apr 16 20:33:59 2012 & 547 & LINE & B3A \\
S~Cep & 21:35:12.83 & 78:37:28.20 & 1342246553 & 1115 & Jun 01 18:47:49 2012 & 3258 & LINE & B2B-R1A-R1B\\
  &    &    & 1342246554 & 1115 & Jun 01 19:22:51 2012 & 547 & LINE & B3A\\
RV~Cyg & 21:43:16.33 & 38:01:03.00 & 1342247466 & 1140 & Jun 27 11:54:06 2012 & 2815 & LINE & B2B-R1A\\
  &    &    & 1342247467 & 1140 & Jun 27 12:34:59 2012 & 2055 & LINE & B2B-R1A-R1B\\
  &    &    & 1342247468 & 1140 & Jun 27 13:00:23 2012 & 955 & LINE & B3A\\
LL~Peg & 23:19:12.39 & 17:11:35.40 & 1342257222 & 1310 & Dec 14 13:16:04 2012 & 547 & LINE & B3A\\
  &    &    & 1342257635 & 1317 & Dec 21 09:38:25 2012 & 2465 & LINE$^\star$ & B2B-R1B\\
  &    &    & 1342257684 & 1319 & Dec 22 16:45:23 2012 & 1261 & LINE & R1A\\\hline
  \end{tabular}
  \end{center}
  }
\end{table*}

The line selection in the OT2~program aimed to include \water lines at wavelengths where confusion due to blending with other molecular emission lines is reduced to a minimum, and is based on the molecular inventory made for CW~Leo \citep{dec2010c}. The circumstellar environment of both CW~Leo and R~Scl (for which line scans were also obtained) are spatially resolved. This severely complicates the data reduction process \citep{dec2010c,deb2012}, especially given our analysis strategy outlined in Sect.~\ref{sect:corr}. We have therefore excluded both sources from the present study. We discuss the spatial extension in Appendix~\ref{sect:extension}. Finally, a detached shell has been detected with the PACS instrument for U~Hya. This detached shell falls outside the central spaxel of PACS, and is located too far from the central source to be important for the CO and \water emission. The central component of U~Hya is essentially a point source and can be safely included in the sample.

\subsection{Data reduction}
The MESS~observations were performed with the standard Astronomical Observing Template (AOT) for SED mode. The OT2~data were taken with the AOT for PACS Line Spectroscopy (chopped / nodded), which allows for deeper observations focusing on a subset of wavelength ranges. In a first iteration of the requested observation scheme, eleven line scans were taken for five OT2 targets. We then optimized our observation scheme to include only nine wavelength ranges for the rest of the OT2 targets. All observations were reduced with the appropriate interactive pipeline in HIPE 11 with calibration set 45. The absolute flux calibration is based on the normalization method, in which the flux is normalized to a model of the telescope background radiation. This is possible since the "off-source", which is almost completely dominated by the telescope background radiation, is measured at every wavelength. Consequently, this method allows us to track the response drifts of every detector during the observation, whereas the standard flux calibration via the calibration block only gives a reference point at the start of the observation. The normalization method and the comparison with the calibration block method will be published in a forthcoming PACS-calibration publication. 

The data have been spectrally rebinned with an oversampling factor of two, i.e.~a Nyquist sampling with respect to the native instrumental resolution. We extracted the spectra from the central spaxel of every observation and applied a point-source correction. Finally, a pointing correction was applied to all MESS~targets, as well as to the OT2 targets that show a continuum flux > 2 Jy. Applying the pointing correction to weaker sources introduces too large an uncertainty. For these sources, we opted instead to add 5\% additional flux across all line scans, which is the average flux increase introduced by the pointing correction in observations with a continuum flux > 2 Jy. The data reduction has an absolute-flux-calibration uncertainty of 20\%. The MESS~spectra and the OT2 line scans are shown in the appendix, in Figs.~\ref{fig:mess1} up to \ref{fig:mess12} and Figs.~\ref{fig:ot2_1} up to \ref{fig:ot2_13}, respectively.
\begin{table*}[!t]
  {
  \setlength{\tabcolsep}{3pt}
  \caption{Properties of the sample of carbon-rich AGB stars observed with \emph{Herschel}{ (see Sect.~\ref{sect:sample})}. The first six sources are covered in the MESS~program; the rest in the OT2~program. Given per source are the IRAS number, variability type (Mira or semiregular), pulsational period ($P$), {6.3 \mic flux ($F_\mathrm{6.3\ \mu m}$, with the suffix \emph{a} for ISO SWS data and \emph{b} for photometric data)}, adopted distance ($d$),{ range of distance estimates in the literature ($\Delta d$)}, stellar velocity with respect to the local standard of rest ($\varv_\mathrm{LSR}$), stellar luminosity ($L_\star$), stellar effective temperature {($T_\star$; with $\circ$ added for assumed values)}, gas mass-loss rate ($\mg$), terminal gas velocity ($\vg$), and {wind density tracer ($\shellmass$).} The source denoted with ($\star$) is of spectral type CJ and is possibly an extrinsic carbon star \citep{abi2010}.}\label{table:sample}
  \vspace{-0.5cm}
  \begin{center}
  \small

\begin{tabular}[c]{lllllllllllll}\hline\hline\rule[0mm]{0mm}{3mm}Star  & IRAS &Var.& $P$ &$F_\mathrm{6.3\ \mu m}$& $\varv_\mathrm{LSR}$& $d$ & $\Delta d$& $L_\star$ &$T_\star$&$\mg$&$\vg$&$\shellmass$ \\
name  & number &type& (days) &(10$^2$ Jy) &(\kms)& (pc) &(pc)  & ($10^{3}$ L$_\odot$)&(K)&($\msunyr$)&(\kms)&(\myrkmsfrac) \vspace{0.05cm}\\\hline
\rule{0pt}{10pt}RW Lmi &10131+3049&SRa &640 $^{(1)}$&25$^\mathrm{b}$ &-1.8 $^{(16)}$& 410 $^{(8)}$ & 320-710 & 8.3 $^{(8)}$ & 2470 $^{(17)}$ & 5.2$\times 10^{-6}\ ^{(18)}$ & 16.5 $^{(18)}$ &3.2$\times 10^{-7}$\\
V~Hya &10491-2059&SR/Mira&531 $^{(1)}$&12$^\mathrm{b}$ &-16.0 $^{(9)}$&340 $^{(6,8)}$ & 330-2160& 8.3 $^{(6)}$ & 2160 $^{(17)}$ & 2.7$\times 10^{-6}\ ^{(15)}$ & 15.0 $^{(20)}$ & 1.8$\times 10^{-7}$\\
II~Lup &15194-5115&Mira&580 $^{(2)}$&6.0$^\mathrm{b}$  &-15.0 $^{(16)}$& 640 $^{(6)}$ & 470-640 & 9.1 $^{(6)}$ & 2000 $^{\circ}$& 1.5$\times 10^{-5}\ ^{(18)}$ & 21.0 $^{(18)}$ & 7.0$\times 10^{-7}$\\
V~Cyg &20396+4757&Mira&421 $^{(1)}$&9.7$^\mathrm{a}$ & 15.0 $^{(16)}$& 420 $^{(7)}$ & 270-740 & 6.6 $^{(7)}$ & 1875 $^{(17)}$ & 1.7$\times 10^{-6}\ ^{(18)}$ & 10.5 $^{(18)}$ & 1.6$\times 10^{-7}$\\
LL~Peg &23166+1655&Mira&696 $^{(2)}$&0.9$^\mathrm{a}$ &-31.0 $^{(16)}$ &1050 $^{(6)}$ & 950-1150&11.0 $^{(6)}$& 2000 $^{\circ}$& 1.1$\times 10^{-5}\ ^{(18)}$ & 13.5 $^{(18)}$ & 8.5$\times 10^{-7}$\\
LP~And &23320+4316&Mira&614 $^{(1)}$&4.0$^\mathrm{a}$ &-17.0 $^{(16)}$& 840 $^{(6,17)}$& 610-870 & 9.7 $^{(6)}$ & 2040 $^{(17)}$ & 2.2$\times 10^{-5}\ ^{(18)}$ & 13.5 $^{(18)}$ & 1.6$\times 10^{-6}$\\\hline
\rule{0pt}{10pt}V384~Per &03229+4721&Mira&535 $^{(1)}$&4.6$^\mathrm{a}$&-16.2 $^{(15)}$ & 720 $^{(6,17)}$& 560-1060& 8.4 $^{(6)}$ &1820 $^{(17)}$ & 4.1$\times 10^{-6}\ ^{(18)}$ &14.5 $^{(18)}$&2.8$\times 10^{-7}$\\
R~Lep &04573-1452&Mira&427 $^{(1)}$&4.0$^\mathrm{b}$    & 18.5 $^{(12)}$ & 413 $^{(5)}$ & 250-480 & 5.2 $^{(5,6)}$ & 2290 $^{(17)}$ & 1.3$\times 10^{-6}\ ^{(18)}$ & 17.0 $^{(12)}$ &8.1$\times 10^{-8}$\\
W~Ori &05028+0106&SRb &212 $^{(1)}$&3.3$^\mathrm{a}$   & 18.8 $^{(16)}$ & 377 $^{(5)}$ & 220-460 & 8.0 $^{(5,8)}$ & 2625 $^{(17)}$ & 2.1$\times 10^{-7}\ ^{(18)}$ & 12.0 $^{(12)}$ & 1.8$\times 10^{-8}$\\
S~Aur &05238+3406&SR/Mira&596 $^{(1)}$&1.7$^\mathrm{b}$  &-21.0 $^{(12)}$ &1010 $^{(6,17)}$&300-1130& 9.4 $^{(6)}$  & 1940 $^{(17)}$& 4.5$\times 10^{-6}\ ^{(18)}$ & 25.0 $^{(12)}$ &1.8$\times 10^{-7}$\\
U~Hya &10350-1307&SRb &450 $^{(1)}$&2.8$^\mathrm{b}$   &-31.0 $^{(16)}$ & 208 $^{(5)}$ & 160-980 & 4.2 $^{(5,8)}$ & 2965 $^{(17)}$ & 1.4$\times 10^{-7}\ ^{(18)}$ & 7.0 $^{(12)}$ & 2.0$\times 10^{-8}$\\
QZ~Mus &11318-7256&Mira&535 $^{(1)}$&4.0$^\mathrm{b}$   &-2.0 $^{(16)}$ & 660 $^{(6)}$ & 620-720 & 8.4 $^{(6)}$  & 2200 $^{\circ}$ & 4.8$\times 10^{-6}\ ^{(15)}$ & 26.5 $^{(15)}$ & 1.8$\times 10^{-7}$\\
Y~CVn$^{\star}$&12427+4542&SRb&157 $^{(1)}$&3.7$^\mathrm{a}$ & 21.0 $^{(16)}$ &320 $^{(5)}$ & 170-340 & 8.7 $^{(5,8)}$  & 2760 $^{(17)}$& 3.2$\times 10^{-7}\ ^{(18)}$ & 8.5 $^{(12)}$ & 3.8$\times 10^{-8}$\\
AFGL~4202&14484-6152&Mira&566 $^{(3)}$&4.4$^\mathrm{b}$ & 24.4 $^{(15)}$ & 611 $^{(6,15)}$& 570-900 & 8.9 $^{(6)}$  & 2200 $^{\circ}$ & 4.5$\times 10^{-6}\ ^{(15)}$ & 19.0 $^{(15)}$ & 2.4$\times 10^{-7}$\\
V821~Her &18397+1738&Mira&511 $^{(4)}$&4.4$^\mathrm{b}$ &-0.5 $^{(16)}$ & 750 $^{(6)}$ & 600-900 & 7.5 $^{(6)}$  & 2200 $^{\circ}$ & 2.8$\times 10^{-6}\ ^{(18)}$ & 13.0 $^{(18)}$ & 2.2$\times 10^{-7}$\\
V1417~Aql&18398-0220&Mira&617 $^{(4)}$&4.2$^\mathrm{a}$ & 3.0 $^{(15)}$ & 870 $^{(6)}$ & 870-950 &10.8 $^{(6)}$  & 2000 $^{\circ}$ & 1.7$\times 10^{-5}\ ^{(19)}$ & 36.0 $^{(15)}$ & 4.7$\times 10^{-7}$\\
S~Cep &21358+7823&Mira&487 $^{(1)}$&7.0$^\mathrm{a}$   &-15.5 $^{(15)}$ & 407 $^{(5)}$ & 380-720 & 6.4 $^{(5,6)}$ & 2095 $^{(17)}$ & 1.4$\times 10^{-6}\ ^{(18)}$ & 21.5 $^{(18)}$ & 6.4$\times 10^{-8}$\\
RV~Cyg &21412+3747&SRb &263 $^{(1)}$&1.1$^\mathrm{b}$ & 17.0 $^{(12)}$ & 640 $^{(8)}$ & 350-850 &13.4 $^{(8)}$  & 2675 $^{(17)}$ & 2.0$\times 10^{-7}\ ^{(12)}$ & 13.0 $^{(12)}$ & 1.5$\times 10^{-8}$\\\hline
  %\multicolumn{12}{l}{\tablefoottext{$\star$}{Spectral type CJ; possibly extrinsic carbon star \citep{abi2010}.}}
  \end{tabular}
  \end{center}
  \vspace{-0.3cm}
  \small $^{(1)}$~\citet{sam2009}, $^{(2)}$~\citet{leb1992}, $^{(3)}$~\citet{pri2010}, $^{(4)}$~\citet{gua2013}, $^{(5)}$~\citet{van2007}, $^{(6)}$~\citet{whi2006}, $^{(7)}$~\citet{whi2008}, $^{(8)}$~\citet{ber2005}, $^{(9)}$~\citet{sah2009}, $^{(10)}$~\citet{epc1990}, $^{(11)}$~\citet{lou1993}, $^{(12)}$~\citet{olo1993}, $^{(13)}$~\citet{gro1998b}, $^{(14)}$~\citet{kna1998}, $^{(15)}$~\citet{gro2002}, $^{(16)}$~\citet{deb2010}, $^{(17)}$~\citet{ber2001}, $^{(18)}$~\citet{sch2013}, $^{(19)}$~\citet{oli2001}, $^{(20)}$~\citet{kna1997}.
  }
\end{table*}
\subsection{Line strengths}\label{sect:ls}
Integrated line strengths, $I_\mathrm{int}$, of CO, $^{13}$CO, ortho-\water, and para-\water are listed in Table~\ref{table:intintmess} for the MESS~targets and in Tables~\ref{table:intintot2old} and \ref{table:intintot2new} for the OT2 targets of the appendix. Tables~\ref{table:unidentified} and \ref{table:unidentified2} list the strengths of emission lines in the OT2 line scans that are not attributed to CO or \water and for which we have not attempted to identify the molecular carrier. Following \citet{lom2013}, the line strengths were measured by fitting a Gaussian on top of a continuum. The reported uncertainties include the fitting uncertainty and the absolute-flux-calibration uncertainty of 20\%. Measured line strengths are flagged as line blends if they fulfill at least one of two criteria: 1) the full width at half maximum (FWHM) of the fitted Gaussian is larger than the FWHM of the PACS spectral resolution by at least 20\%, and 2) multiple CO or \water transitions have a central wavelength within the FWHM of the fitted central wavelength of the emission line. In the latter case, the additional transitions contributing to the emission line are listed in Tables~\ref{table:intintmess}, \ref{table:intintot2old}, and \ref{table:intintot2new} immediately below the first contributing transition. Other molecules were not considered. Because the OT2~program was specifically targeted at unblended lines based on the line survey of CW~Leo, line detections in the OT2 wavelength ranges can be reliably attributed to CO and \water. Similarly, lines detected in the same wavelength ranges in the MESS~data (given in red in Table~\ref{table:intintmess}) have reliable molecular identifications. Outside these wavelength ranges, we point out that the reported line strengths not flagged as line blends may still be affected by emission from other molecules or from \water transitions not included in our line list (see \citeauthor{dec2010a}~\citeyear{dec2010a} for details). 

\subsection{Stellar and circumstellar properties}\label{sect:sample}
Values for several stellar and circumstellar properties were gathered from the literature and are listed in Table~\ref{table:sample}. In Sects.~\ref{sect:comp} and \ref{sect:grad}, we compare our sample of AGB sources to a set of theoretical models with a generalized set of parameters, as opposed to a tailored modeling of each source. To this end, we did not blindly assume literature values for the properties listed in Table~\ref{table:sample}, but instead carefully scaled relevant values to ensure homogeneity and consistency within the sample. In what follows, we describe this procedure where relevant. {Throughout the paper, we refer to three distinct regions in the AGB wind, following \citet{wil1998}: inner, intermediate, and outer. As a guideline, this corresponds to $r<10\ \mathrm{R}_\star$, $10\ \mathrm{R}_\star < r < 100\ \mathrm{R}_\star$, and $r > 100\ \mathrm{R}_\star$, respectively, for an average mass-loss rate of $\sim 10^{-6}\ \msunyr$.}

The pulsational period $P$ is taken from the \emph{General Catalog of Variable Stars} (GCVS; \citeauthor{sam2009}~\citeyear{sam2009}) when available. For the other sources, the period is taken from \citet{leb1992}, \citet{pri2010}, or \citet{gua2013}. We make use of period-luminosity $PL$-relations for both the luminosity $L_\star$ and the distance $d$. For the Miras, $L_\star$ and $d$ are taken from \citet{whi2006,whi2008}. If not available, we use their $PL$-relation in combination with the apparent bolometric magnitude given by \citeauthor{ber2001} (\citeyear{ber2001}; for LP~And and V384~Per) or by \citeauthor{gro2002} (\citeyear{gro2002}; for AFGL~4202). For the SRa/b pulsators, we take $L_\star$ and $d$ from the PL-relation of \citet{ber2005}. If Hipparcos parallax measurements with an uncertainty less than 40\% are available, we rescale the luminosity given by these $PL$-relations to the measured distance \citep{van2007}. The uncertainty on the distance estimate for the other objects is taken to be 40\% owing to the broad range of distance estimates given in the literature; see column six in Table~\ref{table:sample}. %{We assume a 30\% uncertainty on the stellar luminosities.} 
To allow for a direct comparison between measured line strengths, all objects in the sample are placed at an arbitrary distance of 100 pc by rescaling the observed fluxes.

The stellar velocity $\varv_\mathrm{LSR}$ with respect to the local standard of rest is taken from \citet{deb2010}. If not in their sample, it is taken from \citet{olo1993} or \citet{gro2002}. For the stellar effective temperature $T_\star$ we follow \citet{ber2001}, who derived relations for $T_\star$ versus several colors based on a sample of 54 carbon stars. 
%The uncertainty on these values is estimated to be 140 K. 
However, $T_\star$ is notoriously difficult to constrain for stars with {a large infrared (IR) excess}. For this reason, the reddest carbon stars are absent in the sample of \citet{ber2001}. Two of these absent sources, II~Lup and LL~Peg, are included in the classification of cool carbon variables (CVs) of \citet{kna1999b} as CV7 objects, as they have the reddest spectral energy distribution (SED) among carbon stars. The average effective temperature attributed by \citet{ber2001} to the CV7 class is 2000 K, which we adopt for II~Lup and LL~Peg, as well as for V1417~Aql, which has an IR color similar to II~Lup and LL~Peg. 
%For these objects, we assume an uncertainty in $T_\star$ of 200 K. 
{While bluer than II~Lup, LL~Peg, and V1417~Aql, the remaining objects still show relatively red IR colors and have intermediate-to-high mass-loss rates. Hence, we assume they are either CV6 or CV7, to which \citet{ber2001} assign a temperature range of 2000-2400 K.} %We adopt $T_\star = 2200$ K, with an uncertainty of 300 K.
{We do not take into account time-dependent variations in stellar parameters. We therefore assume $R_\star$ to be the stellar radius associated with a blackbody radiator, following the Stefan-Boltzmann relation. Taking into account that $L_\star$ gives an average stellar luminosity scaled with distance through the $PL$-relations, $R_\star$ should give a reasonable estimate of the average stellar radius.} %The uncertainty on $R_\star$ is calculated by the $L_\star$ and $T_\star$ error propagation through the Stefan-Boltzmann relation of.

A broad range of gas mass-loss rates $\mg$ can be found in the literature for all objects in the sample, derived from either low-$J$ CO emission lines or SED modeling. We only use $\mg$ estimates derived from CO modeling because mass-loss rates derived from modeling the thermal dust emission require a conversion using a dust-to-gas ratio, which introduces a large uncertainty. {$\mg$ values derived from CO lines do depend on the CO abundance with respect to H$_2$ (\coabun), a parameter that is also not well constrained. In Sect.~\ref{sect:cols} we show that the impact of the CO abundance is limited in the context of the constraints that we have from chemical models.} {To maintain consistency, we rescale quoted mass-loss rates in the literature based on the distance $d_\mathrm{lit}^2$ for which they were derived to the distance used here (see column 7 in Table~\ref{table:sample} by applying the scaling factor $d^2$/$d_\mathrm{lit}^2$; \citeauthor{ram2008}~\citeyear{ram2008}, \citeauthor{deb2010}~\citeyear{deb2010}).} Most values for $\mg$ were taken from the recent work by \citet{sch2013}. Other values are taken from \citet{gro2002}, \citet{oli2001}, or \citet{olo1993}. The uncertainty on $\mg$ amounts to a factor of three. The gas terminal velocity $\vg$ is taken from \citet{olo1993}, \citet{gro2002}, and \citet{sch2013}. {The uncertainty on $\vg$ is usually not more than 10\%. The final column of Table~\ref{table:sample} lists values for $\shellmass$, which is a quantity that we use as a density tracer (see, e.g., \citeauthor{ram2009}~\citeyear{ram2009}). The uncertainty on $\shellmass$ is dominated by the uncertainty on $\mg$.}
% a column-density proxy \begin{equation}\label{eq:densproxy}\bar{m} = \bar{\rho} R_\star = \frac{\mg}{4\pi\ R_\star^2\ \vg} \times R_\star,\end{equation}where $\bar{\rho}$ is a characteristic density for the stellar wind and $R_\star$ is the stellar radius. Given a wavelength-dependent mass-extinction coefficient, which only depends on dust and gas characteristics, this proxy can be translated into an optical thickness of the wind. 

{To have an indicator for the dust content of the stellar wind and because of its relevance for \water excitation (see Sect.~\ref{sect:63flux}), we list the measured 6.3 \mic flux for each source in Jansky (not distance scaled). These are taken from ISO-SWS spectra if available. In all other cases, the values are derived from an interpolation of photometric measurements at shorter and longer wavelengths.}

\subsection{{V~Hya and S~Aur}}
A special note is warranted for V~Hya, which is suggested to be in transition between the AGB stage and the planetary nebula stage (e.g.,~\citeauthor{kna1997}~\citeyear{kna1997}, \citeauthor{sah2003}~\citeyear{sah2003}, \citeauthor{sah2009}~\citeyear{sah2009}). Clearly, V~Hya does not necessarily follow the general trends observed in other semiregular AGB stars. An indication for this is a stellar luminosity of $17.9 \times 10^{3}$ L$_\odot$ derived from the $PL$-relation of \citet{ber2005}, which is unusually high for a carbon AGB star (see, e.g., the overview in Fig.~C.2.~of \citeauthor{deb2010}~\citeyear{deb2010}, and the luminosity function in Fig.~4 of \citeauthor{gua2013}~\citeyear{gua2013}). The Mira $PL$-relation of \citet{whi2006}, which we adopted, instead leads to $L_\star = 8.3\times10^{3}$ L$_\odot$, in agreement with many other studies dedicated to the peculiar kinematic structure of this source. The use of the Mira $PL$-relation is further supported by the findings of \citet{kna1999}, who suggest V~Hya may be a Mira. Additionally, most of the kinematic complexity in V~Hya occurs in the outer circumstellar wind where multiple components in the kinematic structure are observed in the low-$J$ CO emission lines, including a high-velocity bipolar outflow. CS and HC$_3$N {emission lines, which are formed in the inner or intermediate wind, show only one component with an expansion velocity of $\sim15$ \kms \citep{kna1997}, indicating that their formation region behaves more like a normal spherically symmetric AGB wind. Most lines detected in the PACS wavelength range are formed in this region.} We take $\varv_\mathrm{LSR} = -16.0$ \kms from \citet{sah2009}. 

{There is some debate whether S~Aur is a semiregular variable or a Mira. The GCVS catalog lists S~Aur as a semiregular, but the light curve amplitude in $V$ is $>2.5$ mag, categorizing it as a Mira variable. Moreover, the effective temperature of S~Aur ($T_\star = 1940$ K) is extremely low, and therefore more reminiscent of Miras than semiregulars. Because there is no a priori reason to assume that S~Aur is a semiregular, we treat the source as a Mira for the distance and luminosity determination. We note that, in the discussion of the importance of variability in SRa sources, the variability type for both V~Hya and S~Aur is debatable.}
\section{Trend analysis}\label{sect:corr}
To determine dependencies of the \water abundance on stellar and/or circumstellar properties, we combine two methods. In this section, we look for empirical correlations between observed molecular-emission line strengths and mass-loss rate. In Sect.~\ref{sect:comp} and \ref{sect:grad}, we perform a parameter study by calculating a grid of theoretical radiative-transfer models to compare with the measured line strengths. This combined approach allows us to identify model-independent \water emission trends and to disentangle radiative-transfer effects from other effects that contribute to the observed correlations.
\subsection{The observed CO line strength as an H$_2$ density tracer}\label{sect:coash2}
{Because one of the goals of this study is to constrain the \water abundance with respect to H$_2$ ($n_{\mathrm{H}_2\mathrm{O}}/n_{\mathrm{H}_2}$) in the sample sources, the ratio $I_{\mathrm{H}_2\mathrm{O}}/\mg$ is of interest as the \water number density is proportional to $I_{\mathrm{H}_2\mathrm{O}}$ and the H$_2$ number density to $\mg$. However, large uncertainties affect this ratio, owing to the uncertainties on the mass-loss rate itself and to the distance scaling that is necessary to compare the measurements within the sample. As such, considering line-strength ratios rather than line strengths is preferred, {as these are distance independent}. An interesting line-strength ratio is \water/CO, which provides an \water abundance proxy via $$I_{\mathrm{H}_2\mathrm{O}}/I_\mathrm{CO} \sim n_{\mathrm{H}_2\mathrm{O}}/n_\mathrm{CO} = (n_{\mathrm{H}_2\mathrm{O}}/n_{\mathrm{H}_2}) \times (n_{\mathrm{H}_2}/n_{\mathrm{CO}}),$$ assuming that CO has a constant molecular abundance with respect to H$_2$ throughout the entire wind, up to the photodissociation radius, and in the absence of optical-depth effects. 

\begin{table}[!t]
  {%\renewcommand{\arraystretch}{1.2}
  \setlength{\tabcolsep}{5pt}
  \caption{CO transitions selected for this study based on the wavelength ranges of the OT2 line scans. Given are the central wavelength ($\lambda_0$), upper-level energy ($E_\mathrm{u}$), and number of targets with a detection ($n$) of the emission line.}\label{table:co}\vspace{-0.4cm}
\begin{center}
\begin{tabular}[c]{lllll}\hline\hline\rule[0mm]{0mm}{3mm}
\rule{0pt}{10pt}Molecule & Transition &$\lambda_0$ (\mic)&$E_\mathrm{u}$ (cm$^{-1}$)&$n$\\\hline
\rule{0pt}{10pt}$^{12}$CO &$J=15-14$ &173.6&461.1& 18\\
&$J=18-17$ &144.8&656.8&18 \\
&$J=24-23$ &108.8&1151&18 \\
&$J=29-28$ &90.2&1668&18 \\
& $J=30-29$ &87.2&1783&17 \\
&$J=36-35$ &72.8&2550&10 \\
&$J=38-37$ &69.1&2836&8 \\\hline
\rule{0pt}{10pt}$^{13}$CO&$J=19-18$ &143.5&697.6&8 \\\hline
\end{tabular}
  \end{center}
  }
\end{table}
Seven CO transitions and one $^{13}$CO transition have been observed in the wavelength ranges of the OT2 line scans. We limit our study to these transitions because a maximum of only six detections are available for the other CO transitions from the MESS~data, and some of those lines may be affected by line blending. An overview of the relevant CO transitions is given in Table~\ref{table:co}. Fig.~\ref{fig:covmdot} shows the measured line strengths of CO $J=15-14$, {scaled to a distance of 100 pc. A correlation between line strength and mass-loss rate is present, which is expected considering that the mass-loss rates listed in Table~\ref{table:sample} are exclusively derived from CO emission lines. Because CO is predominantly excited through collisions with H$_2$, {CO is a reliable tracer of $\mg$ and, hence, of $n_{\mathrm{H}_2}$. At the high end of the range of mass-loss rate, the trend flattens off where the lines become optically thick. We show in Sect.~\ref{sect:cols} that theoretical models recover this behavior.} For higher-$J$ levels the flattening of the slope sets in at a lower mass loss because the lines are formed closer to the stellar surface, where the gas density is higher.} Therefore, the $J=15-14$ transition is best suited to act as an H$_2$ density tracer. CO $J=15-14$ has been detected in all objects in the sample, and none of them are flagged as a line blend.

As shown in Fig~\ref{fig:covmdot}, the Miras and SRa sources cannot be distinguished based on CO line strength. The SRb sources cluster at the low end of the range of mass-loss rate, but still seem to follow the linear trend set by the Miras and SRa sources. Studies on large populations have shown that Miras are considered to be fundamental-mode pulsators, while semiregulars are overtone pulsators or short-period fundamental-mode pulsators \citep{woo1999,woo2010}. The differentiation between SRa and SRb variables is based on the regularity of the light curves of these sources, but no definite conclusion can be drawn about the pulsational mode they exhibit. As shown by \citet{bow1988}, overtone pulsators are significantly less efficient at driving a stellar wind than fundamental-mode pulsators. If one assumes that SRa sources pulsate in a short-period fundamental mode, and SRb sources in a first or second overtone, this could explain the clear difference in terms of mass-loss rate between these two variability classes. Another suggestion is that SRb sources are unstable in more than one pulsation mode, and thus experience more than one pulsation period characteristic of each mode, explaining the lower periodicity of their light curves \citep{sos2013}. This may also decrease the efficiency with which a wind is driven. {We recall that two out of the three SRa sources in our sample have a debatable variability type} and were treated as Miras for the luminosity and distance determination.
\begin{figure}[!t]
\resizebox{9.0cm}{!}{\includegraphics{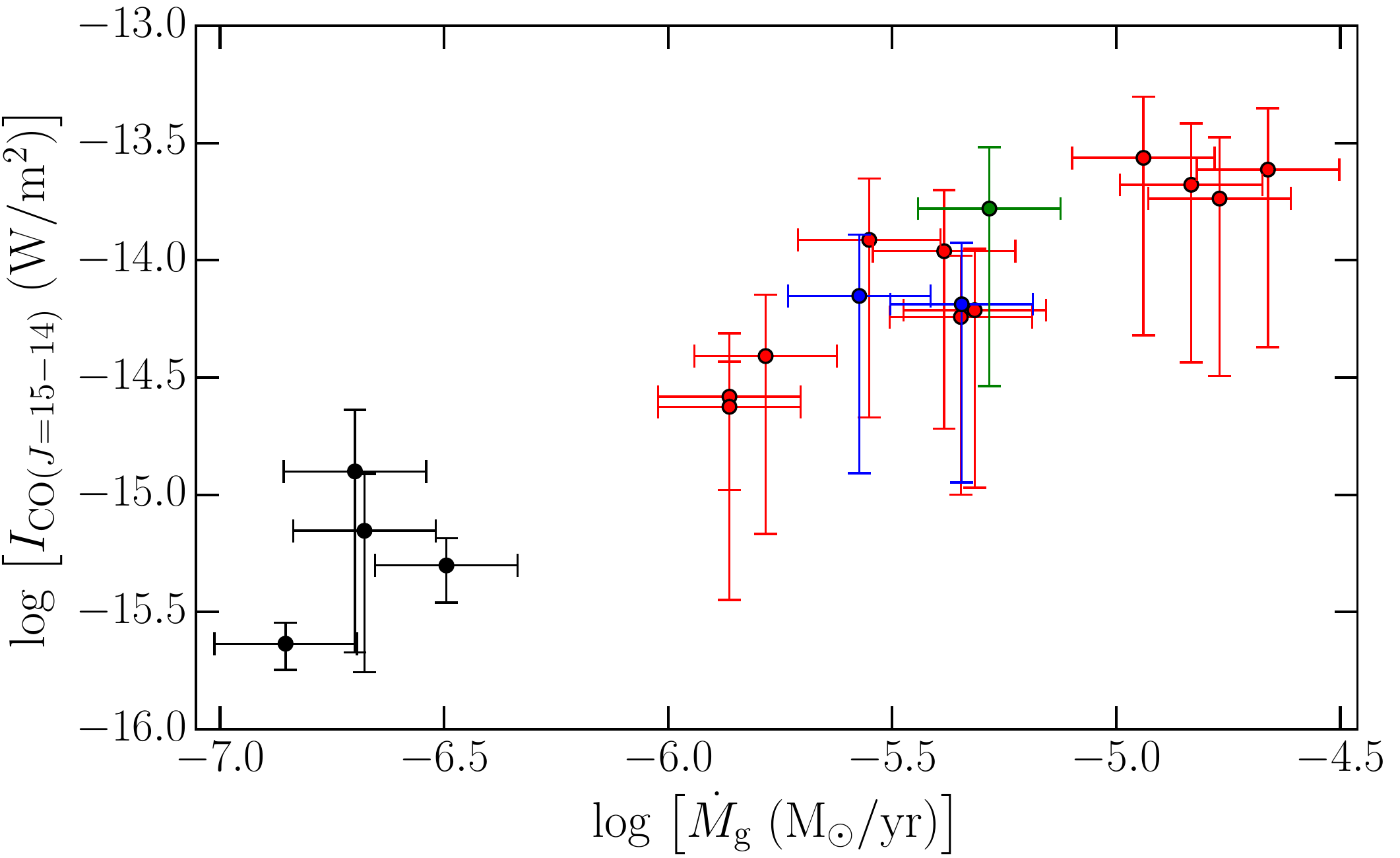}}
\caption{Line strengths of the CO $J=15-14$ transition as a function of the mass-loss rate $\mg$. The data points are color coded according to the variability type: Miras in red, {SR/Mira sources in blue, the SRa source in green,} and SRb sources in black. The line strengths are scaled to a distance of 100 pc.}
\label{fig:covmdot}
\end{figure}

\subsection{The \water/CO line-strength ratio versus $\mg$}\label{sect:obsh2o}
We only take the \water transitions in the wavelength ranges of the OT2 line scans into account. Their central wavelengths and upper-level energies are listed in the first columns of Table~\ref{table:corr}. Two additional transitions, with higher upper-level energies, are included in Table~\ref{table:intintot2old} and \ref{table:intintot2new}, but both occur in a blend with another \water transition listed in Table~\ref{table:corr} and do not contribute significantly to the emission. We do not consider them in the remainder of this study. In what follows, we primarily look at the \water $J_{\mathrm{K}_\mathrm{a}, \mathrm{K}_\mathrm{c}}=2_{1,2} - 1_{0,1}$ line because it is the only \water line detected in the entire sample.

\begin{table*}[!t]
  {
  \setlength{\tabcolsep}{4pt}
  \caption{{\water transitions selected for this study. Given are the central wavelength ($\lambda_0$), upper-level energy, ($E_\mathrm{u}$) and number of targets with a detection ($n$) of the emission line. Also given for each transition are the empirical fitting results of the linear correlation $Y = \bar{a} + \bar{b}X$ between the \water/CO line-strength ratio and the mass-loss rate. $n_\mathrm{inc} = n - n_\mathrm{SRb}$ gives the number of detections included in the fit, $\bar{a}$ and $\bar{b}$ the mean coefficients, $\sigma_\mathrm{\bar{a}}$ and $\sigma_\mathrm{\bar{b}}$ the fitting uncertainties on both coefficients, and $\sigma_\mathrm{\bar{a}\bar{b}}$ the covariance between the two. The middle five columns assume the logarithm of the line-strength ratio as the independent variable ($X$) and the logarithm of the mass-loss rate as the variable ($Y$), while the last five columns give the results for the inverse relation. The trends are valid for the subsample of Miras and SRa sources only; see Sect.~\ref{sect:obsh2o}.}}\label{table:corr}
  \begin{center}
  \begin{tabular}[c]{lllll||rrrrrr|rrrrr}\hline\hline\rule[0mm]{0mm}{3mm}
\rule{0pt}{10pt}Molecule & Transition &$\lambda_0$ (\mic)&$E_\mathrm{u}$ (cm$^{-1}$)&$n$&$n_\mathrm{inc}$&$\bar{a}$ & $\sigma_\mathrm{\bar{a}}$ & $\bar{b}$ & $\sigma_\mathrm{\bar{b}}$ &$\sigma_\mathrm{\bar{a}\bar{b}}$& $\bar{a}$ & $\sigma_\mathrm{\bar{a}}$ & $\bar{b}$ & $\sigma_\mathrm{\bar{b}}$ &$\sigma_\mathrm{\bar{a}\bar{b}}$\\\hline
\rule{0pt}{10pt}o-H$_2$O &$J_{\mathrm{K}_\mathrm{a}, \mathrm{K}_\mathrm{c}}=2_{2,1} - 2_{1,2}$ &180.5&134.9 & 11& 11 & -5.9 & 0.3 & -0.9 & 0.4 & 0.12 & -2.1 & 0.7 & -0.25 & 0.12 & 0.08 \\
&$J_{\mathrm{K}_\mathrm{a}, \mathrm{K}_\mathrm{c}}=2_{1,2} - 1_{0,1}$ &179.5&79.5& 18 &14& -5.51 & 0.07 & -0.8 & 0.2 & 0.012 & -2.2 & 0.6 & -0.38& 0.11 & 0.06 \\
&$J_{\mathrm{K}_\mathrm{a}, \mathrm{K}_\mathrm{c}}=3_{0,3} - 2_{1,2}$ &174.6&136.8& 12 & 9 & -5.70 & 0.13 & -0.8 & 0.3 & 0.03 & -2.8 & 0.8 & -0.45 & 0.16 & 0.13 \\
&$J_{\mathrm{K}_\mathrm{a}, \mathrm{K}_\mathrm{c}}=2_{2,1} - 1_{1,0}$ &108.1&134.9& 17 & 13 & -5.42 & 0.06 & -0.7 & 0.2 & 0.004 & -2.1 & 0.7 & -0.38& 0.12 & 0.08 \\
&$J_{\mathrm{K}_\mathrm{a}, \mathrm{K}_\mathrm{c}}=7_{0,7} - 6_{1,6}$ &72.0&586.2& 13 & 12 & -5.59 & 0.09 & -0.7 & 0.2 & 0.017 & -2.5 & 0.7 & -0.41 & 0.14 & 0.10 \\
&$J_{\mathrm{K}_\mathrm{a}, \mathrm{K}_\mathrm{c}}=3_{3,0} - 2_{2,1}$ &66.4&285.4& 15 & 12 & -5.32 & 0.06 & -0.8 & 0.3 & -0.005 & -1.3 & 0.6 & -0.26 & 0.11 & 0.06 \\\hline
\rule{0pt}{10pt}p-H$_2$O & $J_{\mathrm{K}_\mathrm{a}, \mathrm{K}_\mathrm{c}}=4_{1,3} - 3_{2,2}$ &144.5&275.5& 8 & 8 & -5.9 & 0.3 & -0.5 & 0.3 & 0.09 & -2.8 & 1.2& -0.3& 0.2 & 0.3 \\
& $J_{\mathrm{K}_\mathrm{a}, \mathrm{K}_\mathrm{c}}=3_{1,3} - 2_{0,2}$ &138.5&142.3& 17 & 13 & -5.68 & 0.11 & -0.9 & 0.2 & 0.02 & -2.5 & 0.6 & -0.39 & 0.12 & 0.07 \\
& $J_{\mathrm{K}_\mathrm{a}, \mathrm{K}_\mathrm{c}}=3_{2,2} - 2_{1,1}$ &90.0&206.3& 15 & 14 & -5.60 & 0.11 & -0.5 & 0.2 & 0.02 & -2.3 & 0.8 & -0.34 & 0.15 & 0.12 \\
& $J_{\mathrm{K}_\mathrm{a}, \mathrm{K}_\mathrm{c}}=7_{1,7} - 6_{0,6}$ &71.5&586.4& 8 & 7 & -5.7 & 0.2 & -0.7 & 0.4 & 0.08 & -2.3 & 1.0 & -0.33 & 0.19 & 0.19 \\\hline
 \end{tabular}
  \end{center}
  }
\end{table*}
Fig.~\ref{fig:h2ovmdot} shows the line-strength ratio of \water $J_{\mathrm{K}_\mathrm{a}, \mathrm{K}_\mathrm{c}}=2_{1,2} - 1_{0,1}$ and CO $J=15-14$ as a function of the mass-loss rate. Several qualitative conclusions can be drawn. A downward trend toward higher mass-loss rate is present in the \water/CO line-strength ratios, indicated by the green arrow superimposed on the data points {(see Sect.~\ref{sect:lsfa})}. {Assuming \water is homogeneously distributed within the formation region of a given line, this suggests that the \water abundance also decreases with increasing mass-loss rate in the same fashion.} Fig.~\ref{fig:h2ovmdot} shows the line-strength ratios for only one \water line, but the trend is significant for other \water lines as well (see Sect.~\ref{sect:lsfa}). However, contrary to the CO line strengths, the \water/CO line-strength ratios of the low-$\mg$ SRb sources do not follow the trend set by the Miras and SRa sources. Instead, they group together at the low end of the range of mass-loss rate featuring low line-strength ratios. Though only four sources can be considered, of which one is flagged as a line blend {(see Sect.~\ref{sect:ls} for clarification)}, {a tentative upward trend between the \water/CO line-strength ratio and the mass-loss rate appears present within the SRb sample.}
\begin{figure}[!t]
\resizebox{9.0cm}{!}{\includegraphics{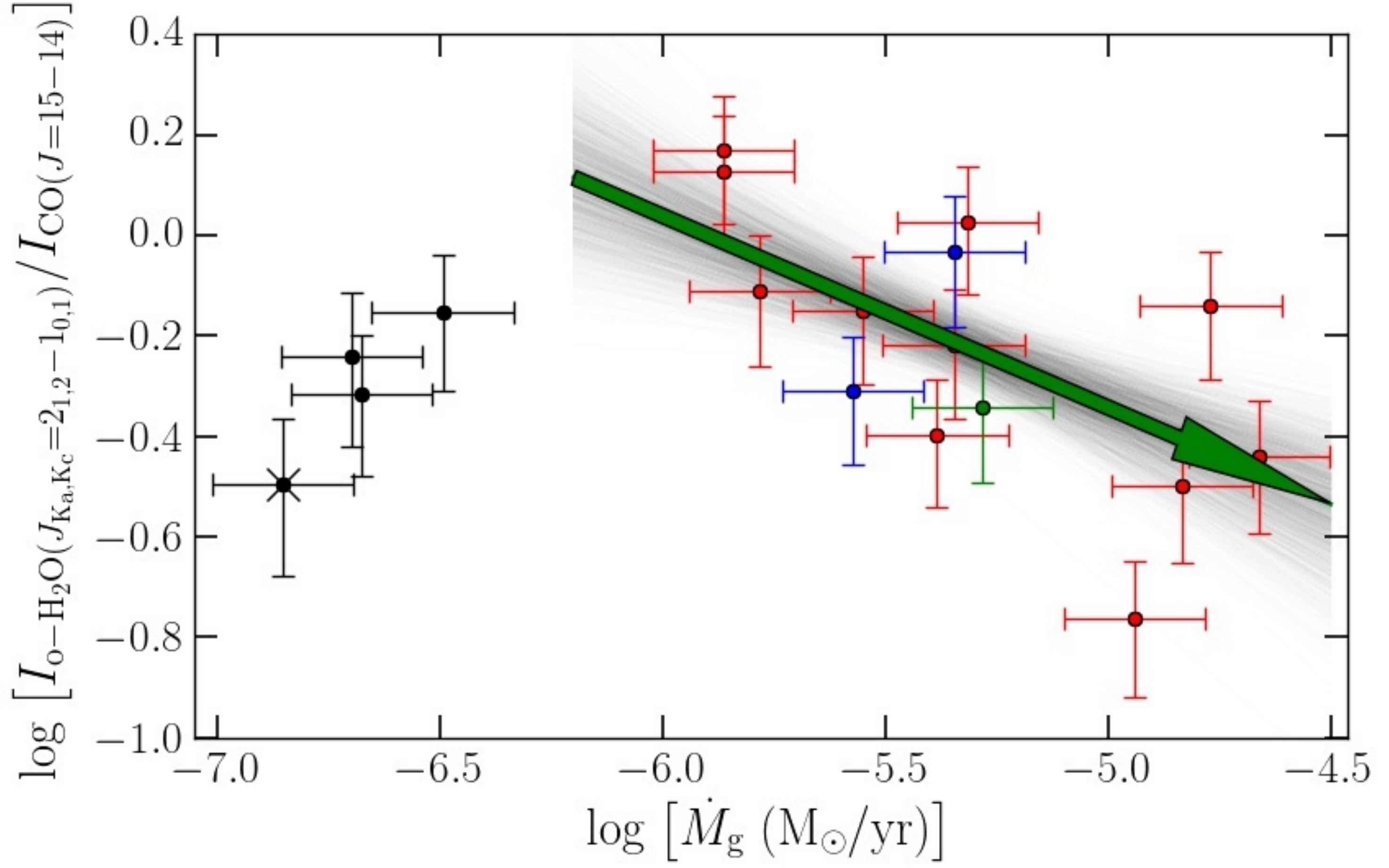}}
\caption{Line-strength ratio of the \water $J_{\mathrm{K}_\mathrm{a}, \mathrm{K}_\mathrm{c}}=2_{1,2} - 1_{0,1}$ transition and CO $J=15-14$ transition as a function of the mass-loss rate $\mg$. The data points are color coded according to the variability type: Miras in red, {SR/Mira sources in blue, the SRa source in green,} and SRb sources in black. A black cross superimposed on a point indicates that the \water line strength is flagged as a blend {(see Sect.~\ref{sect:ls} for clarification). The gray lines show the individual Monte Carlo linear fitting results %of $I_{\mathrm{oH_2O}}/I_\mathrm{CO} = a + b\mg$ to a large number of guesses drawn from the $\mg$ and $I_{\mathrm{H}_2\mathrm{O}}/I_\mathrm{CO}$ distributions of each 
to the data points for Miras and SRa sources. The green arrow indicates the mean linear relation (see Sect.~\ref{sect:lsfa}).} %, for which the coefficients $\bar{a}$ and $\bar{b}$ are given in Table~\ref{table:corr}.
}
\label{fig:h2ovmdot}
\end{figure}

% $^{12}$CO &$J=15-14$ & 18&7 &2 &0.78 &0.15 &0.3 & -10.1&0.6 & 0.75&0.11 &0.07 \\
% &$J=18-17$ &18 &7 &2 &0.89 &0.14 &0.3 & -10.0& 0.6&0.75 & 0.11&0.07 \\
% &$J=24-23$ &18 &7 &2 &0.91 &0.17 &0.4 & -10.5& 0.6&0.67 & 0.11&0.07 \\
% &$J=29-28$ &18 &7 &3 &0.92 &0.19 &0.5 & -10.9& 0.6&0.61 & 0.11&0.07 \\
% & $J=30-29$ &17 &7 &2 &0.88 &0.17 &0.4 & -10.5& 0.6&0.70 & 0.11&0.07 \\
% &$J=36-35$ &10 &8 &4 &1.0 &0.3 &1.0 & -11.6& 0.8&0.53 & 0.13&0.11 \\\hline
% $^{13}$CO&$J=19-18$ &8 &7 &3 &0.86 &0.19 &0.5 & -11.1&0.8 & 0.72&0.12 &0.09 \\\hline
{The difference between the SRb sources, on the one hand, and the Miras and SRa sources, on the other hand, suggests some dependence of \water emission on pulsational properties.} Fig.~\ref{fig:h2ovp} gives the line-strength ratio of the \water $J_{\mathrm{K}_\mathrm{a}, \mathrm{K}_\mathrm{c}}=2_{1,2} - 1_{0,1}$ transition and the CO $J=15-14$ transition as a function of pulsational period. The data points are color coded according to the wind density tracer $\shellmass$. The Miras and SRa sources are shown in blue, red, and green for increasing $\shellmass$ (as indicated in the legend). An increasing outflow density, and thus a decreasing \water/CO line-strength ratio, is associated with an increasing pulsational period. The pulsational period and mass-loss rate were derived independently (see Table~\ref{table:sample} for references.) This supports previous theoretical \citep{bow1988} and observational \citep{woo2007,deb2010} studies that have shown a strong correlation between the mass-loss rate and the pulsational period of AGB stars. {The SRb sources (shown in black in Fig.~\ref{fig:h2ovp}) do not show a clear-cut correlation between wind density and pulsational period.} We note that the \water $J_{\mathrm{K}_\mathrm{a}, \mathrm{K}_\mathrm{c}}=2_{1,2} - 1_{0,1}$ transition detected in U Hya (the right most black point in Fig.~\ref{fig:h2ovp}) is flagged as a line blend, which effectively makes the \water/CO line-strength ratio an upper limit.

\begin{figure}[!t]
\resizebox{\hsize}{!}{\includegraphics{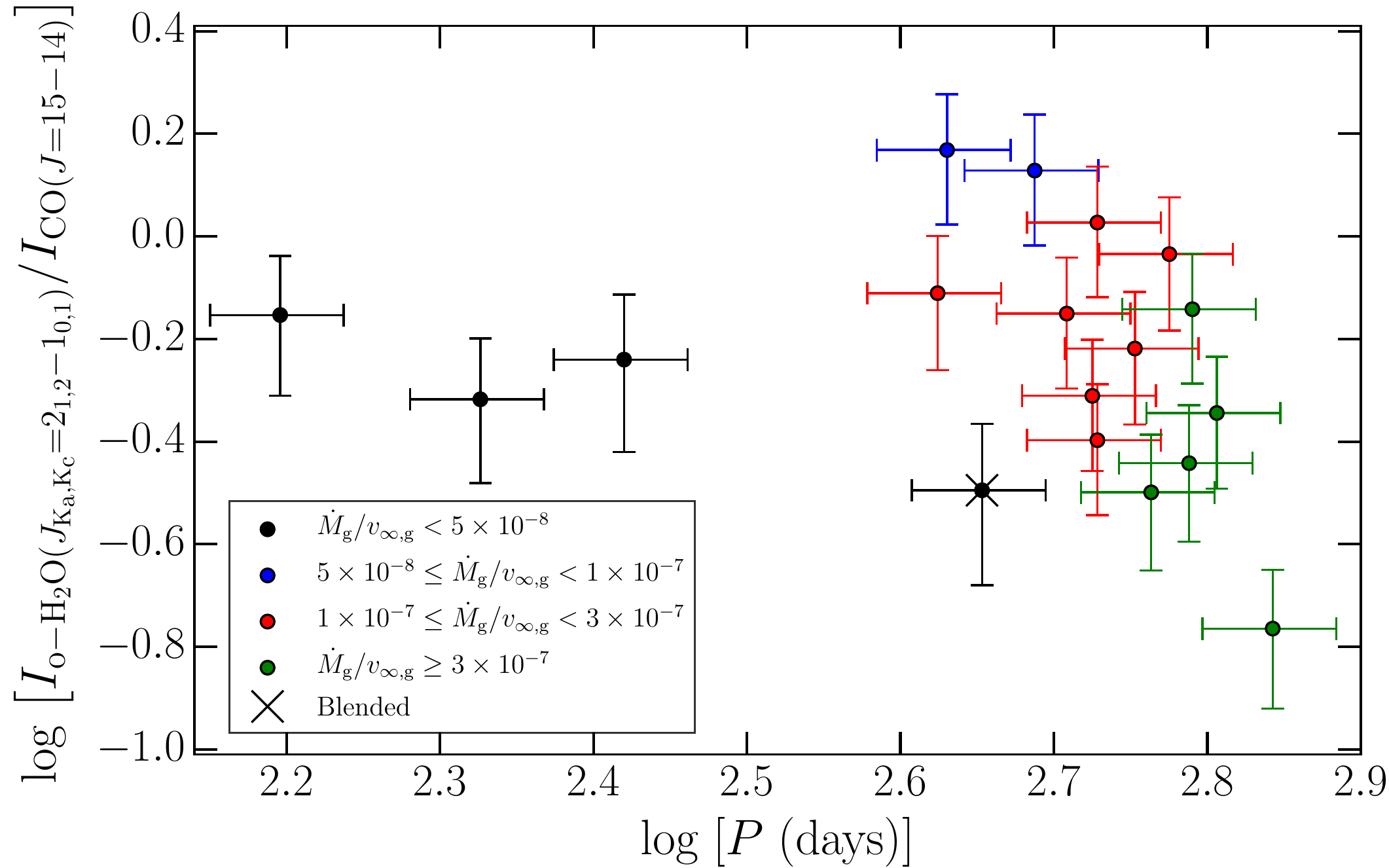}}
\caption{Line-strength ratio of the \water $J_{\mathrm{K}_\mathrm{a}, \mathrm{K}_\mathrm{c}}=2_{1,2} - 1_{0,1}$ transition and the CO $J=15-14$ transition as a function of the pulsational period. The points with error bars give the measured \water/CO line-strength ratios, color coded according to the value of the wind density proxy $\shellmass$ in units of \myrkms (see legend). A black cross superimposed on the data point indicates that the \water line is flagged as a blend.}
\label{fig:h2ovp}
\end{figure}
\subsection{Least-squares fitting approach}\label{sect:lsfa}
To quantify the {negative correlation} between measured \water/CO line-strength ratios and mass-loss rates of the Miras and the SRa sources, we apply a least-squares fitting technique to fit a linear function in logarithmic scale. {Measurements are included only when $\mg > 5 \times 10^{-7}$ $\msunyr$. This removes the four SRb sources from the statistical sample.} We have to take into account the uncertainties on the measured values, which follow a normal distribution in linear space, and the uncertainty on the mass-loss rate to assess the accuracy of the fitted slope and intercept. Studies investigating the mass-loss rate of AGB outflows typically report uncertainties of a factor of three \citep{ram2008,deb2010,lom2013,sch2013}. For our purposes, we assume that the derived $\mg$ values follow a normal distribution in logarithmic scale with the 3$\sigma$-confidence level equal to this factor of three accuracy.

To ensure a proper error propagation, we apply a Monte Carlo-like approach, in which we draw a large number of guesses ($N=10^6$) for the relevant quantities from their respective distributions. Since we fit the observed line-strength ratios in logarithmic scale, we can also apply this approach to the mass-loss rate, for which we draw the guess from the normal distribution of logarithmic values. This results in $N$ linear relations from which we calculate the mean slope and intercept to arrive at a mean relation between the relevant quantities. At the same time, we also determine whether the slope and intercept of the $N$ relations are correlated. This approach is applied to all \water transitions. The number of data points $n$ per transition taken into account for the linear fit is given in column 5 of Table~\ref{table:corr}. The mean coefficients $\bar{a}$ and $\bar{b}$ of the linear relation $Y = \bar{a} + \bar{b}X$ between $I_{\mathrm{H}_2\mathrm{O}}/I_\mathrm{CO}$ and $\mg$ are listed in the next ten columns of Table~\ref{table:corr}; their uncertainties and the covariance between them are also listed. We give the results for $I_{\mathrm{H}_2\mathrm{O}}/I_\mathrm{CO}$ as independent variable $X$ in columns 6 through 10 and the results for the inverse relation in columns 11 through 15. Taking the reciprocal of one relation does not necessarily result in the coefficients of the inverse relation because the least-squares minimization only takes the vertical residuals between the data points and the best linear fit into account. The $N$ individual linear fit results in the Monte Carlo approach are shown in gray-black in Fig.~\ref{fig:h2ovmdot}. The green arrow indicates the mean linear relation according to the coefficients given in Table~\ref{table:corr} for the \water $J_{\mathrm{K}_\mathrm{a}, \mathrm{K}_\mathrm{c}}=2_{1,2} - 1_{0,1}$ transition.

Notably, within the fitting uncertainties, the slope of the linear relation {is similar} for all ortho- and para-\water lines. We list the covariance between the slope and the intercept of the linear relation as well, which is a measure of how closely correlated the slope and the intercept are. With the exception of one, all relations listed in Table~\ref{table:corr} show a strong correlation between the slope and the intercept, meaning that a larger intercept must be associated with a steeper slope. This is evidenced by the gray lines in Fig.~\ref{fig:h2ovmdot}, which seem to knot together in the intermediate $\mg$ region, while spreading out for more extreme values of $\mg$. The \water/CO line-strength ratio for the $J_{\mathrm{K}_\mathrm{a}, \mathrm{K}_\mathrm{c}}=3_{3,0} - 2_{2,1}$ transition is attributed to a small negative covariance when taking $I_{\mathrm{H}_2\mathrm{O}}/I_\mathrm{CO}$ as the independent variable $X$. This suggests that the slope and intercept of the linear relation are weakly correlated, hence the negative value. However, the slope-intercept correlation is very weak for this particular transition because of a large scatter between the data points. As such, the linear fit to this \water/CO line-strength ratio and the mass-loss rate is less reliable, but still confirms the observed downward trend based on the negative slope $\bar{b} = -0.8$.

The relations in columns 6 through 10 can serve as a mass-loss indicator as long as measurements for the relevant \water and CO line strengths are available. The relations in columns 11 through 15 are helpful in predicting the \water/CO line-strength ratio, given a mass-loss rate. When using these relations to estimate a mass-loss rate or predict a line-strength ratio, the uncertainty on the result can be determined from the relation $$\sigma_{\rm Y} = \sqrt{\sigma_{\rm \bar{a}}^2 + \bar{b}^2\sigma_{\rm X}^2 + X^2\sigma_{\rm \bar{b}}^2 +X\sigma_{\rm \bar{a}\bar{b}}^2}.$$ Barring systematic effects in the assumed $\mg$ values for our sample, this leads to an uncertainty of about 0.3 dex on the logarithmic values.
\begin{table}[!t]
  {
  \setlength{\tabcolsep}{5pt}
  \caption{Stellar and circumstellar parameters of the model grid described in Sect~\ref{sect:theomodel}. The first and second column list the parameter and its unit, the third column lists the adopted value in the {standard} model grid, the fourth column indicates the sampling range in which an individual parameter is allowed to vary, and the last column gives the step size with which the parameter was probed. Listed are the gas mass-loss rate ($\mg$), \water abundance with respect to molecular hydrogen (\waterabun), power of the adopted radial gas kinetic-temperature profile given in Eq.~\ref{eq:tkin}~($\epsilon$), effective temperature ($T_\star$), luminosity ($L_\star$), gas terminal velocity ($\vg$), dust-to-gas ratio ($\psi$), and CO abundance with respect to molecular hydrogen (\coabun).}\label{table:modelpars}
  \begin{center}
  \begin{tabular}[c]{llllll}\hline\hline\rule[0mm]{0mm}{3mm}
 Parameter & Unit & Standard & Range & Step size \\\hline
\rule{0pt}{10pt}$\log(\mg)$ & $\msunyr$ & & $[-8.0,-4.5]$ & 0.5 \\
$\log($\waterabun)& & & $[-10,-4]$ & 1\\
$\epsilon$& & 0.4 & $[0.3,0.9]$ & 0.1\\
$T_\star$& $10^3$ K & 2.4 & $[2.4,3.0]$ & 0.3\\
$L_\star$& 10$^3$ L$_\odot$ & 8 & $[4,12]$ & 4 \\
$\vg$& \kms & 10 & $[10,25]$ & 5\\
$\log(\psi)$& & -3 & $[-3.3,-2.7]$ & 0.3 \\
\coabun & $10^{-3}$ & 0.8 & $[0.6,1.2]$ & 0.2 \\\hline
\end{tabular}
  \end{center}
  }
\end{table}

\section{Sample-wide \water abundance}\label{sect:comp}
A negative correlation between the \water/CO line-strength ratio and the mass-loss rate is evident for the Miras and SRa sources. We compute a set of radiative-transfer models to investigate the role of optical-depth effects and to establish whether or not this points to a negative correlation between the \water abundance and mass-loss rate. Because modeling the line strengths for each source individually is beyond the scope of this study, we opt for an approach in which we calculate these line strengths for models covering the parameter range appropriate for Miras, SRa, and SRb sources.

\subsection{The model grid}\label{sect:theomodel}
We set up a model grid with a fine sampling of the \water abundance\footnote{All values for \waterabun are given for ortho-\water only.}, the mass-loss rate $\mg$, and the gas temperature profile $T_\mathrm{g}(r)$, and with a coarse sampling of the other stellar and circumstellar properties: the gas terminal velocity $\vg$, the stellar effective temperature $T_\star$, the stellar luminosity $L_\star$, the dust-to-gas ratio $\psi$, and the CO abundance with respect to molecular hydrogen. {We refer to a single set of values for the latter set of properties as the {standard} model grid, for which the values are listed in Table.~\ref{table:modelpars}{, and we represent it by a black curve in the figures in Sect.~\ref{sect:comp} for clarity.} In this grid, the mass-loss rate and the \water abundance are allowed to vary between $1 \times 10^{-8}\ \msunyr$ and $3\times10^{-5}\ \msunyr$, and $10^{-10}$ and $10^{-4}$, respectively. To probe the sensitivity of the observed \water emission to the other stellar and circumstellar properties, we created secondary model grids in which at most one additional fixed parameter from the {standard} grid was allowed to vary. We consider each grid separately in Sects.~\ref{sect:cols}~and~\ref{sect:h2o}.} Table~\ref{table:modelpars} lists both the adopted value for the {standard} model grid as well as the sampling range and step size of the parameters. Beam effects or other telescope-related properties have been corrected for during the PACS data reduction, such that measured line strengths can be directly compared with the intrinsic line strengths of theoretical predictions. This assumes that the PACS observations are not spatially resolved, which has been one of our target selection criteria (see Sect.~\ref{sect:targetsel}). Unfortunately, even though it is the prototypical carbon-rich AGB star, CW~Leo has to be excluded from the sample as a result of its spatial extent as observed by the PACS instrument. We refer to {the work by \cite{cer2014}} for typical CO and \water line strengths, but {we caution that these values must be rescaled to 100 pc to facilitate a comparison with our results, which is not straightforward given CW~Leo's extension.}

We calculate spectral line profiles using \gastronoom \citep{dec2006,dec2010a,lom2013}. In these calculations, the density distribution of the outflow is assumed to be smooth and spherically symmetric, i.e.~we do not take a small-scale structure in the form of clumps or a large-scale structure in the form of a disk or polar outflows into account. {We do not take masing into account in our modeling.} We use a COMARCS synthetic spectrum for the central star \citep{ari2009} with $log(g) = 0.0$, a C/O ratio of 1.4, $M_\star = 1.0$ M$_\odot$, a microturbulent velocity of 2.5 \kms, and solar metallicity. For CO, we take transitions in the ground- and first-vibrational state up to $J = 60$ into account. The energy levels, transition frequencies, and Einstein A coefficients were taken from \citet{goo1994} and the CO-H$_2$ collision rates from \citet{lar2002} (see Appendix A in \citeauthor{dec2010a}~\citeyear{dec2010a} for more details). For \water, we take into account the 45 lowest levels of the ground state and the $\nu_2=1$ and $\nu_3=1$ vibrationally excited states. Level energies, frequencies, and Einstein A coefficients were taken from the HITRAN water line list \citep{rot2009}, and \water-H$_2$ collisional rates from \citet{fau2007} (see \citeauthor{dec2010a}~\citeyear{dec2010a}, and Appendix B in \citeauthor{dec2010b}~\citeyear{dec2010b} for more details). {Recently, \citet{dub2009} and \citet{dan2011} published new \water-H$_2$ collisional rates. \citet{dan2012} compared these collision rates to those from \citet{fau2007} and found that the line strengths can be affected by up to a factor of 3 for low \water abundance (\waterabun$\sim 10^{-8}$) and low density ($n_{\mathrm{H}_2} < 10^7$) regimes. They also note that when \water excitation is dominated by pumping via the dust radiation field, these differences are attenuated. Hence, we do not expect this to affect our results significantly.} 

The molecular abundances with respect to H$_2$ of both CO and \water are assumed to be constant throughout the wind up to the photodissociation radius where interstellar UV photons destroy the molecules. The CO photodissociation radius is set by the formalism of \citet{mam1988}. For \water we use the analytic formula from \citet{gro1994}. The acceleration of the wind to the terminal expansion velocity $\vg$ of the gas is set by momentum transfer from dust to gas, assuming full momentum coupling between the two components \citep{kwo1975}. The gas turbulent velocity $\vs$ is fixed at 1.5 \kms. Because the cooling contribution from HCN is not well constrained \citep{dec2010a,deb2012}, we approximate the gas kinetic-temperature structure with a power law of the form \begin{eqnarray}\label{eq:tkin}T_\mathrm{g}(r) = T_\star\ \left(\frac{r}{R_\star}\right)^{-\epsilon},\end{eqnarray} where $r$ is the distance to the center of the star. As shown by \citet{lom2013}, dust can play an important role in \water excitation. Following \citet{lom2012}, we use a distribution of hollow spheres (DHS, \citeauthor{min2003}~\citeyear{min2003}) with filling factor 0.8 to represent the dust extinction properties, a dust composition that is 75\% amorphous carbon, 10\% silicon carbide, and 15\% magnesium sulfide, {and assume composite dust grains, leading to thermal equilibrium between all three dust species}. The optical properties used to calculate the extinction contribution from these species are taken from \citet{jag1998b}, \citet{pit2008}, and \citet{beg1994}, respectively. We take the inner radius of the dusty circumstellar envelope to match the dust condensation radius, which is determined following \citet{kam2009} with use of the dust radiative-transfer code \mcmax \citep{min2009a}. Typical inner-radius values lie between 2 and 2.5 \rstar.}

\subsection{The 6.3 \mic flux}\label{sect:63flux}
\begin{figure}[!t]
\resizebox{9.0cm}{!}{\includegraphics{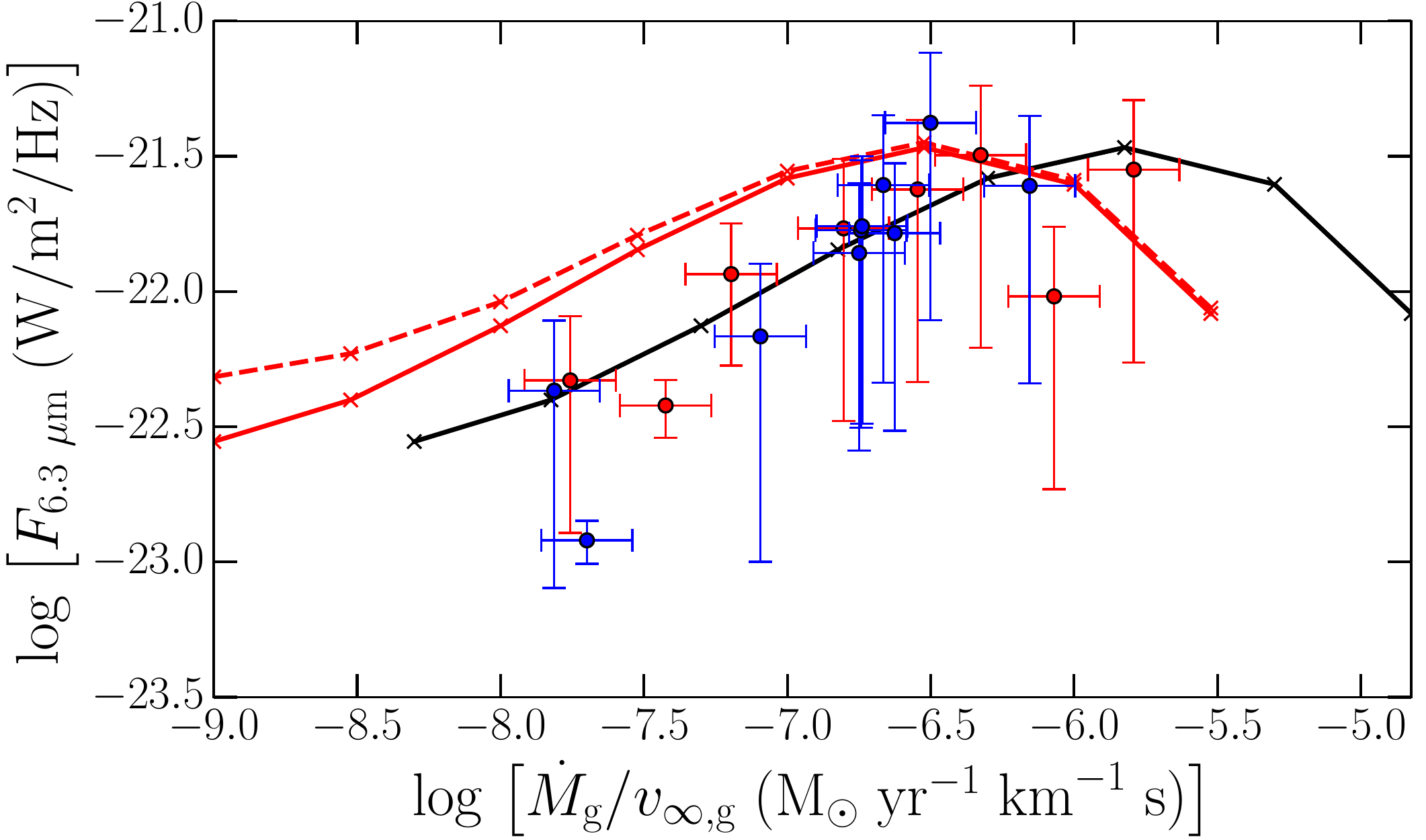}}
\caption{Fluxes at 6.3 \mic as a function of $\shellmass$. The data points are color coded according to the method with which the 6.3 \mic flux was measured: from ISO data in red, and interpolation of photometric data points in blue. The full and dashed lines are fluxes calculated from models. The red lines represent models with $\psi = 0.005$, while the black line shows models with $\psi = 0.001$. The dashed line makes use of a blackbody spectrum of 2400 K for the central star, while the full lines use a COMARCS spectrum. The 6.3 \mic fluxes are scaled to a distance of 100 pc.}
\label{fig:f63vshellmass}
\end{figure}
The excitation analysis of \water is important when considering \water emission from any type of source. {We refer to \citet{gon2007}, \citet{mae2008}, and \citet{lom2013} for examples of overviews of the most important excitation channels for \water. These include: 1) collisional excitation; 2) radiative vibrational excitation in the near- and mid-IR; and 3) radiative rotational excitation in the mid- and far-IR.} The $\nu_2 = 1$, $\nu_1 = 1$, and $\nu_3=1$ vibrational states can be accessed by absorption of radiation at about 6.3 \mic and 2.7 \mic, respectively. Especially the $\nu_2=1$ state was shown to have a strong impact on the excitation of \water molecules by \citet{gon2007}. We therefore carefully consider whether our modeling approach correctly reproduces the observed flux at 6.3 \mic for our sample.

The stellar spectrum and the presence of dust primarily determine the flux at 6.3 \mic. Atmospheric absorption bands can have a significant impact on the near-IR flux. For this reason, we make use of a COMARCS synthetic spectrum as opposed to a blackbody spectrum for a more reliable estimate of the stellar flux at 6.3 \mic. This flux depends on the pulsational phase of the star, which is not taken into account in the COMARCS models (e.g.,~\citeauthor{deb2012}~\citeyear{deb2012} for CW~Leo). Time-dependent modeling of the atmosphere and inner wind is beyond the scope of this work.

\begin{figure*}[!ht]
$\begin{array}{cc}
\resizebox{9.0cm}{!}{\includegraphics{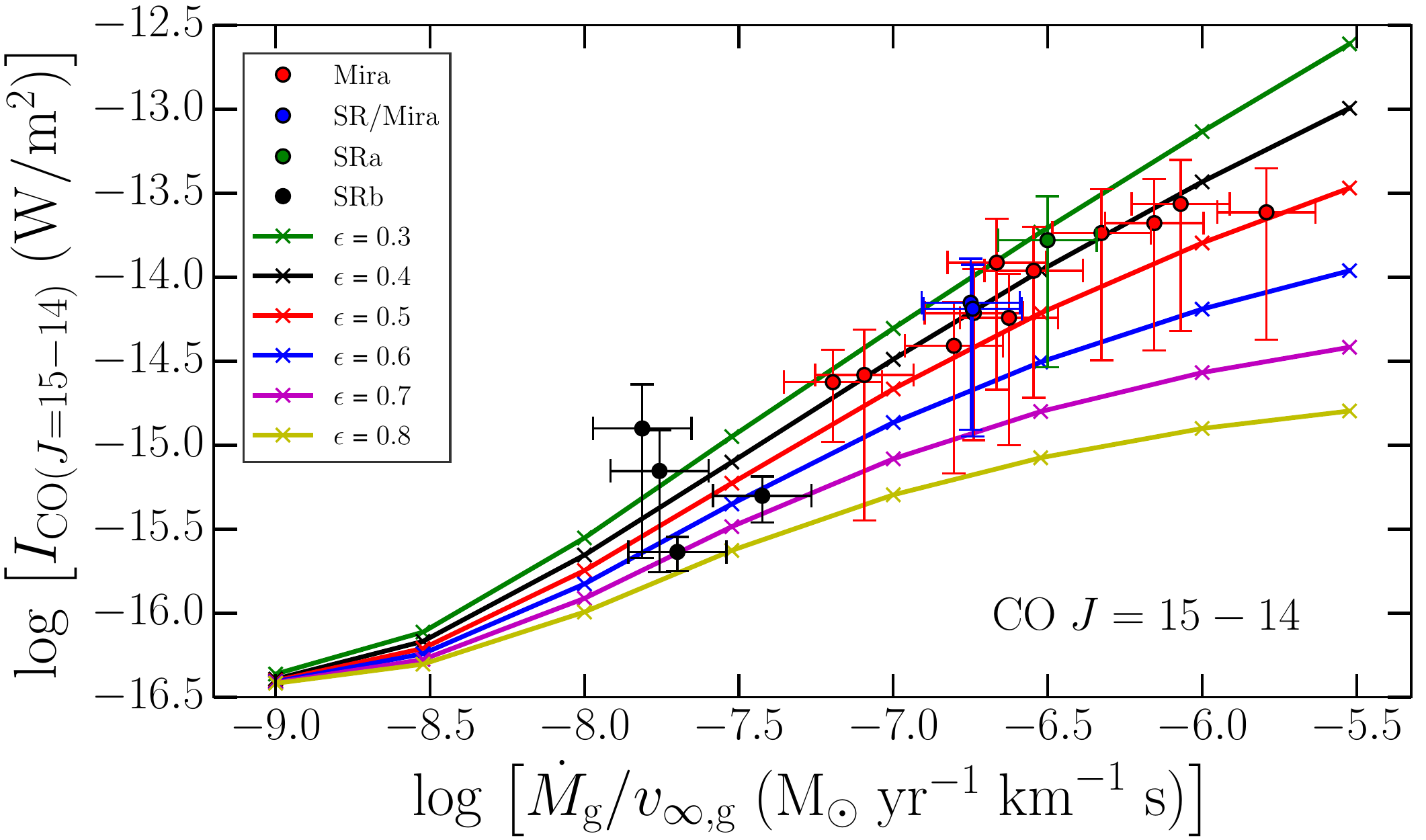}} & \resizebox{9.0cm}{!}{\includegraphics{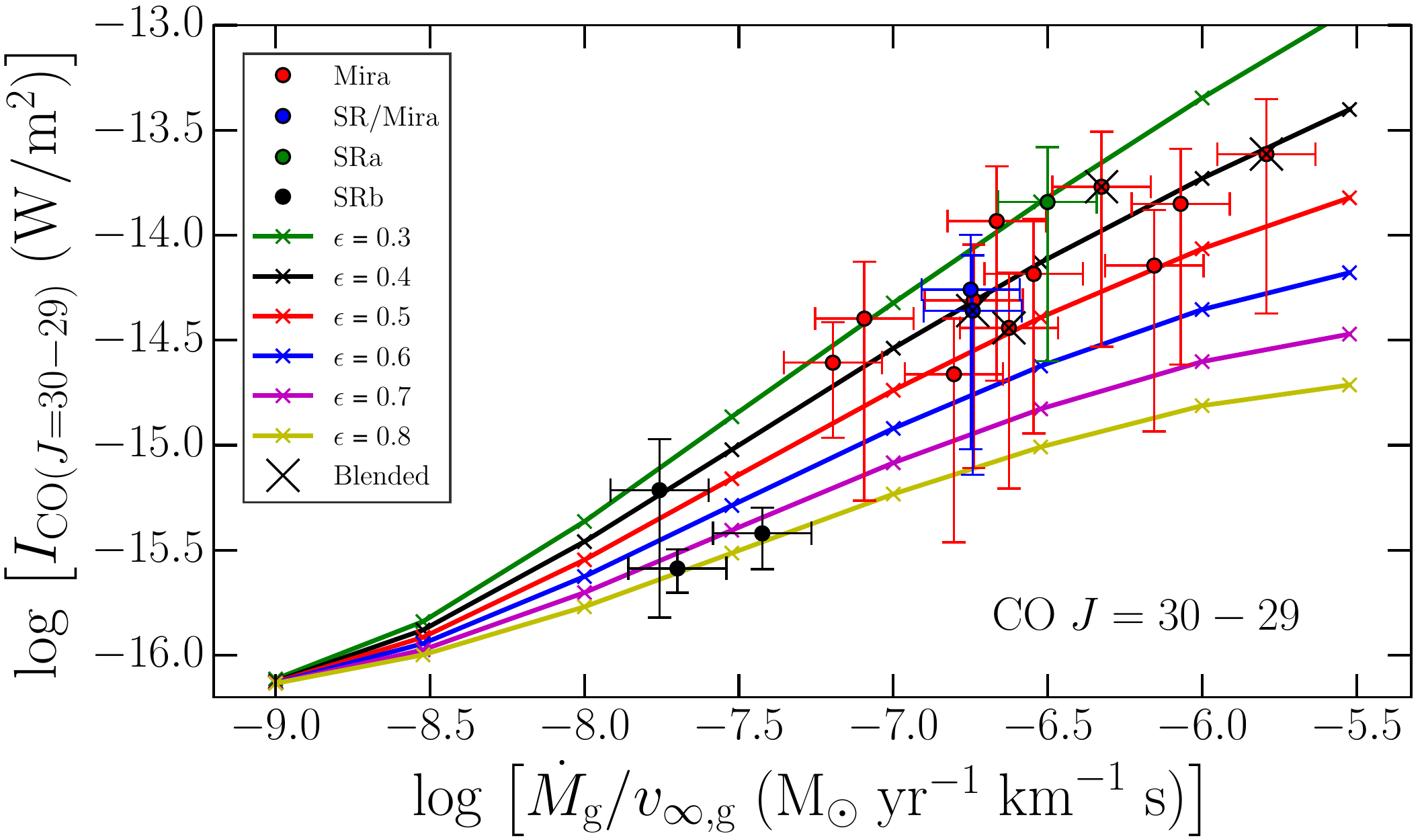}} 
\end{array}$
\caption{Line strengths of two CO transitions as a function of $\shellmass$: CO $J=15-14$ on the left and CO $J=30-29$ on the right. The points with error bars give the measured CO line strengths, color coded according to the variability type. A black cross superimposed on the data point indicates that the CO line is flagged as a blend. The colored curves show the predicted CO line strength for various values of the temperature power law exponent $\epsilon$. Adopted values for other parameters are listed in Table~\ref{table:modelpars}. The line strengths are scaled to a distance of 100 pc.}
\label{fig:covshellmass_eps}
\end{figure*}

The presence of dust reddens the stellar spectrum and affects the radiation field that \water is subjected to. The amount of reddening depends critically on the optical depth in the dust continuum. Reddening has two major effects. Firstly, a higher dust content smooths out the stellar spectrum. In other words, using a synthetic spectrum rather than a blackbody spectrum becomes irrelevant for high mass-loss rates. Secondly, the spectral reddening shifts a large portion of the emitted photons away from the near-IR to the mid-IR. In first order, the 2.7 \mic \water vibrational excitation channels become less relevant for higher mass-loss rates. {Once $\mg$ is high enough to turn the star into an extreme carbon star (e.g., in the case of LL~Peg, where the 11-\mic SiC feature is in absorption; see, for instance, \citeauthor{lom2012}~\citeyear{lom2012}) the 6.3 \mic \water vibrational excitation channel loses importance as well, in favor of the far-IR rotational excitation channels of \water.}

{To illustrate these effects, Fig.~\ref{fig:f63vshellmass} shows the predicted and measured 6.3 \mic fluxes for our sample of carbon stars {scaled to a distance of 100 pc. The uncertainties on the observed fluxes are predominated by the uncertainty on the distance.} The models are calculated for a blackbody and a COMARCS stellar spectrum of $T_\star = 2400$ K, for two different dust-to-gas ratios: the canonical value of 0.005 and a value of 0.001. The 6.3 \mic flux is only weakly dependent on the stellar spectrum in the low mass-loss rate regime. The dust-to-gas ratio has a much more pronounced effect across all densities. The overall trend supports $\psi = 0.001$. For reference, the dust opacity at 6.3 $\mu$m is $4\times10^3$ cm$^2$~g$^{-1}$. \citet{erik2014} find similar low dust-to-gas ratios from their wind model calculations in line with our findings.}

Hence, in what follows, we do not calculate models to fit every source individually, and instead make assumptions to reproduce the 6.3 \mic flux on average for the whole sample. We use a COMARCS synthetic spectrum of 2400 K (synthetic spectra for even lower temperatures are not available) and a dust-to-gas ratio of 0.001 for the standard model grid. However, we vary these parameters to probe their effect on the \water line strengths, if needed. {Many sources in our sample, all of which probe the upper range of $\shellmass$, are predicted to have a lower effective temperature than the 2400 K used here. Because the 6.3 \mic flux of the high-$\shellmass$ sources is insensitive to direct stellar light, the adopted effective temperature does not affect the \water excitation. We therefore have a preference in the model grid for a higher effective temperature, which better represents the low-$\shellmass$ sources.}
\begin{figure*}[!t]
$\begin{array}{cc}
\resizebox{9.0cm}{!}{\includegraphics{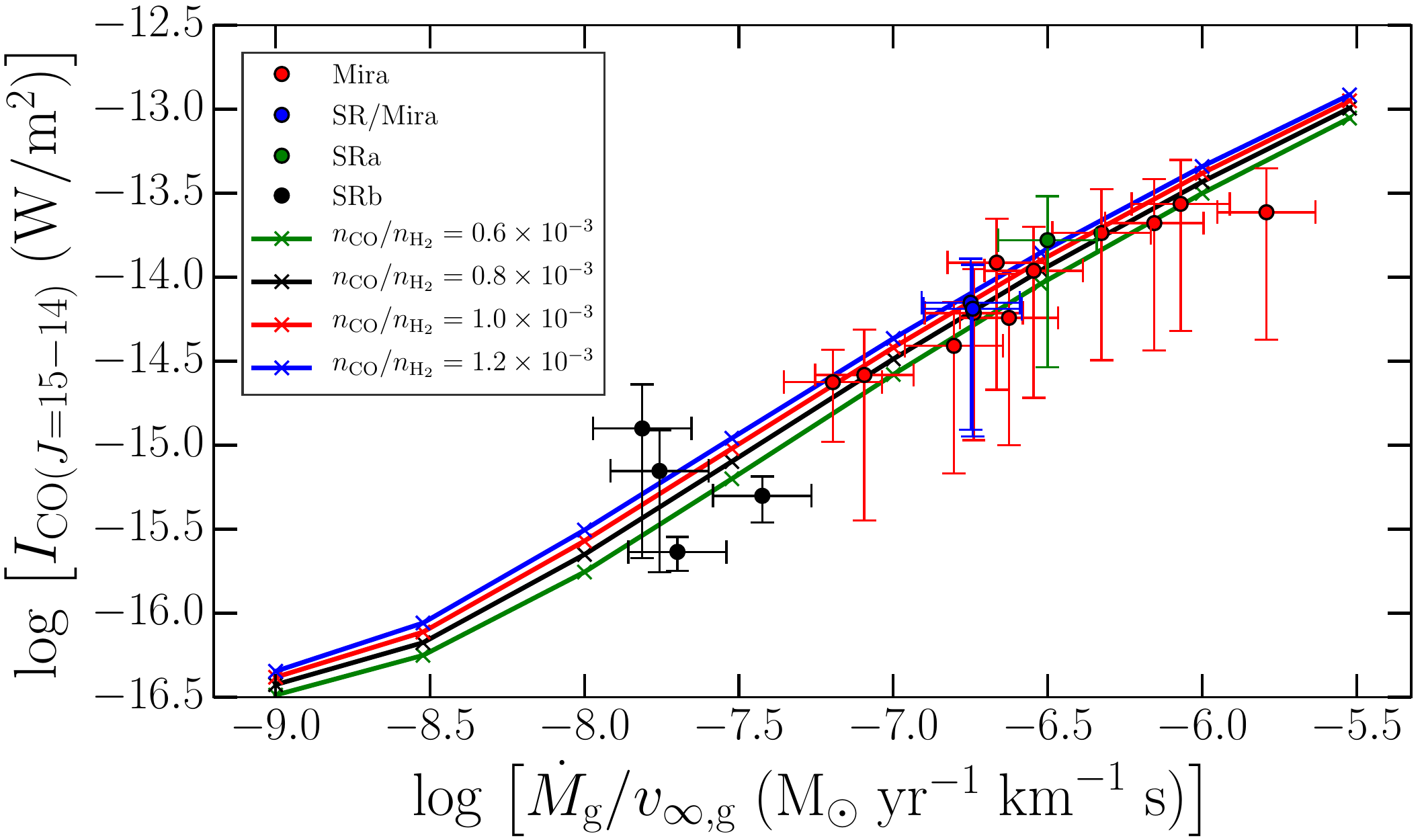}} & \resizebox{9.0cm}{!}{\includegraphics{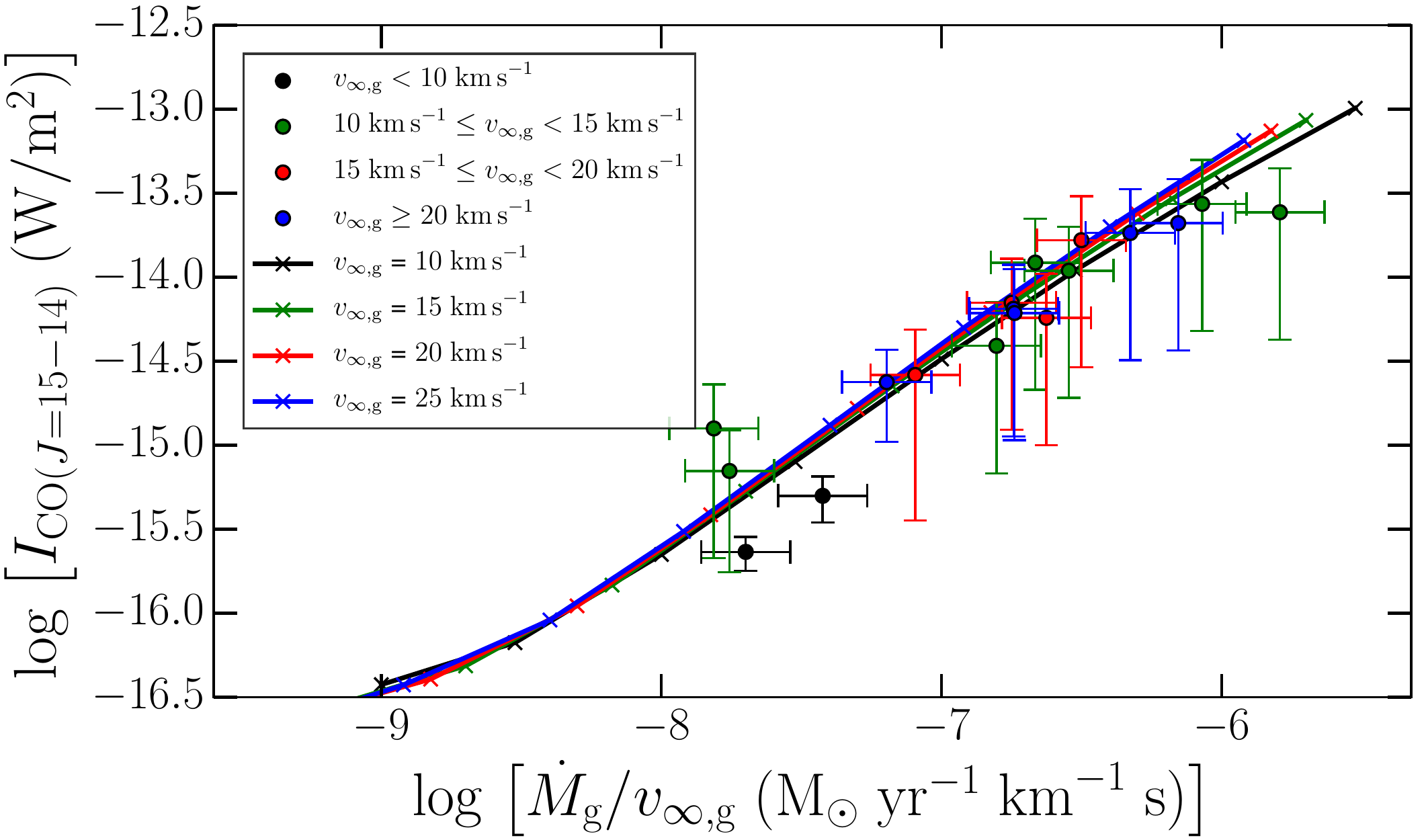}} \\
\resizebox{9.0cm}{!}{\includegraphics{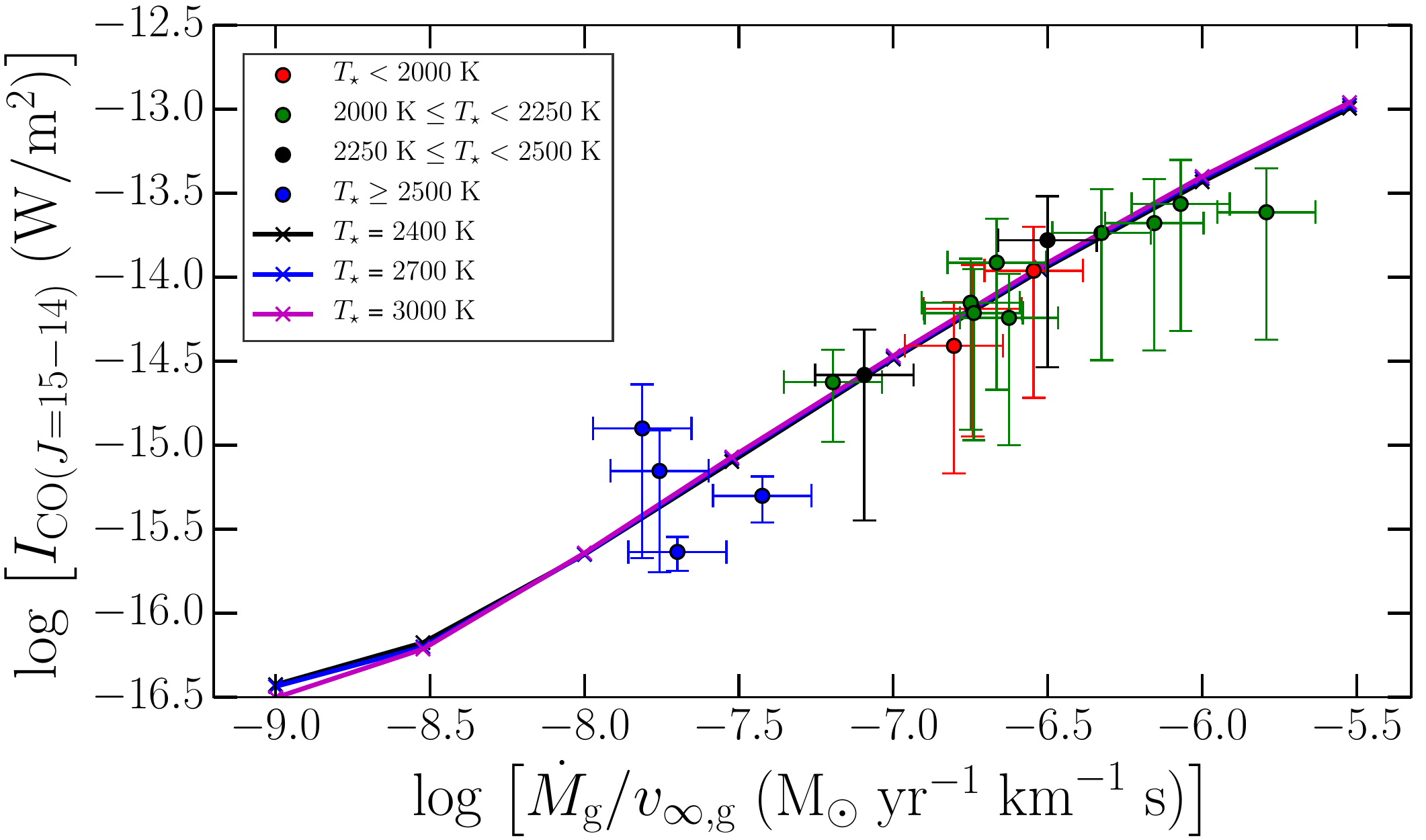}} & \resizebox{9.0cm}{!}{\includegraphics{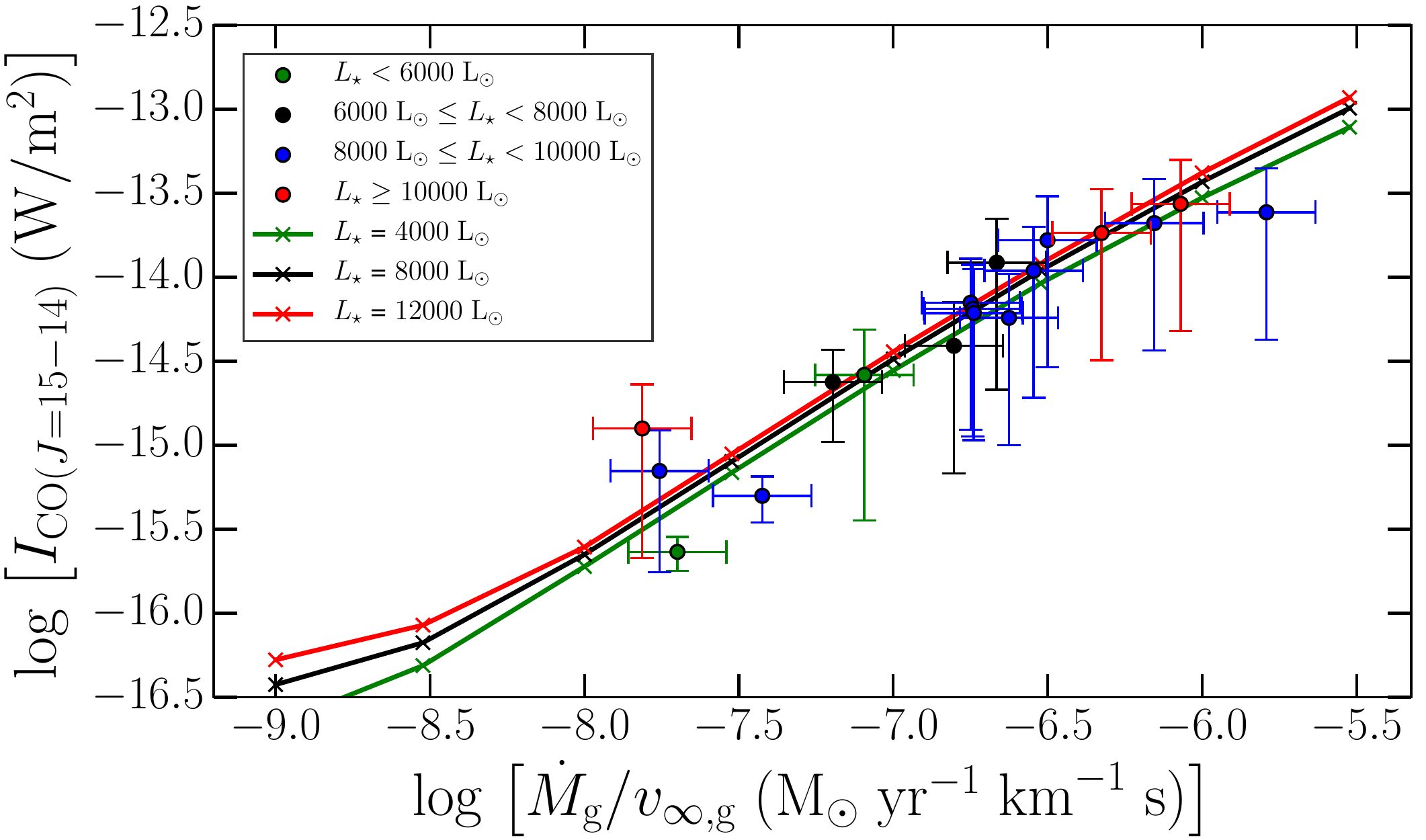}} 
\end{array}$
\caption{Line strengths of the CO $J=15-14$ transition as a function of $\shellmass$. The points with error bars give the measured CO line strengths, color coded according to the pulsational type or according to the values of the relevant quantity indicated in the legend. Each panel shows curves for different values of a given parameter in the model grid with similar color coding as the data points, if applicable{: the CO abundance with respect to H$_2$ (\coabun; upper left panel), gas terminal velocity ($\vg$; upper right panel), stellar effective temperature ($T_\star$; lower left panel), and stellar luminosity ($L_\star$; lower right panel).} Adopted values for other parameters are listed in Table~\ref{table:modelpars}. The line strengths are scaled to a distance of 100 pc.}
\label{fig:covshellmass_pars}
\end{figure*}

\subsection{CO line strengths}\label{sect:cols}
To allow for a direct comparison between measured and predicted \water/CO line-strength ratios, it is important that the {standard} model grid predicts the observed CO line strengths well. {We show CO line strengths as a function of the circumstellar density tracer given by $\shellmass$, probing a broad range of values for the mass-loss rate but keeping the terminal expansion velocity constant unless noted otherwise. The most influential property that affects CO emission other than the circumstellar density is the gas kinetic temperature. In Fig.~\ref{fig:covshellmass_eps}, we consider CO emission calculated in the {standard} model grid for various values of the exponent of the temperature power law.}

Because the distances to many sources are uncertain, it is difficult to constrain the exponent of the temperature law, as shown in Fig.~\ref{fig:covshellmass_eps} for $J=15-14$ on the left, and for $J=30-29$ on the right. The most probable value taking into account both these CO transitions as well as others (not shown here) is $\epsilon = 0.4$. As discussed in Sect.~\ref{sect:coash2}, the trend in the observed CO line strengths as a function of the mass-loss rate flattens off at higher values. The theoretical predictions confirm this observed trend, but mainly for higher temperature exponents. The effect on the CO $J=15-14$ transition appears to be limited, which confirms this line to be a suitable H$_2$ density tracer. {There seems to be a larger spread in CO line strengths among the SRb sources, though given the small sample size and the uncertainties on the distance it is premature to conclude that this points to a temperature law that is deviant from that of Mira and SRa sources.}

Recent findings by {\citet{cer2014} point to a possible time variability in the high-$J$ CO line strengths for the high-$\mg$ carbon star CW~Leo. Significant line variability is detected in $^{13}$CO $J=18-17$ with an amplitude of up to 30\%. {The variability is most likely caused by the variation in stellar luminosity with pulsational phase.} The main isotopologue of CO is more optically thick than $^{13}$CO, so time variability is expected to have a smaller effect and, if present, seems well within the uncertainties on the observed CO line strengths. Nevertheless, we must be careful in interpreting model-to-data comparisons of higher excitation CO lines, as we do not take time variability into account. We have two observations for several CO lines at different phases for the high-$\mg$ source LL~Peg, of which the integrated line strengths are given in Tables~\ref{table:intintmess}~and~\ref{table:intintot2old}. None of these CO lines convincingly show any variability, except for the $J=29-28$ transition. One of its line detections, though, is flagged as a blend, which invalidates the line as a reliable variability tracer. {The winds of SRb sources are the least opaque, implying that the CO lines of these stars likely suffer the most from temporal effects. The spread in CO line strengths in the SRb sources, as shown in Fig.~\ref{fig:covshellmass_eps}, could be related to this. The CO $J=15-14$ line is formed in the intermediate wind even in SRb sources, so circumstellar density variations due to stellar pulsations do not affect the line directly. The CO $J=30-29$ line, however, is formed in the inner wind in SRb sources and one should be cautious when comparing predicted and observed line strengths.} 
%The observed time variability also prohibits the use of a CO/6.3-\mic line-to-continuum flux ratio to cancel out the distance dependence. It is, at this point, extremely difficult to correct for the phase dependence between the CO line strengths and the 6.3-\mic flux for the entire sample, which would introduce an uncertainty of the same order as caused by the poorly known distances.
%Because CO excitation is dominated by collisions, the CO line strength depends on the total CO density and the temperature structure throughout the wind. However, several quantities affect both the density and temperature structure. Whereas the mass-loss rate is useful in terms of predictive power and in characterizing the wind, too many other factors play a role in determining the physical structure of the wind. Hence, owing to the spread in parameter space inherent to the sample, determining general trends in the line strengths is difficult solely with respect to the mass-loss rate. To make meaningful conclusions, we consider the line-strength trends in terms of the density-proxy $\bar{m}$, which according to Eq.~\ref{eq:densproxy} also takes into account the gas terminal velocity, the stellar luminosity and the stellar effective temperature, all of which are given in Table~\ref{table:sample}. 

Other stellar or circumstellar properties are less important for CO emission. Fig.~\ref{fig:covshellmass_pars} presents an overview of {standard} theoretical models for the CO $J=15-14$ transition with $\epsilon = 0.4$, in which only one additional parameter is allowed to vary. The top left panel shows that \coabun does not have a significant effect on the CO line strengths relative to the effect of the explored range of mass-loss rates. The CO abundance is notoriously difficult to constrain from CO observations alone because it is completely degenerate with respect to the gas mass-loss rate. We therefore keep it fixed at $0.8 \times 10^{-3}$ in our {standard} model grid. From chemical network calculations, \citet{che2012} found \coabun $= 0.9 \times 10^{-3}$ for CW~Leo.

{The top right, bottom left, and bottom right panels in Fig.~\ref{fig:covshellmass_pars} show predictions for several values of $\vg$, $T_\star$ and $L_\star$, respectively. The gas terminal velocity has only a minor effect on the CO line-strength predictions. Variations in terminal velocity are equivalent to variations in mass-loss rate when comparing line strengths to the density tracer $\shellmass$, so this behavior is expected. The stellar temperature and luminosity both have no significant effect on the CO line strengths, given the uncertainties on the measured values. CO excitation primarily happens through collisions with H$_2$, so the gas temperature distribution is the most important factor. The stellar temperature in our models essentially shifts the temperature profile up or down by an absolute amount but does not change the gradient throughout the wind. If the stellar temperature increases, it implies that the CO $J=15-14$ line is formed in a region slightly further out. In a first approximation, the width of the line formation region increases with the square of the distance from the stellar surface, while the circumstellar density decreases with the square of the distance and the CO abundance remains constant. {As a result, for a given density profile, the CO line strengths do not change significantly depending on the radial distance at which the lines are formed.} The stellar luminosity also does not contribute directly to CO excitation unless the circumstellar density reaches very low values. This explains the low sensitivity of the CO line strengths. \citet{deb2010} show similar low sensitivities to stellar properties for lower-$J$ CO lines from large model grid calculations.}

%The observed spread in CO line strengths versus $\bar{m}$ with respect to the theoretical predictions can be attributed mainly to the difference in the stellar luminosity and effective temperature. {As an example, the SRb objects in particular deviate somewhat from the {standard} theoretical predictions given by the black curve in the bottom two figures. Two of these objects have large distance uncertainties, so not much can be done to further constrain stellar properties. The other two SRb objects have more reliable distance measurements. A combination of a decrease in $L_\star$ and an increase in $T_\star$ in the model predictions fixes the issue for U Hya (the lowermost black data point), for which $L_\star$ was estimated to be $4.2 \times 10^3$ L$_\odot$ and $T_\star$ was estimated to be $3000$ K, see Table~\ref{table:sample}. To best represent the bulk of the sample, we adopt $\vg = 10$ \kms, $T_\star = 2400$ K, and $L_\star = 8000$ L$_\odot$ for the {standard} model grid. The temperature profile is the main determinant of the CO line strength, as CO molecules are primarily excited through collisions with H$_2$.}

\begin{figure*}[!t]
$\begin{array}{cc}
\resizebox{9.0cm}{!}{\includegraphics{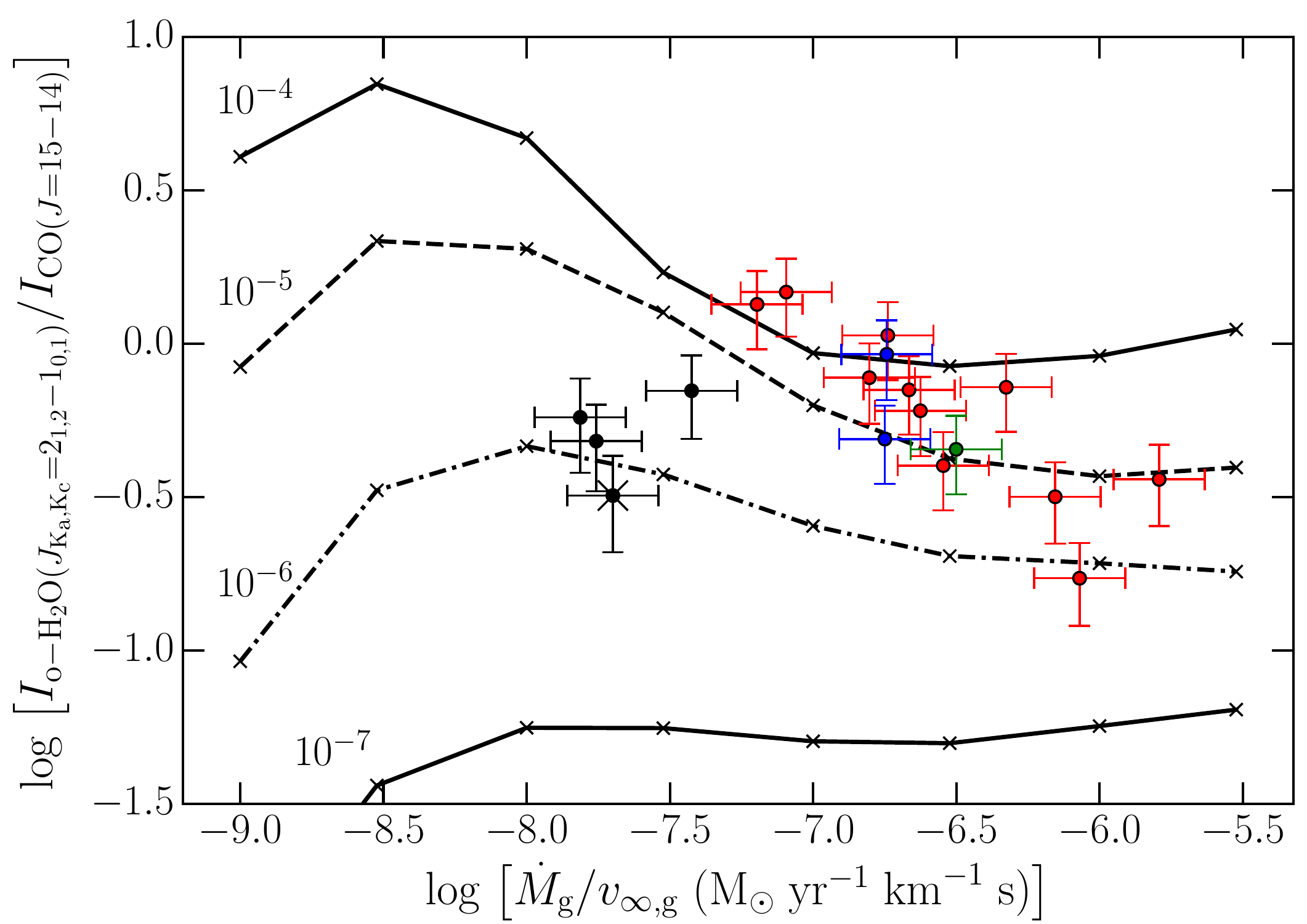}} & \resizebox{9.0cm}{!}{\includegraphics{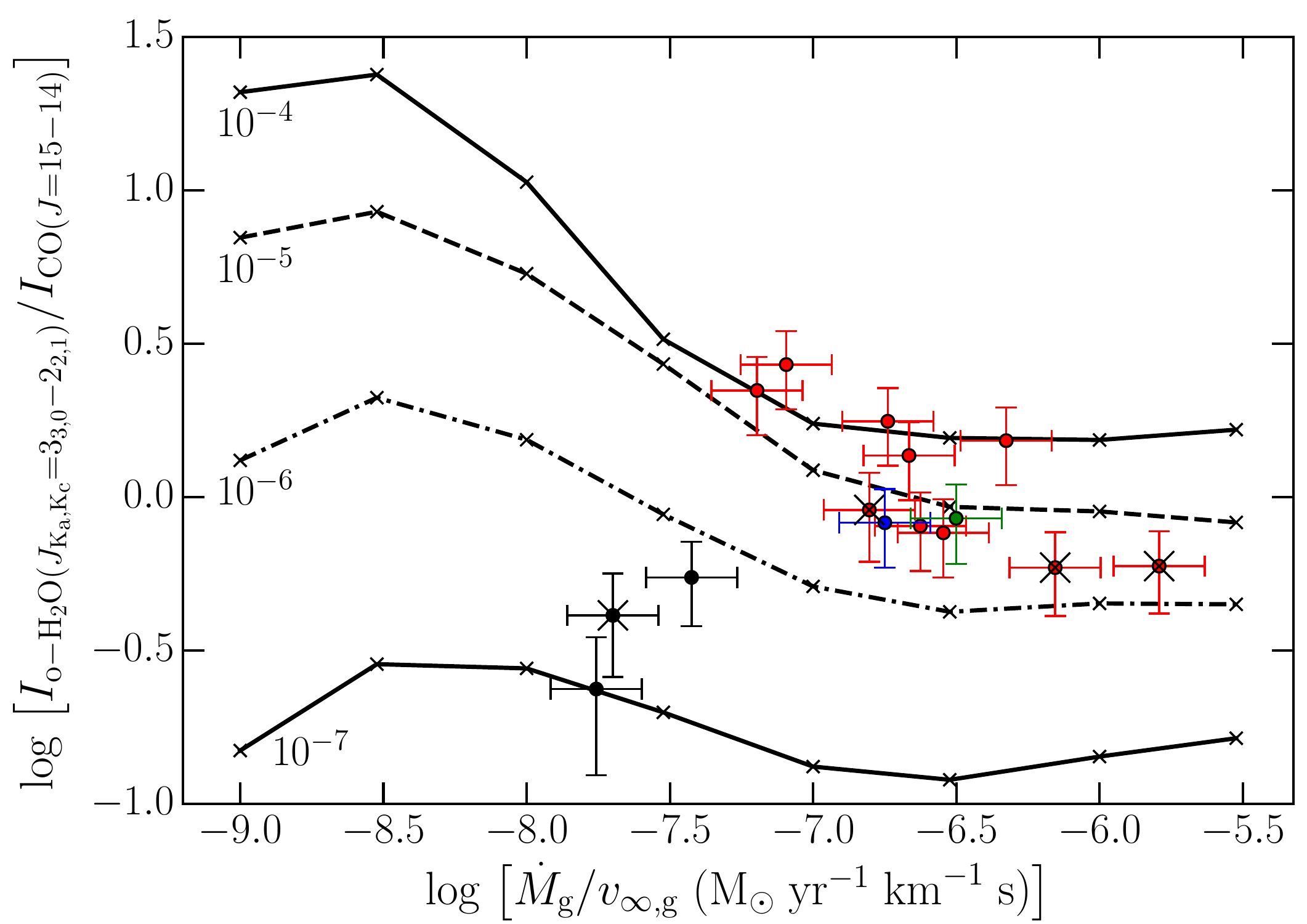}} 
\end{array}$
\caption{Line-strength ratio of two \water transitions and the CO $J=15-14$ transition as a function of $\shellmass$: {the cold $J_{\mathrm{K}_\mathrm{a}, \mathrm{K}_\mathrm{c}}=2_{1,2} - 1_{0,1}$ line on the left, and the warm $J_{\mathrm{K}_\mathrm{a}, \mathrm{K}_\mathrm{c}}=3_{3,0} - 2_{2,1}$ line on the right.} The points with error bars give the measured \water/CO line-strength ratios, color coded according to the variability type: Miras in red, SR/Mira sources in blue, the SRa source in green, and SRb sources in black. A black cross superimposed on the data point indicates that the \water line is flagged as a blend. The curves show the predicted \water/CO line-strength ratios for various values of \waterabun as indicated in the figures. Adopted values for other parameters are listed in Table~\ref{table:modelpars}.}
\label{fig:h2ovshellmass}
\end{figure*}
\begin{figure*}[!t]
$\begin{array}{cc}
\resizebox{9.0cm}{!}{\includegraphics{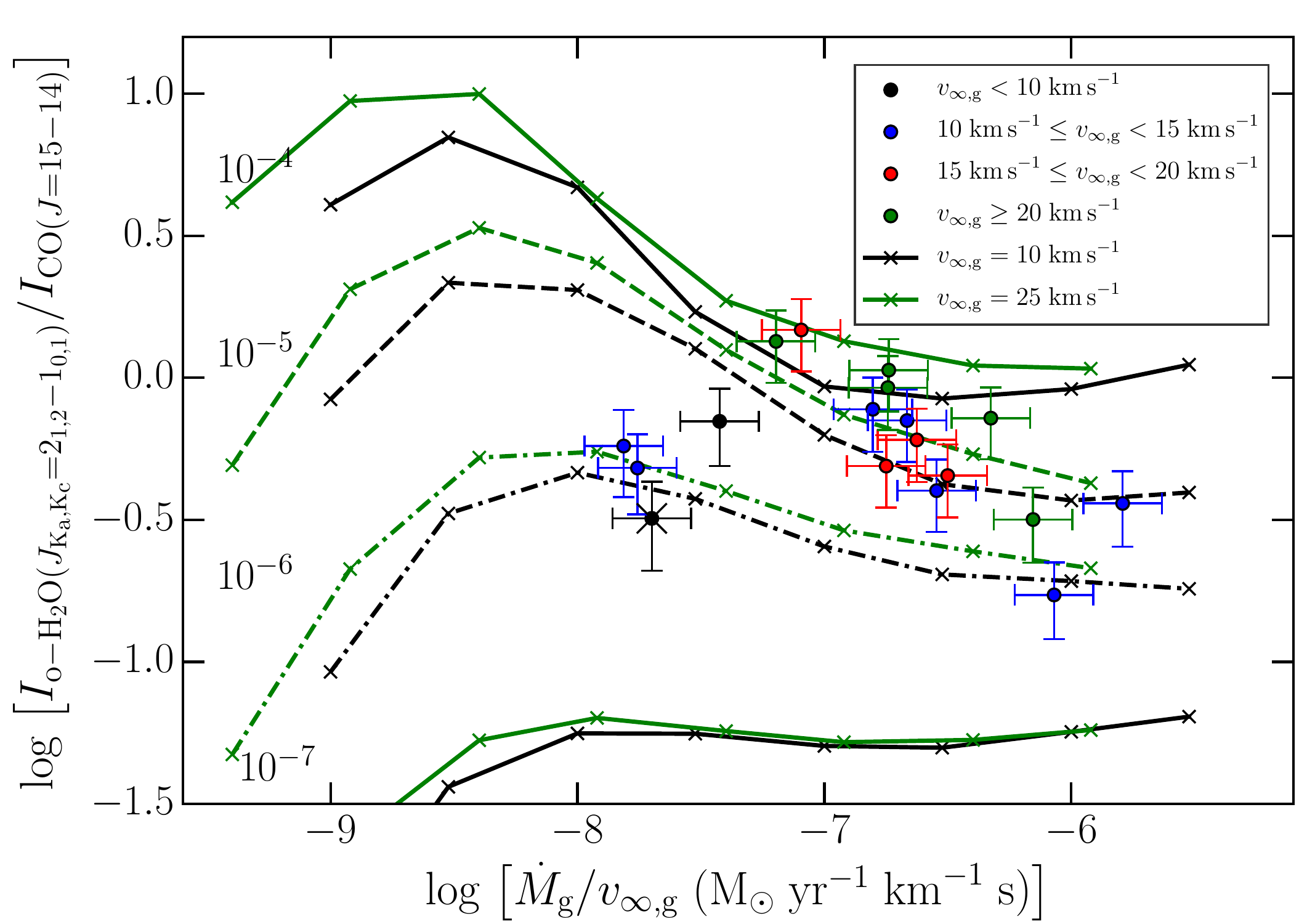}} & \resizebox{9.0cm}{!}{\includegraphics{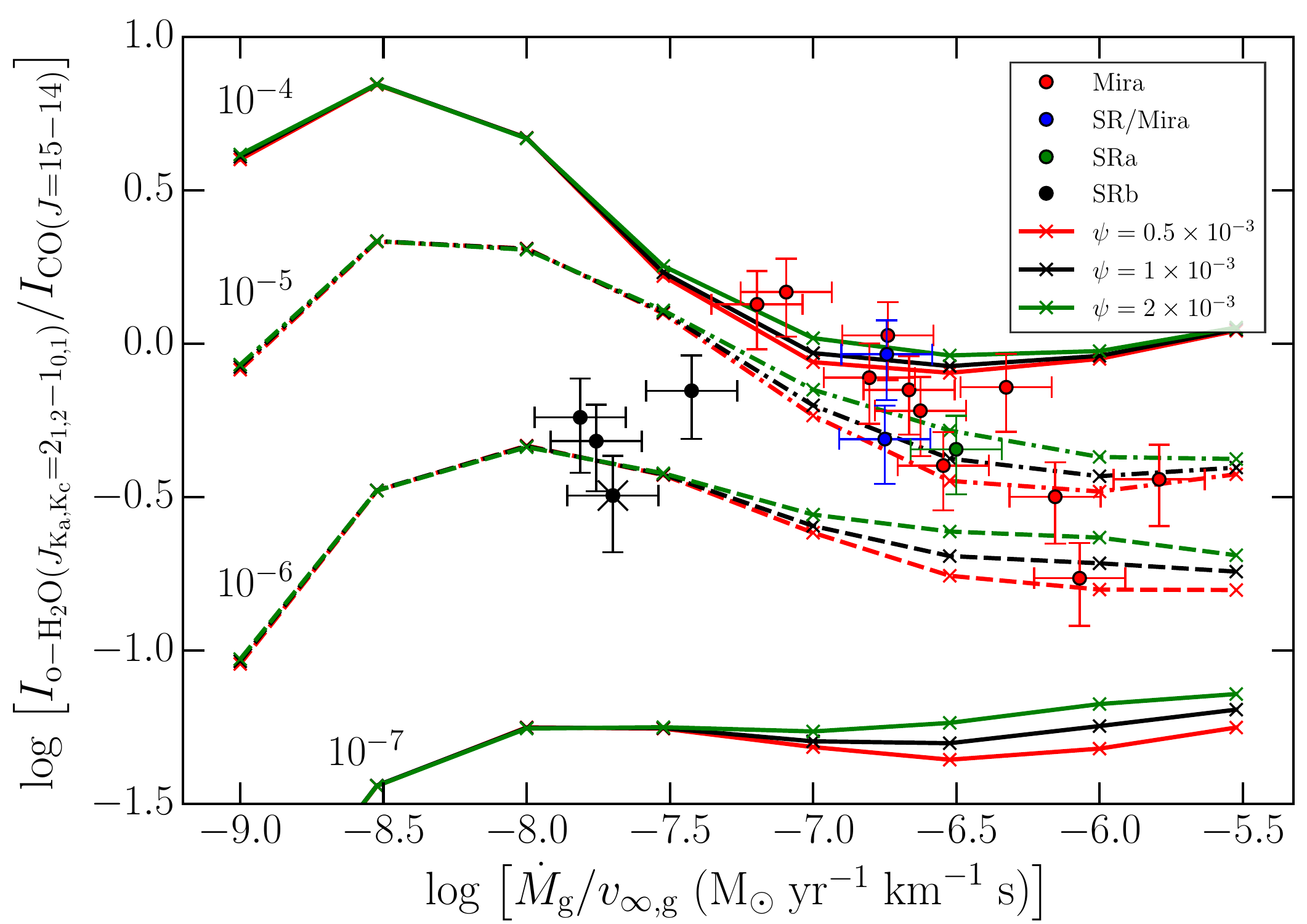}} 
\end{array}$
\caption{Line-strength ratio of the cold \water $J_{\mathrm{K}_\mathrm{a}, \mathrm{K}_\mathrm{c}}=2_{1,2} - 1_{0,1}$ transition and the CO $J=15-14$ transition as a function of $\shellmass$. The points with error bars give the measured \water/CO line-strength ratios, color coded according to the values of $\vg$ (left panel) or according to the variability type (right panel). A black cross superimposed on the data point indicates that the \water line is flagged as a blend. The black curves show the same predicted \water/CO line-strength ratios as Fig.~\ref{fig:h2ovshellmass} with \waterabun as indicated in the figures, $\vg = 10$ \kms and $\psi = 1\times10^{-3}$. In the left panel, the green curves show the predicted \water/CO line-strength ratios for $\vg = 25$ \kms. In the right panel, predictions for $\psi = 0.5\times10^{-3}$ in red and $\psi = 2\times10^{-3}$ in green are shown. Adopted values for other parameters are listed in Table~\ref{table:modelpars}.}
\label{fig:h2ovshellmass_pars}
\end{figure*}

\subsection{\water/CO line-strength ratios}\label{sect:h2o}
Following the approach for CO lines from the previous section, we now use the {standard} model grid in Fig.~\ref{fig:h2ovshellmass} to probe the influence of $\mg$ and \waterabun on the \water/CO line-strength ratio. Fig.~\ref{fig:h2ovshellmass_pars} shows the model grids in which the gas expansion velocity and the dust-to-gas ratio are allowed to vary in addition to the mass-loss rate. Varying the gas expansion velocity implies that changes in $\shellmass$ are not exclusively due to the mass-loss rate. Fig.~\ref{fig:h2ovshellmass} shows the measured \water/CO line-strength ratios for a cold ortho-\water transition ($J_{\mathrm{K}_\mathrm{a}, \mathrm{K}_\mathrm{c}}=2_{1,2} - 1_{0,1}$ with $E_\mathrm{u} = 114.4$ K) on the left and a warm ortho-\water transition ($J_{\mathrm{K}_\mathrm{a}, \mathrm{K}_\mathrm{c}}=3_{3,0} - 2_{2,1}$ with $E_\mathrm{u} = 410.6$ K) on the right. Additionally, predicted line-strength ratios from the {standard} model grid with adopted parameters given in Table~\ref{table:modelpars} are superimposed on the data points. The observed line-strength ratios span more than two orders of magnitude in \water vapor abundance. This is the case for all \water lines in the sample, i.e.~for both cold and warm \water emission. {For the cold emission line, the \water abundances range from $10^{-6}$ up to $10^{-4}$ for the Mira and SRa sources, and cluster around $10^{-6}$ for the SRb sources with the exception of Y~CVn, which requires an abundance of $\sim 5\times 10^{-6}$. For the warm emission line, the same range of \water abundances is found for the Mira and SRa sources, while the abundance is an order of magnitude lower for the SRb sources.}

{As discussed in Sect.~\ref{sect:obsh2o}, the model predictions confirm that SRb sources show lower \water abundances overall. The absolute values should be considered tentatively because the CO line strengths of the SRb sample are not very well reproduced by our chosen model (see Sect.~\ref{sect:cols}), but the difference between the SRb sample and the Mira/SRa sample is large enough to be significant. The tentative upward trend with respect to $\mg$ revealed in Sect.~\ref{sect:obsh2o} is less convincing with respect to $\shellmass$, which is really a testament to the small sample size and the uncertainties. Hence, we cannot be conclusive about the trend. In addition, a significantly different abundance for cold and warm \water is derived for the SRb sample. Such differences are not prominent for the Miras and SRa sources. However, the $J_{\mathrm{K}_\mathrm{a}, \mathrm{K}_\mathrm{c}}=3_{3,0} - 2_{2,1}$ \water transition is formed in the inner wind in SRb stars. As discussed before for the CO lines, our model predictions do not represent the inner wind well, so they are not reliable. The $J_{\mathrm{K}_\mathrm{a}, \mathrm{K}_\mathrm{c}}=2_{1,2} - 1_{0,1}$ \water transition is primarily formed in the intermediate wind, even in SRb stars, so predictions for that line are robust. }

In terms of sensitivity to the assumptions of the {standard} model grid, {only the gas terminal velocity and the dust-to-gas ratio have a noticeable impact on the calculated \water/CO line-strength ratios under the important assumption that our CO line-strength predictions are accurate.} Fig.~\ref{fig:h2ovshellmass_pars} shows \water/CO line-strength-ratio predictions for the {standard} model grid, in which either $\vg$ or $\psi$ is allowed to vary. The left panel gives the results for $\vg = 10$ \kms in black ({standard} model-grid value) and $\vg = 25$ \kms in green. In the optically thin regime, a change in $\vg$, and therefore in the density tracer $\shellmass$, does not substantially affect the \water emission, as shown by the models for \waterabun $ = 10^{-7}$. For higher \water abundances, the lines become optically thick, so that a change in $\vg$ affects \water line strengths significantly. The differences are however well within the uncertainty in the observed line-strength ratios. 

{The right panel in Fig.~\ref{fig:h2ovshellmass_pars} gives the lowest (in red) and highest $\psi$ value (in green) in the grid compared to $\psi = 0.001$ ({standard} model-grid value) in black. The \water/CO line-strength-ratio sensitivity to the dust-to-gas ratio arises because \water is primarily excited radiatively by IR photons emitted by dust in high-density environments (e.g.,~the right panel of Fig.~\ref{fig:h2ovshellmass_pars} for $\log(\shellmass) > -7.0$). Here, the higher $\psi$ results in stronger \water emission, while CO line strengths remain mostly unaffected \citep{lom2013}. In low-density environments, direct stellar light dominates \water excitation and the sensitivity of \water line strengths to the dust-to-gas ratio is lost. Again, the differences are well within the uncertainty on the observed line-strength ratios. We conclude that effects of both $\vg$ and $\psi$ cannot explain the observed trend in the Miras and SRa sources.}

The comparison between the observed \water/CO line-strength ratios and the theoretical predictions excludes radiative-transfer effects as the sole cause of the downward trend between the \water/CO line-strength ratio and $\shellmass$. This confirms that the \water/CO line-strength ratio can be treated as an \water abundance proxy and that {the \water abundance correlates negatively with the circumstellar density} {in the Miras and SRa sources}. Because the downward trend exists for all \water transitions regardless of the energy levels involved, {it is the \water formation mechanism itself that becomes less efficient with increasing circumstellar density.}

\subsection{{Model reliability}}
{We mention a few caveats regarding the conclusion concerning the \water/CO line-strength ratios.} The assumed exponent of the temperature law $\epsilon = 0.4$ has a significant impact on the \water/CO line-strength ratios because of its importance for the CO line strength, emphasizing the need to predict the observed CO line strengths accurately. Collisions play a minor role in \water excitation, so the temperature law does not directly influence the \water line strengths (e.g.,~\citeauthor{lom2013}~\citeyear{lom2013} for the high-$\mg$ case). {This also corroborates the use of older \water-H$_2$ collision rates, as discussed in Sect.~\ref{sect:theomodel}.}

Time variability can be an issue in the \water lines. Recent CW~Leo results derived from \emph{Herschel}-PACS data show variability in \water line strengths up to 50\% {\citep{cer2014}}. {While CW~Leo shows this for the high-$\mg$ case, a similar behavior may occur at low $\mg$. Assuming 50\% to be the norm, this variability is within the uncertainty on our line-strength ratios. \water line variability primarily arises from changes in the radiation field, i.e.~in the efficiency of radiative pumping.}

Finally, the predicted \water abundances are noticeably higher than reported in previous studies for carbon-rich AGB winds, e.g.,~$0.2-0.5\times 10^{-5}$ for V~Cyg \citep{neu2010}, while we predict $1-12\times 10^{-5}$ for the $J_{\mathrm{K}_\mathrm{a}, \mathrm{K}_\mathrm{c}}=2_{1,2} - 1_{0,1}$ line and $0.1-1\times 10^{-5}$ for the $J_{\mathrm{K}_\mathrm{a}, \mathrm{K}_\mathrm{c}}=3_{3,0} - 2_{2,1}$ line. We must proceed with caution in comparing \water abundances found here with \water abundances derived from an in-depth modeling for individual sources. We do not take into account source-specific deviations from the model grid (e.g.,we underestimate the 6.3 \mic flux for V~Cyg specifically; see Fig.~\ref{fig:f63vshellmass}), nor do we consider in-depth all of the available \water lines for each source. We therefore do not list estimates of \water abundances for individual sources in our sample. The results presented here serve a different goal: constraining the dependence of the \water abundance on the circumstellar density and, thus, the mass-loss rate. The absolute values may shift up or down somewhat depending on the model assumptions, but the relative difference between sources with different mass-loss rate is robust. In the case of V~Cyg, a higher model prediction for the 6.3 \mic flux would decrease the \water abundance and bring the result more in line with that of \citet{neu2010}. Moreover, \citeauthor{neu2010}~use a significantly higher mass-loss rate. Overall, we arrive at a similar \water outflow rate as they do.
%However, as pointed out in Sect.~\ref{sect:63flux}, we require a low dust-to-gas ratio and make use of a COMARCS synthetic atmosphere spectrum to reproduce the 6.3-\mic fluxes for the whole sample. Moreover, we take into account a nonzero drift velocity by calculating the momentum transfer from dust to gas. All these differences add up and make it difficult to compare abundances derived in previous studies. As long as the 6.3-\mic flux is reproduced on average for the whole sample, the differences between this and previous studies in derived abundances boil down to , which directly results in higher prediction of the \water abundances for the whole sample.

\section{\water abundance gradients within single sources}\label{sect:grad}
{In this section, we look for trends in the radial dependence of the \water abundance within individual sources to help constrain the \water formation mechanism in carbon-rich winds. To this end, \water transitions formed in different regions in the wind are compared to trace the radial profile of the \water abundance. A similar strategy was followed by \citet{kho2014}.}
\begin{figure}[!t]
$\begin{array}{l}
\resizebox{8.8cm}{!}{\includegraphics{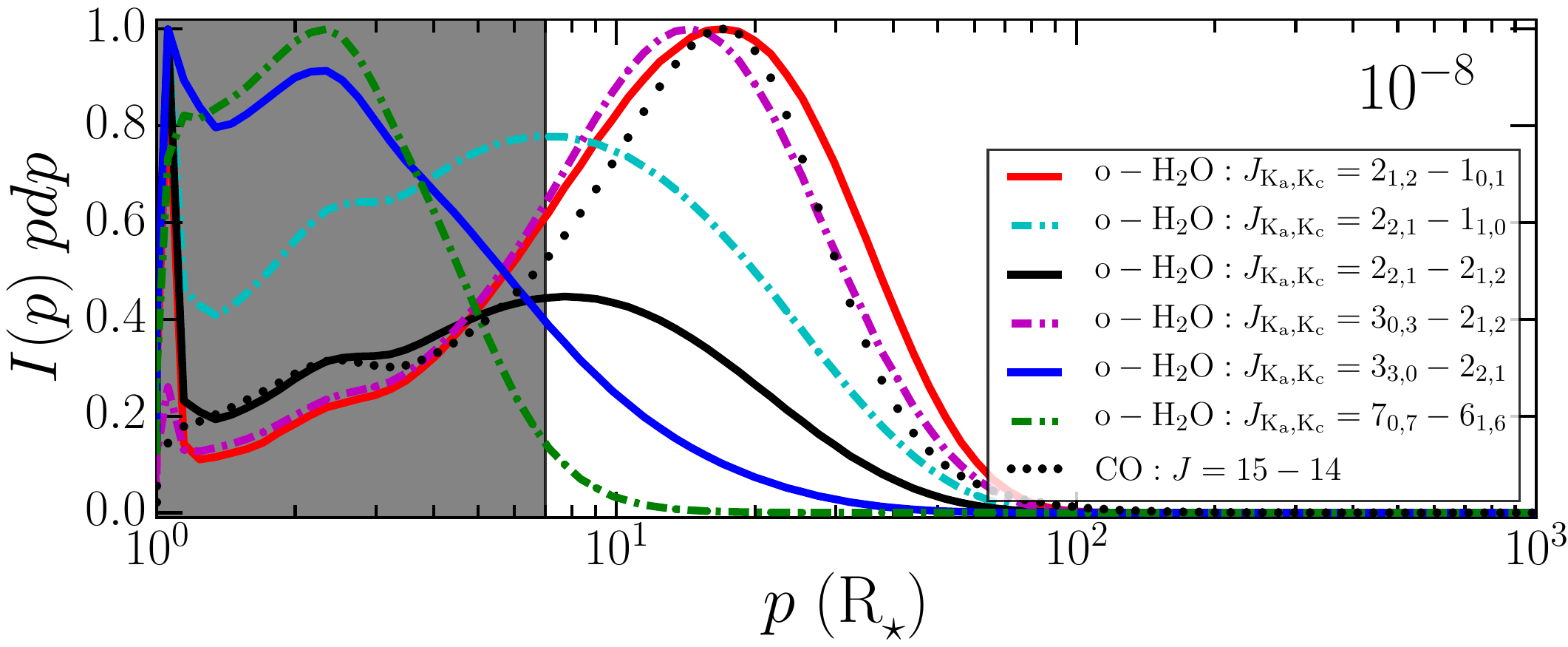}} \\
\resizebox{8.8cm}{!}{\includegraphics{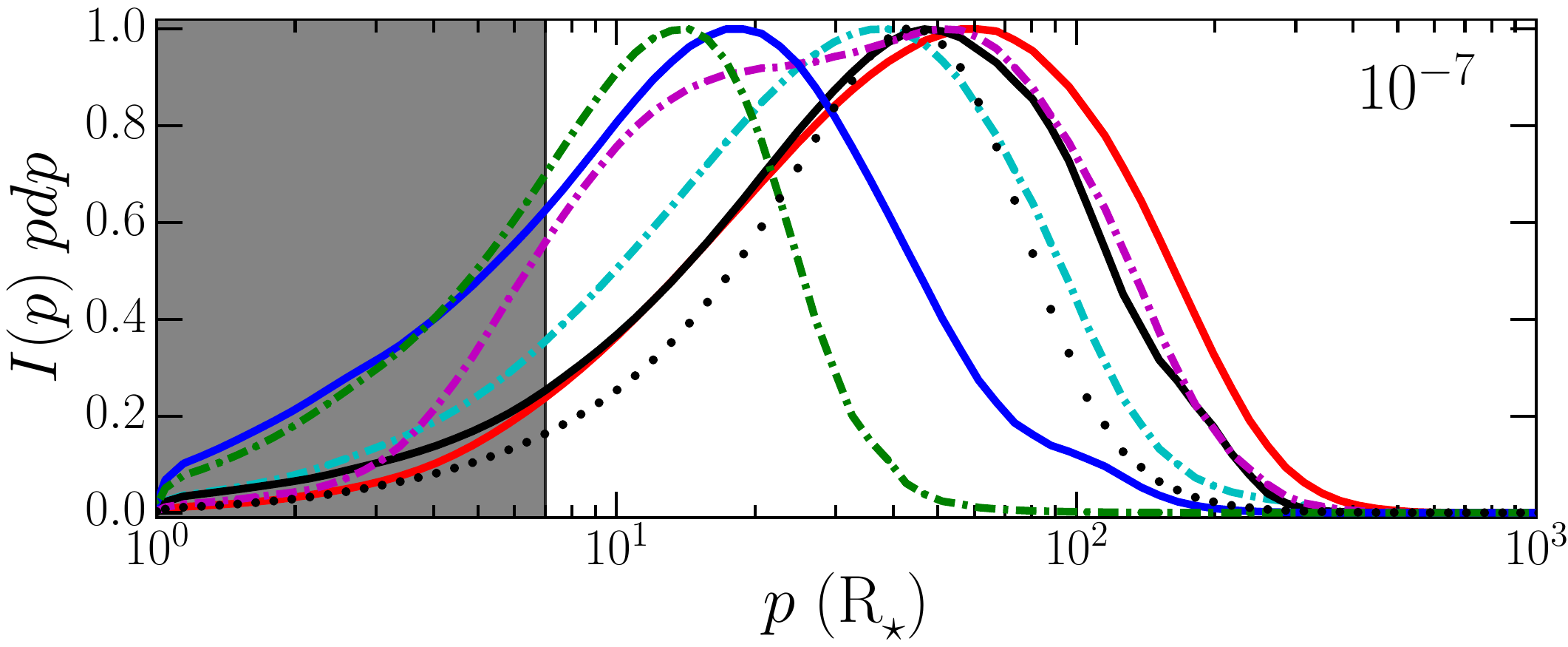}} \\
\resizebox{8.8cm}{!}{\includegraphics{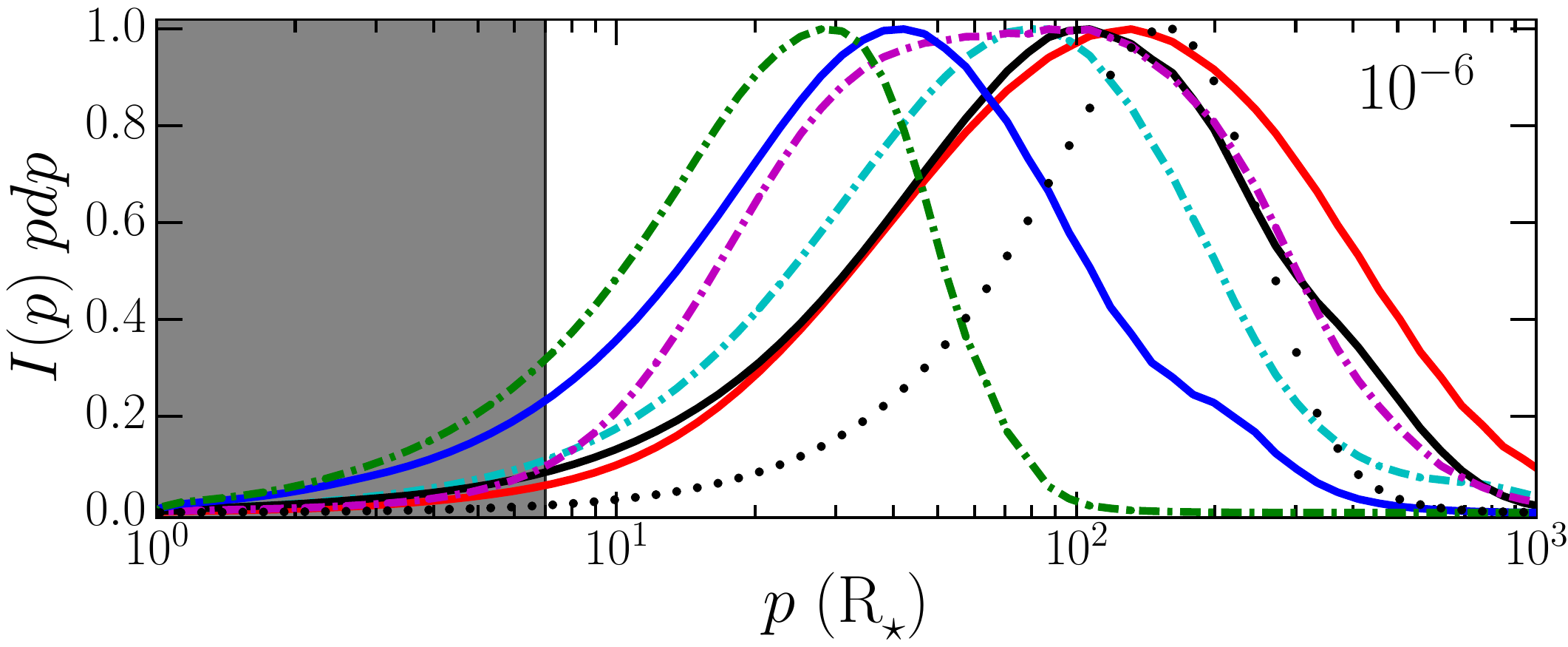}} 
\end{array}$
\caption{Line contribution regions for ortho-\water. The normalized quantity $I(p) \ pdp$ is shown as a function of the impact parameter $p$ for six transitions identified in the legend in the top panel. The different panels show the line contributions for models with $\shellmass = 10^{-8}$ \myrkms and \waterabun$ = 10^{-6}$ in the top panel, $\shellmass = 10^{-7}$ \myrkms and \waterabun$ = 10^{-4}$ in the middle panel, and $\shellmass = 10^{-6}$ \myrkms and \waterabun$ = 10^{-5}$ in the bottom panel. The gray area indicates the wind acceleration zone.}
\label{fig:lcp}
\end{figure}
\subsection{Molecular line contribution regions}
Radial abundance gradients of a molecular species are probed by emission lines formed in different regions of the outflow. For CO, the excitation occurs primarily through collisions with H$_{2}$ and is thus coupled to the gas kinetic temperature. %CO excitation is quantified in terms of the rotational quantum number $J$, which follows a simple ladder structure. 
A high-$J$ CO transition forms closer to the stellar surface than does a low-$J$ transition because the former is populated in a zone where the temperature is higher. {Hence, assuming the gas temperature profile is known, it is possible to identify a radial gradient simply by studying the CO abundance as a function of $J$.} For \water, the situation is different as the levels are mainly radiatively excited and \water excitation does not follow a simple $J$-ladder, like CO. As a result, the line contribution region of a given \water transition cannot be located through a straightforward scheme such as for CO, and requires models to establish which transitions trace which part of the wind.

{Fig.~\ref{fig:lcp} shows the normalized quantity $I_\mathrm{p}\ pdp$ as a function of the impact parameter $p$, where $I_p$ is the predicted intensity at line center.} This quantity indicates from where emission in a given line originates in the wind. From the top panel to the bottom panel in Fig.~\ref{fig:lcp}, $\shellmass$ increases. 
%For each case, a typical \water abundance has been chosen based on the comparison of the measured \water/CO line-strength ratios to the theoretical results in Sect.~\ref{sect:h2o}. 
The top panel assumes $\shellmass = 10^{-8}$ \myrkms, a typical value for the SRb sources, which cluster around the theoretical model with \waterabun $ = 10^{-6}$ in Fig.~\ref{fig:h2ovshellmass}. The middle panel and bottom panel represent the low and high end $\shellmass$ values of the Miras and SRa sources: $\shellmass = 10^{-7}$ \myrkms and $\shellmass = 10^{-6}$ \myrkms, values for which data points cluster around \waterabun $= 10^{-4}$ and $10^{-5}$, respectively, in Fig.~\ref{fig:h2ovshellmass}. In each panel, the gray area indicates the wind acceleration zone. In the $\shellmass= 10^{-8}$ \myrkms model, higher energy emission lines form close to the stellar surface and may be affected by stellar pulsations, ongoing dust formation, or wind acceleration. The CO $J=15-14$ line is shown for comparison. Typically, a CO line forms in a narrower region because of its sensitivity to the temperature profile only, while an \water line forms in a wider region owing to the nonlocal nature of radiative excitation.

{We assume a constant mass-loss rate. A time-variable mass loss can cause changes in the density profile throughout the wind. This would have a similar effect on the line strengths as a nonconstant molecular abundance profile. For instance, a recent decrease in mass loss results in less emission from the region close to the stellar surface. Nevertheless, even though variable mass loss may explain discrepancies between observed and predicted line-strength ratios for specific sources, it is highly unlikely that all sources in our sample suffer from a variable mass loss on a short timescale of a few hundred years.}
 
{As noted previously, our predictions for lines formed at the base of the wind (at $r <7$ \rstar, indicated by the gray area in Fig.~\ref{fig:lcp}) are less reliable. We do not take into account the effects of the periodic shocks moving through the medium, and make assumptions regarding the dust formation, initial acceleration, and temperature profile in the first few stellar radii.}

\subsection{\water/\water line-strength ratios}
{By comparing the strengths of two \water lines formed in different regions of the wind, information on the radial dependence of the \water abundance can be inferred. For this, it is important that the lines included in the comparison are formed outside the acceleration zone. Hence, from here onward, we discuss the SRb sources separately from the SRa sources and the Miras.}
\subsubsection{The SRb sources}
A significant portion of the observed lines in SRb stars form at $r < 7$ \rstar (see the top panel of Fig.~\ref{fig:lcp}). {The right-hand panel in Fig.~\ref{fig:h2ovshellmass_pars} shows that the sensitivity of \water emission to the dust-to-gas ratio in the intermediate wind becomes negligible for $\log(\shellmass) < -7.0$, which includes all the SRb sources. This behavior is also expected to hold for \water lines formed in the inner wind. The mass-loss rate is so low that the contribution of dust emission to the overall radiation field is minor. Hence, \water excitation by dust is irrelevant for determining the \water line strengths at such mass-loss rates.} That said, it remains difficult to gauge the effect of wind acceleration on the strengths of lines formed in the inner wind. Individual differences between the observed sources and the {standard} model grid may have a significant impact on the comparison of the \water lines. Our modeling approach also does not take pulsational shocks or the phase dependence of the pulsation pattern into account. Hence, using our approach and realizing that our sample is small, we cannot derive meaningful constraints for the radial dependence of the \water abundance for SRb sources.

{However, it is clear that the \water line strengths measured in the acceleration zone compared to those measured in the intermediate wind imply vastly different \water abundances for individual sources as predicted by our simplified model of the inner wind. This is evident from the comparison of the two \water lines shown in Fig.~\ref{fig:h2ovshellmass} as black data points for the SRb sample. The line shown in the left panel is formed in the intermediate wind, while the line in the right panel is formed in the inner wind. Two scenarios are possible:}
\begin{enumerate}
 \item {Shocks are important in determining the density and/or abundance profile of \water in the inner wind and directly affect \water excitation. It is likely that shocks are actively contributing to the formation of \water. The pulsation periodicity of SRb stars compared to Mira and SRa sources may affect the efficiency with which \water forms, relating back to the different trends depending on pulsation type reported in Sect.~\ref{sect:obsh2o}.}
 \item {Shocks are not important for these lines. An alternative cause for the different \water abundances between inner and intermediate wind of SRb stars is needed. This implies that the \water formation mechanism is not related to shocks. }
\end{enumerate}
{We cannot distinguish between these two scenarios within the current setup of our modeling strategy.}
\subsubsection{The Miras and SRa sources}\label{sect:gradmiras}
{For the Miras and SRa sources, we aim to distinguish between different abundance profiles based on the \water/\water line-strength ratios. For this purpose, we have calculated additional models to compare with the {standard} model grid, assuming different \water abundance profiles. These predictions are compared to measurements in Fig.~\ref{fig:h2ograd} for the abundance profiles shown in Fig.~\ref{fig:h2oabun}. The profiles are:}
\begin{enumerate}
 \item {Constant}:{ The \water abundance is assumed to be constant throughout the wind, up to the photodissociation region.}
 \item {Decin/Ag\'undez}: {This model is based on inner-wind penetration of interstellar UV. \citet{dec2010c} and \citet{agu2010} show that the \water abundance profile follows roughly the same shape for different mass-loss rates. The profiles show a positive \water abundance gradient in the inner and intermediate wind, increasing quickly to a maximum value between 5 and 20 \rstar. The model with a low mass-loss rate of 10$^{-7}\ \msunyr$ reaches a maximum \water abundance of 10$^{-6}$, while the models with higher mass-loss rates reach $\sim 2\times10^{-7}$. We used the $\mg = 10^{-5}\ \msunyr$ and $10^{-6}\ \msunyr$ cases to compare with the $\shellmass$ range of the Miras and SRa sources, up to the radius at which they find the highest \water abundance. From that radius onward, the abundance is assumed to be constant up to the photodissociation region.}
 \item {Cherchneff}: This model describes the effect of a shock-induced formation mechanism. The \water abundance profile is as predicted by \citet{che2011} for CW~Leo and is representative of the inner wind for about 80\% of the duration of a shock. The profile predicts a high \water abundance near the stellar surface, which then quickly decreases to a freeze-out value about three orders of magnitude lower depending on the phase. {We assume that this abundance then remains constant at the freeze-out value outside the shock zone up to the photodissociation radius.}
\end{enumerate}
\begin{figure}[!t]
\resizebox{9.0cm}{!}{\includegraphics{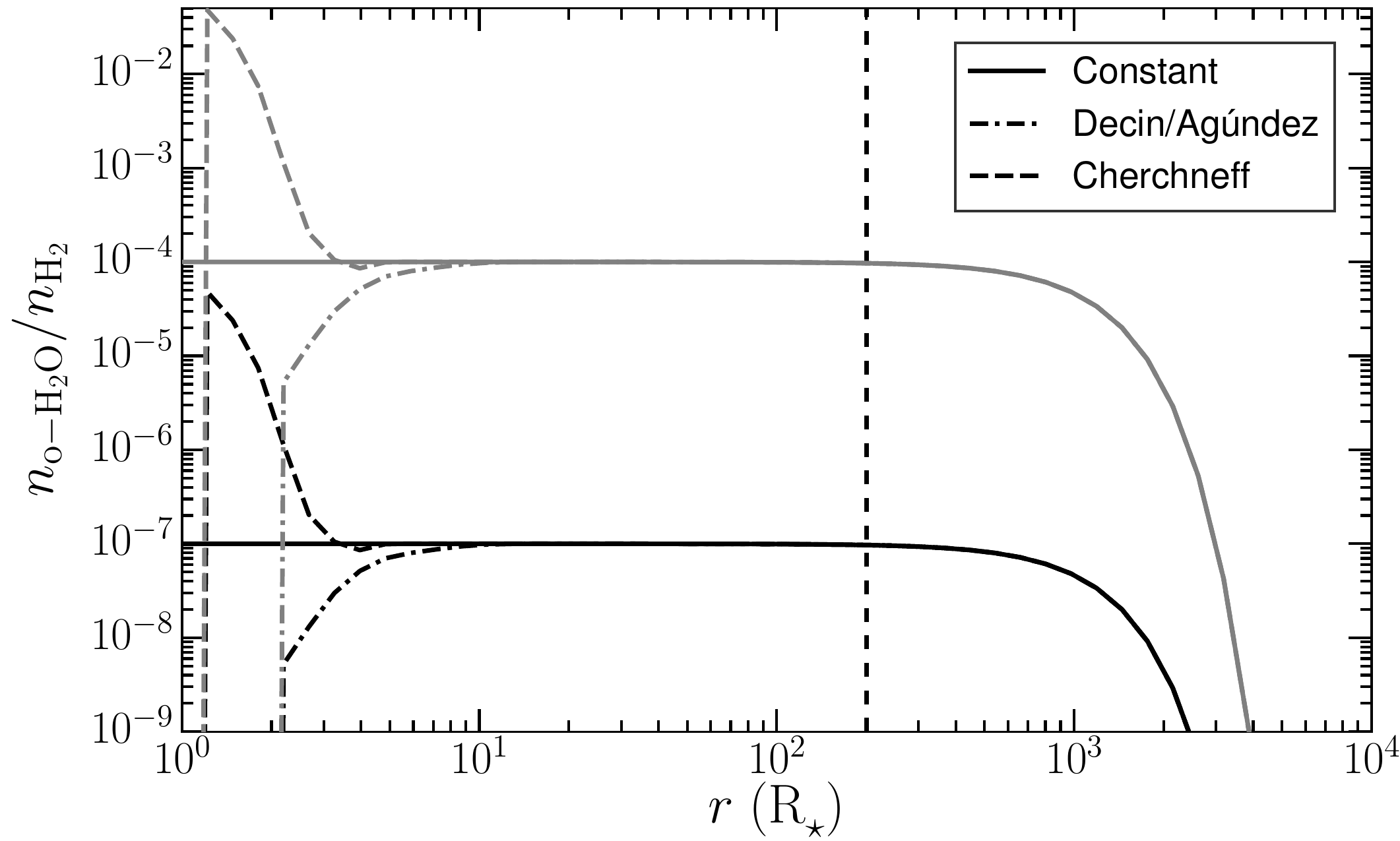}}
\caption{\water abundance profiles for three different chemical models at two different representative abundances at the radius indicated by the dashed vertical line: \waterabun $= 10^{-7}$ in black and \waterabun $= 10^{-4}$ in gray. The Constant abundance profile is a reference model. The Decin/Ag\'undez profile refers to the results of \citet{dec2010c} and \citet{agu2010}. Cherchneff's profile refers to her results of \citeyear{che2011}. See Sect.~\ref{sect:gradmiras} for further details.} 
\label{fig:h2oabun}
\end{figure}

{In all three cases, the photodissociation radius is taken from the analytic formula of \citet{gro1994}. Which photodissociation radius is used here is not important since we want to gauge the sensitivity of \water/\water line-strength ratios to differences in the \water abundance profile caused by different formation mechanisms. Whatever the real photodissociation radius is, it should affect the presented models for different \water abundance profiles in the same way, and hence has no impact on our conclusions. {As our interest lies in the inner and intermediate wind, we assume the same photodissociation law as from \citet{gro1994} in the \citet{dec2010c} and \citet{agu2010} abundance profiles at radii beyond their maximal abundance value. In this way, we can compare the effects of the different formation mechanisms.} To probe the effect of the absolute \water abundance, each of the profiles is scaled to a representative abundance at a radius in the outflow just before photodissociation sets in. In the model grid, this representative \water abundance scales from $10^{-7}$ up to $10^{-4}$ in factors of 10. The abundance profiles associated with the lowest and highest representative abundance are shown in Fig.~\ref{fig:h2oabun}. The higher representative abundances are not necessarily supported by the theories of \citet{agu2010} and \citet{che2011}. These abundance profiles are {not} tailored specifically according to the physical properties of the winds at different $\shellmass$, and only provide an indication of how an inner- and intermediate-wind abundance gradient would affect the \water/\water line-strength ratios.}
\begin{figure*}[!t]
$\begin{array}{cc}
\resizebox{9.0cm}{!}{\includegraphics{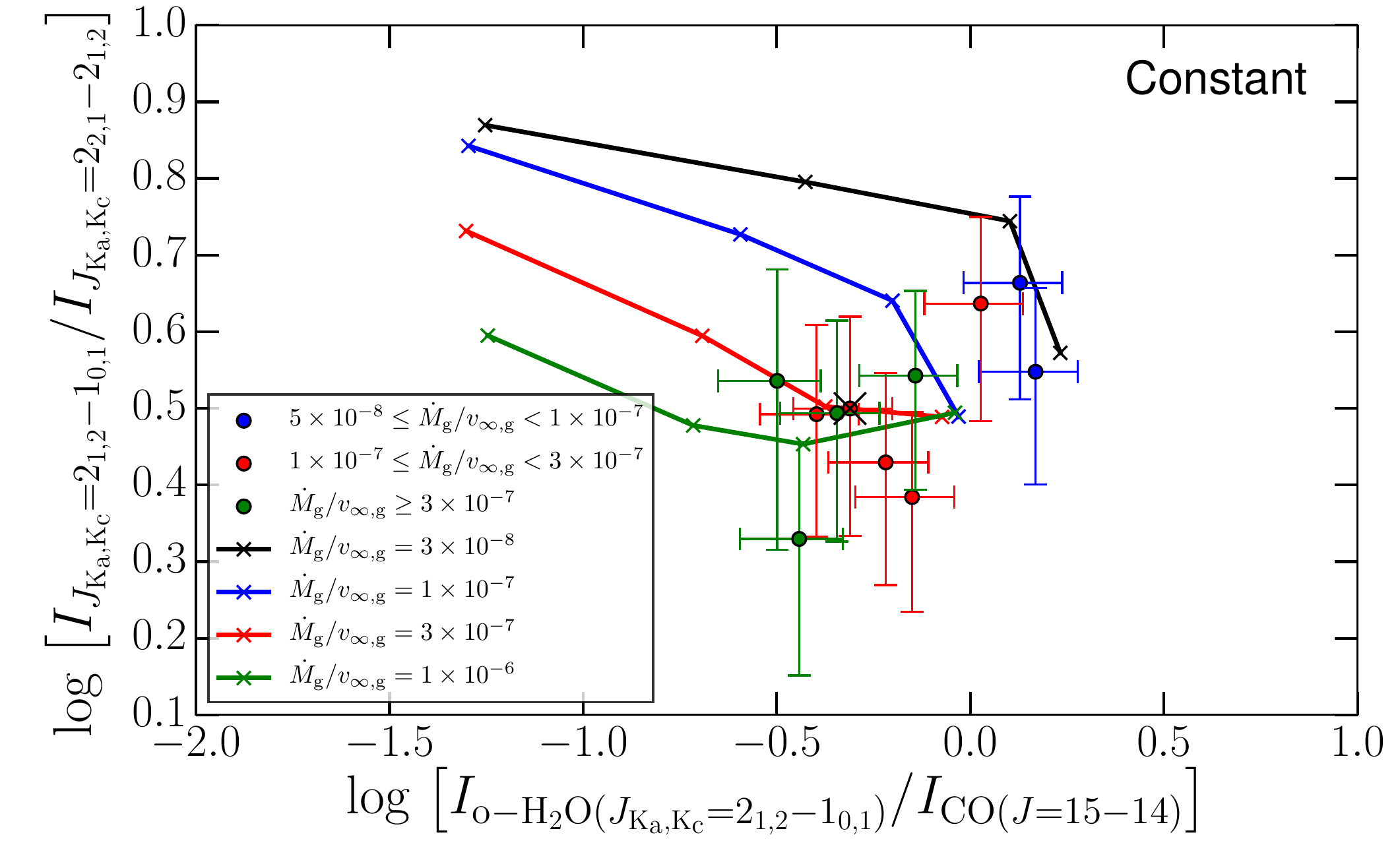}} &\resizebox{9.0cm}{!}{\includegraphics{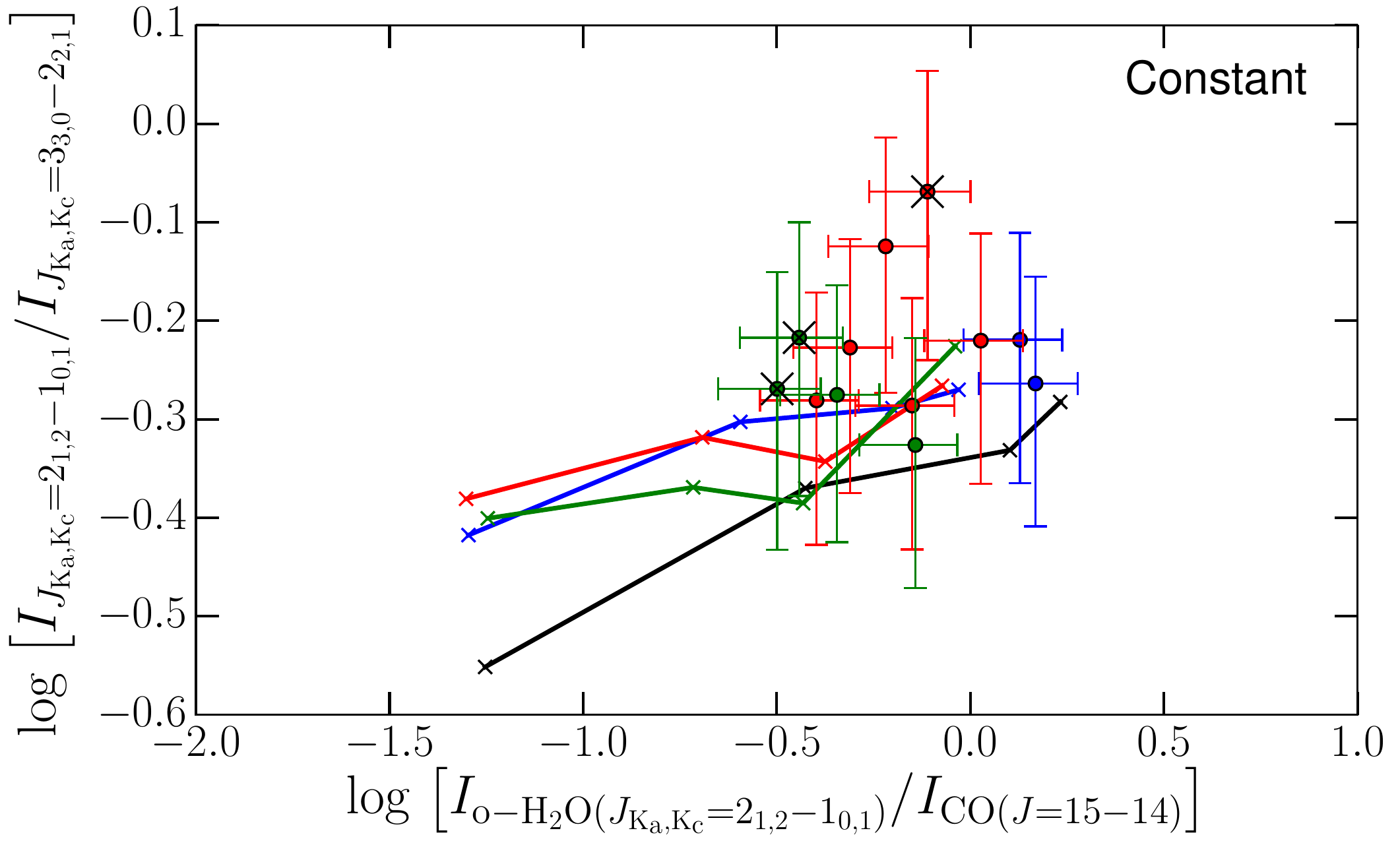}} \\
\resizebox{9.0cm}{!}{\includegraphics{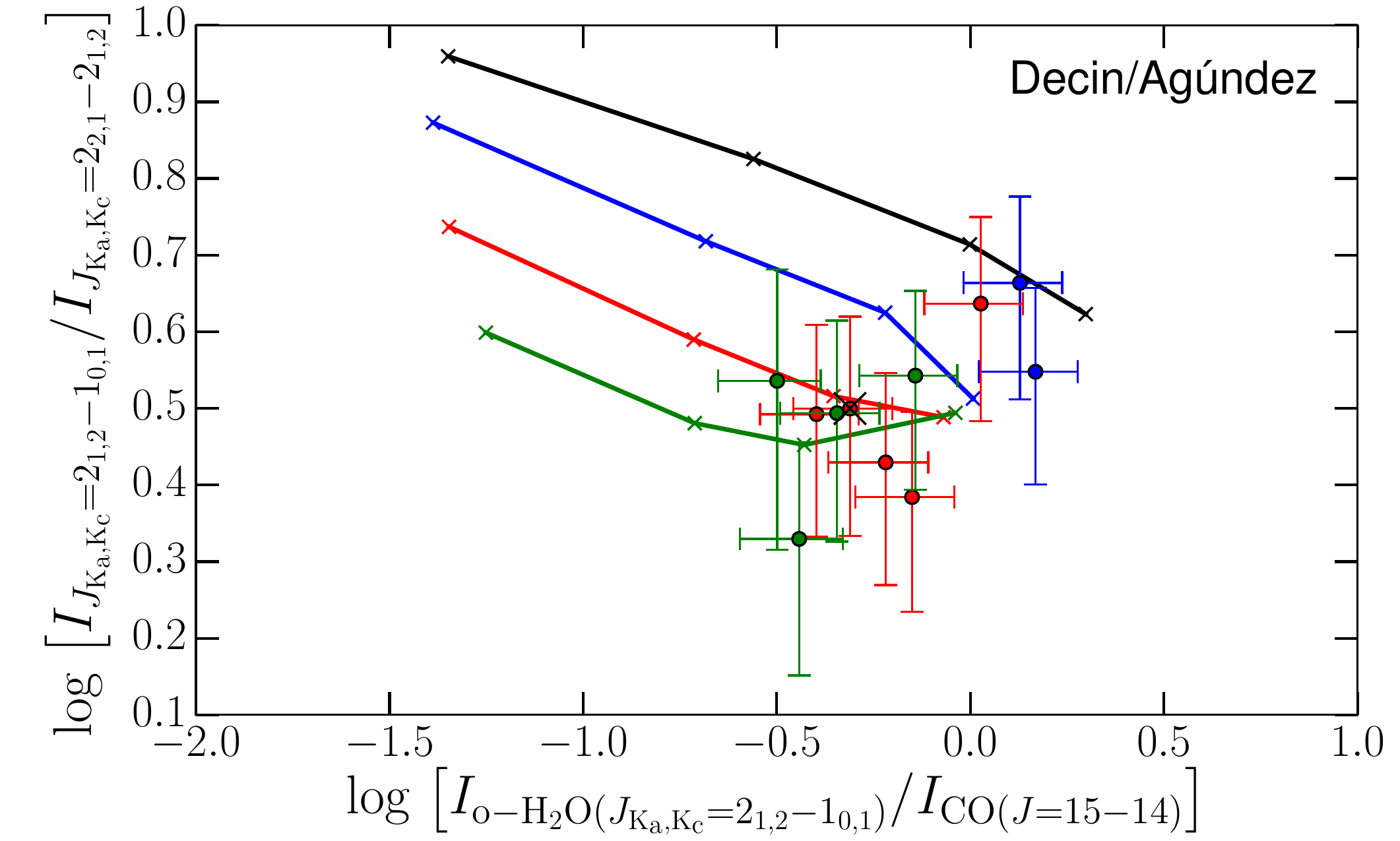}} &\resizebox{9.0cm}{!}{\includegraphics{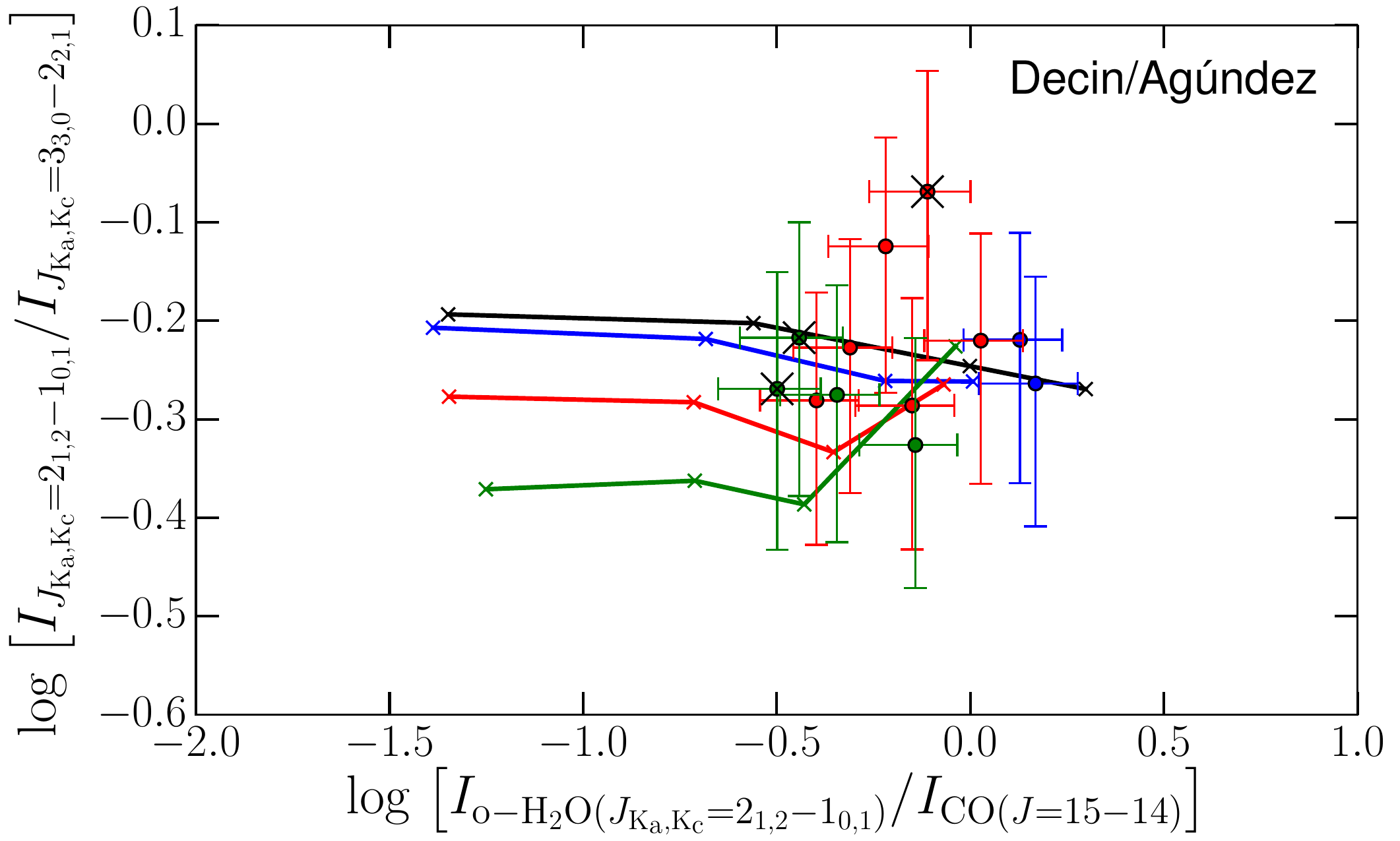}} \\
\resizebox{9.0cm}{!}{\includegraphics{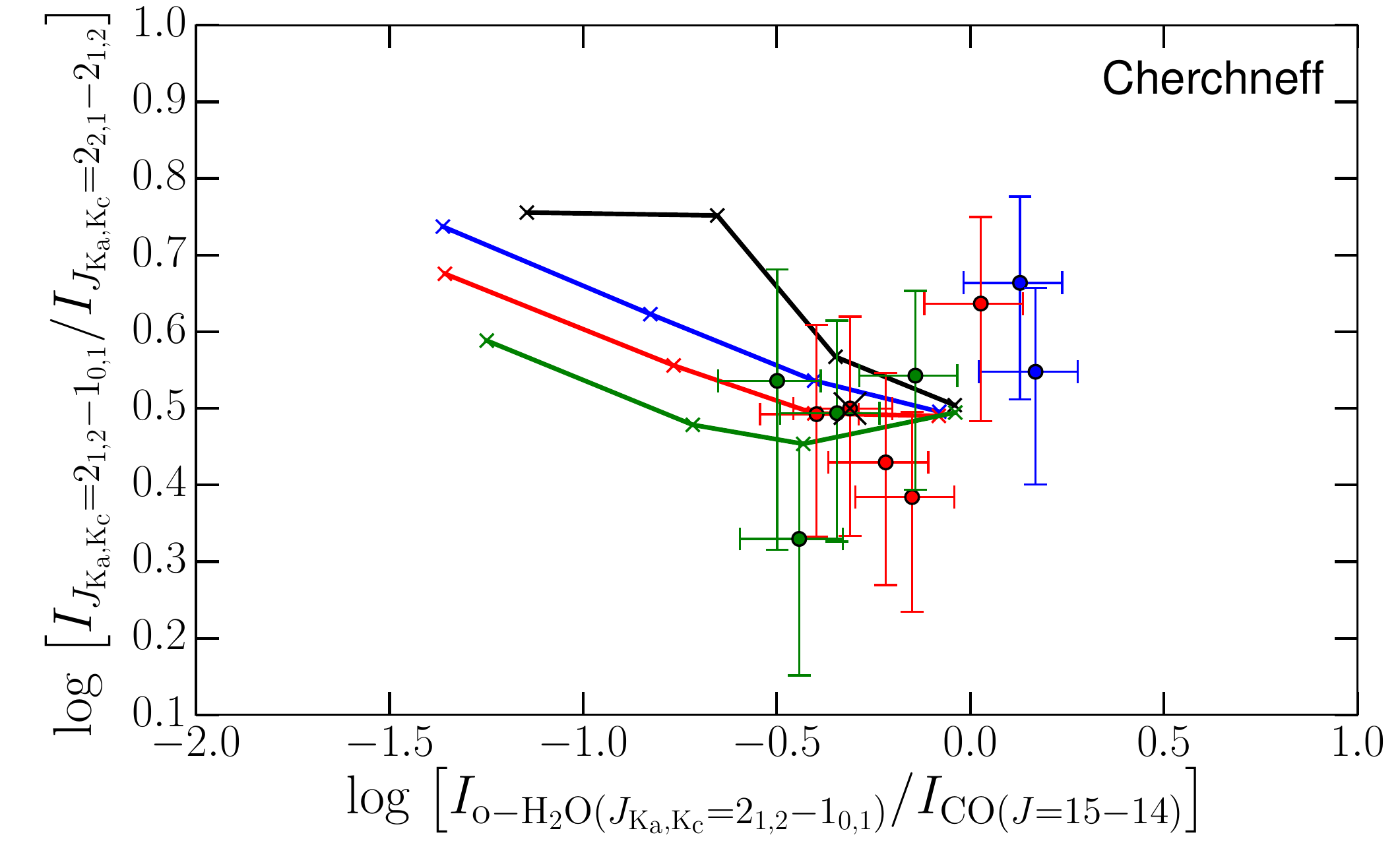}} &\resizebox{9.0cm}{!}{\includegraphics{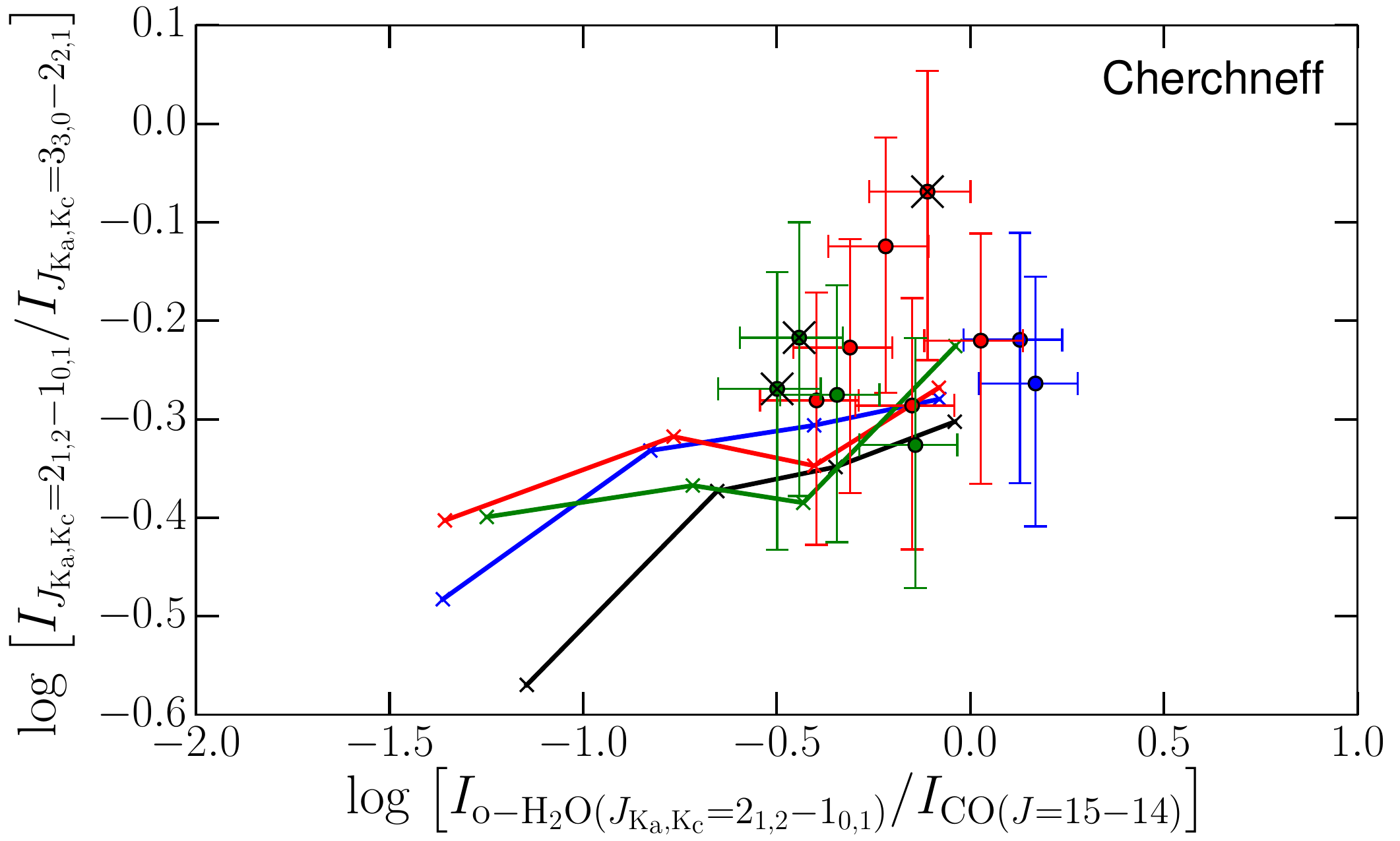}} 
\end{array}$
\caption{ \water/\water line-strength ratio versus the \water/CO line-strength ratio for a selection of \water transitions. The points with error bars give the measured line-strength ratios, color coded according to the $\shellmass$ range to which the sources belong (in units of \myrkms), as indicated in the first panel. Undetected lines are not included in the figures. A black cross superimposed on the data point indicates that one of the \water lines is flagged as a blend. The colored curves show the predicted line-strength ratios for various values of $\shellmass$, in the same range as the data points for each color. Each row of two panels shows results for different \water abundance profiles (see Sect.~\ref{sect:gradmiras} for more details.) The crosses superimposed on the curves indicate the models for an increasing \water abundance from left to right in factors of 10, with the highest maximum value being $10^{-4}$. Adopted values for other parameters are listed in Table~\ref{table:modelpars}.}
\label{fig:h2ograd}
\end{figure*}

{Two \water/\water line-strength ratios are shown in Fig.~\ref{fig:h2ograd} for each of the \water abundance profiles. The first column compares the $J_{\mathrm{K}_\mathrm{a}, \mathrm{K}_\mathrm{c}}=2_{2,1} - 2_{1,2}$ line in the denominator to the $J_{\mathrm{K}_\mathrm{a}, \mathrm{K}_\mathrm{c}}=2_{1,2} - 1_{0,1}$ line in the numerator. The formation regions of these two lines differ slightly. The second column compares the $J_{\mathrm{K}_\mathrm{a}, \mathrm{K}_\mathrm{c}}=3_{3,0} - 2_{2,1}$ line in the denominator to the $J_{\mathrm{K}_\mathrm{a}, \mathrm{K}_\mathrm{c}}=2_{1,2} - 1_{0,1}$ line in the numerator, the former originating much deeper in the outflow than the latter. The \water/\water line-strength ratios are shown as a function of the \water/CO line-strength ratios on the horizontal axis for the \water line in common between both cases. The theoretical predictions are superimposed as full curves on the data points. The color coding is such that the same colors between data points and theoretical predictions have a similar $\shellmass$ value. The points on the theoretical curves represent \water abundance values, increasing from left to right (as expected from the \water/CO line-strength ratio).}

The major differences between the \water abundance profiles occur in the inner wind up to $r \sim 10$ \rstar. We would expect to see the most profound effect on lines formed in the inner wind, but this is precisely where our line formation predictions are less reliable. {That does not mean that lines formed primarily outside this zone remain unaffected. The nonlocal nature of radiative pumping implies that a high or low amount of \water in the inner wind can still affect emission lines formed further out owing to radiative pumping effects. Moreover, a different \water abundance profile may shift the line formation regions inward or outward in the wind. It is therefore worth checking how differences in the \water abundance profile in the intermediate as well as the inner wind affect the line strengths.} Both columns in Fig.~\ref{fig:h2ograd} compare the $J_{\mathrm{K}_\mathrm{a}, \mathrm{K}_\mathrm{c}}=2_{1,2} - 1_{0,1}$ line with a line that is formed deeper in the wind, but mostly at radii larger than $\sim 10$ \rstar. All in all, the three abundance profiles predict subtle differences in line-strength ratios. We find that, in the case of optically thin lines (e.g.,~for $\shellmass = 3 \times 10^{-8}$ \myrkms and $\shellmass = 1 \times 10^{-7}$ \myrkms, the black and blue curves in Fig.~\ref{fig:h2ograd}), the Decin/Ag\'undez \water abundance profile systematically increases the \water/\water line-strength ratios with respect to the constant abundance models. This should come as no surprise because the Decin/Ag\'undez \water abundance profile results in a {lower} abundance closer to the stellar surface, {which in turn implies that the lines forming deeper in the wind decrease in strength relative to the lines forming further out.} At high $\shellmass$ the lines saturate and there is no noticeable difference between the constant and Decin/Ag\'undez cases. Compared to the constant abundance profile, the line strengths saturate more quickly for the Cherchneff abundance profile. When reaching representative abundances in the intermediate wind on the order of $10^{-4}$, there is no noticeable difference between different $\shellmass$ values. 

Comparing the predictions with the measurements, we are immediately confronted with the limitations of the PACS data, which have uncertainties that are too large to distinguish between the different abundance profiles. Nevertheless, the \water abundance is consistently predicted to be on the order of $10^{-6}-10^{-4}$ for the entire $\shellmass$ range. This range is up to three orders of magnitude larger than what is predicted by \citet{dec2010c} and \citet{agu2010}, as well as by \citet{che2011} for the case of CW~Leo. The measurements for the \water/\water line-strength ratio involving the $J_{\mathrm{K}_\mathrm{a}, \mathrm{K}_\mathrm{c}}=2_{2,1} - 2_{1,2}$ line (first column in Fig.~\ref{fig:h2ograd}) tentatively suggest a trend where the ratio increases as $\shellmass$ decreases (compare the green data points with the blue). {Finally, our results for the Cherchneff abundance profile should not be taken at face value. While the profile for the low \water abundance value of $10^{-7}$ in the intermediate wind (as shown in Fig.~\ref{fig:h2oabun}) is representative of the chemical models calculated for CW~Leo, the high abundance profiles are not. We artificially introduced extremely high abundance values in the inner wind (at $r<3\ \mathrm{R}_\star$) when scaling the profile up or down in representative intermediate-wind abundance. Of course, if we correct for such unreasonably high values, the \water abundance profile would flatten out and become more similar to the constant \water abundance profile. We included the experimental profiles to probe the effect of an increased inner-wind \water abundance on the \water/\water line-strength ratios. They do not represent a realistic view of what a chemical model following \citet{che2011} might look like if it was made to produce higher \water abundances.}

{It is clear that we have reached the limitations of a grid/sample-based approach. In-depth modeling of individual sources is required to rule out unique differences between observed sources and the model grid. This would allow us to derive further meaningful constraints on the \water abundance profile.}

\begin{table*}[!t]
  {
  \setlength{\tabcolsep}{2.7pt}
  \caption{Proposed \water formation mechanisms and the type of \water produced. Typical values for a set of stellar and circumstellar parameters are given: $\shellmass$ the predicted \water abundance, radius $R_\mathrm{i}$ at which \water formation is expected to begin, radius $R_\mathrm{o}$ from which the \water abundance is expected to remain constant, a typical temperature range $T_\mathrm{f}$ for each mechanism, the original literature references (given below the table), and finally additional comments.} \label{table:formation}\vspace{-0.4cm}
  \begin{center}
  \begin{tabular}[c]{llllllllll}\hline\hline\rule[0mm]{0mm}{3mm}
\water formation mechanism& Type  & $\shellmass$& \waterabun  & $R_\star$  & $R_\mathrm{i}$& $R_\mathrm{o}$& T$_\mathrm{f}$& Ref.   & Comments\\
      &  &(\myrkmsfrac) & (10$^{-7}$) & ($10^{13}$ cm)& (R$_\star$)  & (R$_\star$) & (K)   &    &  \vspace{0.05cm}\\\hline
\rule{0pt}{10pt}Evaporation of icy bodies& cold & 1.4-3.4$\times 10^{-6}$ & 4-24 & 7.65   & 15   & /    & $<500$  & 1,2,10 & \emph{Ruled out}  \\
Radiative association H$_2$ + O &cold & 2.1$\times 10^{-6}$ & 1    & 6.5    & 150   & 750   & $<200$  & 4,9   & \emph{Ruled out}  \\
Fischer-Tropsch catalysis &intermediate & 3.4$\times 10^{-6}$  & 1-100  & 7.0    & 15   & 45   & $<500$  & 3    & $\sim$ Fe-grain density\\
Shock chemistry &warm & 1.0-4.0$\times 10^{-6}$   & 1-7  & 6.5    & 1.2   & 3.0   & $>850$  & 7,8   & $\sim$ shock strength \\
UV photodissociation& warm& 1.4$\times 10^{-6}$ & 2   & 5.1   & 2    & 20   & $>300$  & 5    & $\sim$ degree of clumping  \\
        && 6.7$\times 10^{-7}$  & 2   & 5.0    & 2    & 12   & $>300$  & 6    & $\sim$ degree of clumping  \\
        && 6.7$\times 10^{-8}$  & 2   & 5.0    & 2    & 8    & $>300$  & 6    & $\sim$ degree of clumping  \\
        && 6.7$\times 10^{-9}$  & 10   & 5.0    & 2    & 5    & $>300$  & 6    & $\sim$ degree of clumping  \\\hline
\end{tabular}
  \end{center}
  \vspace{-0.3cm}
  \small $^{(1)}$~\citet{mel2001}, $^{(2)}$~\citet{for2001}, $^{(3)}$~\citet{wil2004}, $^{(4)}$~\citet{agu2006}, $^{(5)}$~\citet{dec2010c}, $^{(6)}$~\citet{agu2010}, $^{(7)}$~\citet{che2011}, $^{(8)}$~\citet{che2012}, $^{(9)}$~\citet{tal2010}, $^{(10)}$~\citet{neu2011b}.
  }
\end{table*}

\section{Discussion}\label{sect:disc}
Different \water formation mechanisms lead to different properties of the \water abundance profile in the wind of carbon stars. In Table~\ref{table:formation}, we summarize these properties for five proposed mechanisms, although most of these predictions are model dependent and have been tailored to explain the \water observations of CW~Leo, which is the prototypical high mass-loss rate carbon star. Hence a straightforward comparison of predicted values with the \water observations reported in this study is not feasible, unless the model assumptions of the \water formation mechanism agree with the properties of our sample stars. In the table we list the $\shellmass$ value for which the \water abundance was derived and typical radii and temperatures associated with the formation mechanism. The mechanisms based on the evaporation of icy bodies and radiative association of H$_2$ and O are listed for completeness, but have been firmly ruled out as viable production mechanisms by previous studies \citep{neu2011b,tal2010}. That leaves one mechanism capable of producing {cold} \water in the intermediate wind from $\sim 15$ R$_\star$ onward, and two mechanisms for producing {warm} \water in regions closer to the stellar surface. Our observations place four constraints on the \water formation mechanism.
\begin{enumerate}
 \item As shown by previous studies for singular sources, and now confirmed to hold for {all} stars in a sample of 18 sources, \water exists in the inner and intermediate wind. For high mass-loss rate objects, we confirm the presence of \water at least {as close to the stellar surface as} $\sim$ 10 R$_\star$, just outside the acceleration zone. For low mass-loss rate objects, \water is present around $\sim 2$ R$_\star$.
 \item {The \water abundance is in the range of $10^{-6}-10^{-4}$. This is significantly higher than the predictions of state-of-the-art formation mechanisms.}
 \item The \water formation mechanism becomes less efficient with increasing mass-loss rate.
 \item {This negative correlation between \water abundance and mass-loss rate} is observed for mass-loss rates higher than $\sim 5 \times 10^{-7}\ \msunyr$. The SRb sources in our sample do not follow the trend.
\end{enumerate}
We now discuss properties of \water formation mechanisms that so far have not been disproved and relate them to the suggested criteria.

\subsection{Fischer-Tropsch catalysis}
Fischer-Tropsch catalysis allows for a broad range of \water abundances to be produced by tweaking the Fe-grain number density, but it is unclear how circumstellar column density affects the Fe-grain number density. However, in terms of the other requirements, the Fischer-Tropsch mechanism cannot be reconciled with our observations. Firstly, the presence of warm \water in the inner wind cannot be explained. Secondly, the mechanism would have to become more efficient at lower wind densities. This is counterintuitive for a mechanism based on dust grains acting as a catalyst because lower densities reduce the amount of interaction between dust and gas that is needed to produce \water. Thirdly, the mechanism cannot explain the presence of \water in SRb objects. In these sources, \water is located close to the stellar surface in too hot of an environment for the mechanism to operate. Even though Fischer-Tropsch catalysis may contribute to \water formation in carbon-rich environments, it seems very unlikely that it is universally active. Further modeling of this production mechanism for low mass-loss-rate objects needs to be performed to see if it still functions in low-density regions and whether or not it becomes more efficient.

\subsection{UV photodissociation in the inner wind}
\citet{agu2010} have looked into a range of mass-loss rates for the mechanism of UV photodissociation, allowing a comparison with our results. \citet{dec2010c} report results for the same mechanism for CW~Leo. Table~\ref{table:formation} summarizes the results for the different $\shellmass$ values. In short, the photodissociation mechanism relies heavily on the degree of clumping in the wind for interstellar UV photons to be able to penetrate deeply into the wind. Therefore, the mechanism provides a natural way to explain a broad range of \water abundances.

The model results shown by \citet{agu2010} also predict a decreasing \water abundance for objects with high mass-loss rates. This model predicts similar \water abundances for high and intermediate mass loss, but a sharp increase in \water abundance for low mass-loss rates. The discontinuity occurs when the major UV-shielded component, i.e.,~the clumps, becomes transparent. Our observations do not show such a sharp increase at a given mass-loss rate, but the model still provides enough flexibility in terms of the clump properties to allow for a more gradual dependence between the \water abundance and $\shellmass$. Moreover, once both the UV-shielded and UV-exposed components become optically thin, the \water abundance can be expected to flatten off. This would explain why the SRb sources at $\mg < 5 \times 10^{-7}\ \msunyr$ do not follow the {negative correlation} between the \water/CO line-strength ratio and the mass-loss rate. {However, the \water abundance toward lower mass-loss rates does not flatten off, and instead decreases (as shown in Fig.~\ref{fig:h2ovmdot}). Hence, there appears to be a dependence on the stellar pulsation type, which is difficult to reconcile with this mechanism.}

{We cannot constrain the radial \water abundance gradient derived by \citet{agu2010} given the uncertainties on the measured line strengths and the low sensitivity of the model predictions to density changes. \citeauthor{agu2010} also predict a maximum abundance of $2\times10^{-7}$ to $10^{-6}$, depending on the mass-loss rate. However, w}e require these profiles to reach a maximum abundance up to three orders of magnitude greater, if they are to explain our \water/\water line-strength ratios. There might still be a reasonable degree of flexibility in the formalism to allow for this increase, but this would require further investigation.

The UV-photodissociation scenario suggests that the C$^{17}$O and C$^{18}$O isotopologues also provide atomic oxygen to produce the minor isotopologues \wateril and \waterih, while the main CO isotopologue shields itself from UV radiation. As a result, one expects an isotope-selective enhancement of the \wateril and \waterih abundances with respect to the main \water isotopologue. Recently, \citet{neu2013} have shown for CW~Leo that this isotope-selective enhancement is less than expected. They suggest that dissociation of C$^{16}$O must contribute a significant number of oxygen atoms as well, if UV photodissociation serves as a basis for \water formation. {Alternatively, if self-shielding of C$^{16}$O proves to be too efficient, another mechanism that is indiscriminate of CO isotopologues, should contribute to \water formation in addition to UV photodissociation in the inner wind.}

\subsection{Shock-induced NLTE chemistry}
% ?????? If the variability type and pulsation pattern have no impact on \water formation, the correlation seen for the other objects might be expected to continue to mass-loss rates below $5 \times 10^{-7}\ \msunyr$ or to flatten off if \water formation saturates. Saturation could occur because of limited availability of the reactants required to form \water. If such saturation does not occur, the pulsation pattern may play a role in \water formation.
{As first proposed by \citet{wil1998} and \citet{che2006}, shock-induced} NLTE chemistry provides a universal method to produce \water in carbon-rich AGB stars: all of them show regular or semiregular pulsational variability, providing the shockwaves that are needed to break up CO and allow \water to form. \water is thus expected close to the stellar surface and a dependence on the variability type and pulsation amplitude could be explained in this framework. Important aspects of our \water analysis concern the similar \water line strengths between Miras and SRa sources, and the breakdown of the {negative correlation} between the \water/CO line-strength ratio and $\shellmass$ at the low end of the range of mass-loss rate that is populated by SRb sources. It could be that SRa sources pulsate in a short-period fundamental mode and SRb sources in a first or second overtone mode. This could affect the shock strengths and densities, which in turn could influence \water formation. Indeed, \citet{bow1988} found that the overtone pulsational modes experience smaller amplitude shocks. This could lead to a clear differentiation between Miras/SRa sources and SRb sources in terms of \water formation. Alternatively,{ the lesser regularity of the pulsations, a.k.a.~periodicity,} of SRb sources may also point to instabilities in multiple pulsation modes (e.g.,~\citeauthor{sos2013}~\citeyear{sos2013}), which could result in weaker shocks as well. In contrast, SRa sources are only unstable in one pulsation mode. 
%Note that two out of three SRa sources in our sample have a debatable variability type and were treated as a Mira for the luminosity and distance determination.

Because \citet{che2011,che2012} has focused on CW\ Leo, a source with a high mass-loss rate that has a period of 650 days, it is difficult to predict how her results would translate to the case of lower or multiple periods. \citet{che2012} states that similar trends can be expected in carbon-rich AGB stars other than CW~Leo. She explains that a lower shock strength can result in a higher \water abundance due to the complex interplay between the consumption of free oxygen by both \water and SiO formation processes. As \citet{che2011,che2012} notes, these results rely heavily on the interplay between \water and SiO production, of which some involved reaction rates are not well constrained. If this process proves viable, it may explain why shorter-period pulsators, and thus lower shock strengths \citep{bow1988}, show higher \water abundances. {Even though they have a shorter pulsational period, SRb sources instead show lower \water abundances. However, they also pulsate less regularly. \citet{che2011,che2012} does not consider less regular shocks of lower strength, so it is unclear what their effect would be.}

\citet{che2012} predicts a strong line variability with time for lines formed within $\sim 3$ R$_\star$, i.e.,~where the shocks are strong. \water abundances can vary several orders of magnitude in this region, and for up to $\sim 80\%$ of one pulsational phase they are significantly higher than at larger distances from the stellar surface {(see also Fig.~\ref{fig:h2oabun})}. Outside this region, {the \water abundance chemically freezes out to its final value over the course of one period, at $\sim 7 \times 10^{-7}$.} As shown in Sect.~\ref{sect:gradmiras} for Mira and SRa sources, we require much larger abundances to explain the observed \water/\water line-strength ratios. Hence, the freeze-out over the course of one pulsation phase should occur at much larger abundances instead. It is at this point unclear whether that is possible in \citeauthor{che2011}'s chemical model.

For SRb sources, we cannot draw any firm conclusions owing to the lower reliability of our models in the shocked region. The time variability for lines formed close to the stellar surface could provide an explanation for the erratic behavior of the \water/\water line-strength ratios observed in these sources. 

{Interestingly, a shock-induced formation mechanism would not discriminate between isotopologues when breaking up CO. As discussed before when considering \water formation by interstellar UV photodissociation, additional \water formation with $^{16}$O is needed \citep{neu2013}. The shock mechanism readily provides this. Additionally, the shock mechanism and the UV photodissociation mechanism both predict significantly lower \water abundances than required. Therefore, our findings suggest that both mechanisms should contribute to warm \water formation in carbon-rich environments. Further studies expanding upon the parameter space of both chemical models are required to probe what range of \water abundances can be produced.}

\section{Conclusions}\label{sect:conc}
We report on new \water observations made with the PACS instrument onboard the \emph{Herschel} space observatory for a sample of 18 carbon-rich AGB stars in the framework of the MESS guaranteed-time key project (P.I.:~M.~Groenewegen) and an OT2~project (P.I.:~L.~Decin). \water has been detected in all sample stars, spanning a broad range of mass-loss rates and several variability types. The \water emission lines include both warm and cold \water and trace the inner and intermediate wind, providing an unprecedented data set that contributes to solving the issue of \water formation in carbon-rich environments. We present line-strength measurements for CO, $^{13}$CO, ortho-\water, and para-\water between 60 \mic and 190 \mic. 

{For Miras and SRa sources, we find that the observed \water/CO line-strength ratios decrease as a function of the circumstellar density. A comparison of the CO line strengths with a model grid suggests} that a single temperature power law with an exponent $\epsilon = 0.4$ explains all CO observations with a moderate sensitivity to other parameters. As such, CO line measurements can be used as a reliable H$_2$ density tracer. We provide linear fitting coefficients for \water/CO line-strength ratios versus mass-loss rate, which can be used as either a {distance-independent} mass-loss indicator or to predict line-strength ratios if an estimate of the mass-loss rate is available. 

A clear {negative correlation} is evident between the \water/CO line-strength ratios and the mass-loss rate for $\mg > 5 \times 10^{-7}\ \msunyr$, regardless of the upper excitation level of the \water transitions or the variability type. The low mass-loss-rate SRb sources in our sample deviate from this trend. Only the gas terminal velocity and the dust-to-gas ratio noticeably impact the \water/CO line-strength ratios, but not enough to explain the {negative correlation}. This confirms that the \water/CO line-strength ratio is a valid distance-independent \water abundance tracer. As a result, the \water abundance needed to explain the observed line strengths depends on $\shellmass$. {When comparing \water/\water line-strength ratios with our model grid, we find that the measurements are not sensitive enough to distinguish between different \water abundance profiles.}

Until now, five \water formation mechanisms have been suggested for carbon stars. Three of these mechanisms explain the presence of cold \water and two predict warm \water close to the stellar surface. Two cold-\water formation mechanisms have already been ruled out on the basis of previous studies. This leaves a \water formation mechanism based on Fischer-Tropsch catalysis on Fe grains in the intermediate wind, and two warm-\water formation mechanisms: one induced by pulsational shocks just outside the stellar surface, and one by photodissociation of molecules such as $^{13}$CO and SiO in the inner wind by interstellar UV photons. {We derive four constraints that must be fulfilled by an \water formation mechanism: 1) warm \water is present close to or inside the acceleration zone in {all} 18 sources in our sample, 2) \water abundances are significantly higher than predicted by chemical formation mechanisms, 3) \water formation becomes less efficient with increasing mass loss regardless of the \water formation zone, and 4) the \water properties of the SRb sources are disparate from those of Miras and SRa sources.}

{The Fischer-Tropsch catalysis scenario fails to fully explain up to three of these criteria. Of the two warm-\water formation mechanisms, shock-induced NLTE chemistry looks the most promising, as the mechanism has the potential to fulfill all formation criteria. A mechanism based on interstellar UV photons cannot easily explain the peculiar behavior of the SRb sources in terms of \water emission, nor the absence of an isotope-selective enhancement of the \water isotopologues.}

{Both mechanisms currently fail to predict the high \water abundances required to reproduce the observed line strengths. This warrants further investigation of the chemical models.}
%How flexible is photodissociation by interstellar UV photons in terms of how much \water can be produced and why is there no isotope-selective enhancement of \water? How does \water formation depend on shock strengths and periodicity in shock-induced NLTE chemistry? }

\begin{acknowledgements}
We would like to thank the anonymous referee for the extensive and detailed comments on the manuscript. We would also like to thank R.~Guandalini for his contribution to the study. RL acknowledges financial support from the Fund for Scientific Research - Flanders (FWO) under grant number ZKB5757-04-W01, from the Department of Physics and Astronomy of the KU~Leuven, and from the Belgian Federal Science Policy Office via the PRODEX Program of ESA under grant number C90371. LD acknowledges financial support from the FWO. PR, NC, JD, JB, MG, and BV acknowledge support from the Belgian Federal Science Policy Office via the PRODEX Programme of ESA. FK is supported by the FWF project P23586 and the ffg ASAP project HIL. E.G-A is a Research Associate at the Harvard-Smithsonian CfA, and thanks the Spanish Ministerio de Econom\'{\i}a y Competitividad for support under projects AYA2010-21697-C05-0 and FIS2012-39162-C06-01 and partial support from NHSC/JPL RSA 1455432. PACS has been developed by a consortium of institutes led by MPE (Germany) and including UVIE (Austria); KUL, CSL, IMEC (Belgium); CEA, OAMP (France); MPIA (Germany); IFSI, OAP/AOT, OAA/CAISMI, LENS, SISSA (Italy); IAC (Spain). This development has been supported by the funding agencies BMVIT (Austria), ESA-PRODEX (Belgium), CEA/CNES (France), DLR (Germany), ASI (Italy), and CICT/MCT (Spain). For the computations we used the infrastructure of the VSC - Flemish Supercomputer Center, funded by the Hercules Foundation and the Flemish Government - department EWI. 
\end{acknowledgements}
\bibliographystyle{aa}
\bibliography{allreferences}

\begin{thebibliography}{86}
\expandafter\ifx\csname natexlab\endcsname\relax\def\natexlab#1{#1}\fi

\bibitem[{{Abia} {et~al.}(2010){Abia}, {Cunha}, {Cristallo}, {de Laverny},
  {Dom{\'{\i}}nguez}, {Eriksson}, {Gialanella}, {Hinkle}, {Imbriani},
  {Recio-Blanco}, {Smith}, {Straniero}, \& {Wahlin}}]{abi2010}
{Abia}, C., {Cunha}, K., {Cristallo}, S., {et~al.} 2010, \apjl, 715, L94

\bibitem[{{Ag{\'u}ndez} \& {Cernicharo}(2006)}]{agu2006}
{Ag{\'u}ndez}, M. \& {Cernicharo}, J. 2006, \apj, 650, 374

\bibitem[{{Ag{\'u}ndez} {et~al.}(2010){Ag{\'u}ndez}, {Cernicharo}, \&
  {Gu{\'e}lin}}]{agu2010}
{Ag{\'u}ndez}, M., {Cernicharo}, J., \& {Gu{\'e}lin}, M. 2010, \apjl, 724, L133

\bibitem[{{Aringer} {et~al.}(2009){Aringer}, {Girardi}, {Nowotny}, {Marigo}, \&
  {Lederer}}]{ari2009}
{Aringer}, B., {Girardi}, L., {Nowotny}, W., {Marigo}, P., \& {Lederer}, M.~T.
  2009, \aap, 503, 913

\bibitem[{{Begemann} {et~al.}(1994){Begemann}, {Dorschner}, {Henning},
  {Mutschke}, \& {Thamm}}]{beg1994}
{Begemann}, B., {Dorschner}, J., {Henning}, T., {Mutschke}, H., \& {Thamm}, E.
  1994, \apjl, 423, L71

\bibitem[{{Bergeat} \& {Chevallier}(2005)}]{ber2005}
{Bergeat}, J. \& {Chevallier}, L. 2005, \aap, 429, 235

\bibitem[{{Bergeat} {et~al.}(2001){Bergeat}, {Knapik}, \& {Rutily}}]{ber2001}
{Bergeat}, J., {Knapik}, A., \& {Rutily}, B. 2001, \aap, 369, 178

\bibitem[{{Bowen}(1988)}]{bow1988}
{Bowen}, G.~H. 1988, \apj, 329, 299

\bibitem[{{Cernicharo} {et~al.}(2014){Cernicharo}, {Teyssier},
  {Quintana-Lacaci}, {Daniel}, {Agundez}, {Velilla-Prieto}, {Decin}, {Guelin},
  {Encrenaz}, {Garcia-Lario}, {de}, {Barlow}, {Groenewegen}, {Neufeld}, \&
  {Pearson}}]{cer2014}
{Cernicharo}, J., {Teyssier}, D., {Quintana-Lacaci}, G., {et~al.} 2014, \apjl,
  796, L21

\bibitem[{{Cherchneff}(2006)}]{che2006}
{Cherchneff}, I. 2006, \aap, 456, 1001

\bibitem[{{Cherchneff}(2011)}]{che2011}
{Cherchneff}, I. 2011, \aap, 526, L11

\bibitem[{{Cherchneff}(2012)}]{che2012}
{Cherchneff}, I. 2012, \aap, 545, A12

\bibitem[{{Daniel} {et~al.}(2011){Daniel}, {Dubernet}, \& {Grosjean}}]{dan2011}
{Daniel}, F., {Dubernet}, M.-L., \& {Grosjean}, A. 2011, \aap, 536, A76

\bibitem[{{Daniel} {et~al.}(2012){Daniel}, {Goicoechea}, {Cernicharo},
  {Dubernet}, \& {Faure}}]{dan2012}
{Daniel}, F., {Goicoechea}, J.~R., {Cernicharo}, J., {Dubernet}, M.-L., \&
  {Faure}, A. 2012, \aap, 547, A81

\bibitem[{{De Beck} {et~al.}(2010){De Beck}, {Decin}, {de Koter}, {Justtanont},
  {Verhoelst}, {Kemper}, \& {Menten}}]{deb2010}
{De Beck}, E., {Decin}, L., {de Koter}, A., {et~al.} 2010, \aap, 523, A18

\bibitem[{{De Beck} {et~al.}(2012){De Beck}, {Lombaert}, {Ag{\'u}ndez},
  {Daniel}, {Decin}, {Cernicharo}, {M{\"u}ller}, {Min}, {Royer},
  {Vandenbussche}, {de Koter}, {Waters}, {Groenewegen}, {Barlow}, {Gu{\'e}lin},
  {Kahane}, {Pearson}, {Encrenaz}, {Szczerba}, \& {Schmidt}}]{deb2012}
{De Beck}, E., {Lombaert}, R., {Ag{\'u}ndez}, M., {et~al.} 2012, \aap, 539,
  A108

\bibitem[{{Decin}(2012)}]{dec2012}
{Decin}, L. 2012, Advances in Space Research, 50, 843

\bibitem[{{Decin} {et~al.}(2010{\natexlab{a}}){Decin}, {Ag{\'u}ndez}, {Barlow},
  {Daniel}, {Cernicharo}, {Lombaert}, {De Beck}, {Royer}, {Vandenbussche},
  {Wesson}, {Polehampton}, {Blommaert}, {De Meester}, {Exter}, {Feuchtgruber},
  {Gear}, {Gomez}, {Groenewegen}, {Gu{\'e}lin}, {Hargrave}, {Huygen}, {Imhof},
  {Ivison}, {Jean}, {Kahane}, {Kerschbaum}, {Leeks}, {Lim}, {Matsuura},
  {Olofsson}, {Posch}, {Regibo}, {Savini}, {Sibthorpe}, {Swinyard}, {Yates}, \&
  {Waelkens}}]{dec2010c}
{Decin}, L., {Ag{\'u}ndez}, M., {Barlow}, M.~J., {et~al.} 2010{\natexlab{a}},
  \nat, 467, 64

\bibitem[{{Decin} {et~al.}(2010{\natexlab{b}}){Decin}, {De Beck},
  {Br{\"u}nken}, {M{\"u}ller}, {Menten}, {Kim}, {Willacy}, {de Koter}, \&
  {Wyrowski}}]{dec2010a}
{Decin}, L., {De Beck}, E., {Br{\"u}nken}, S., {et~al.} 2010{\natexlab{b}},
  \aap, 516, A69

\bibitem[{{Decin} {et~al.}(2006){Decin}, {Hony}, {de Koter}, {Justtanont},
  {Tielens}, \& {Waters}}]{dec2006}
{Decin}, L., {Hony}, S., {de Koter}, A., {et~al.} 2006, \aap, 456, 549

\bibitem[{{Decin} {et~al.}(2010{\natexlab{c}}){Decin}, {Justtanont}, {De Beck},
  {Lombaert}, {de Koter}, {Waters}, {Marston}, {Teyssier}, {Sch{\"o}ier},
  {Bujarrabal}, {Alcolea}, {Cernicharo}, {Dominik}, {Melnick}, {Menten},
  {Neufeld}, {Olofsson}, {Planesas}, {Schmidt}, {Szczerba}, {de Graauw},
  {Helmich}, {Roelfsema}, {Dieleman}, {Morris}, {Gallego},
  {D{\'{\i}}ez-Gonz{\'a}lez}, \& {Caux}}]{dec2010b}
{Decin}, L., {Justtanont}, K., {De Beck}, E., {et~al.} 2010{\natexlab{c}},
  \aap, 521, L4

\bibitem[{{Dubernet} {et~al.}(2009){Dubernet}, {Daniel}, {Grosjean}, \&
  {Lin}}]{dub2009}
{Dubernet}, M.-L., {Daniel}, F., {Grosjean}, A., \& {Lin}, C.~Y. 2009, \aap,
  497, 911

\bibitem[{{Epchtein} {et~al.}(1990){Epchtein}, {Le Bertre}, \&
  {Lepine}}]{epc1990}
{Epchtein}, N., {Le Bertre}, T., \& {Lepine}, J.~R.~D. 1990, \aap, 227, 82

\bibitem[{{Eriksson} {et~al.}(2014){Eriksson}, {Nowotny}, {H{\"o}fner},
  {Aringer}, \& {Wachter}}]{erik2014}
{Eriksson}, K., {Nowotny}, W., {H{\"o}fner}, S., {Aringer}, B., \& {Wachter},
  A. 2014, \aap, 566, A95

\bibitem[{{Faure} {et~al.}(2007){Faure}, {Crimier}, {Ceccarelli}, {Valiron},
  {Wiesenfeld}, \& {Dubernet}}]{fau2007}
{Faure}, A., {Crimier}, N., {Ceccarelli}, C., {et~al.} 2007, \aap, 472, 1029

\bibitem[{{Ford} {et~al.}(2003){Ford}, {Neufeld}, {Goldsmith}, \&
  {Melnick}}]{for2003}
{Ford}, K.~E.~S., {Neufeld}, D.~A., {Goldsmith}, P.~F., \& {Melnick}, G.~J.
  2003, \apj, 589, 430

\bibitem[{{Gonz{\'a}lez-Alfonso} {et~al.}(2007){Gonz{\'a}lez-Alfonso},
  {Neufeld}, \& {Melnick}}]{gon2007}
{Gonz{\'a}lez-Alfonso}, E., {Neufeld}, D.~A., \& {Melnick}, G.~J. 2007, \apj,
  669, 412

\bibitem[{{Goorvitch} \& {Chackerian}(1994)}]{goo1994}
{Goorvitch}, D. \& {Chackerian}, Jr., C. 1994, \apjs, 91, 483

\bibitem[{{Groenewegen}(1994)}]{gro1994}
{Groenewegen}, M.~A.~T. 1994, \aap, 290, 531

\bibitem[{{Groenewegen} {et~al.}(2002){Groenewegen}, {Sevenster}, {Spoon}, \&
  {P{\'e}rez}}]{gro2002}
{Groenewegen}, M.~A.~T., {Sevenster}, M., {Spoon}, H.~W.~W., \& {P{\'e}rez}, I.
  2002, \aap, 390, 511

\bibitem[{{Groenewegen} {et~al.}(2011){Groenewegen}, {Waelkens}, {Barlow},
  {Kerschbaum}, {Garcia-Lario}, {Cernicharo}, {Blommaert}, {Bouwman}, {Cohen},
  {Cox}, {Decin}, {Exter}, {Gear}, {Gomez}, {Hargrave}, {Henning},
  {Hutsem{\'e}kers}, {Ivison}, {Jorissen}, {Krause}, {Ladjal}, {Leeks}, {Lim},
  {Matsuura}, {Naz{\'e}}, {Olofsson}, {Ottensamer}, {Polehampton}, {Posch},
  {Rauw}, {Royer}, {Sibthorpe}, {Swinyard}, {Ueta}, {Vamvatira-Nakou},
  {Vandenbussche}, {van de Steene}, {van Eck}, {van Hoof}, {van Winckel},
  {Verdugo}, \& {Wesson}}]{gro2011}
{Groenewegen}, M.~A.~T., {Waelkens}, C., {Barlow}, M.~J., {et~al.} 2011, \aap,
  526, A162

\bibitem[{{Groenewegen} {et~al.}(1998){Groenewegen}, {Whitelock}, {Smith}, \&
  {Kerschbaum}}]{gro1998b}
{Groenewegen}, M.~A.~T., {Whitelock}, P.~A., {Smith}, C.~H., \& {Kerschbaum},
  F. 1998, \mnras, 293, 18

\bibitem[{{Guandalini} \& {Cristallo}(2013)}]{gua2013}
{Guandalini}, R. \& {Cristallo}, S. 2013, \aap, 555, A120

\bibitem[{{Habing} \& {Olofsson}(2003)}]{hab2003}
{Habing}, H.~J. \& {Olofsson}, H., eds. 2003, {Asymptotic giant branch stars}

\bibitem[{{Hasegawa} {et~al.}(2006){Hasegawa}, {Kwok}, {Koning}, {Volk},
  {Justtanont}, {Olofsson}, {Sch{\"o}ier}, {Sandqvist}, {Hjalmarson}, {Olberg},
  {Winnberg}, {Nyman}, \& {Frisk}}]{has2006}
{Hasegawa}, T.~I., {Kwok}, S., {Koning}, N., {et~al.} 2006, \apj, 637, 791

\bibitem[{{J{\"a}ger} {et~al.}(1998){J{\"a}ger}, {Mutschke}, \&
  {Henning}}]{jag1998b}
{J{\"a}ger}, C., {Mutschke}, H., \& {Henning}, T. 1998, \aap, 332, 291

\bibitem[{{Kama} {et~al.}(2009){Kama}, {Min}, \& {Dominik}}]{kam2009}
{Kama}, M., {Min}, M., \& {Dominik}, C. 2009, \aap, 506, 1199

\bibitem[{{Khouri} {et~al.}(2014){Khouri}, {de Koter}, {Decin}, {Waters},
  {Lombaert}, {Royer}, {Swinyard}, {Barlow}, {Alcolea}, {Blommaert},
  {Bujarrabal}, {Cernicharo}, {Groenewegen}, {Justtanont}, {Kerschbaum},
  {Maercker}, {Marston}, {Matsuura}, {Melnick}, {Menten}, {Olofsson},
  {Planesas}, {Polehampton}, {Posch}, {Schmidt}, {Szczerba}, {Vandenbussche},
  \& {Yates}}]{kho2014}
{Khouri}, T., {de Koter}, A., {Decin}, L., {et~al.} 2014, \aap, 561, A5

\bibitem[{{Knapik} {et~al.}(1999){Knapik}, {Bergeat}, \& {Rutily}}]{kna1999b}
{Knapik}, A., {Bergeat}, J., \& {Rutily}, B. 1999, \aap, 344, 263

\bibitem[{{Knapp} {et~al.}(1999){Knapp}, {Dobrovolsky}, {Ivezi{\'c} }, {Young},
  {Crosas}, {Mattei}, \& {Rupen}}]{kna1999}
{Knapp}, G.~R., {Dobrovolsky}, S.~I., {Ivezi{\'c} }, Z., {et~al.} 1999, \aap,
  351, 97

\bibitem[{{Knapp} {et~al.}(1997){Knapp}, {Jorissen}, \& {Young}}]{kna1997}
{Knapp}, G.~R., {Jorissen}, A., \& {Young}, K. 1997, \aap, 326, 318

\bibitem[{{Knapp} {et~al.}(1998){Knapp}, {Young}, {Lee}, \&
  {Jorissen}}]{kna1998}
{Knapp}, G.~R., {Young}, K., {Lee}, E., \& {Jorissen}, A. 1998, \apjs, 117, 209

\bibitem[{{Kwok}(1975)}]{kwo1975}
{Kwok}, S. 1975, \apj, 198, 583

\bibitem[{{Larsson} {et~al.}(2002){Larsson}, {Liseau}, \&
  {Men'shchikov}}]{lar2002}
{Larsson}, B., {Liseau}, R., \& {Men'shchikov}, A.~B. 2002, \aap, 386, 1055

\bibitem[{{Le Bertre}(1992)}]{leb1992}
{Le Bertre}, T. 1992, \aaps, 94, 377

\bibitem[{{Lombaert} {et~al.}(2012){Lombaert}, {de Vries}, {de Koter}, {Decin},
  {Min}, {Smolders}, {Mutschke}, \& {Waters}}]{lom2012}
{Lombaert}, R., {de Vries}, B.~L., {de Koter}, A., {et~al.} 2012, \aap, 544,
  L18

\bibitem[{{Lombaert} {et~al.}(2013){Lombaert}, {Decin}, {de Koter},
  {Blommaert}, {Royer}, {De Beck}, {de Vries}, {Khouri}, \& {Min}}]{lom2013}
{Lombaert}, R., {Decin}, L., {de Koter}, A., {et~al.} 2013, \aap, 554, A142

\bibitem[{{Loup} {et~al.}(1993){Loup}, {Forveille}, {Omont}, \&
  {Paul}}]{lou1993}
{Loup}, C., {Forveille}, T., {Omont}, A., \& {Paul}, J.~F. 1993, \aaps, 99, 291

\bibitem[{{Maercker} {et~al.}(2012){Maercker}, {Mohamed}, {Vlemmings},
  {Ramstedt}, {Groenewegen}, {Humphreys}, {Kerschbaum}, {Lindqvist},
  {Olofsson}, {Paladini}, {Wittkowski}, {de Gregorio-Monsalvo}, \&
  {Nyman}}]{mae2012}
{Maercker}, M., {Mohamed}, S., {Vlemmings}, W.~H.~T., {et~al.} 2012, \nat, 490,
  232

\bibitem[{{Maercker} {et~al.}(2008){Maercker}, {Sch{\"o}ier}, {Olofsson},
  {Bergman}, \& {Ramstedt}}]{mae2008}
{Maercker}, M., {Sch{\"o}ier}, F.~L., {Olofsson}, H., {Bergman}, P., \&
  {Ramstedt}, S. 2008, \aap, 479, 779

\bibitem[{{Mamon} {et~al.}(1988){Mamon}, {Glassgold}, \& {Huggins}}]{mam1988}
{Mamon}, G.~A., {Glassgold}, A.~E., \& {Huggins}, P.~J. 1988, \apj, 328, 797

\bibitem[{{Melnick} {et~al.}(2001){Melnick}, {Neufeld}, {Ford}, {Hollenbach},
  \& {Ashby}}]{mel2001}
{Melnick}, G.~J., {Neufeld}, D.~A., {Ford}, K.~E.~S., {Hollenbach}, D.~J., \&
  {Ashby}, M.~L.~N. 2001, \nat, 412, 160

\bibitem[{{Melnick} {et~al.}(2000){Melnick}, {Stauffer}, {Ashby}, {Bergin},
  {Chin}, {Erickson}, {Goldsmith}, {Harwit}, {Howe}, {Kleiner}, {Koch},
  {Neufeld}, {Patten}, {Plume}, {Schieder}, {Snell}, {Tolls}, {Wang},
  {Winnewisser}, \& {Zhang}}]{mel2000}
{Melnick}, G.~J., {Stauffer}, J.~R., {Ashby}, M.~L.~N., {et~al.} 2000, \apjl,
  539, L77

\bibitem[{{Millar}(2003)}]{mil2003}
{Millar}, T.~J. 2003, in Asymptotic giant branch stars. Astronomy and
  astrophysics library, New York, Berlin: Springer, ed. H.~J. {Habing} \&
  H.~{Olofsson}, 247p

\bibitem[{{Millar}(2015)}]{mil2015}
{Millar}, T.~J. 2015, Plasma Sources Sci. Technol., 24 043001, 31pp

\bibitem[{{Min} {et~al.}(2009){Min}, {Dullemond}, {Dominik}, {de Koter}, \&
  {Hovenier}}]{min2009a}
{Min}, M., {Dullemond}, C.~P., {Dominik}, C., {de Koter}, A., \& {Hovenier},
  J.~W. 2009, \aap, 497, 155

\bibitem[{{Min} {et~al.}(2003){Min}, {Hovenier}, \& {de Koter}}]{min2003}
{Min}, M., {Hovenier}, J.~W., \& {de Koter}, A. 2003, \aap, 404, 35

\bibitem[{{Neufeld} {et~al.}(2010){Neufeld}, {Gonz{\'a}lez-Alfonso}, {Melnick},
  {Pu{\l}ecka}, {Schmidt}, {Szczerba}, {Bujarrabal}, {Alcolea}, {Cernicharo},
  {Decin}, {Dominik}, {Justtanont}, {de Koter}, {Marston}, {Menten},
  {Olofsson}, {Planesas}, {Sch{\"o}ier}, {Teyssier}, {Waters}, {Edwards},
  {McCoey}, {Shipman}, {Jellema}, {de Graauw}, {Ossenkopf}, {Schieder}, \&
  {Philipp}}]{neu2010}
{Neufeld}, D.~A., {Gonz{\'a}lez-Alfonso}, E., {Melnick}, G., {et~al.} 2010,
  \aap, 521, L5

\bibitem[{{Neufeld} {et~al.}(2011{\natexlab{a}}){Neufeld},
  {Gonz{\'a}lez-Alfonso}, {Melnick}, {Szczerba}, {Schmidt}, {Decin}, {Alcolea},
  {de Koter}, {Sch{\"o}ier}, {Bujarrabal}, {Cernicharo}, {Dominik},
  {Justtanont}, {Marston}, {Menten}, {Olofsson}, {Planesas}, {Teyssier}, \&
  {Waters}}]{neu2011b}
{Neufeld}, D.~A., {Gonz{\'a}lez-Alfonso}, E., {Melnick}, G., {et~al.}
  2011{\natexlab{a}}, \apjl, 727, L29

\bibitem[{{Neufeld} {et~al.}(2011{\natexlab{b}}){Neufeld},
  {Gonz{\'a}lez-Alfonso}, {Melnick}, {Szczerba}, {Schmidt}, {Decin}, {de
  Koter}, {Sch{\"o}ier}, \& {Cernicharo}}]{neu2011a}
{Neufeld}, D.~A., {Gonz{\'a}lez-Alfonso}, E., {Melnick}, G., {et~al.}
  2011{\natexlab{b}}, \apjl, 727, L28

\bibitem[{{Neufeld} {et~al.}(2013){Neufeld}, {Tolls}, {Ag{\'u}ndez},
  {Gonz{\'a}lez-Alfonso}, {Decin}, {Daniel}, {Cernicharo}, {Melnick},
  {Schmidt}, \& {Szczerba}}]{neu2013}
{Neufeld}, D.~A., {Tolls}, V., {Ag{\'u}ndez}, M., {et~al.} 2013, \apjl, 767, L3

\bibitem[{{Nordh} {et~al.}(2003){Nordh}, {von Sch{\'e}ele}, {Frisk}, {Ahola},
  {Booth}, {Encrenaz}, {Hjalmarson}, {Kendall}, {Kyr{\"o}l{\"a}}, {Kwok},
  {Lecacheux}, {Leppelmeier}, {Llewellyn}, {Mattila}, {M{\'e}gie}, {Murtagh},
  {Rougeron}, \& {Witt}}]{nor2003}
{Nordh}, H.~L., {von Sch{\'e}ele}, F., {Frisk}, U., {et~al.} 2003, \aap, 402,
  L21

\bibitem[{{Olivier} {et~al.}(2001){Olivier}, {Whitelock}, \&
  {Marang}}]{oli2001}
{Olivier}, E.~A., {Whitelock}, P., \& {Marang}, F. 2001, \mnras, 326, 490

\bibitem[{{Olofsson} {et~al.}(1993){Olofsson}, {Eriksson}, {Gustafsson}, \&
  {Carlstrom}}]{olo1993}
{Olofsson}, H., {Eriksson}, K., {Gustafsson}, B., \& {Carlstrom}, U. 1993,
  \apjs, 87, 267

\bibitem[{{Pilbratt} {et~al.}(2010){Pilbratt}, {Riedinger}, {Passvogel},
  {Crone}, {Doyle}, {Gageur}, {Heras}, {Jewell}, {Metcalfe}, {Ott}, \&
  {Schmidt}}]{pil2010}
{Pilbratt}, G.~L., {Riedinger}, J.~R., {Passvogel}, T., {et~al.} 2010, \aap,
  518, L1

\bibitem[{{Pitman} {et~al.}(2008){Pitman}, {Hofmeister}, {Corman}, \&
  {Speck}}]{pit2008}
{Pitman}, K.~M., {Hofmeister}, A.~M., {Corman}, A.~B., \& {Speck}, A.~K. 2008,
  \aap, 483, 661

\bibitem[{{Poglitsch} {et~al.}(2010){Poglitsch}, {Waelkens}, {Geis},
  {Feuchtgruber}, {Vandenbussche}, {Rodriguez}, {Krause}, {Renotte}, {van
  Hoof}, {Saraceno}, {Cepa}, {Kerschbaum}, {Agn{\`e}se}, {Ali}, {Altieri},
  {Andreani}, {Augueres}, {Balog}, {Barl}, {Bauer}, {Belbachir}, {Benedettini},
  {Billot}, {Boulade}, {Bischof}, {Blommaert}, {Callut}, {Cara}, {Cerulli},
  {Cesarsky}, {Contursi}, {Creten}, {De Meester}, {Doublier}, {Doumayrou},
  {Duband}, {Exter}, {Genzel}, {Gillis}, {Gr{\"o}zinger}, {Henning},
  {Herreros}, {Huygen}, {Inguscio}, {Jakob}, {Jamar}, {Jean}, {de Jong},
  {Katterloher}, {Kiss}, {Klaas}, {Lemke}, {Lutz}, {Madden}, {Marquet},
  {Martignac}, {Mazy}, {Merken}, {Montfort}, {Morbidelli}, {M{\"u}ller},
  {Nielbock}, {Okumura}, {Orfei}, {Ottensamer}, {Pezzuto}, {Popesso},
  {Putzeys}, {Regibo}, {Reveret}, {Royer}, {Sauvage}, {Schreiber}, {Stegmaier},
  {Schmitt}, {Schubert}, {Sturm}, {Thiel}, {Tofani}, {Vavrek}, {Wetzstein},
  {Wieprecht}, \& {Wiezorrek}}]{pog2010}
{Poglitsch}, A., {Waelkens}, C., {Geis}, N., {et~al.} 2010, \aap, 518, L2

\bibitem[{{Price} {et~al.}(2010){Price}, {Smith}, {Kuchar}, {Mizuno}, \&
  {Kraemer}}]{pri2010}
{Price}, S.~D., {Smith}, B.~J., {Kuchar}, T.~A., {Mizuno}, D.~R., \& {Kraemer},
  K.~E. 2010, \apjs, 190, 203

\bibitem[{{Ramstedt} {et~al.}(2009){Ramstedt}, {Sch{\"o}ier}, \&
  {Olofsson}}]{ram2009}
{Ramstedt}, S., {Sch{\"o}ier}, F.~L., \& {Olofsson}, H. 2009, \aap, 499, 515

\bibitem[{{Ramstedt} {et~al.}(2008){Ramstedt}, {Sch{\"o}ier}, {Olofsson}, \&
  {Lundgren}}]{ram2008}
{Ramstedt}, S., {Sch{\"o}ier}, F.~L., {Olofsson}, H., \& {Lundgren}, A.~A.
  2008, \aap, 487, 645

\bibitem[{Rothman {et~al.}(2009)Rothman, Gordon, Barbe, Benner, Bernath, Birk,
  Boudon, Brown, Campargue, Champion, Chance, Coudert, Dana, Devi, Fally,
  Flaud, Gamache, Goldman, Jacquemart, Kleiner, Lacome, Lafferty, Mandin,
  Massie, Mikhailenko, Miller, Moazzen-Ahmadi, Naumenko, Nikitin, Orphal,
  Perevalov, Perrin, Predoi-Cross, Rinsland, Rotger, Šimečková, Smith, Sung,
  Tashkun, Tennyson, Toth, Vandaele, \& Auwera}]{rot2009}
Rothman, L., Gordon, I., Barbe, A., {et~al.} 2009, Journal of Quantitative
  Spectroscopy and Radiative Transfer, 110, 533 , \{HITRAN\}

\bibitem[{{Saavik Ford} \& {Neufeld}(2001)}]{for2001}
{Saavik Ford}, K.~E. \& {Neufeld}, D.~A. 2001, \apjl, 557, L113

\bibitem[{{Sahai} {et~al.}(2003){Sahai}, {Morris}, {Knapp}, {Young}, \&
  {Barnbaum}}]{sah2003}
{Sahai}, R., {Morris}, M., {Knapp}, G.~R., {Young}, K., \& {Barnbaum}, C. 2003,
  \nat, 426, 261

\bibitem[{{Sahai} {et~al.}(2009){Sahai}, {Sugerman}, \& {Hinkle}}]{sah2009}
{Sahai}, R., {Sugerman}, B.~E.~K., \& {Hinkle}, K. 2009, \apj, 699, 1015

\bibitem[{{Samus} {et~al.}(2009){Samus}, {Durlevich}, \& {et al.}}]{sam2009}
{Samus}, N.~N., {Durlevich}, O.~V., \& {et al.} 2009, VizieR Online Data
  Catalog, 1, 2025

\bibitem[{{Sch{\"o}ier} {et~al.}(2013){Sch{\"o}ier}, {Ramstedt}, {Olofsson},
  {Lindqvist}, {Bieging}, \& {Marvel}}]{sch2013}
{Sch{\"o}ier}, F.~L., {Ramstedt}, S., {Olofsson}, H., {et~al.} 2013, \aap, 550,
  A78

\bibitem[{{Soszy{\'n}ski} \& {Wood}(2013)}]{sos2013}
{Soszy{\'n}ski}, I. \& {Wood}, P.~R. 2013, \apj, 763, 103

\bibitem[{{Talbi} \& {Bacchus-Montabonel}(2010)}]{tal2010}
{Talbi}, D. \& {Bacchus-Montabonel}, M. 2010, Chemical Physics Letters, 485, 56

\bibitem[{{van Leeuwen}(2007)}]{van2007}
{van Leeuwen}, F. 2007, \aap, 474, 653

\bibitem[{{Whitelock} {et~al.}(2006){Whitelock}, {Feast}, {Marang}, \&
  {Groenewegen}}]{whi2006}
{Whitelock}, P.~A., {Feast}, M.~W., {Marang}, F., \& {Groenewegen}, M.~A.~T.
  2006, \mnras, 369, 751

\bibitem[{{Whitelock} {et~al.}(2008){Whitelock}, {Feast}, \& {van
  Leeuwen}}]{whi2008}
{Whitelock}, P.~A., {Feast}, M.~W., \& {van Leeuwen}, F. 2008, \mnras, 386, 313

\bibitem[{{Willacy}(2004)}]{wil2004}
{Willacy}, K. 2004, \apjl, 600, L87

\bibitem[{{Willacy} \& {Cherchneff}(1998)}]{wil1998}
{Willacy}, K. \& {Cherchneff}, I. 1998, \aap, 330, 676

\bibitem[{{Wood} {et~al.}(2007){Wood}, {Groenewegen}, {Sloan}, {Blommaert},
  {Cioni}, {Feast}, {Habing}, {Hony}, {Lagadec}, {Loup}, {Matsuura}, {Menzies},
  {Olivier}, {Vanhollebeke}, {van Loon}, {Waters}, {Whitelock}, \&
  {Zijlstra}}]{woo2007}
{Wood}, P., {Groenewegen}, M.~A.~T., {Sloan}, G.~C., {et~al.} 2007, in
  Astronomical Society of the Pacific Conference Series, Vol. 378, Why Galaxies
  Care About AGB Stars: Their Importance as Actors and Probes, ed.
  F.~{Kerschbaum}, C.~{Charbonnel}, \& R.~F. {Wing}, 251

\bibitem[{{Wood}(2010)}]{woo2010}
{Wood}, P.~R. 2010, \memsai, 81, 883

\bibitem[{{Wood} {et~al.}(1999){Wood}, {Alcock}, {Allsman}, {Alves}, {Axelrod},
  {Becker}, {Bennett}, {Cook}, {Drake}, {Freeman}, {Griest}, {King}, {Lehner},
  {Marshall}, {Minniti}, {Peterson}, {Pratt}, {Quinn}, {Stubbs}, {Sutherland},
  {Tomaney}, {Vandehei}, \& {Welch}}]{woo1999}
{Wood}, P.~R., {Alcock}, C., {Allsman}, R.~A., {et~al.} 1999, in IAU Symposium,
  Vol. 191, Asymptotic Giant Branch Stars, ed. T.~{Le Bertre}, A.~{Lebre}, \&
  C.~{Waelkens}, 151

\end{thebibliography}
\newpage
\newpage
\appendix
\section{Radial profiles of 70-$\mu$m and 160-$\mu$m far-IR broadband emission}\label{sect:extension}
Figure~\ref{fig:rad} shows the radial profiles for two different carbon-rich AGB stars (see Table~\ref{table:profiles}) observed with PACS at 70~$\mu$m. The top panel shows a point source, LL~Peg, while the bottom panel shows an extended source, R~Scl. In each figure the radial profile of Vesta, the PACS point spread function (PSF) calibration source, is also shown for comparison. For each object the full width at half maximum (FWHM) given in Table~\ref{table:profiles} is derived from a 2D-Gaussian fit to the bright, central object. The radial profiles\footnote{Azimuthally averaged profiles give similar result (not shown here).} are derived from aperture photometry using circular annuli up to 45''. The subtracted sky background is measured between annuli at 45 and 65''.

The FWHM at 70~$\mu$m is consistently $\sim$7-8''\ for eleven carbon-rich objects included in our sample. This is slightly larger than the $\sim$6'' found for VESTA. Differences in, e.g., observing mode and data processing (HIPE photproject vs. scanamorphos) may underlie this small difference. The brightness of the central object and any extended emission also affect the Gaussian fit. Furthermore, particularly below 10\% flux intensity levels the PSF deviates from a Gaussian shape showing a complicated tri-lobal PSF structure with diffraction spikes. At 160~$\mu$m the FHWM ranges from 12 to 14.5'', compared to the Vesta FWHM of 11.2''. At both wavelengths, RW~LMi appears to be the most extended 'point' source.

For the previously known extended sources R~Scl and CW~Leo we derive larger values for the Gaussian FWHM. However, these should not at all be taken to represent the observed shape for either R~Scl (central point source with a small disk or shell; \citeauthor{mae2012}~\citeyear{mae2012}) or CW~Leo (central, bright point source with a smooth wind and additional density enhancements; \citeauthor{deb2012}~\citeyear{deb2012}). For completeness, the FWHM values of these objects are included in Table~\ref{table:profiles} as well, but they are not included in the present study. PACS photometric data were not available for the objects in the sample not listed in Table~\ref{table:profiles}. However, based on the data reduction of the spectroscopic data, these sources behave like a point source as well, similar to the sources listed here. 
\begin{table}[!h]
\caption{ In the sample for which PACS photometric data were available, we show FWHM of stars, as well as Vesta (PSF calibrator), R~Scl, and CW~Leo.}
\label{table:profiles}
\begin{center}
\begin{tabular}{lll}\hline\hline
Star name      & \multicolumn{2}{c}{FWHM (\arcsec)}  \\
      & 70~$\mu$m    & 160~$\mu$m  \\\hline
RW Lmi      & 7.7      & 14.5    \\
V~Hya      & 7.0      & 13.2    \\
II~Lup    & 7.3      & 14.4    \\
V~Cyg      & 6.9      & 14.3    \\
LL~Peg    & 6.6      & 12.2    \\
LP~And     & 7.1      & 13.8    \\\hline
S~Cep     & 7.1      & 12.7    \\
Y~CVn     &6.7      &12.1    \\
R~Lep     & 7.1      & 12.8    \\
U~Hya     &7.1      & 13.1    \\
W~Ori     &7.0      &12.3    \\\hline
Vesta      & 5.6      & 11.2    \\\hline
R~Scl      & 25.9      & 29.7    \\
CW~Leo      & 10.1      & 15.4    \\\hline
\end{tabular}
\end{center}
\end{table}

\begin{figure}[!t]$
\begin{array}{c}
\resizebox{9cm}{!}{\includegraphics{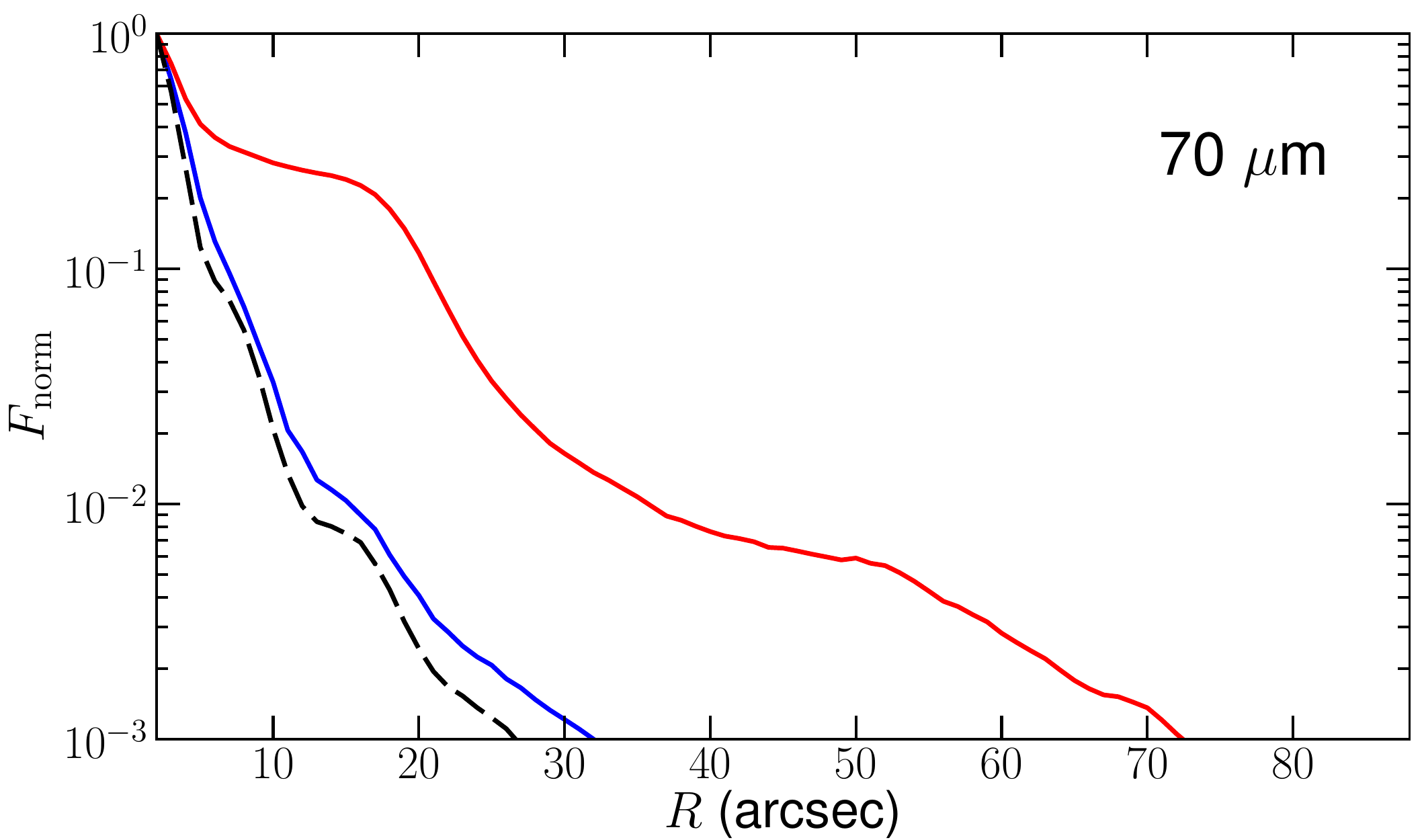}} \\
\resizebox{9cm}{!}{\includegraphics{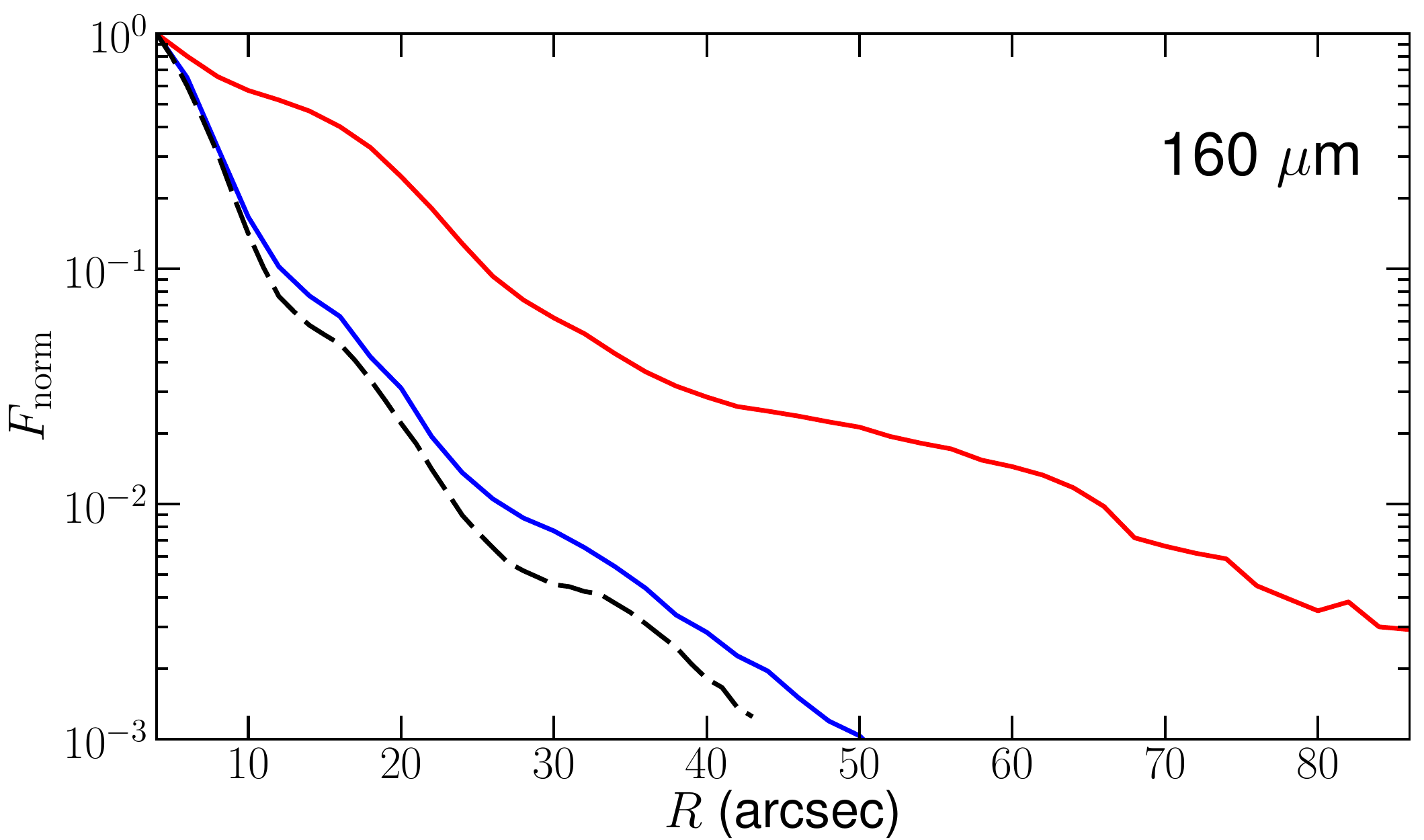}}
\end{array}$
\caption{The normalized flux as a function of the radial distance in arcseconds observed with PACS at 70~$\mu$m in the left panel, and at 160~$\mu$m in the right panel. Shown are Vesta (black dashed), LL~Peg (blue), and R~Scl (red).}
\label{fig:rad}
\end{figure}

\section{The PACS data}
\subsection{The spectra}
Figures~\ref{fig:mess1} through \ref{fig:mess12} show spectra of the sample sources observed in the framework of the MESS~program. We subtracted the continuum of the spectra to improve readability (following \citeauthor{lom2013}~\citeyear{lom2013}) and indicate the identified CO and \water emission lines. Other molecules are not included. The line strengths reported in Table.~\ref{table:intintmess} have been measured before continuum subtraction was performed. Figures~\ref{fig:ot2_1} up to \ref{fig:ot2_13} show the line scans of the sample sources observed in the framework of an OT2~program (P.I.:~L.~Decin). The six sources for which the line strengths are reported in Table~\ref{table:intintot2old} were observed according to an old observation template, resulting in some overlapping wavelength regions between the line scans. For the next set of spectra, the observation scheme was optimized.
\subsection{Integrated line strengths}
Tables~\ref{table:intintmess}, \ref{table:intintot2old}, and \ref{table:intintot2new} list the measured strengths of CO and \water lines in the MESS~spectra, the extended setup of the OT2 line scans, and the optimized setup of the OT2 line scans, respectively. Other molecules have not been taken into account. See Sect.~\ref{sect:ls} for more details on the measurement process, and a few caveats. Finally, Tables~\ref{table:unidentified} and \ref{table:unidentified2} list the significantly detected emission lines in the OT2~data that are not attributed to CO or \water and for which we have not attempted to identify the molecular carrier.
\newpage
\begin{figure*}$
\begin{array}{cc}
\resizebox{8.8cm}{23cm}{\includegraphics[angle=90]{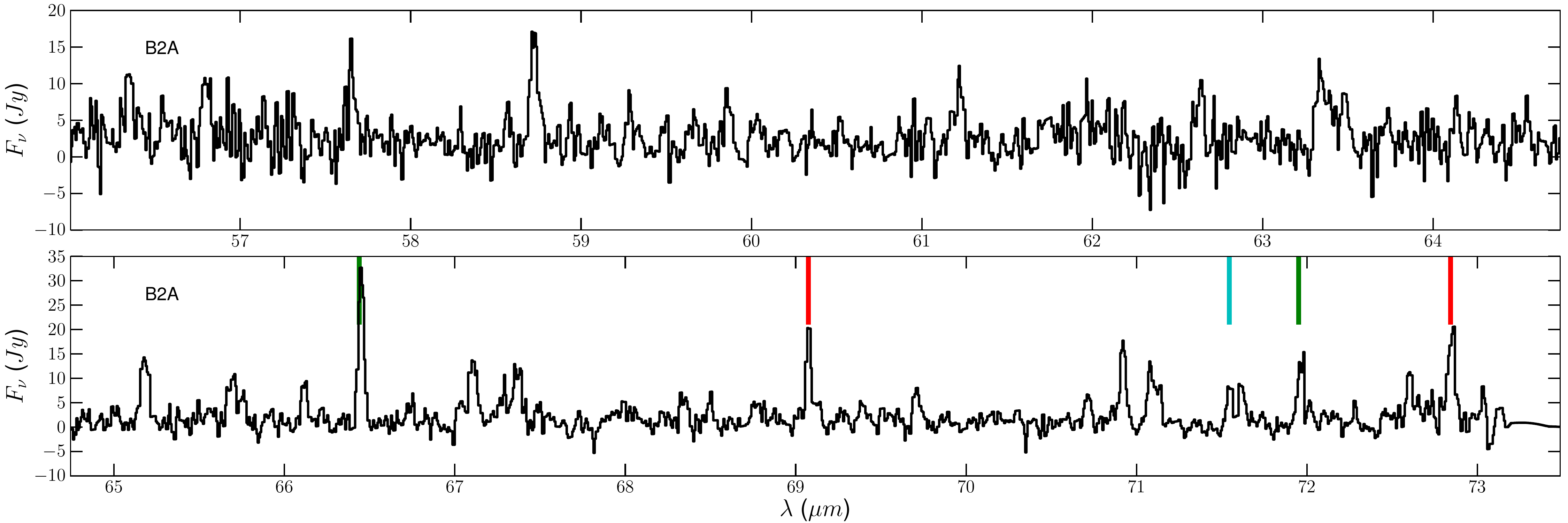}} & \resizebox{8.8cm}{23cm}{\includegraphics[angle=90]{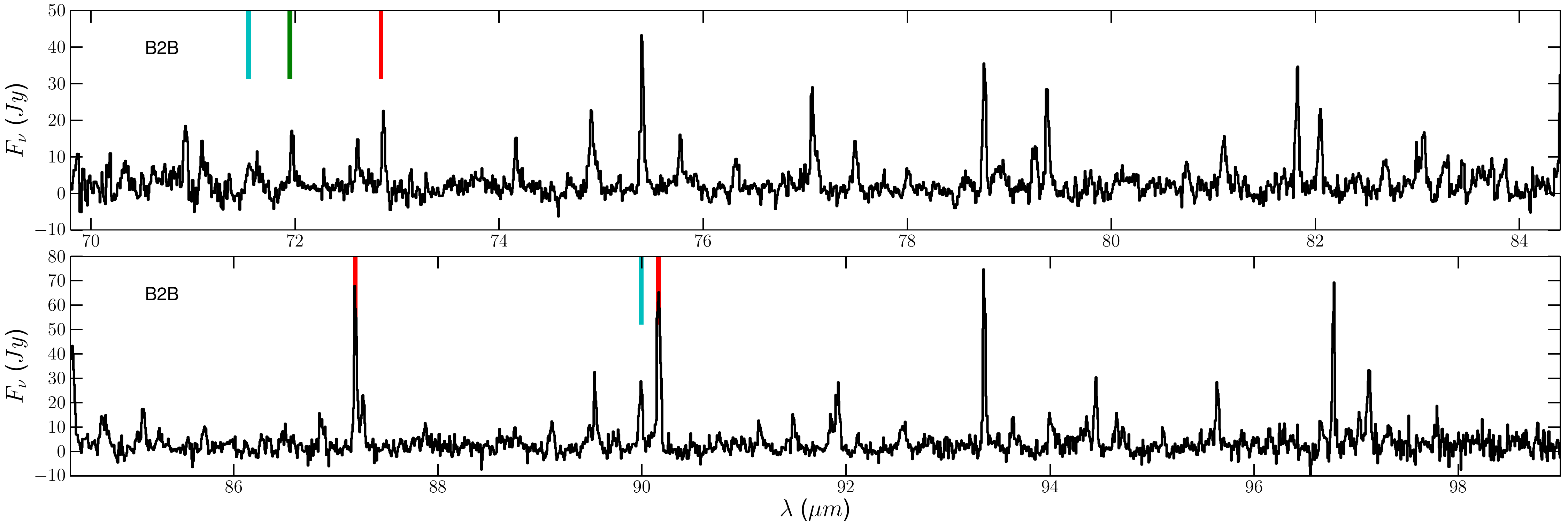}} \\
\end{array}$
\caption{Continuum-subtracted PACS spectrum of RW~LMi is shown in black for the blue bands. The vertical lines indicate molecular identifications according to Table~\ref{table:intintmess}: CO in red, $^{13}$CO in magenta, ortho-\water in green, and para-\water in cyan. If a black dashed line is superimposed over the identification line, the transition was not detected by our line-fitting algorithm. Lines have been indicated only if they occur in the wavelength ranges shared with the OT2 line scans (indicated in red in Table~\ref{table:intintmess}) because the other identifications are less reliable (see Sect.~\ref{sect:ls}). The PACS band is indicated in the upper left corner of each spectrum.}
\label{fig:mess1}
\end{figure*}
\begin{figure*}$
\begin{array}{cc}
\resizebox{8.8cm}{24cm}{\includegraphics[angle=90]{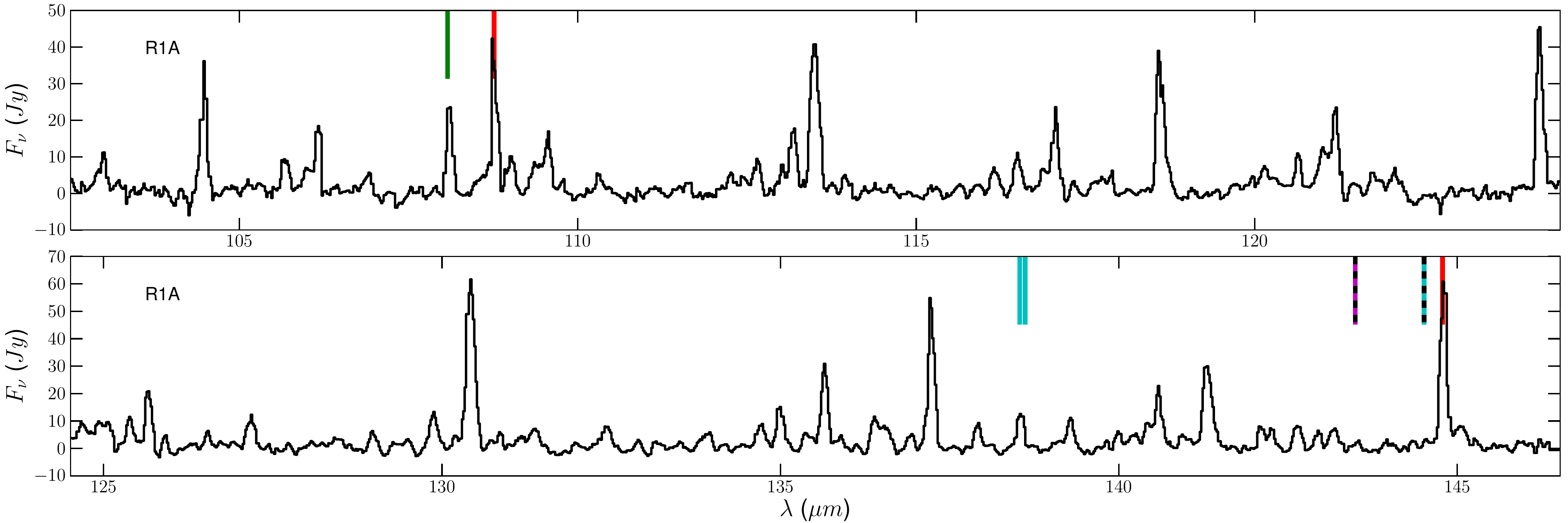}} & \resizebox{8.8cm}{24cm}{\includegraphics[angle=90]{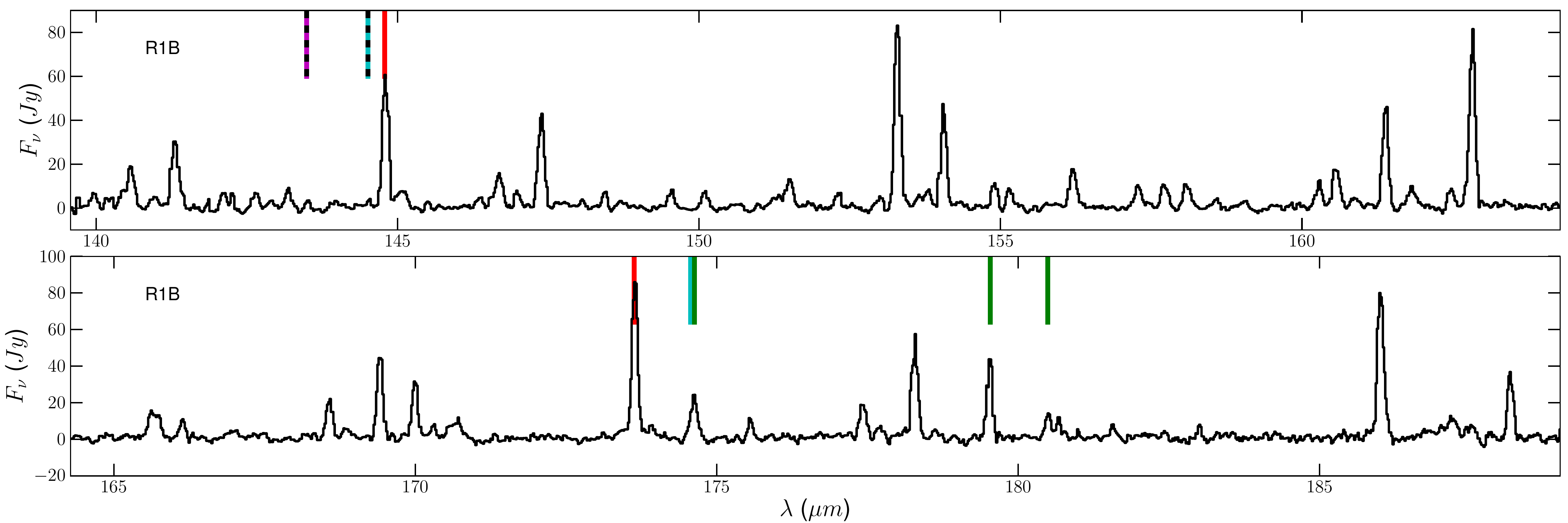}} \\
\end{array}$
\caption{Continuum-subtracted PACS spectrum of RW~LMi is shown for the red bands. The line types are the same as Fig.~\ref{fig:mess1}.}
\label{fig:mess2}
\end{figure*}

\begin{figure*}$
\begin{array}{cc}
\resizebox{8.8cm}{24cm}{\includegraphics[angle=90]{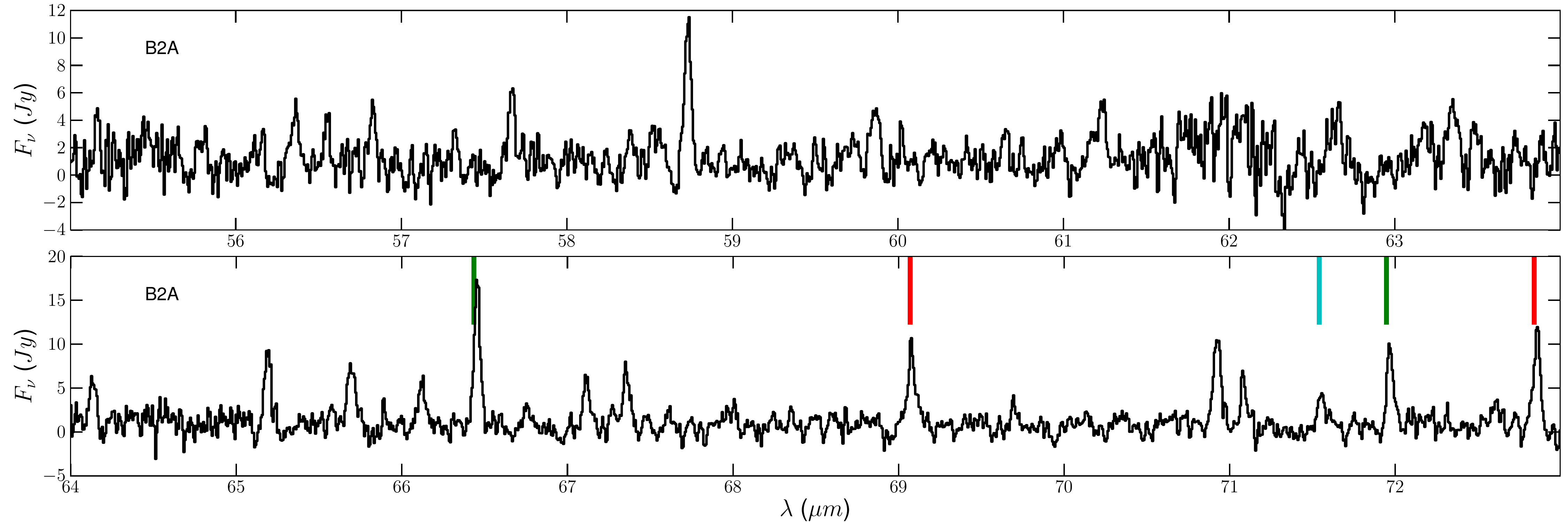}} & \resizebox{8.8cm}{24cm}{\includegraphics[angle=90]{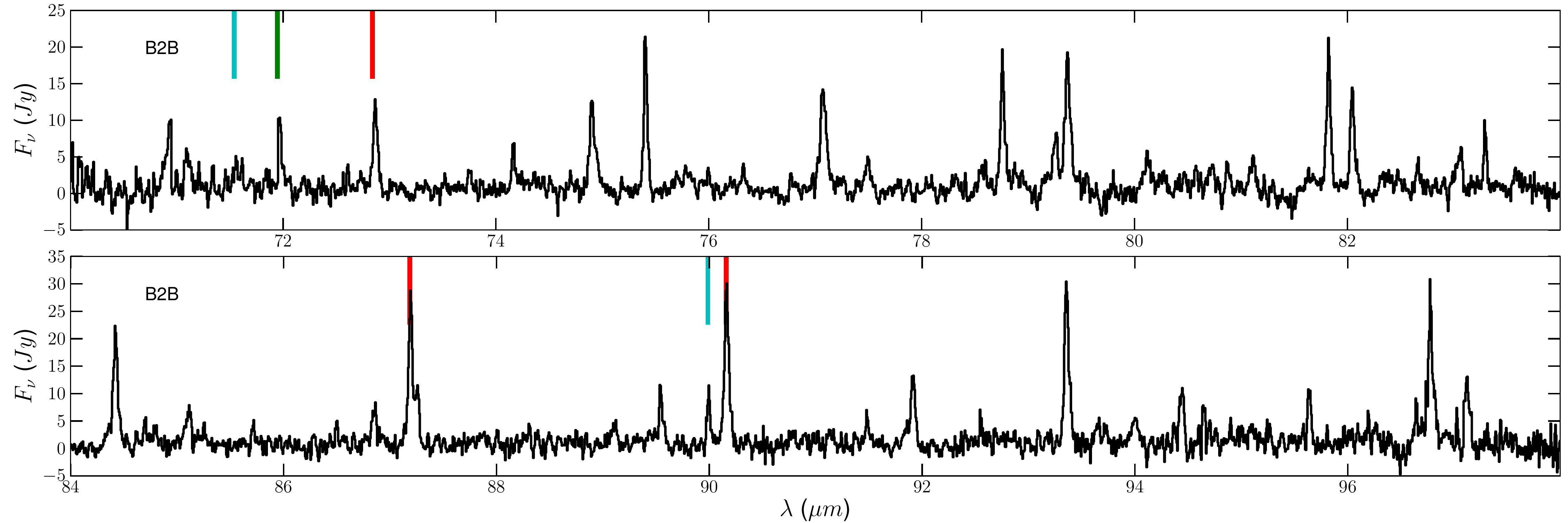}} \\
\end{array}$
\caption{Continuum-subtracted PACS spectrum of V~Hya is shown for the blue bands. The line types are the same as Fig.~\ref{fig:mess1}.}
\label{fig:mess3}
\end{figure*}
\begin{figure*}$
\begin{array}{cc}
\resizebox{8.8cm}{24cm}{\includegraphics[angle=90]{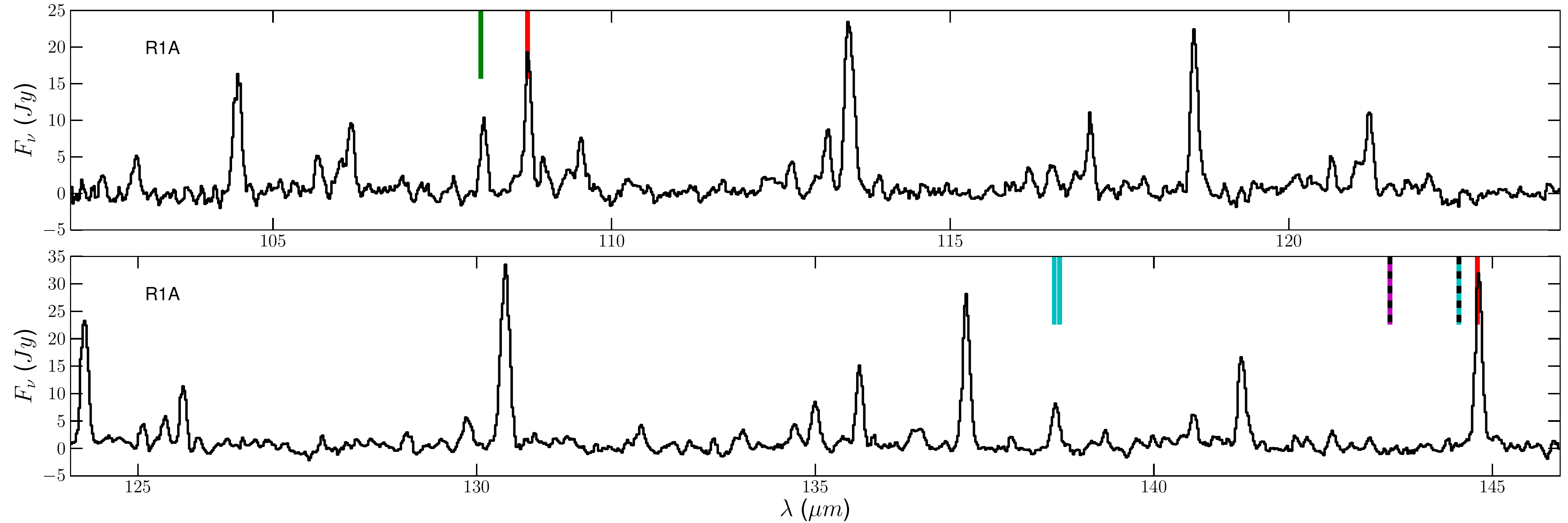}} & \resizebox{8.8cm}{24cm}{\includegraphics[angle=90]{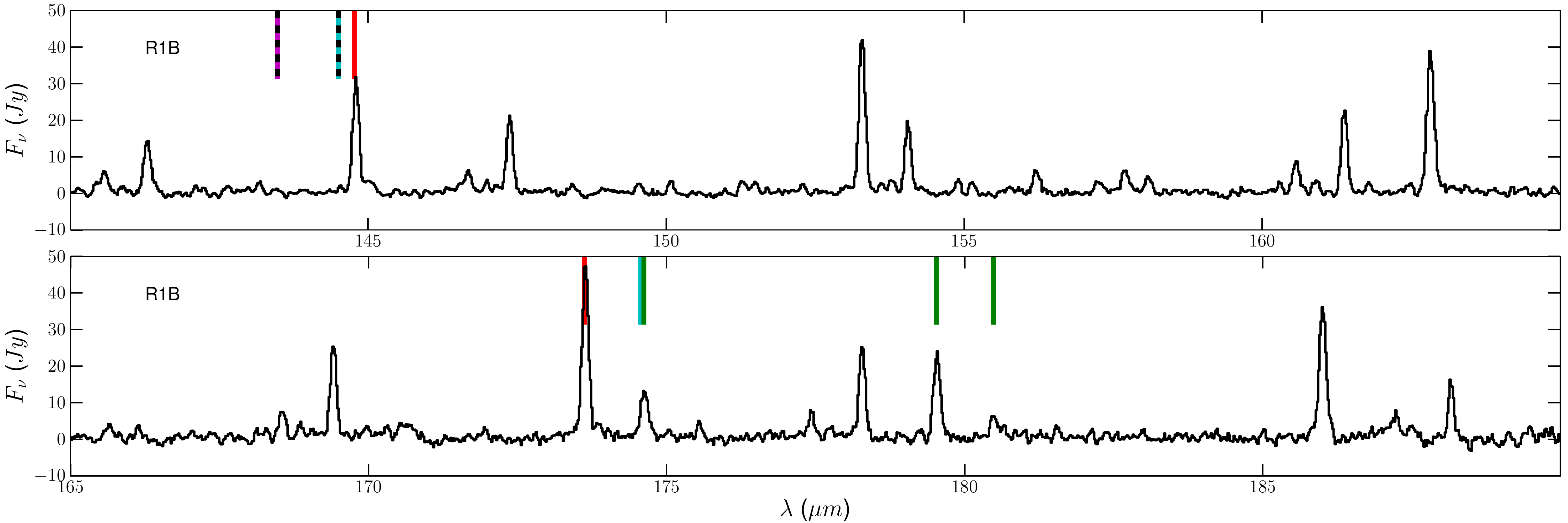}} \\
\end{array}$
\caption{Continuum-subtracted PACS spectrum of V~Hya is shown for the red bands. The line types are the same as Fig.~\ref{fig:mess1}.}
\label{fig:mess4}
\end{figure*}

\begin{figure*}$
\begin{array}{cc}
\resizebox{8.8cm}{24cm}{\includegraphics[angle=90]{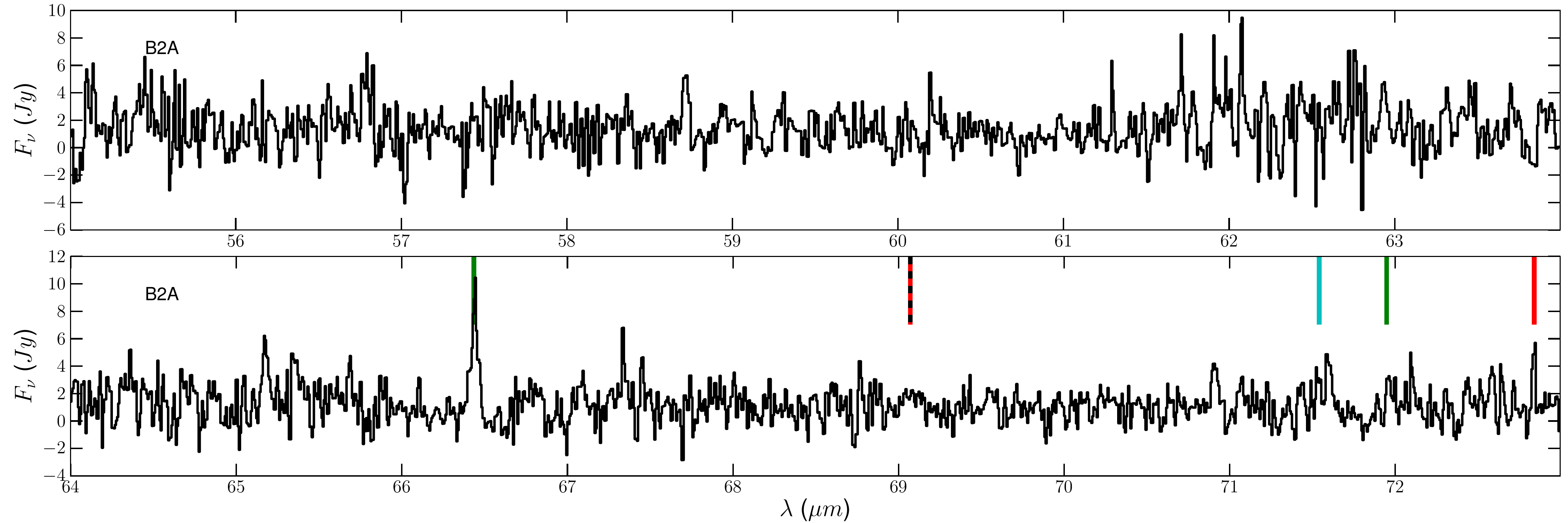}} & \resizebox{8.8cm}{24cm}{\includegraphics[angle=90]{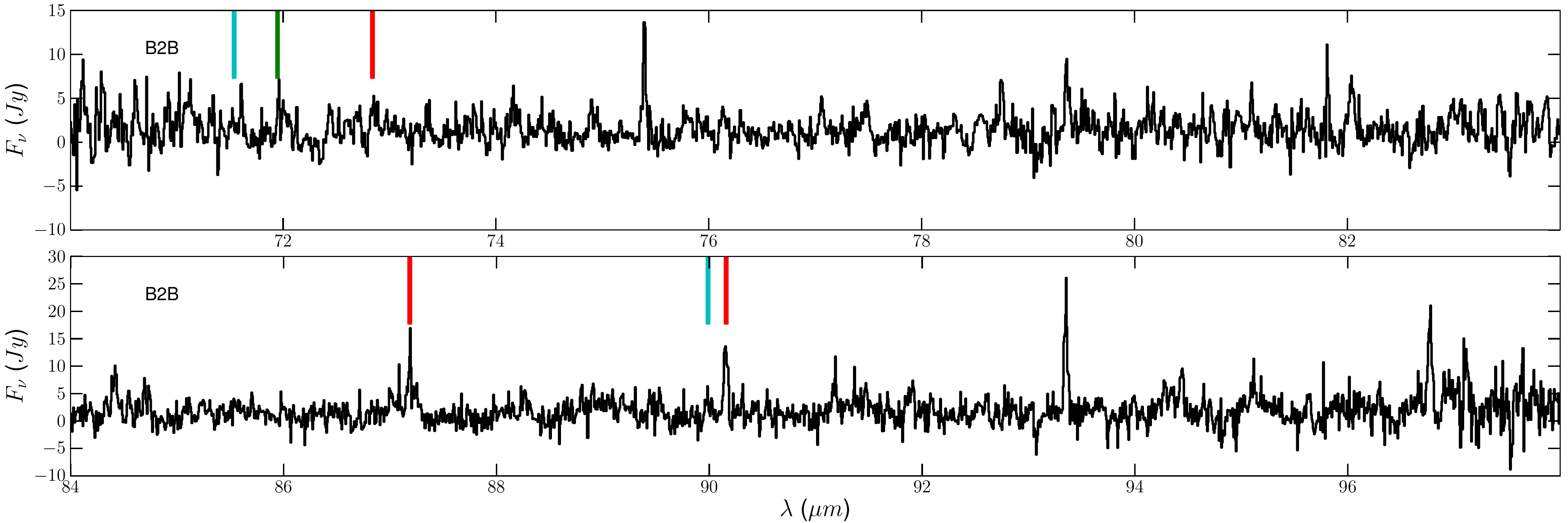}} \\
\end{array}$
\caption{Continuum-subtracted PACS spectrum of II~Lup is shown for the blue bands. The line types are the same as Fig.~\ref{fig:mess1}.}
\label{fig:mess5}
\end{figure*}
\begin{figure*}$
\begin{array}{cc}
\resizebox{8.8cm}{24cm}{\includegraphics[angle=90]{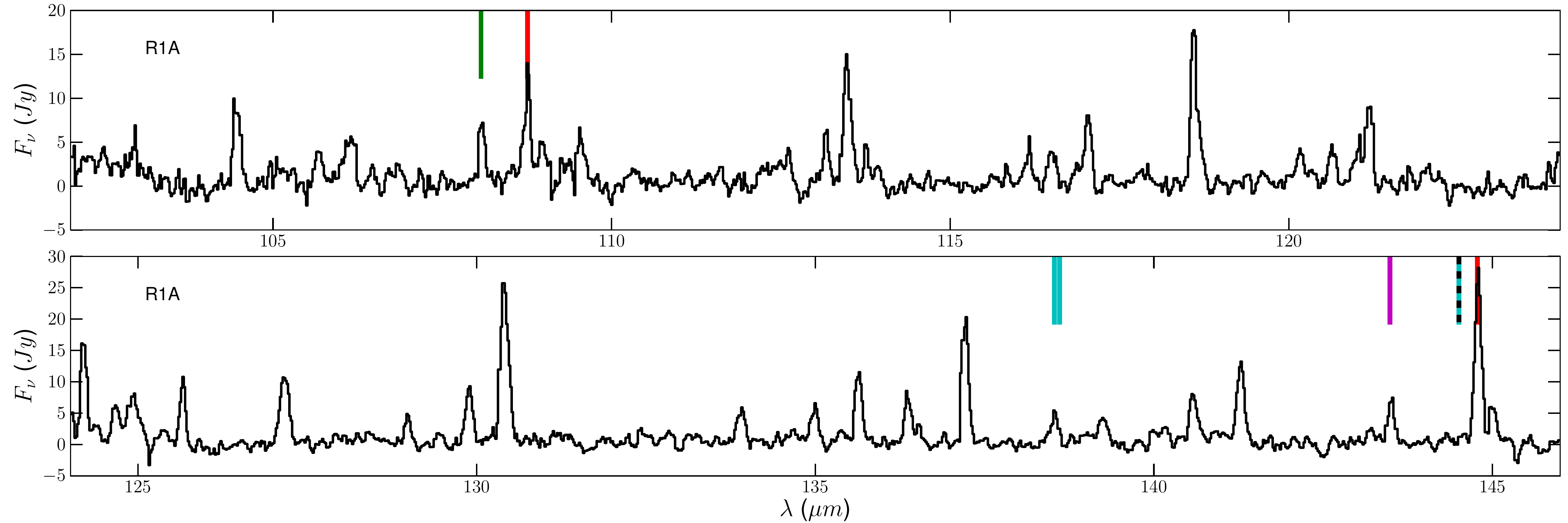}} & \resizebox{8.8cm}{24cm}{\includegraphics[angle=90]{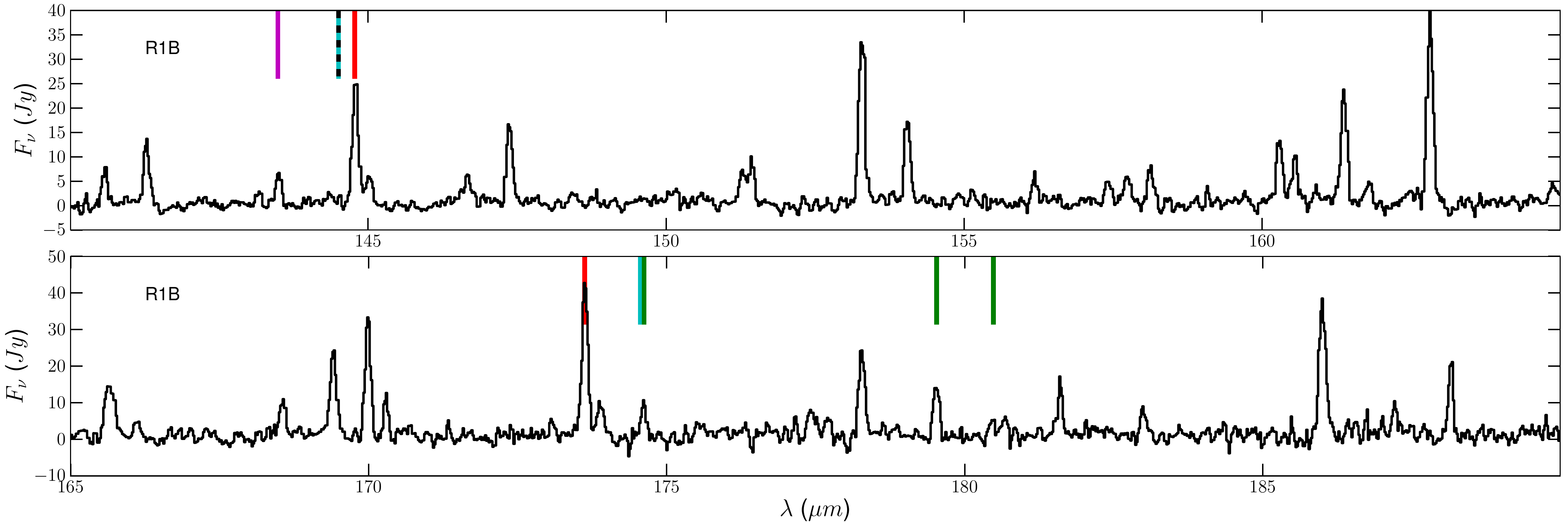}} \\
\end{array}$
\caption{Continuum-subtracted PACS spectrum of II~Lup is shown for the red bands. The line types are the same as Fig.~\ref{fig:mess1}.}
\label{fig:mess6}
\end{figure*}

\begin{figure*}$
\begin{array}{cc}
\resizebox{8.8cm}{24cm}{\includegraphics[angle=90]{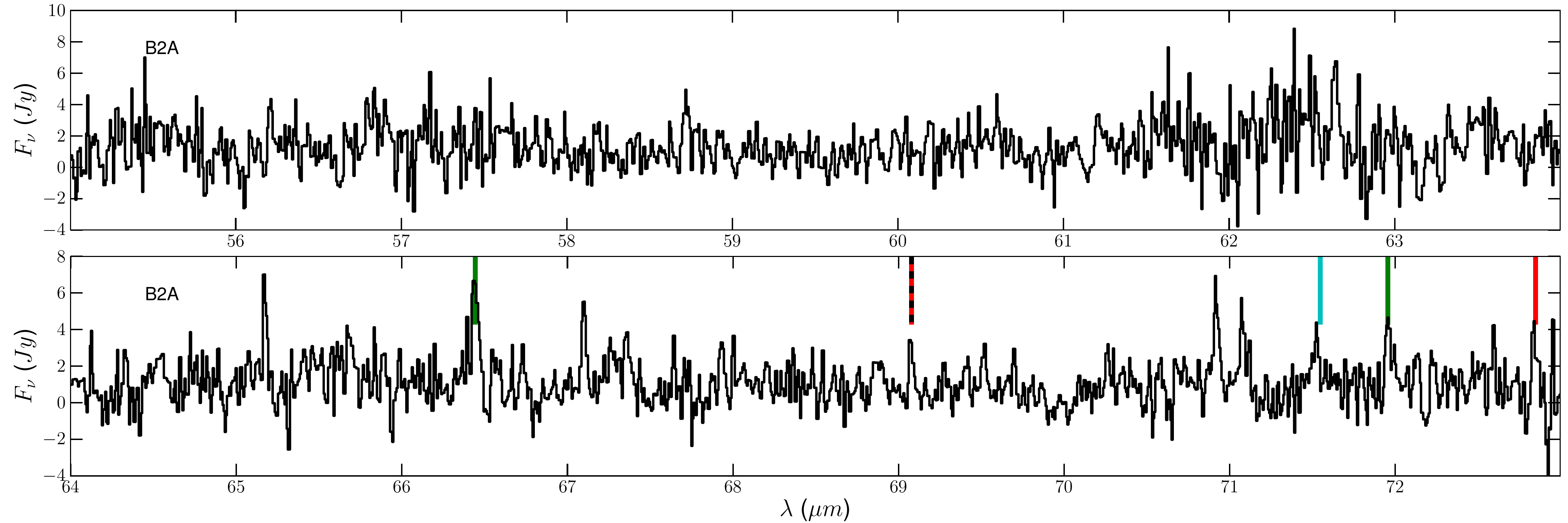}} & \resizebox{8.8cm}{24cm}{\includegraphics[angle=90]{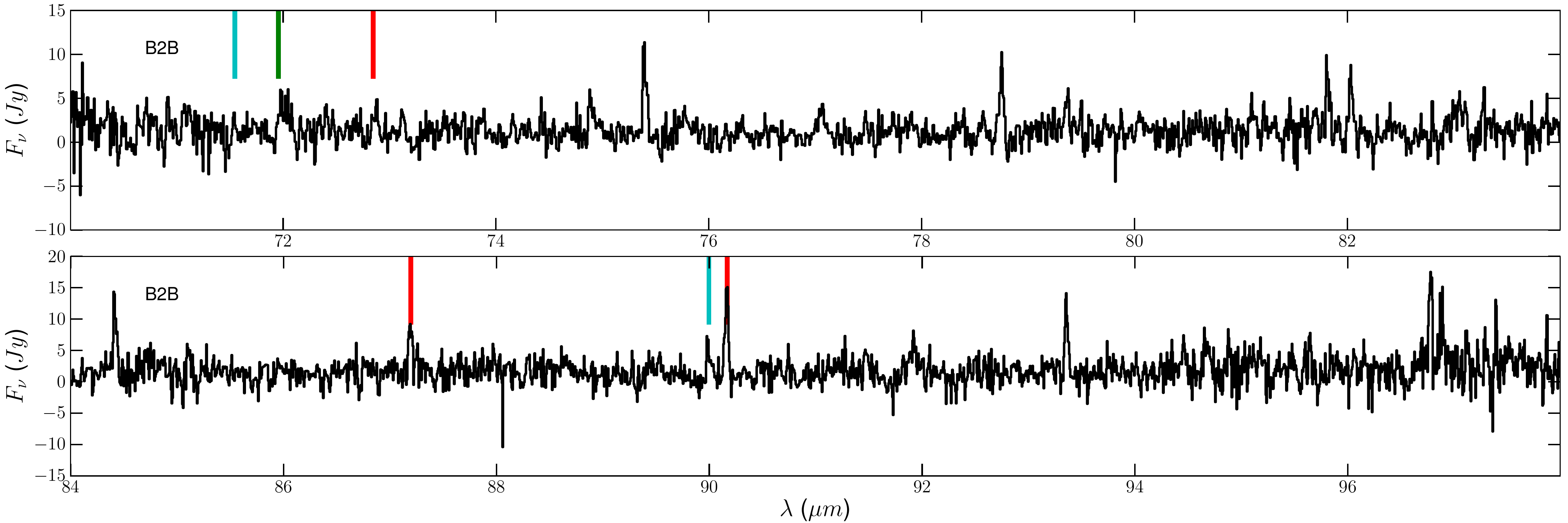}} \\
\end{array}$
\caption{Continuum-subtracted PACS spectrum of V~Cyg is shown for the blue bands. The line types are the same as Fig.~\ref{fig:mess1}.}
\label{fig:mess7}
\end{figure*}
\begin{figure*}$
\begin{array}{cc}
\resizebox{8.8cm}{24cm}{\includegraphics[angle=90]{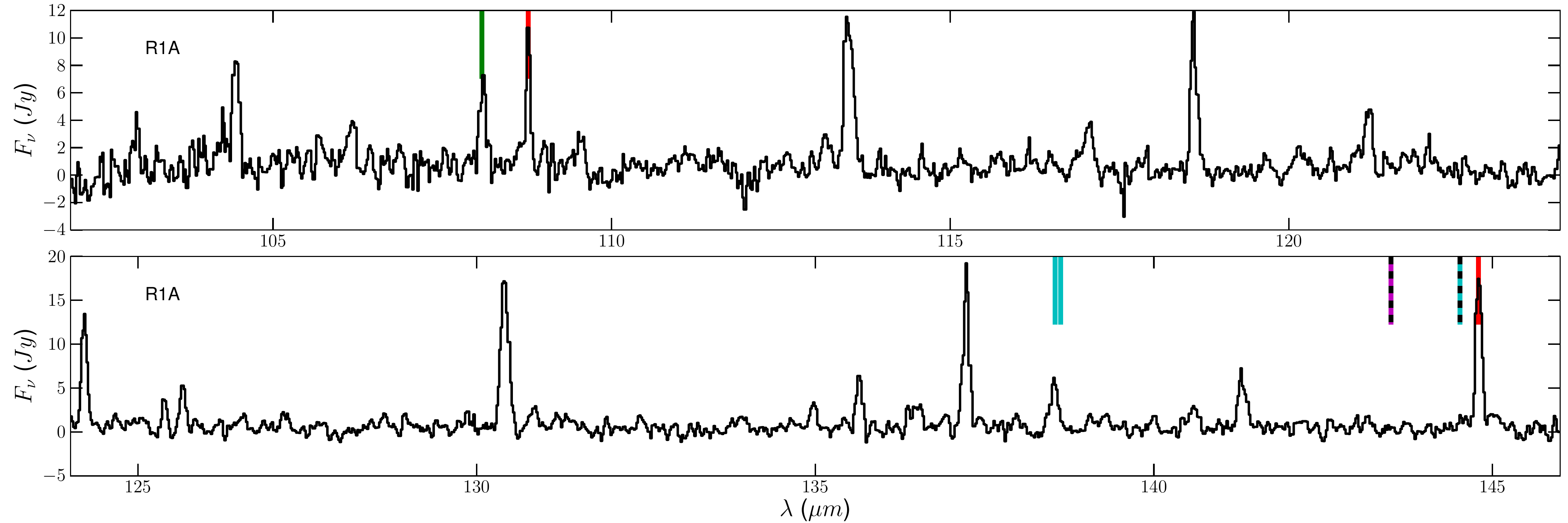}} & \resizebox{8.8cm}{24cm}{\includegraphics[angle=90]{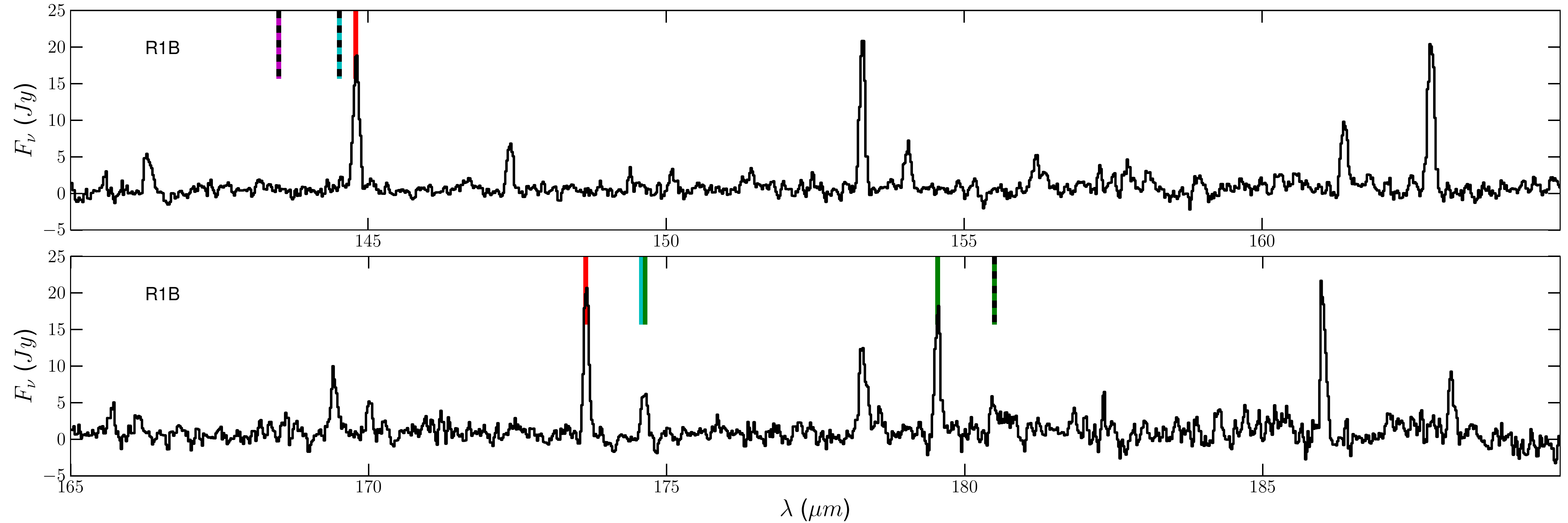}} \\
\end{array}$
\caption{Continuum-subtracted PACS spectrum of V~Cyg is shown for the red bands. The line types are the same as Fig.~\ref{fig:mess1}.}
\label{fig:mess8}
\end{figure*}

\begin{figure*}$
\begin{array}{cc}
\resizebox{8.8cm}{24cm}{\includegraphics[angle=90]{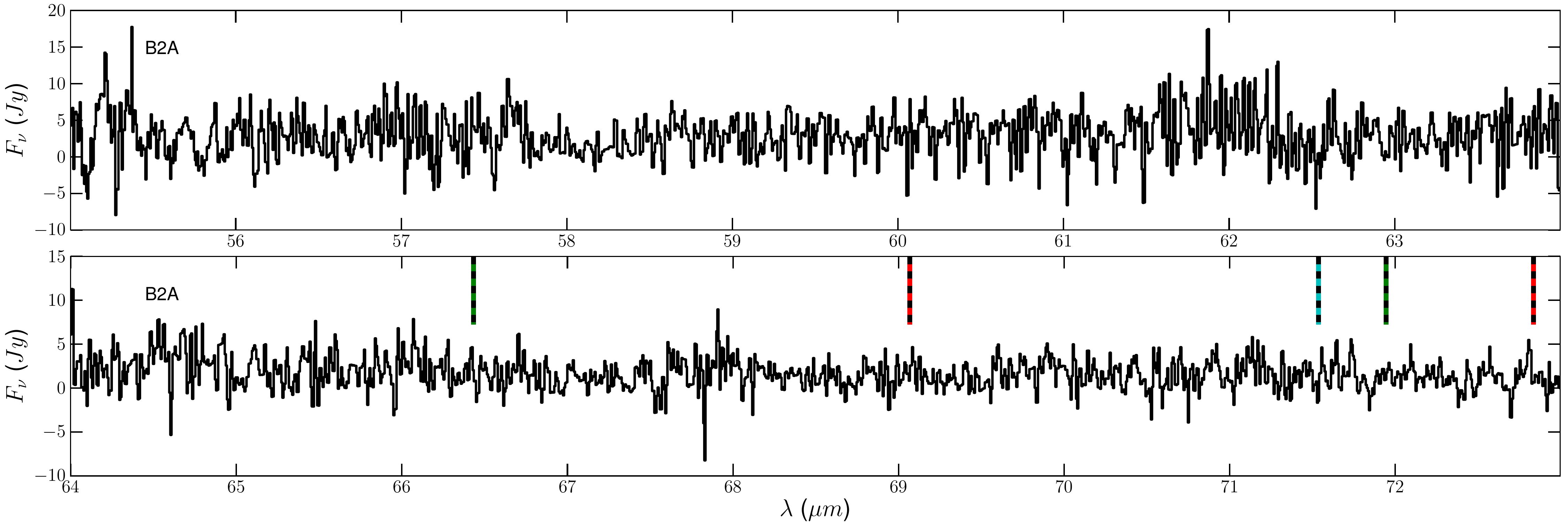}} & \resizebox{8.8cm}{24cm}{\includegraphics[angle=90]{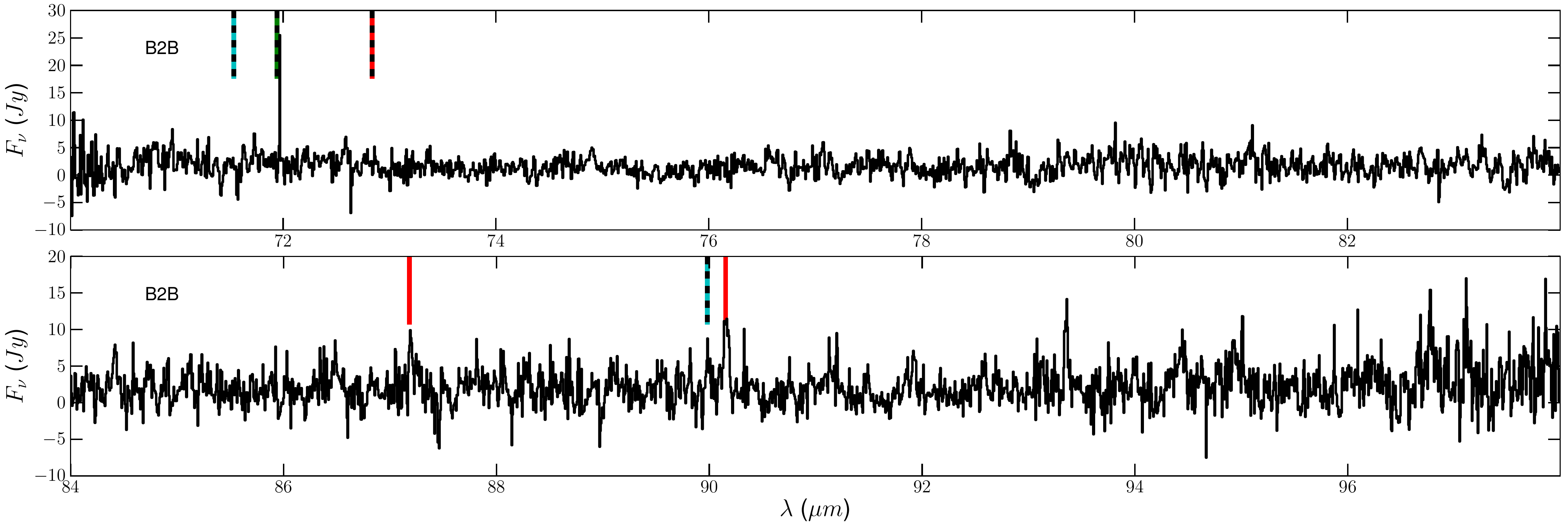}} \\
\end{array}$
\caption{Continuum-subtracted PACS spectrum of LL~Peg is shown for the blue bands. The line types are the same as Fig.~\ref{fig:mess1}.}
\label{fig:mess9}
\end{figure*}
\begin{figure*}$
\begin{array}{cc}
\resizebox{8.8cm}{24cm}{\includegraphics[angle=90]{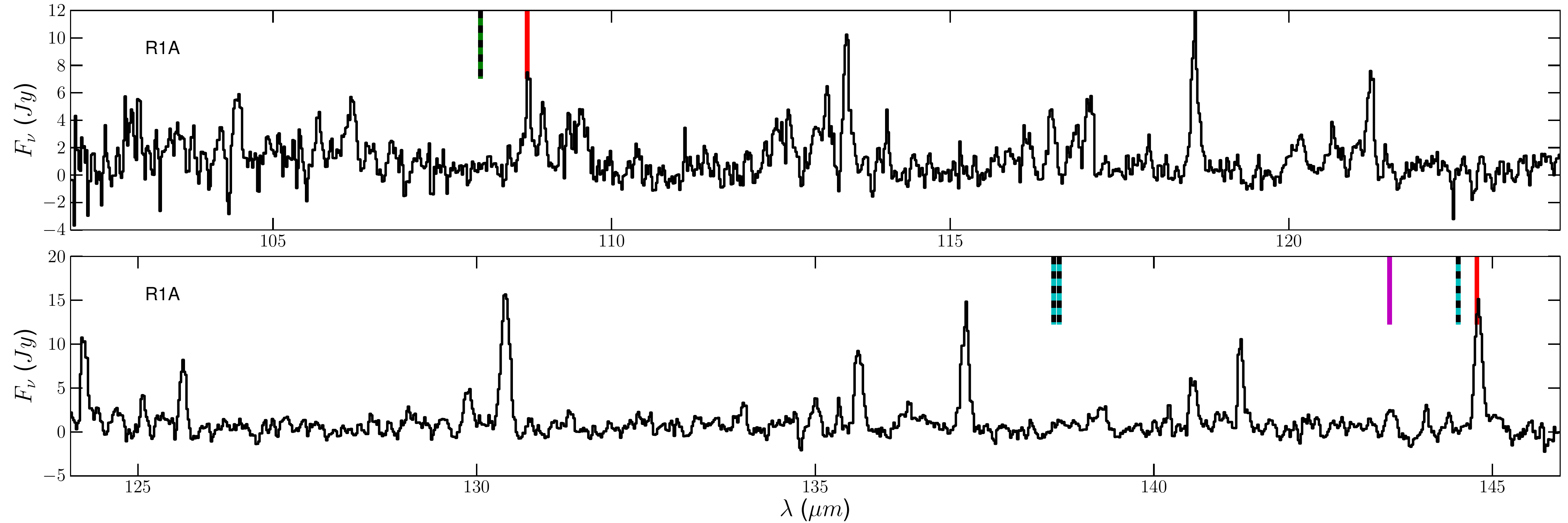}} & \resizebox{8.8cm}{24cm}{\includegraphics[angle=90]{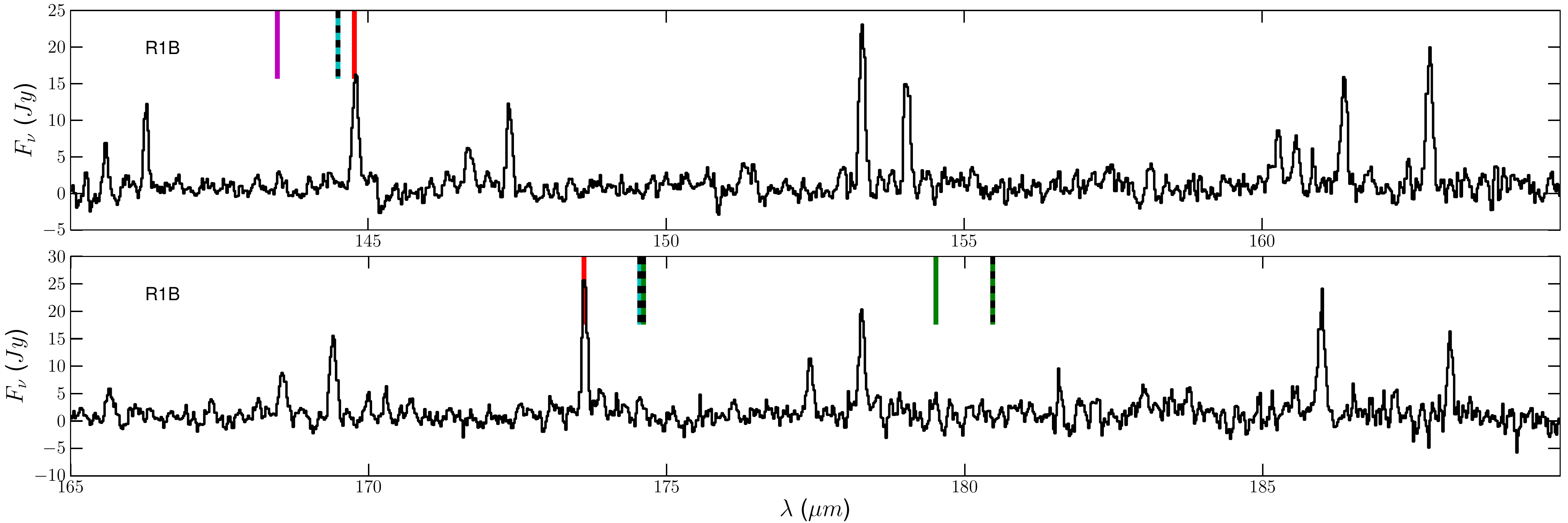}} \\
\end{array}$
\caption{Continuum-subtracted PACS spectrum of LL~Peg is shown for the red bands. The line types are the same as Fig.~\ref{fig:mess1}.}
\label{fig:mess10}
\end{figure*}

\begin{figure*}$
\begin{array}{cc}
\resizebox{8.8cm}{24cm}{\includegraphics[angle=90]{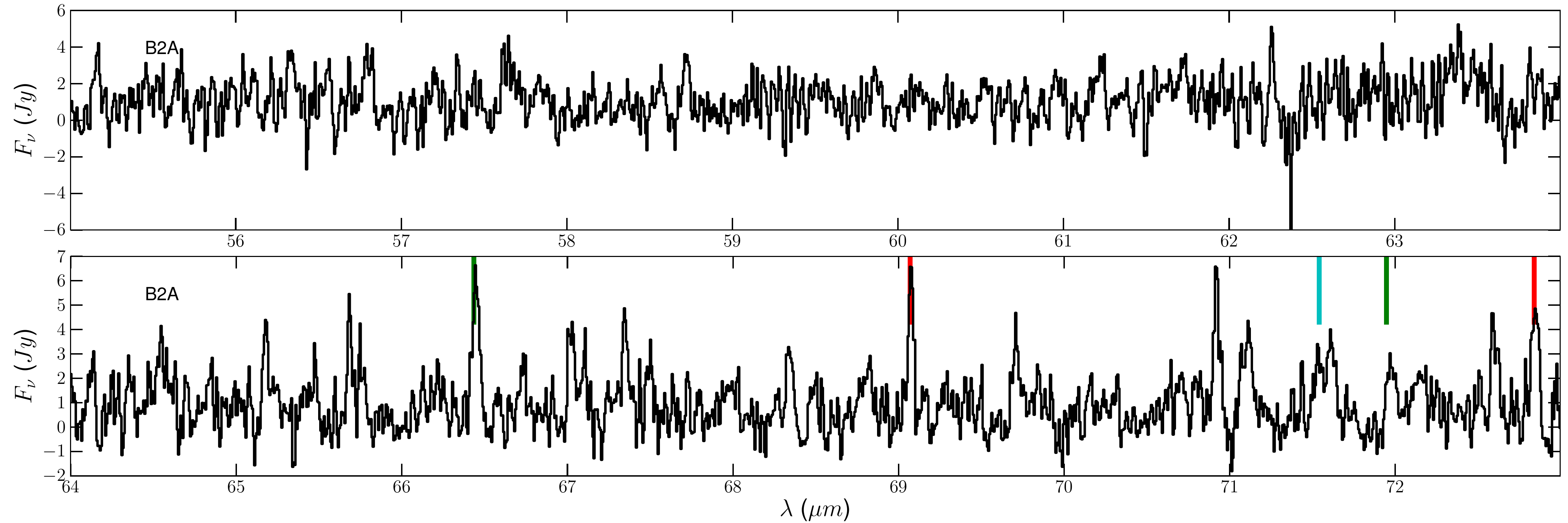}} & \resizebox{8.8cm}{24cm}{\includegraphics[angle=90]{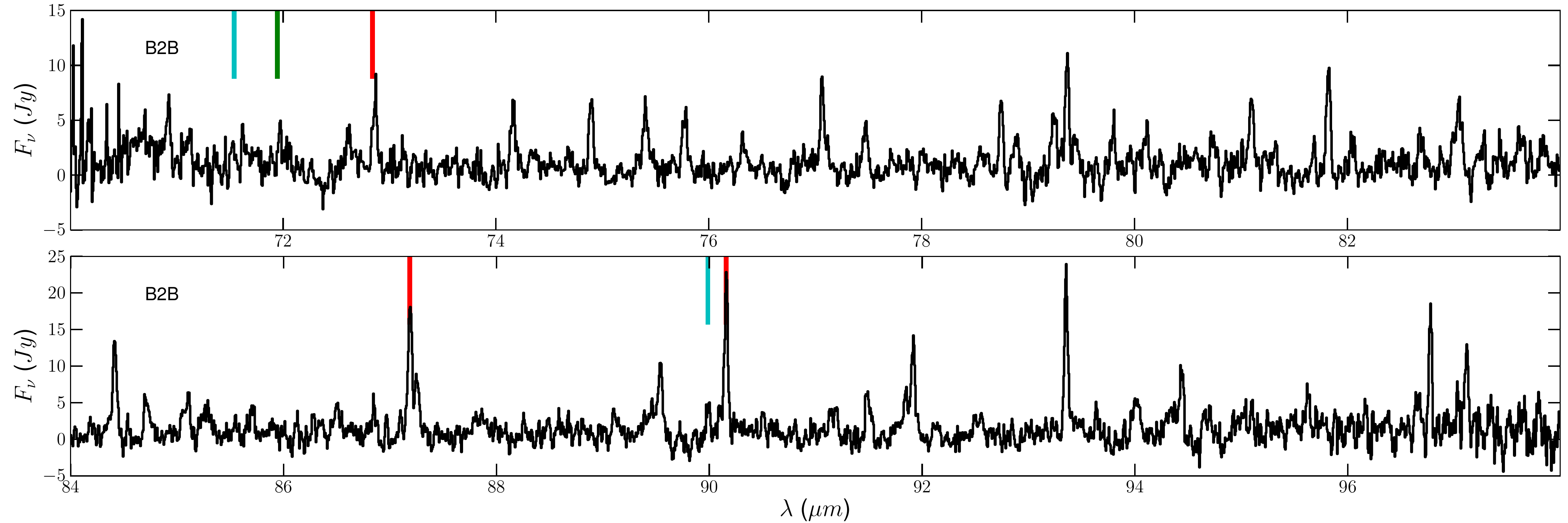}} \\
\end{array}$
\caption{Continuum-subtracted PACS spectrum of LP~And is shown for the blue bands. The line types are the same as Fig.~\ref{fig:mess1}.}
\label{fig:mess11}
\end{figure*}
\begin{figure*}$
\begin{array}{cc}
\resizebox{8.8cm}{24cm}{\includegraphics[angle=90]{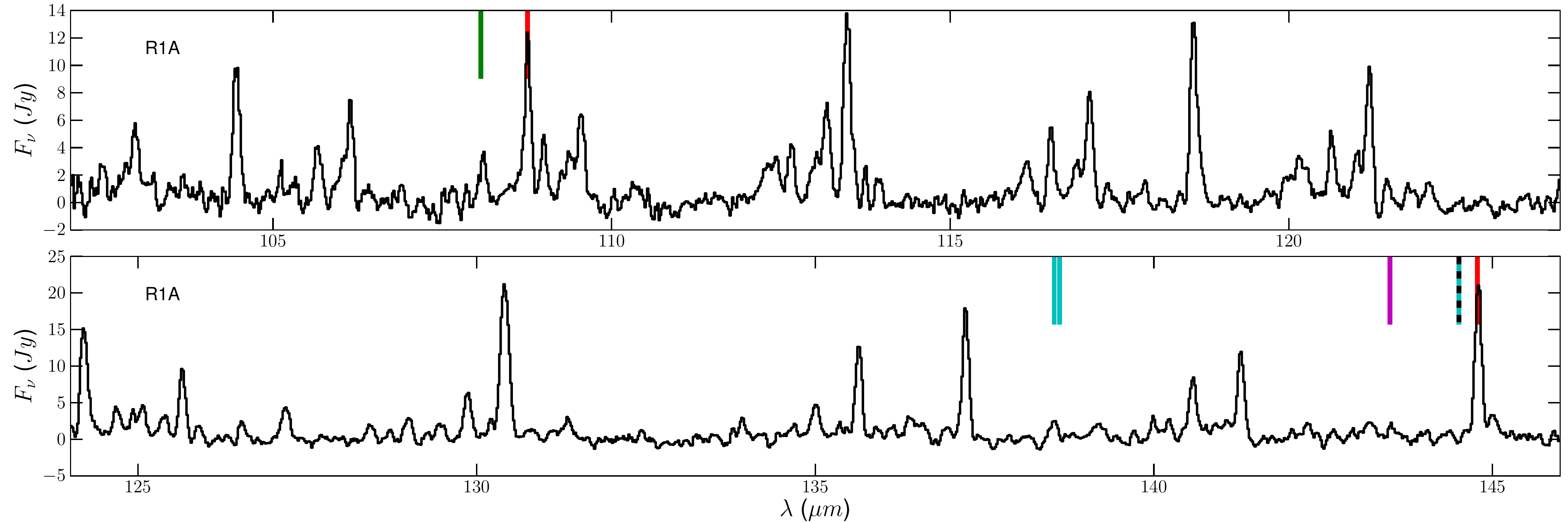}} & \resizebox{8.8cm}{24cm}{\includegraphics[angle=90]{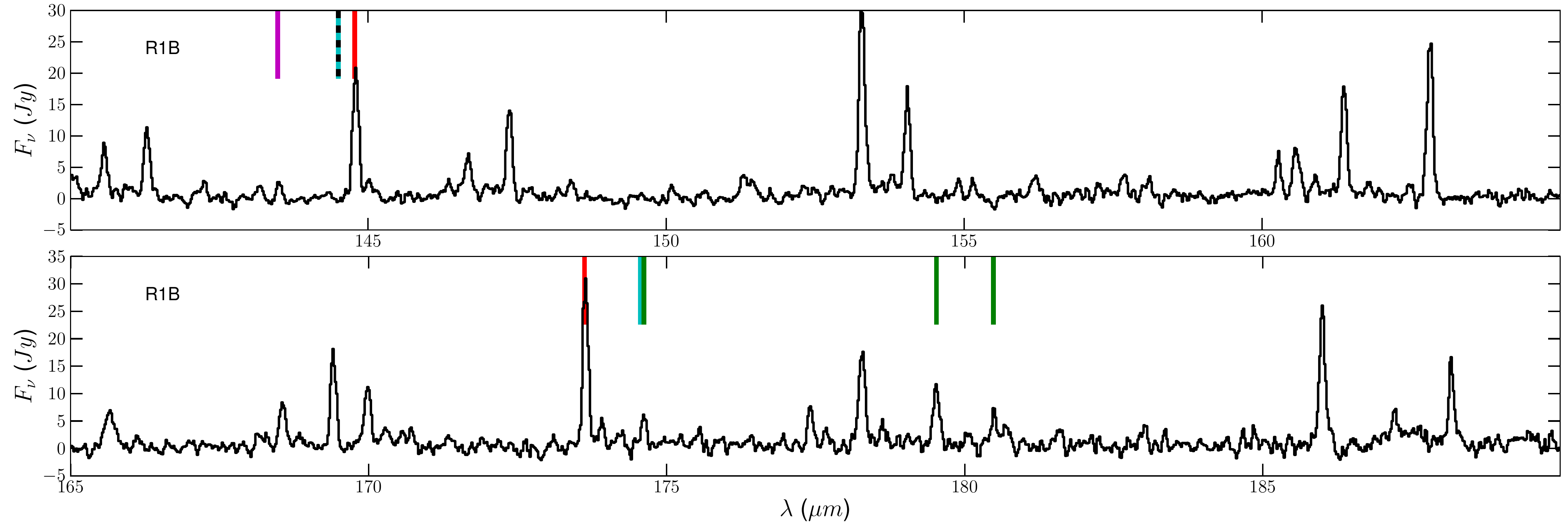}} \\
\end{array}$
\caption{Continuum-subtracted PACS spectrum of LP~And is shown for the red bands. The line types are the same as Fig.~\ref{fig:mess1}.}
\label{fig:mess12}
\end{figure*}
\clearpage
\begin{figure*}[!t]
\resizebox{\hsize}{12cm}{\includegraphics{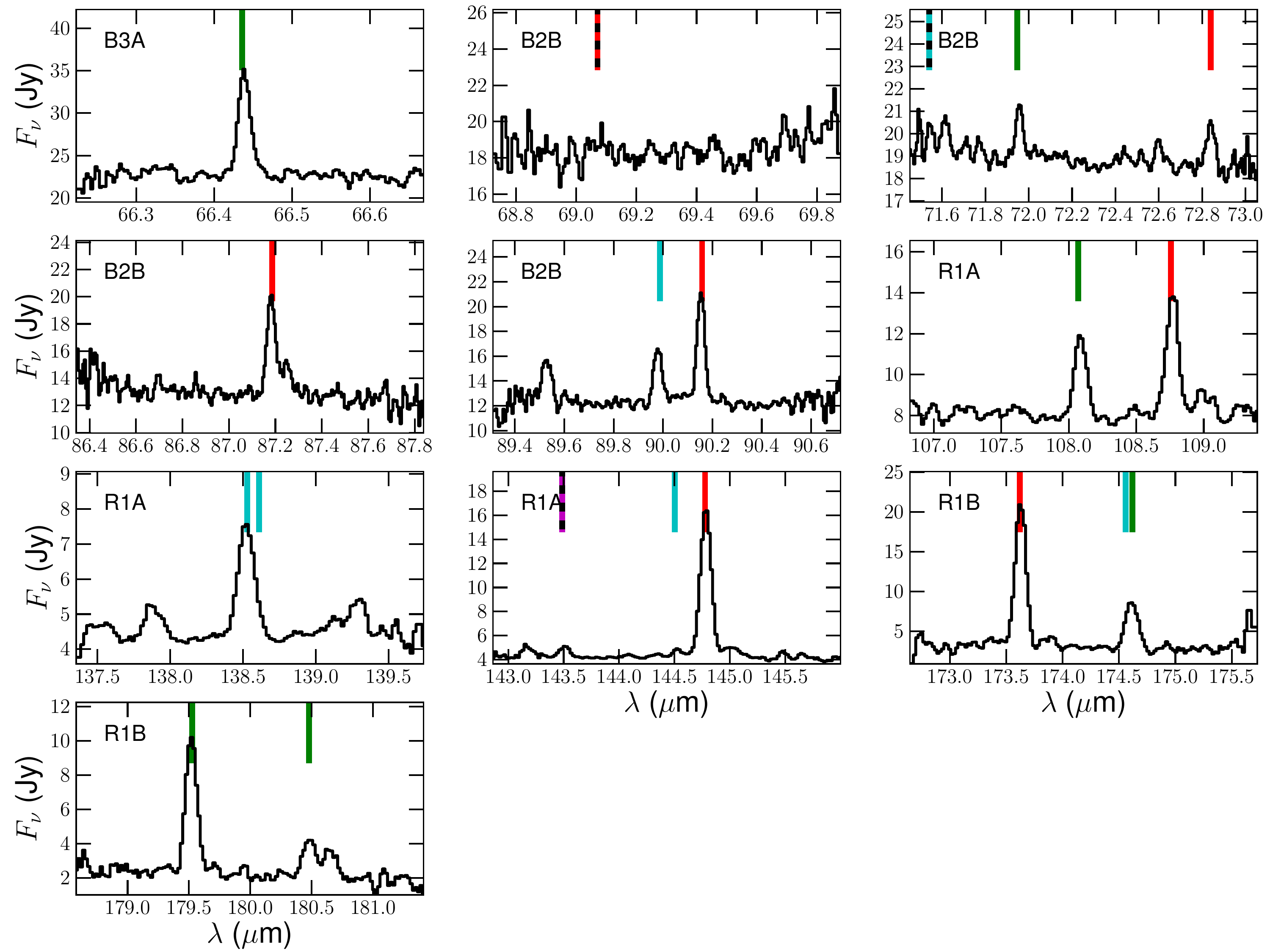}} 
\caption{Line scans of V384~Per are shown in black. The vertical lines indicate molecular identifications according to Table~\ref{table:intintot2old} and \ref{table:intintot2new}: CO in red, $^{13}$CO in magenta, ortho-\water in green, and para-\water in cyan. If a black dashed line is superimposed over the identification line, the transition was not detected by our line-fitting algorithm. The PACS band is indicated in the top left.}
\vspace{1.0cm}
\label{fig:ot2_1}
\resizebox{\hsize}{12cm}{\includegraphics{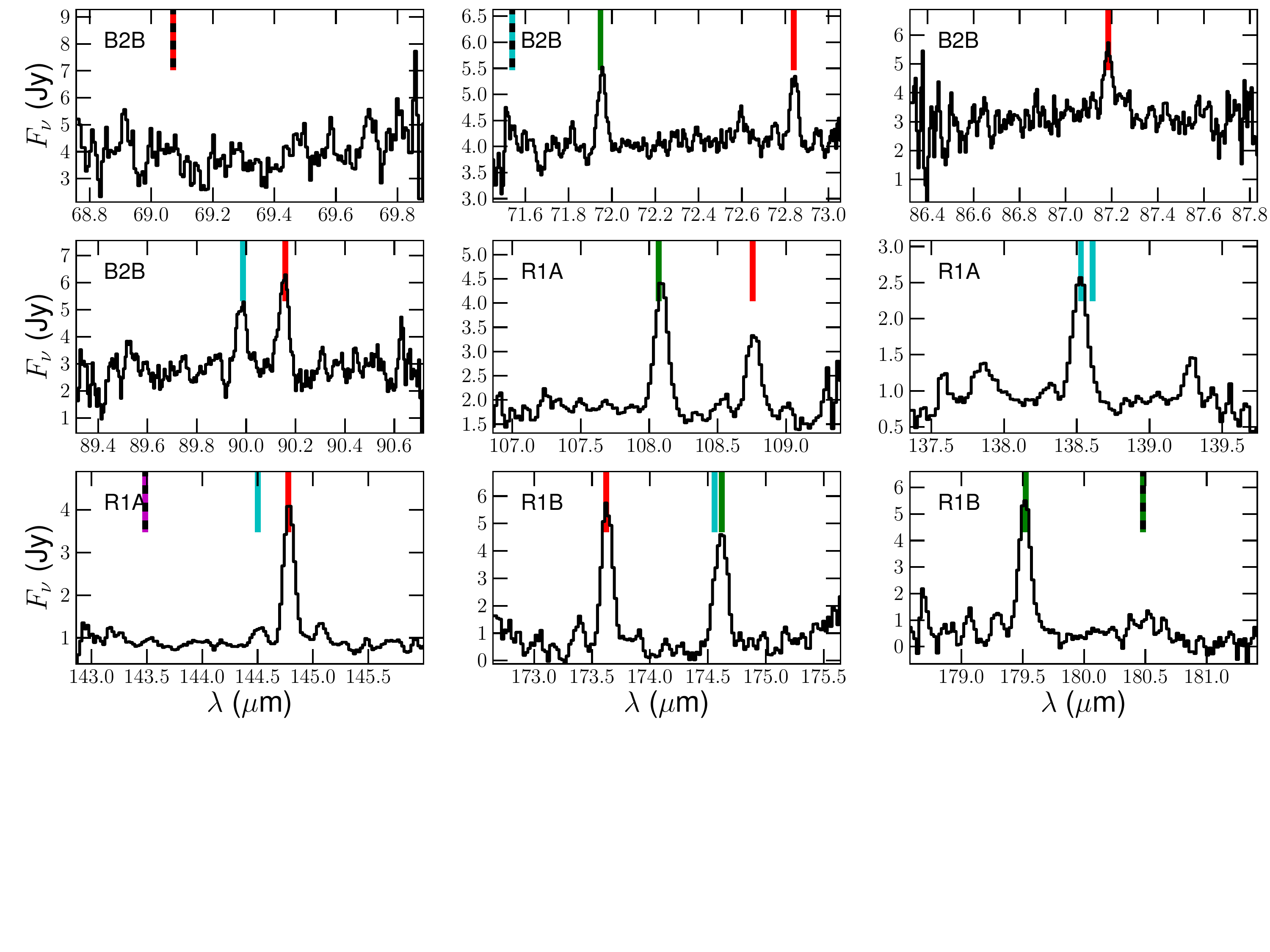}} 
%\vspace{-3.3cm}
\caption{Line scans of S~Aur. The line types are the same as Fig.~\ref{fig:ot2_1}.}
\label{fig:ot2_2}
\end{figure*}

\begin{figure*}[!t]
%\vspace{-0.3cm}
\resizebox{\hsize}{12cm}{\includegraphics{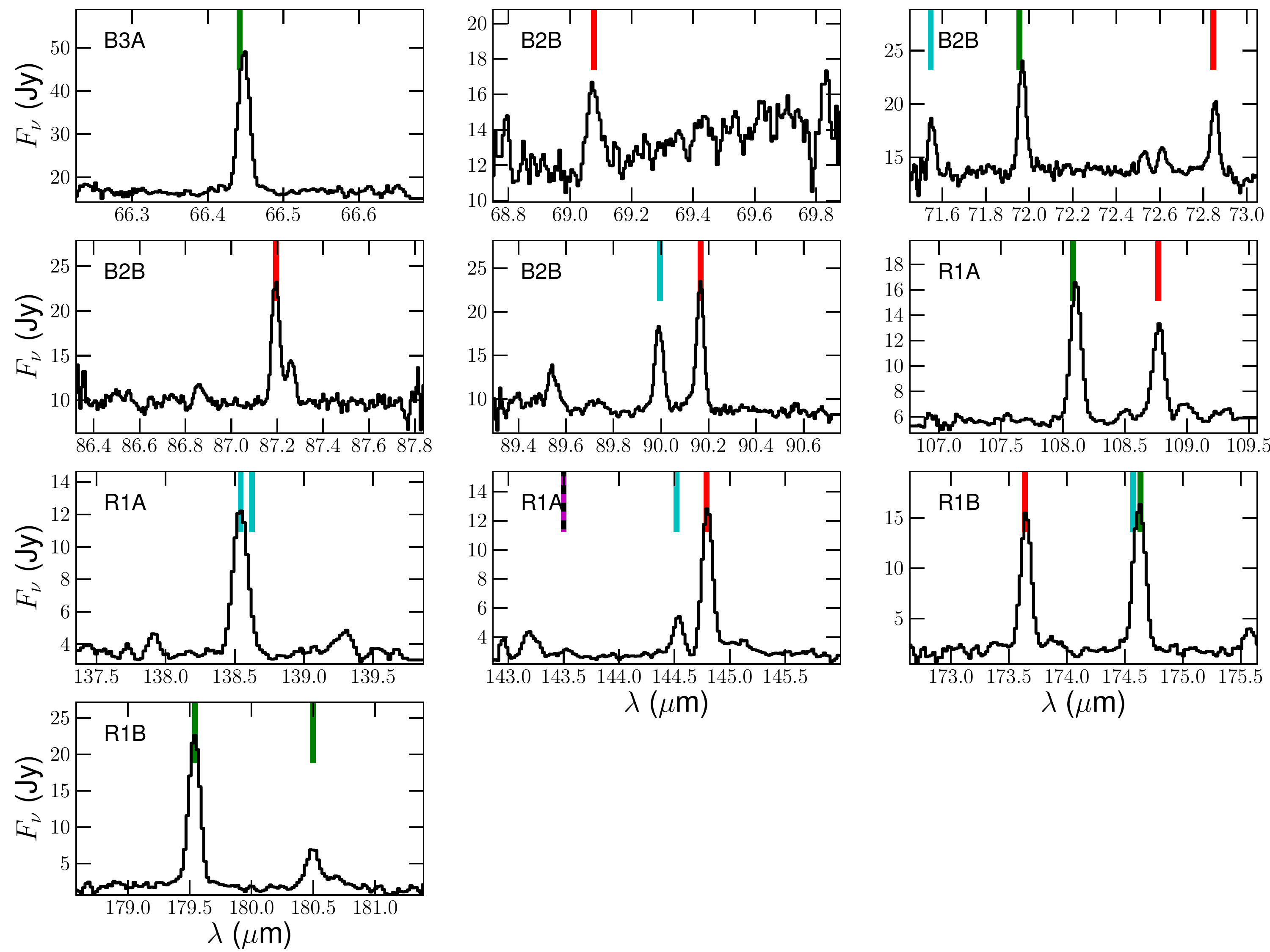}} 
%\vspace{-0.5cm}
\caption{Line scans of R~Lep. The line types are the same as Fig.~\ref{fig:ot2_1}.}
\label{fig:ot2_3}
\vspace{0.3cm}
\resizebox{\hsize}{12cm}{\includegraphics{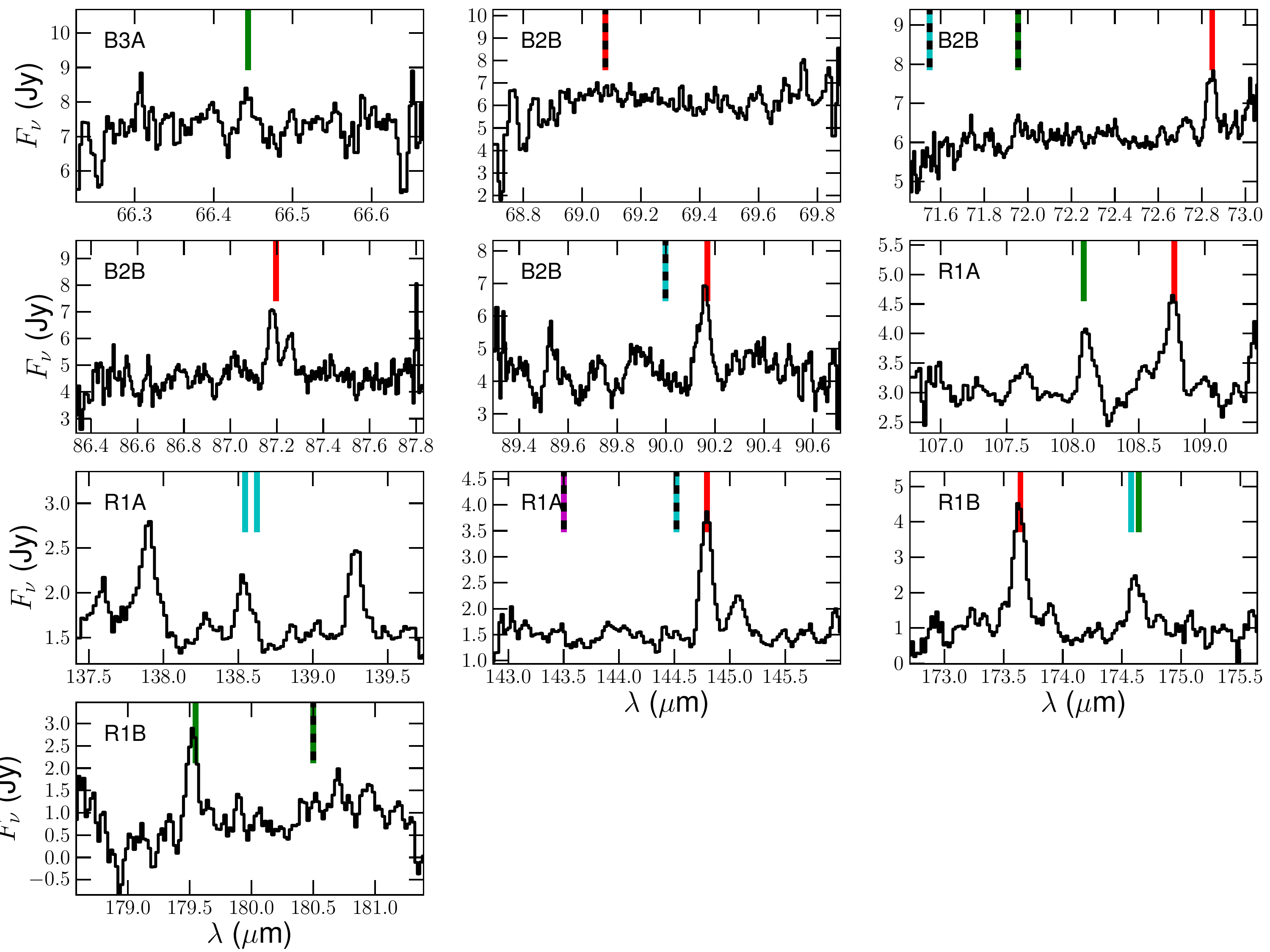}} 
%\vspace{-0.5cm}
\caption{Line scans of W~Ori. The line types are the same as Fig.~\ref{fig:ot2_1}.}
%\vspace{-0.3cm}
\label{fig:ot2_4}
\end{figure*}

\begin{figure*}[!t]
%\vspace{-0.3cm}
\resizebox{\hsize}{12cm}{\includegraphics{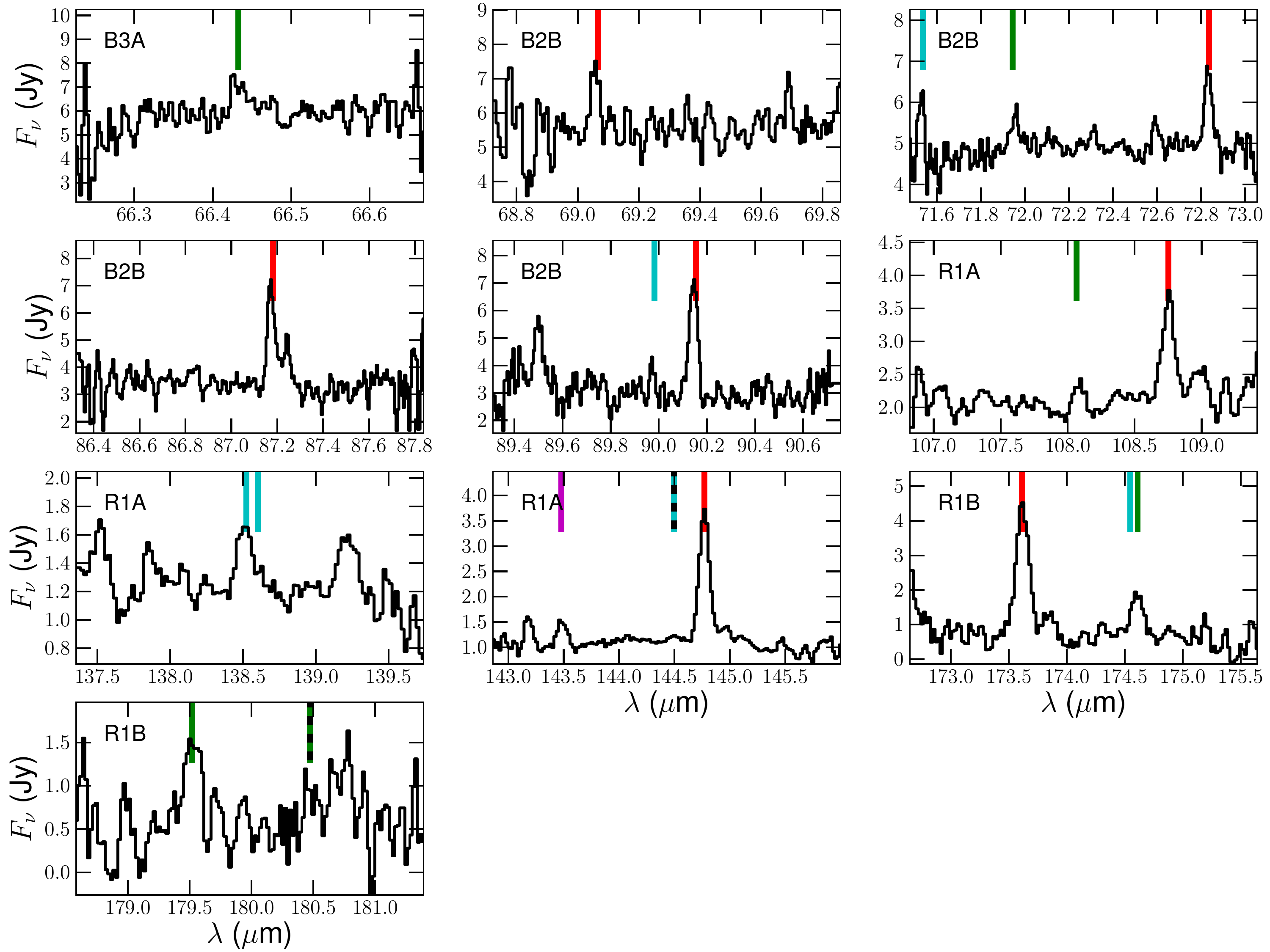}} 
%\vspace{-0.5cm}
\caption{Line scans of U~Hya. The line types are the same as Fig.~\ref{fig:ot2_1}.}
\label{fig:ot2_5}
\vspace{0.3cm}
\resizebox{\hsize}{12cm}{\includegraphics{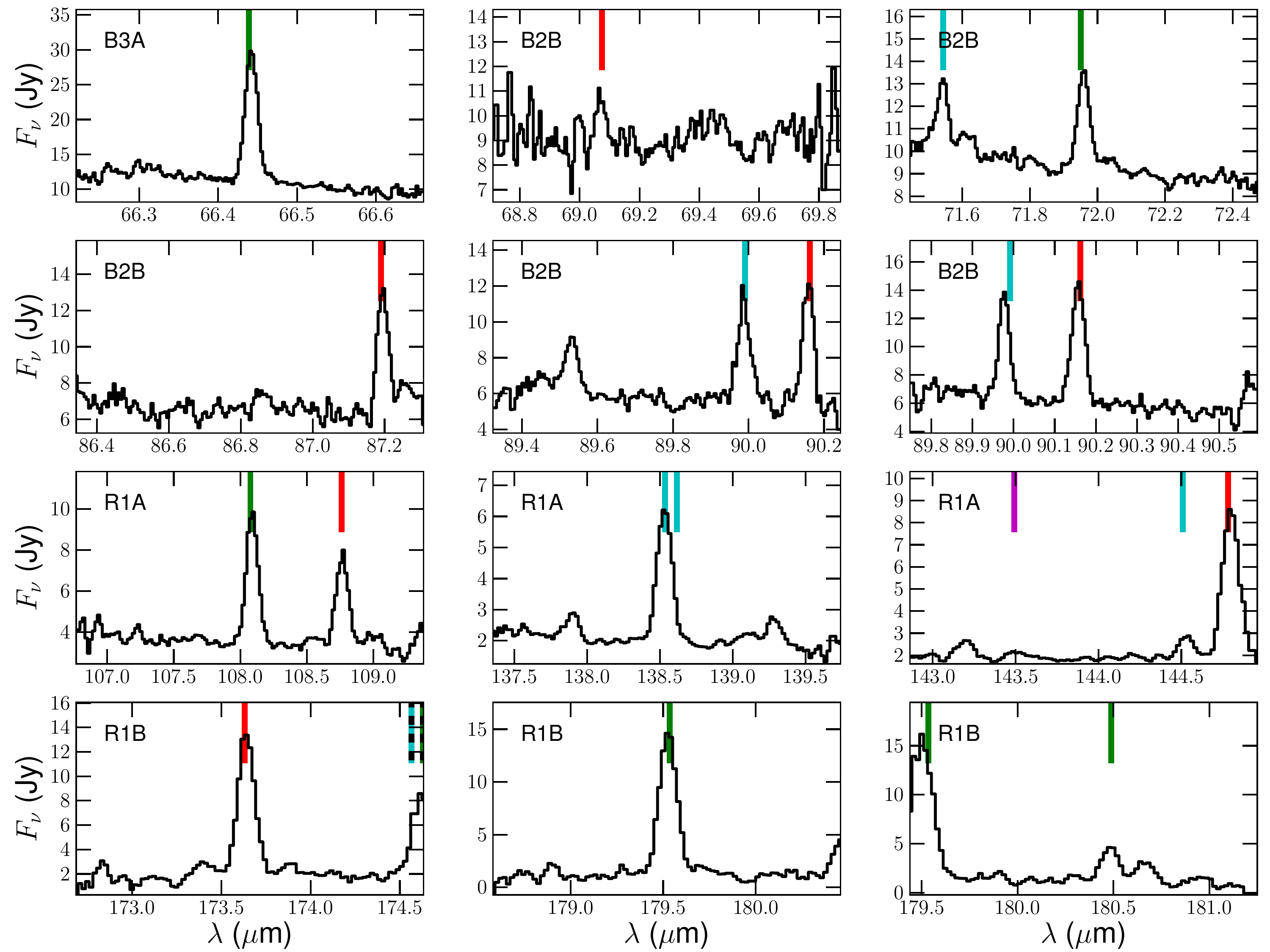}} 
%\vspace{-0.5cm}
\caption{Line scans of QZ~Mus. The line types are the same as Fig.~\ref{fig:ot2_1}.}
%\vspace{-0.3cm}
\label{fig:ot2_6}
\end{figure*}

\begin{figure*}[!t]
%\vspace{-0.3cm}
\resizebox{\hsize}{12cm}{\includegraphics{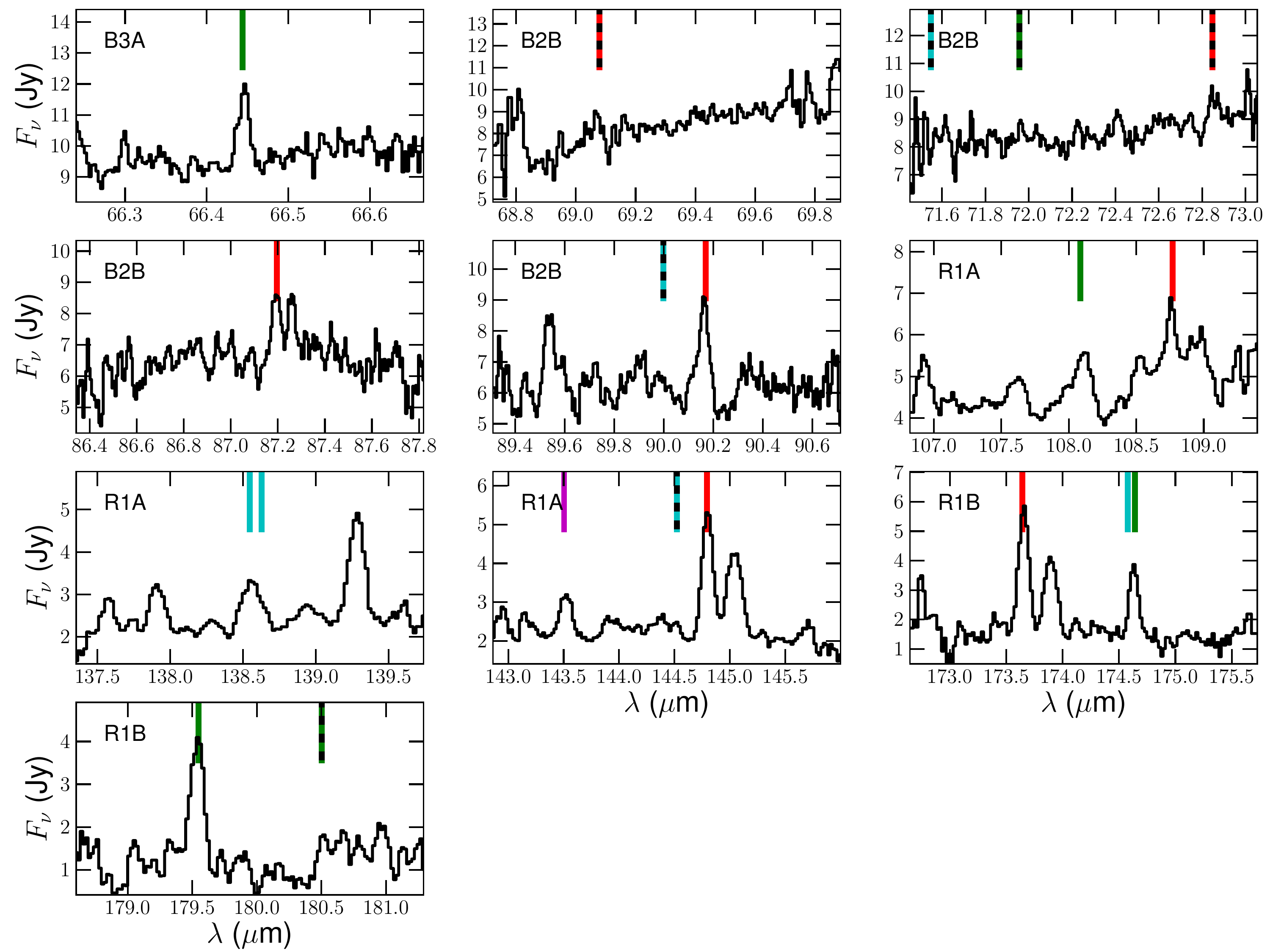}} 
%\vspace{-0.5cm}
\caption{Line scans of Y~CVn. The line types are the same as Fig.~\ref{fig:ot2_1}.}
\label{fig:ot2_7}
\vspace{0.3cm}
\resizebox{\hsize}{12cm}{\includegraphics{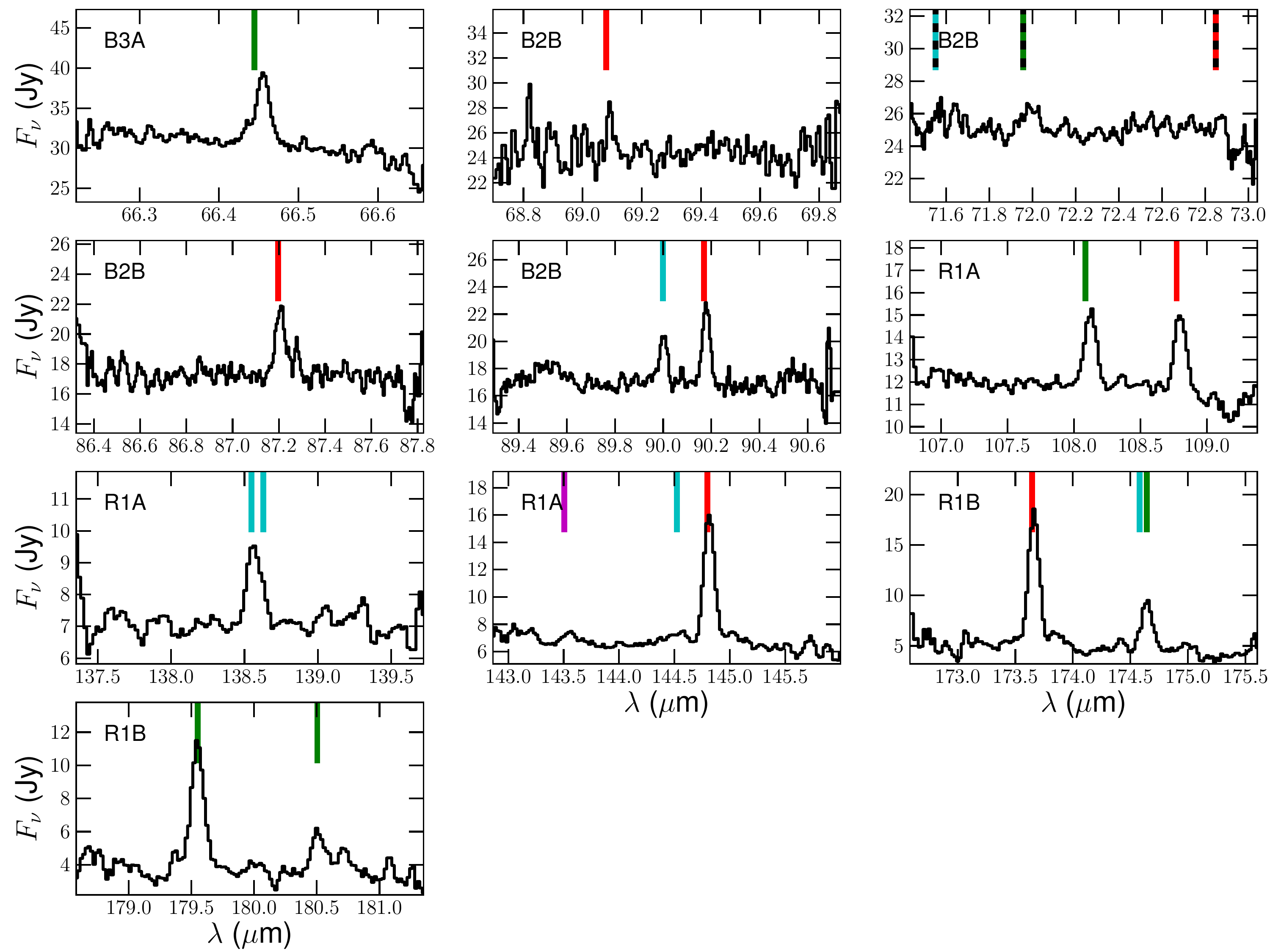}} 
%\vspace{-0.5cm}
\caption{Line scans of AFGL~4202. The line types are the same as Fig.~\ref{fig:ot2_1}.}
%\vspace{-0.3cm}
\label{fig:ot2_8}
\end{figure*}

\begin{figure*}[!t]
%\vspace{-0.3cm}
\resizebox{\hsize}{12cm}{\includegraphics{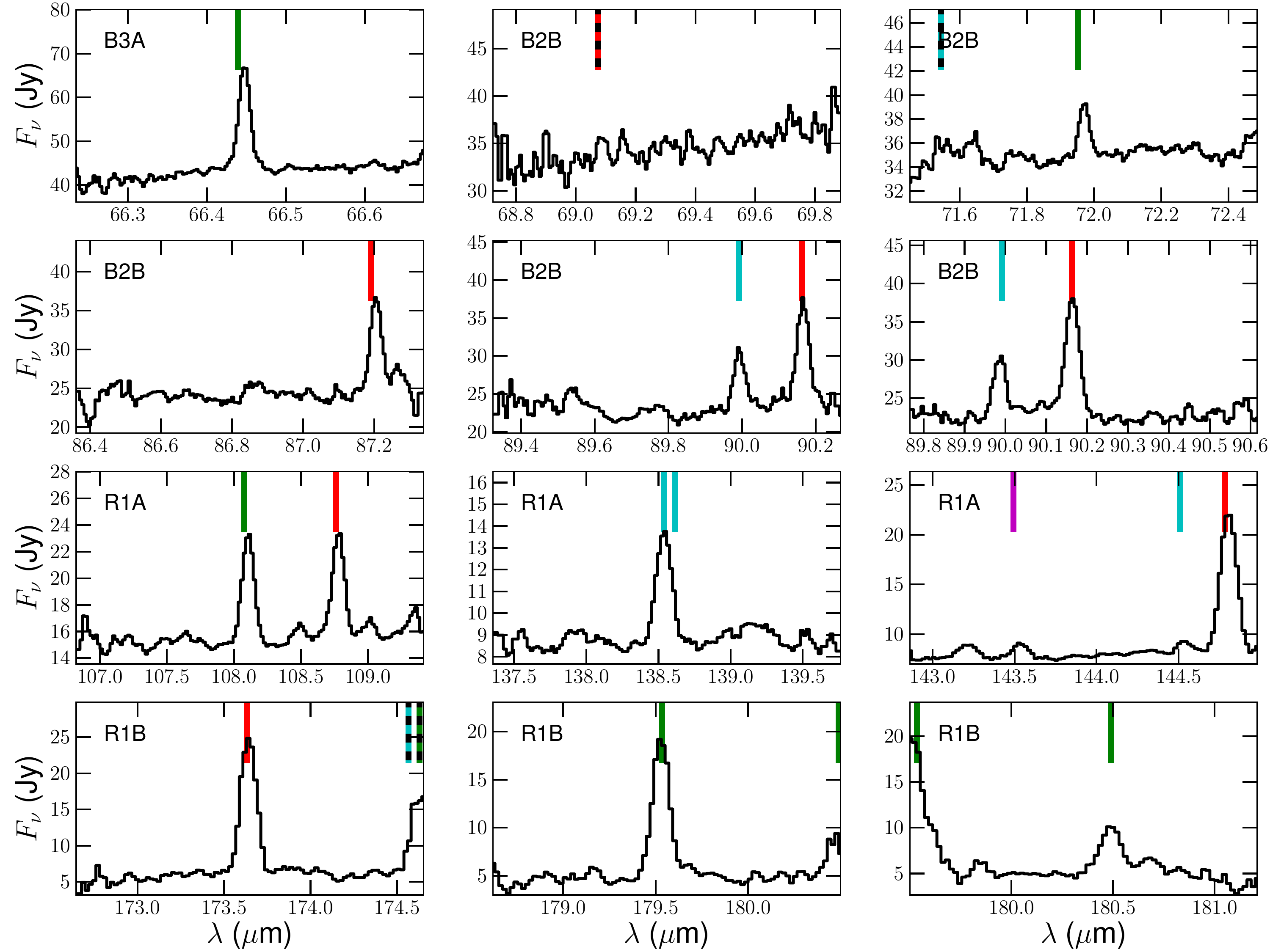}} 
%\vspace{-0.5cm}
\caption{Line scans of V821~Her. The line types are the same as Fig.~\ref{fig:ot2_1}.}
\label{fig:ot2_9}
\vspace{0.3cm}
\resizebox{\hsize}{12cm}{\includegraphics{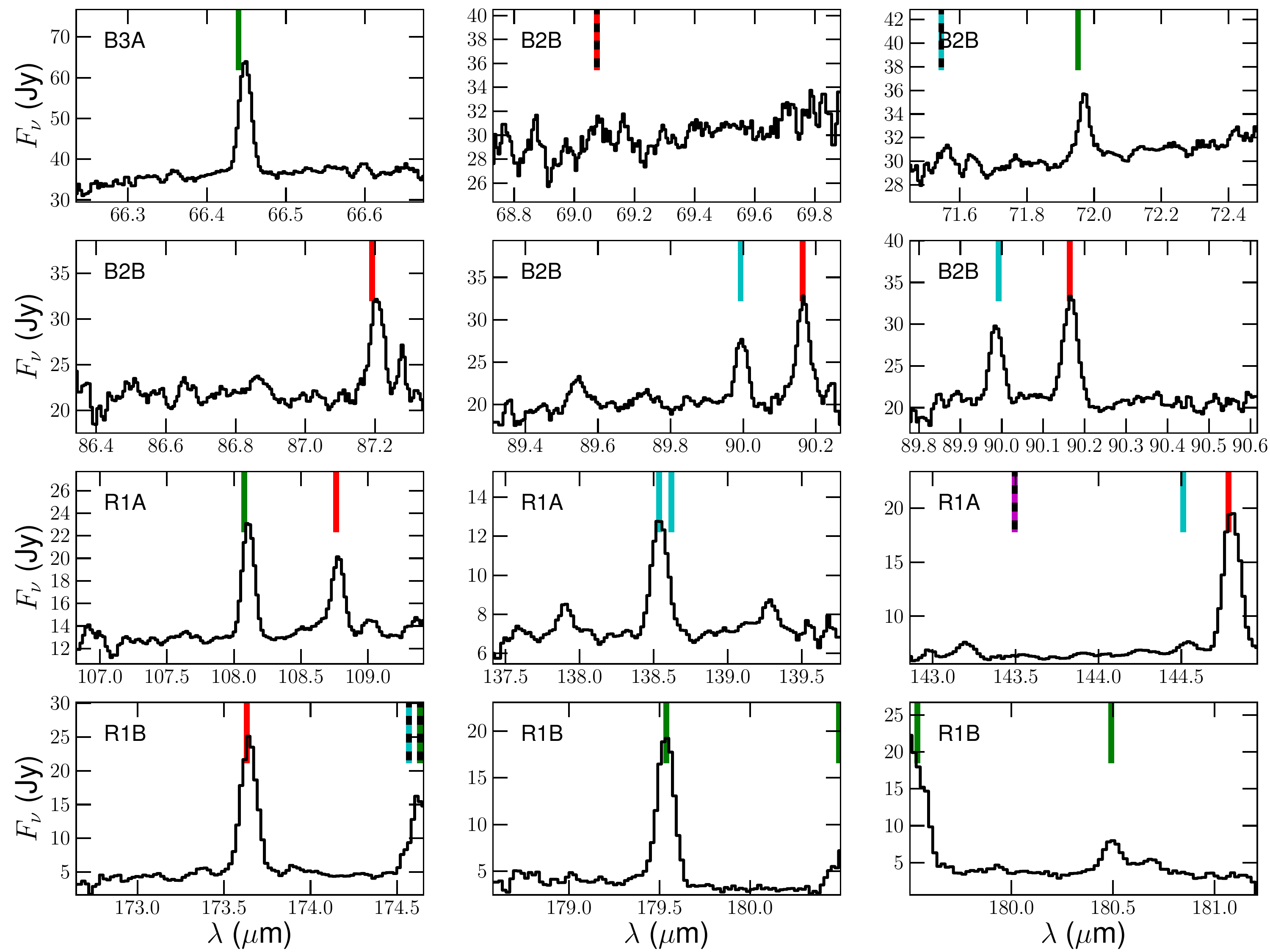}} 
%\vspace{-0.5cm}
\caption{Line scans of V1417~Aql. The line types are the same as Fig.~\ref{fig:ot2_1}.}
%\vspace{-0.3cm}
\label{fig:ot2_10}
\end{figure*}

\begin{figure*}[!t]
%\vspace{-0.3cm}
\resizebox{\hsize}{12cm}{\includegraphics{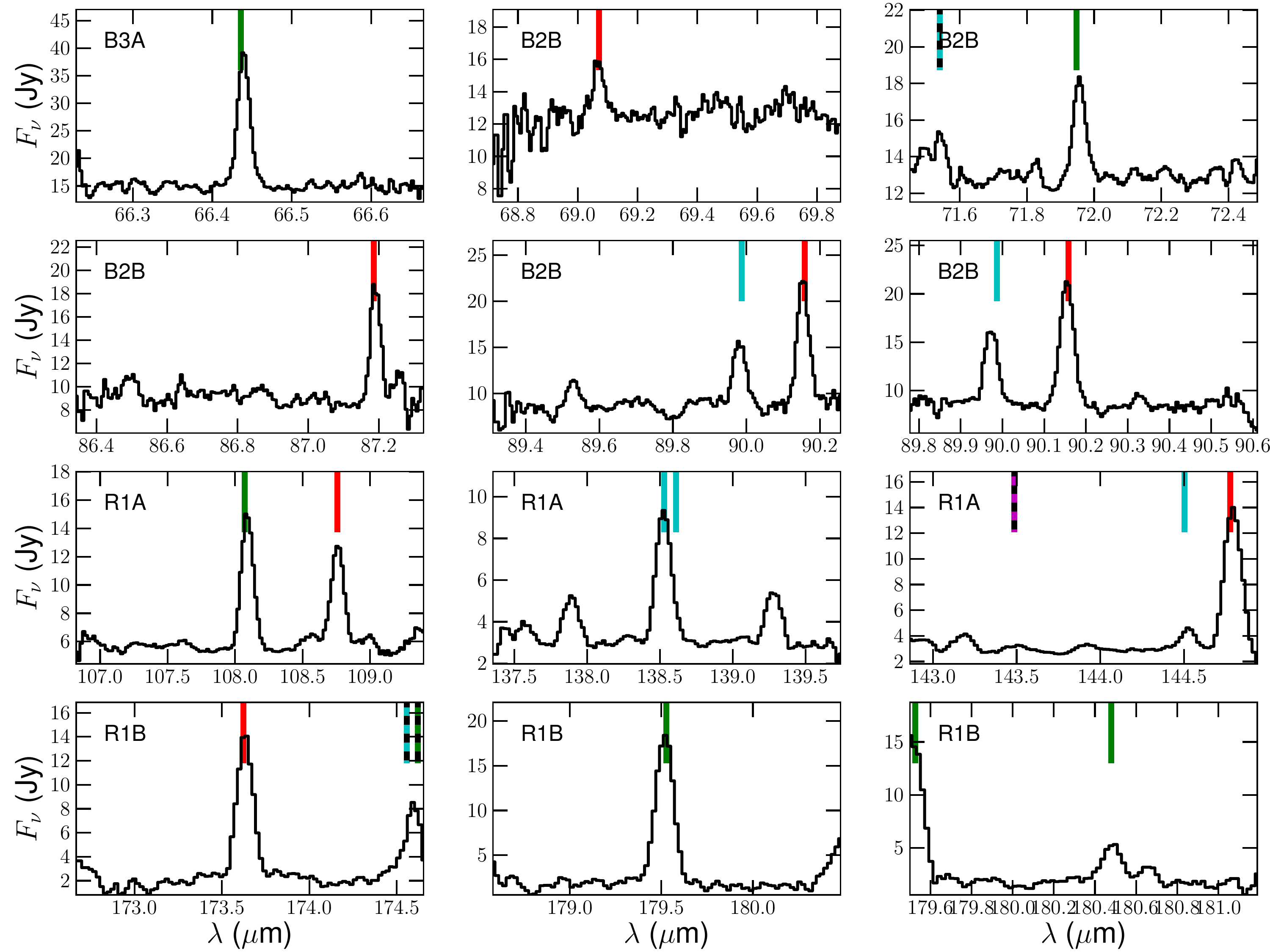}} 
%\vspace{-0.5cm}
\caption{Line scans of S~Cep. The line types are the same as Fig.~\ref{fig:ot2_1}.}
\label{fig:ot2_11}
\vspace{0.3cm}
\resizebox{\hsize}{12cm}{\includegraphics{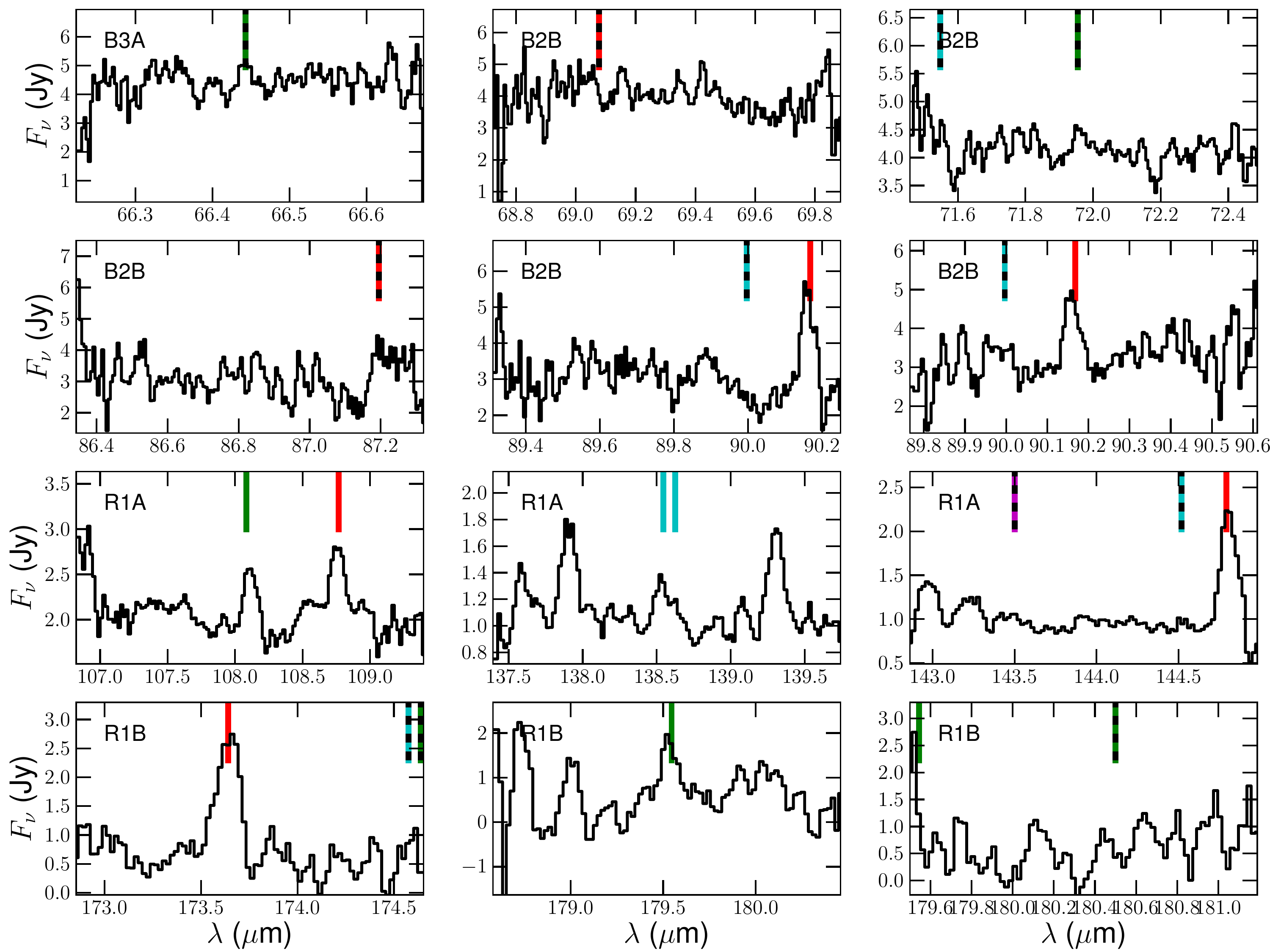}} 
%\vspace{-0.5cm}
\caption{Line scans of RV~Cyg. The line types are the same as Fig.~\ref{fig:ot2_1}.}
%\vspace{-0.3cm}
\label{fig:ot2_12}
\end{figure*}
\newpage
\begin{figure*}[!t]
%\vspace{-0.3cm}
\resizebox{\hsize}{12cm}{\includegraphics{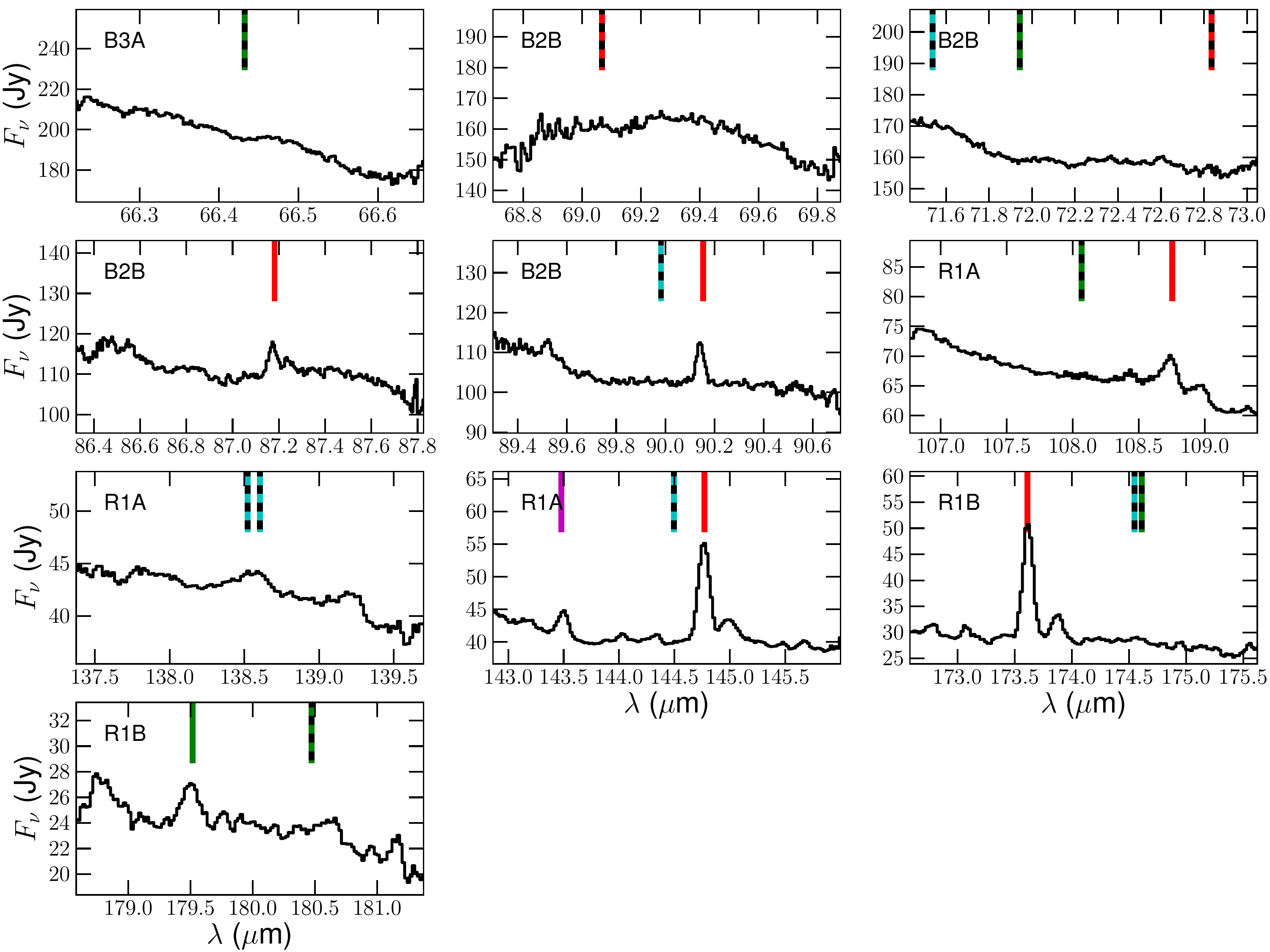}} 
%\vspace{-0.5cm}
\caption{Line scans of LL~Peg. The line types are the same as Fig.~\ref{fig:ot2_1}.}
\label{fig:ot2_13}
\end{figure*}

\onecolumn
%\begin{longtab}
\begin{landscape}
\tiny
\begin{longtable}{lllrrrrrrr}
\caption{{Integrated line strengths $I_\mathrm{int}$ (W m$^{-2}$) for a selection of lines in the PACS spectra of carbon stars in the MESS~sample. The rest wavelength $\lambda_0$ (\mic) of the transition is indicated. The percentages between brackets give the uncertainty on $I_\mathrm{int}$, which includes both the fitting uncertainty and the PACS absolute-flux-calibration uncertainty of 20\%. Line strengths indicated with $\star$ are flagged for potential line blends. Transitions that might cause the line blend are mentioned immediately below the flagged transition. Molecular transitions indicated in red coincide with the wavelength range of the line scans in the OT2~program. These detections are not expected to be affected by blends with molecules other than CO and \water. Detections listed in black have to be treated with caution, as they have not been checked for potential blending with other molecules. }}\label{table:intintmess}\\
\hline\hline
&&&&RW~LMi&V~Hya&II~Lup&V~Cyg&LL~Peg&LP~And \\\hline
PACS&Molecule&Rotational&$\lambda_0$&\multicolumn{6}{c}{$I_\mathrm{int}$} \\
band&&transition&$\mu$m&\multicolumn{6}{c}{(W m$^{-2}$)} \\\hline
\endfirsthead
\caption{continued.}\\
\hline\hline
&&&&RW~LMi&V~Hya&II~Lup&V~Cyg&LL~Peg&LP~And \\\hline
PACS&Molecule&Rotational&$\lambda_0$&\multicolumn{6}{c}{$I_\mathrm{int}$} \\
band&&transition&$\mu$m&\multicolumn{6}{c}{(W m$^{-2}$)} \\\hline
\endhead
\hline 
% \multicolumn{10}{c}{\tablefoot{\tablefoottext{$\star$}{\parbox{\LTcapwidth}{Line strengths flagged for potential line blends. Transitions that might cause the line blend are mentioned immediately below the flagged transition.}}
% }}
\endfoot
R1B&p-H$_2$O&$J_{\mathrm{K}_\mathrm{a}, \mathrm{K}_\mathrm{c}}=4_{1,3} - 4_{0,4}$&187.11&/&$^\star$~7.71e-17 (26.2\%)&5.83e-17 (30.5\%)&/&/&4.23e-17 (25.0\%)\\
&CO&$J=14 - 13$&186.00&9.19e-16 (20.3\%)&4.61e-16 (20.2\%)&4.57e-16 (20.4\%)&1.98e-16 (21.7\%)&2.29e-16 (21.1\%)&2.64e-16 (20.7\%)\\
&$^{13}$CO&$J=15 - 14$&181.61&6.11e-17 (26.1\%)&/&1.28e-16 (21.9\%)&/&5.41e-17 (28.8\%)&/\\
&{\color{red}o-H$_2$O}&{\color{red}$J_{\mathrm{K}_\mathrm{a}, \mathrm{K}_\mathrm{c}}=2_{2,1} - 2_{1,2}$}&{\color{red}180.49}&{\color{red}1.43e-16 (24.6\%)}&{\color{red}$^\star$~9.43e-17 (24.5\%)}&{\color{red}4.73e-17 (33.4\%)}&{\color{red}/}&{\color{red}/}&{\color{red}5.84e-17 (25.8\%)}\\
&{\color{red}o-H$_2$O}&{\color{red}$J_{\mathrm{K}_\mathrm{a}, \mathrm{K}_\mathrm{c}}=2_{1,2} - 1_{0,1}$}&{\color{red}179.53}&{\color{red}4.47e-16 (20.5\%)}&{\color{red}2.98e-16 (20.3\%)}&{\color{red}1.62e-16 (21.6\%)}&{\color{red}1.72e-16 (20.9\%)}&{\color{red}/}&{\color{red}1.25e-16 (21.7\%)}\\
&{\color{red}o-H$_2$O}&{\color{red}$J_{\mathrm{K}_\mathrm{a}, \mathrm{K}_\mathrm{c}}=3_{0,3} - 2_{1,2}$}&{\color{red}174.63}&{\color{red}$^\star$~3.01e-16 (22.0\%)}&{\color{red}$^\star$~1.86e-16 (21.1\%)}&{\color{red}$^\star$~1.08e-16 (24.1\%)}&{\color{red}$^\star$~8.27e-17 (23.6\%)}&{\color{red}/}&{\color{red}$^\star$~5.49e-17 (26.9\%)}\\
&{\color{red}p-H$_2$O}&{\color{red}$J_{\mathrm{K}_\mathrm{a}, \mathrm{K}_\mathrm{c}}=5_{3,3} - 6_{0,6}$}&{\color{red}174.61}&{\color{red}Blended}&{\color{red}Blended}&{\color{red}Blended}&{\color{red}Blended}&{\color{red}/}&{\color{red}Blended}\\
&{\color{red}CO}&{\color{red}$J=15 - 14$}&{\color{red}173.63}&{\color{red}9.88e-16 (20.2\%)}&{\color{red}6.10e-16 (20.1\%)}&{\color{red}5.12e-16 (20.2\%)}&{\color{red}2.21e-16 (20.4\%)}&{\color{red}2.56e-16 (20.8\%)}&{\color{red}3.45e-16 (20.3\%)}\\
&$^{13}$CO&$J=16 - 15$&170.29&1.01e-16 (28.2\%)&/&9.28e-17 (24.8\%)&$^\star$~1.81e-17 (48.5\%)&4.60e-17 (29.4\%)&$^\star$~6.49e-17 (24.5\%)\\
&p-H$_2$O&$J_{\mathrm{K}_\mathrm{a}, \mathrm{K}_\mathrm{c}}=6_{3,3} - 6_{2,4}$&170.14&/&/&/&Blended&/&/\\
&CO&$J=16 - 15$&162.81&1.05e-15 (20.1\%)&6.35e-16 (20.1\%)&5.45e-16 (20.3\%)&3.22e-16 (20.3\%)&2.83e-16 (20.6\%)&3.42e-16 (20.1\%)\\
&o-H$_2$O&$J_{\mathrm{K}_\mathrm{a}, \mathrm{K}_\mathrm{c}}=5_{3,2} - 5_{2,3}$&160.51&2.65e-16 (21.3\%)&1.20e-16 (23.4\%)&1.20e-16 (23.4\%)&/&9.01e-17 (25.3\%)&1.17e-16 (21.5\%)\\
&$^{13}$CO&$J=17 - 16$&160.30&1.36e-16 (24.0\%)&/&1.89e-16 (21.6\%)&/&1.04e-16 (24.4\%)&6.79e-17 (22.8\%)\\
&o-H$_2$O&$J_{\mathrm{K}_\mathrm{a}, \mathrm{K}_\mathrm{c}}=8_{4,5} - 7_{5,2}$&159.05&$^\star$~6.87e-17 (29.5\%)&/&2.24e-17 (37.1\%)&/&/&/\\
&o-H$_2$O&$J_{\mathrm{K}_\mathrm{a}, \mathrm{K}_\mathrm{c}}=5_{2,3} - 4_{3,2}$&156.27&$^\star$~2.95e-16 (20.8\%)&$^\star$~9.70e-17 (22.0\%)&$^\star$~6.93e-17 (25.5\%)&$^\star$~7.01e-17 (28.9\%)&/&$^\star$~5.49e-17 (28.8\%)\\
&p-H$_2$O&$J_{\mathrm{K}_\mathrm{a}, \mathrm{K}_\mathrm{c}}=3_{2,2} - 3_{1,3}$&156.19&Blended&Blended&Blended&Blended&/&Blended\\
&CO&$J=17 - 16$&153.27&1.30e-15 (20.2\%)&7.15e-16 (20.1\%)&5.61e-16 (20.4\%)&2.90e-16 (20.3\%)&3.51e-16 (20.7\%)&4.55e-16 (20.3\%)\\
&$^{13}$CO&$J=18 - 17$&151.43&2.23e-16 (23.1\%)&4.84e-17 (27.9\%)&1.24e-16 (22.7\%)&$^\star$~5.95e-17 (28.7\%)&3.61e-17 (40.8\%)&$^\star$~5.68e-17 (31.6\%)\\
&p-H$_2$O&$J_{\mathrm{K}_\mathrm{a}, \mathrm{K}_\mathrm{c}}=4_{3,1} - 4_{2,2}$&146.92&9.49e-17 (27.0\%)&2.87e-17 (32.7\%)&/&/&$^\star$~5.42e-17 (34.3\%)&/\\
&{\color{red}CO}&{\color{red}$J=18 - 17$}&{\color{red}144.78}&{\color{red}1.07e-15 (20.4\%)}&{\color{red}5.83e-16 (20.2\%)}&{\color{red}4.27e-16 (20.4\%)}&{\color{red}3.07e-16 (20.3\%)}&{\color{red}2.73e-16 (21.5\%)}&{\color{red}3.67e-16 (20.2\%)}\\
&{\color{red}$^{13}$CO}&{\color{red}$J=19 - 18$}&{\color{red}143.49}&{\color{red}/}&{\color{red}/}&{\color{red}1.08e-16 (21.9\%)}&{\color{red}/}&{\color{red}/}&{\color{red}4.72e-17 (26.2\%)}\\\hline
R1A&{\color{red}CO}&{\color{red}$J=18 - 17$}&{\color{red}144.78}&{\color{red}9.84e-16 (20.4\%)}&{\color{red}5.52e-16 (20.1\%)}&{\color{red}4.59e-16 (20.3\%)}&{\color{red}2.66e-16 (20.6\%)}&{\color{red}2.53e-16 (20.5\%)}&{\color{red}3.56e-16 (20.2\%)}\\
&{\color{red}$^{13}$CO}&{\color{red}$J=19 - 18$}&{\color{red}143.49}&{\color{red}/}&{\color{red}/}&{\color{red}1.19e-16 (21.3\%)}&{\color{red}/}&{\color{red}/}&{\color{red}2.62e-17 (36.1\%)}\\
&{\color{red}p-H$_2$O}&{\color{red}$J_{\mathrm{K}_\mathrm{a}, \mathrm{K}_\mathrm{c}}=8_{4,4} - 7_{5,3}$}&{\color{red}138.64}&{\color{red}$^\star$~2.58e-16 (22.0\%)}&{\color{red}$^\star$~1.74e-16 (21.0\%)}&{\color{red}$^\star$~9.89e-17 (25.9\%)}&{\color{red}$^\star$~1.14e-16 (22.9\%)}&{\color{red}/}&{\color{red}$^\star$~5.62e-17 (27.7\%)}\\
&{\color{red}p-H$_2$O}&{\color{red}$J_{\mathrm{K}_\mathrm{a}, \mathrm{K}_\mathrm{c}}=3_{1,3} - 2_{0,2}$}&{\color{red}138.53}&{\color{red}Blended}&{\color{red}Blended}&{\color{red}Blended}&{\color{red}Blended}&{\color{red}/}&{\color{red}Blended}\\
&CO&$J=19 - 18$&137.20&9.56e-16 (20.1\%)&5.64e-16 (20.1\%)&3.97e-16 (20.4\%)&2.91e-16 (20.4\%)&2.83e-16 (20.7\%)&3.22e-16 (20.2\%)\\
&o-H$_2$O&$J_{\mathrm{K}_\mathrm{a}, \mathrm{K}_\mathrm{c}}=3_{3,0} - 3_{2,1}$&136.50&$^\star$~2.44e-16 (27.8\%)&$^\star$~5.84e-17 (43.2\%)&3.45e-17 (36.8\%)&$^\star$~6.86e-17 (30.3\%)&/&3.53e-17 (44.0\%)\\
&$^{13}$CO&$J=20 - 19$&136.35&2.02e-16 (26.2\%)&Blended&1.45e-16 (22.2\%)&2.12e-17 (43.6\%)&/&5.50e-17 (32.5\%)\\
&o-H$_2$O&$J_{\mathrm{K}_\mathrm{a}, \mathrm{K}_\mathrm{c}}=5_{1,4} - 5_{0,5}$&134.94&3.36e-16 (21.5\%)&1.78e-16 (21.1\%)&1.18e-16 (23.3\%)&5.50e-17 (29.2\%)&6.48e-17 (30.9\%)&9.22e-17 (23.2\%)\\
&o-H$_2$O&$J_{\mathrm{K}_\mathrm{a}, \mathrm{K}_\mathrm{c}}=4_{2,3} - 4_{1,4}$&132.41&$^\star$~2.22e-16 (27.4\%)&9.81e-17 (25.2\%)&/&/&/&/\\
&CO&$J=20 - 19$&130.37&$^\star$~1.64e-15 (20.1\%)&$^\star$~9.09e-16 (20.1\%)&$^\star$~6.55e-16 (20.2\%)&$^\star$~4.29e-16 (20.3\%)&$^\star$~4.19e-16 (20.4\%)&$^\star$~5.89e-16 (20.1\%)\\
&p-H$_2$O&$J_{\mathrm{K}_\mathrm{a}, \mathrm{K}_\mathrm{c}}=7_{5,3} - 8_{2,6}$&130.32&Blended&Blended&Blended&Blended&Blended&Blended\\
&$^{13}$CO&$J=21 - 20$&129.89&3.00e-16 (23.6\%)&1.35e-16 (23.8\%)&1.93e-16 (21.9\%)&/&1.12e-16 (24.4\%)&1.50e-16 (21.0\%)\\
&p-H$_2$O&$J_{\mathrm{K}_\mathrm{a}, \mathrm{K}_\mathrm{c}}=4_{0,4} - 3_{1,3}$&125.35&2.84e-16 (23.7\%)&1.27e-16 (25.2\%)&$^\star$~9.16e-17 (31.3\%)&5.07e-17 (28.5\%)&/&$^\star$~9.55e-17 (25.2\%)\\
&CO&$J=21 - 20$&124.19&1.02e-15 (20.3\%)&6.02e-16 (20.3\%)&3.96e-16 (20.5\%)&2.56e-16 (20.6\%)&2.16e-16 (22.0\%)&4.07e-16 (20.3\%)\\
&$^{13}$CO&$J=22 - 21$&124.02&/&/&1.12e-16 (25.8\%)&/&/&2.15e-17 (46.6\%)\\
&o-H$_2$O&$J_{\mathrm{K}_\mathrm{a}, \mathrm{K}_\mathrm{c}}=4_{3,2} - 4_{2,3}$&121.72&/&/&/&/&/&$^\star$~4.94e-17 (29.6\%)\\
&$^{13}$CO&$J=23 - 22$&118.66&$^\star$~1.12e-15 (20.9\%)&$^\star$~6.17e-16 (20.3\%)&$^\star$~4.57e-16 (21.2\%)&$^\star$~2.65e-16 (21.0\%)&$^\star$~2.65e-16 (20.9\%)&$^\star$~3.71e-16 (20.5\%)\\
&CO&$J=22 - 21$&118.58&Blended&Blended&Blended&Blended&Blended&Blended\\
&o-H$_2$O&$J_{\mathrm{K}_\mathrm{a}, \mathrm{K}_\mathrm{c}}=7_{3,4} - 6_{4,3}$&116.78&/&$^\star$~9.63e-17 (28.2\%)&5.66e-17 (37.2\%)&3.14e-17 (51.1\%)&$^\star$~1.13e-16 (27.0\%)&$^\star$~1.00e-16 (24.0\%)\\
&p-H$_2$O&$J_{\mathrm{K}_\mathrm{a}, \mathrm{K}_\mathrm{c}}=5_{3,3} - 5_{2,4}$&113.95&/&/&$^\star$~1.26e-16 (35.5\%)&/&/&4.49e-17 (31.4\%)\\
&$^{13}$CO&$J=24 - 23$&113.75&/&/&5.40e-17 (36.1\%)&/&/&3.83e-17 (29.0\%)\\
&o-H$_2$O&$J_{\mathrm{K}_\mathrm{a}, \mathrm{K}_\mathrm{c}}=4_{1,4} - 3_{0,3}$&113.54&$^\star$~1.49e-15 (20.3\%)&$^\star$~8.92e-16 (20.2\%)&$^\star$~4.79e-16 (20.5\%)&$^\star$~4.48e-16 (20.6\%)&$^\star$~2.64e-16 (21.7\%)&$^\star$~4.35e-16 (20.2\%)\\
&CO&$J=23 - 22$&113.46&Blended&Blended&Blended&Blended&Blended&Blended\\
&o-H$_2$O&$J_{\mathrm{K}_\mathrm{a}, \mathrm{K}_\mathrm{c}}=7_{4,3} - 7_{3,4}$&112.51&1.30e-16 (47.6\%)&/&$^\star$~1.49e-16 (53.7\%)&/&$^\star$~1.27e-16 (35.2\%)&/\\
&{\color{red}CO}&{\color{red}$J=24 - 23$}&{\color{red}108.76}&{\color{red}1.03e-15 (20.9\%)}&{\color{red}5.71e-16 (20.5\%)}&{\color{red}3.24e-16 (21.7\%)}&{\color{red}2.00e-16 (21.7\%)}&{\color{red}1.96e-16 (23.9\%)}&{\color{red}3.42e-16 (20.6\%)}\\
&{\color{red}o-H$_2$O}&{\color{red}$J_{\mathrm{K}_\mathrm{a}, \mathrm{K}_\mathrm{c}}=2_{2,1} - 1_{1,0}$}&{\color{red}108.07}&{\color{red}6.63e-16 (21.6\%)}&{\color{red}2.84e-16 (21.7\%)}&{\color{red}1.74e-16 (24.8\%)}&{\color{red}1.73e-16 (23.4\%)}&{\color{red}/}&{\color{red}8.01e-17 (27.9\%)}\\
&$^{13}$CO&$J=26 - 25$&105.06&/&/&/&/&/&4.41e-17 (34.0\%)\\
&CO&$J=25 - 24$&104.44&9.97e-16 (21.5\%)&$^\star$~6.75e-16 (20.4\%)&3.36e-16 (22.2\%)&2.79e-16 (22.5\%)&1.77e-16 (30.4\%)&3.38e-16 (20.8\%)\\\hline
B2B&CO&$J=27 - 26$&96.77&5.41e-16 (20.8\%)&$^\star$~4.00e-16 (21.5\%)&$^\star$~2.44e-16 (23.5\%)&1.67e-16 (25.9\%)&8.39e-17 (38.9\%)&1.82e-16 (21.7\%)\\
&p-H$_2$O&$J_{\mathrm{K}_\mathrm{a}, \mathrm{K}_\mathrm{c}}=5_{1,5} - 4_{0,4}$&95.63&2.82e-16 (24.0\%)&1.13e-16 (24.5\%)&5.61e-17 (54.8\%)&/&/&/\\
&o-H$_2$O&$J_{\mathrm{K}_\mathrm{a}, \mathrm{K}_\mathrm{c}}=4_{4,1} - 4_{3,2}$&94.71&9.47e-17 (38.0\%)&/&/&/&/&/\\
&o-H$_2$O&$J_{\mathrm{K}_\mathrm{a}, \mathrm{K}_\mathrm{c}}=6_{2,5} - 6_{1,6}$&94.64&1.60e-16 (29.2\%)&6.74e-17 (29.3\%)&/&/&/&/\\
&p-H$_2$O&$J_{\mathrm{K}_\mathrm{a}, \mathrm{K}_\mathrm{c}}=7_{3,5} - 7_{2,6}$&93.38&$^\star$~7.03e-16 (21.1\%)&$^\star$~4.98e-16 (20.7\%)&$^\star$~2.61e-16 (22.1\%)&$^\star$~1.42e-16 (23.0\%)&$^\star$~1.21e-16 (25.7\%)&$^\star$~2.70e-16 (20.5\%)\\
&CO&$J=28 - 27$&93.35&Blended&Blended&Blended&Blended&Blended&Blended\\
&$^{13}$CO&$J=30 - 29$&91.18&$^\star$~1.64e-16 (24.9\%)&/&$^\star$~8.52e-17 (37.7\%)&/&/&5.15e-17 (28.6\%)\\
&{\color{red}CO}&{\color{red}$J=29 - 28$}&{\color{red}90.16}&{\color{red}1.01e-15 (20.6\%)}&{\color{red}$^\star$~5.17e-16 (20.4\%)}&{\color{red}2.11e-16 (22.2\%)}&{\color{red}2.14e-16 (21.2\%)}&{\color{red}$^\star$~2.31e-16 (23.8\%)}&{\color{red}2.89e-16 (20.9\%)}\\
&{\color{red}p-H$_2$O}&{\color{red}$J_{\mathrm{K}_\mathrm{a}, \mathrm{K}_\mathrm{c}}=3_{2,2} - 2_{1,1}$}&{\color{red}89.99}&{\color{red}3.71e-16 (24.1\%)}&{\color{red}1.17e-16 (24.1\%)}&{\color{red}5.49e-17 (41.3\%)}&{\color{red}5.09e-17 (40.6\%)}&{\color{red}9.45e-17 (33.4\%)}&{\color{red}5.53e-17 (37.3\%)}\\
&$^{13}$CO&$J=31 - 30$&88.27&/&/&6.66e-17 (29.5\%)&/&/&/\\
&{\color{red}CO}&{\color{red}$J=30 - 29$}&{\color{red}87.19}&{\color{red}8.55e-16 (20.4\%)}&{\color{red}4.76e-16 (20.5\%)}&{\color{red}1.75e-16 (24.9\%)}&{\color{red}1.24e-16 (26.1\%)}&{\color{red}/}&{\color{red}$^\star$~3.45e-16 (20.6\%)}\\
&o-H$_2$O&$J_{\mathrm{K}_\mathrm{a}, \mathrm{K}_\mathrm{c}}=7_{1,6} - 7_{0,7}$&84.77&/&/&6.22e-17 (36.2\%)&/&/&/\\
&CO&$J=31 - 30$&84.41&$^\star$~8.60e-16 (20.6\%)&$^\star$~4.55e-16 (20.7\%)&$^\star$~2.12e-16 (24.2\%)&1.82e-16 (25.3\%)&/&2.55e-16 (21.1\%)\\
&p-H$_2$O&$J_{\mathrm{K}_\mathrm{a}, \mathrm{K}_\mathrm{c}}=6_{0,6} - 5_{1,5}$&83.28&$^\star$~2.13e-16 (32.1\%)&1.14e-16 (22.5\%)&/&/&/&/\\
&$^{13}$CO&$J=33 - 32$&82.99&/&/&/&/&/&$^\star$~3.62e-17 (39.9\%)\\
&o-H$_2$O&$J_{\mathrm{K}_\mathrm{a}, \mathrm{K}_\mathrm{c}}=8_{3,6} - 8_{2,7}$&82.98&/&/&/&/&/&Blended\\
&o-H$_2$O&$J_{\mathrm{K}_\mathrm{a}, \mathrm{K}_\mathrm{c}}=6_{1,6} - 5_{0,5}$&82.03&4.06e-16 (21.7\%)&$^\star$~2.91e-16 (20.8\%)&$^\star$~1.84e-16 (25.7\%)&1.25e-16 (27.1\%)&/&$^\star$~7.73e-17 (26.0\%)\\
&CO&$J=32 - 31$&81.81&4.94e-16 (21.0\%)&3.55e-16 (20.5\%)&8.47e-17 (27.0\%)&1.12e-16 (27.8\%)&/&2.07e-16 (20.8\%)\\
&p-H$_2$O&$J_{\mathrm{K}_\mathrm{a}, \mathrm{K}_\mathrm{c}}=8_{3,5} - 7_{4,4}$&81.69&/&/&/&/&/&5.02e-17 (28.3\%)\\
&p-H$_2$O&$J_{\mathrm{K}_\mathrm{a}, \mathrm{K}_\mathrm{c}}=7_{2,6} - 7_{1,7}$&81.22&7.07e-17 (39.8\%)&/&/&/&/&/\\
&CO&$J=33 - 32$&79.36&5.52e-16 (21.1\%)&$^\star$~4.83e-16 (20.7\%)&1.65e-16 (28.1\%)&8.18e-17 (33.3\%)&/&2.20e-16 (21.8\%)\\
&p-H$_2$O&$J_{\mathrm{K}_\mathrm{a}, \mathrm{K}_\mathrm{c}}=6_{1,5} - 5_{2,4}$&78.93&$^\star$~3.45e-16 (27.0\%)&/&/&/&/&$^\star$~1.22e-16 (28.4\%)\\
&o-H$_2$O&$J_{\mathrm{K}_\mathrm{a}, \mathrm{K}_\mathrm{c}}=4_{2,3} - 3_{1,2}$&78.74&6.37e-16 (20.7\%)&3.23e-16 (21.3\%)&1.25e-16 (31.3\%)&1.84e-16 (23.6\%)&/&1.53e-16 (23.6\%)\\
&o-H$_2$O&$J_{\mathrm{K}_\mathrm{a}, \mathrm{K}_\mathrm{c}}=7_{5,2} - 7_{4,3}$&77.76&4.78e-17 (33.5\%)&/&/&/&/&/\\
&CO&$J=34 - 33$&77.06&$^\star$~5.89e-16 (21.8\%)&$^\star$~4.48e-16 (20.6\%)&8.66e-17 (29.4\%)&$^\star$~1.15e-16 (25.6\%)&8.88e-17 (31.6\%)&1.76e-16 (21.6\%)\\
&$^{13}$CO&$J=36 - 35$&76.17&/&/&$^\star$~9.11e-17 (30.8\%)&/&/&/\\
&o-H$_2$O&$J_{\mathrm{K}_\mathrm{a}, \mathrm{K}_\mathrm{c}}=5_{5,0} - 5_{4,1}$&75.91&/&/&5.54e-17 (36.7\%)&/&/&/\\
&o-H$_2$O&$J_{\mathrm{K}_\mathrm{a}, \mathrm{K}_\mathrm{c}}=6_{5,2} - 6_{4,3}$&75.83&/&/&$^\star$~1.72e-17 (89.9\%)&/&/&/\\
&p-H$_2$O&$J_{\mathrm{K}_\mathrm{a}, \mathrm{K}_\mathrm{c}}=7_{5,3} - 7_{4,4}$&75.81&$^\star$~2.24e-16 (25.9\%)&/&Blended&/&/&$^\star$~1.58e-16 (22.6\%)\\
&p-H$_2$O&$J_{\mathrm{K}_\mathrm{a}, \mathrm{K}_\mathrm{c}}=5_{5,1} - 5_{4,2}$&75.78&Blended&/&$^\star$~1.19e-16 (34.8\%)&/&/&Blended\\
&o-H$_2$O&$J_{\mathrm{K}_\mathrm{a}, \mathrm{K}_\mathrm{c}}=3_{2,1} - 2_{1,2}$&75.38&7.81e-16 (20.6\%)&4.47e-16 (20.4\%)&2.33e-16 (21.4\%)&2.34e-16 (22.3\%)&/&$^\star$~1.90e-16 (22.0\%)\\
&o-H$_2$O&$J_{\mathrm{K}_\mathrm{a}, \mathrm{K}_\mathrm{c}}=7_{2,5} - 6_{3,4}$&74.95&$^\star$~2.22e-16 (39.3\%)&$^\star$~1.14e-16 (41.1\%)&5.39e-17 (37.2\%)&$^\star$~9.41e-17 (58.7\%)&/&/\\
&CO&$J=35 - 34$&74.89&$^\star$~5.46e-16 (23.2\%)&3.05e-16 (23.8\%)&7.94e-17 (27.8\%)&6.76e-17 (44.1\%)&$^\star$~1.36e-16 (25.0\%)&1.83e-16 (21.7\%)\\
&$^{13}$CO&$J=37 - 36$&74.14&2.67e-16 (22.1\%)&1.09e-16 (23.3\%)&/&/&/&$^\star$~1.61e-16 (23.4\%)\\
&{\color{red}CO}&{\color{red}$J=36 - 35$}&{\color{red}72.84}&{\color{red}3.82e-16 (22.4\%)}&{\color{red}$^\star$~3.47e-16 (20.9\%)}&{\color{red}/}&{\color{red}/}&{\color{red}/}&{\color{red}$^\star$~1.91e-16 (22.8\%)}\\
&{\color{red}o-H$_2$O}&{\color{red}$J_{\mathrm{K}_\mathrm{a}, \mathrm{K}_\mathrm{c}}=7_{0,7} - 6_{1,6}$}&{\color{red}71.95}&{\color{red}3.22e-16 (23.5\%)}&{\color{red}/}&{\color{red}/}&{\color{red}/}&{\color{red}/}&{\color{red}/}\\
&{\color{red}p-H$_2$O}&{\color{red}$J_{\mathrm{K}_\mathrm{a}, \mathrm{K}_\mathrm{c}}=7_{1,7} - 6_{0,6}$}&{\color{red}71.54}&{\color{red}$^\star$~2.90e-16 (28.0\%)}&{\color{red}/}&{\color{red}/}&{\color{red}/}&{\color{red}/}&{\color{red}/}\\
&p-H$_2$O&$J_{\mathrm{K}_\mathrm{a}, \mathrm{K}_\mathrm{c}}=5_{2,4} - 4_{1,3}$&71.07&2.12e-16 (38.2\%)&$^\star$~1.40e-16 (36.9\%)&/&/&/&/\\
&CO&$J=37 - 36$&70.91&4.33e-16 (23.6\%)&$^\star$~2.80e-16 (25.7\%)&/&/&/&/\\
&o-H$_2$O&$J_{\mathrm{K}_\mathrm{a}, \mathrm{K}_\mathrm{c}}=8_{2,7} - 8_{1,8}$&70.70&5.65e-17 (70.5\%)&/&/&/&/&/\\\hline
B2A&{\color{red}CO}&{\color{red}$J=36 - 35$}&{\color{red}72.84}&{\color{red}4.80e-16 (22.3\%)}&{\color{red}$^\star$~3.49e-16 (21.2\%)}&{\color{red}7.06e-17 (34.7\%)}&{\color{red}7.48e-17 (35.7\%)}&{\color{red}/}&{\color{red}$^\star$~1.38e-16 (25.9\%)}\\
&{\color{red}o-H$_2$O}&{\color{red}$J_{\mathrm{K}_\mathrm{a}, \mathrm{K}_\mathrm{c}}=7_{0,7} - 6_{1,6}$}&{\color{red}71.95}&{\color{red}3.47e-16 (24.4\%)}&{\color{red}2.40e-16 (21.2\%)}&{\color{red}5.97e-17 (35.5\%)}&{\color{red}8.00e-17 (34.7\%)}&{\color{red}/}&{\color{red}$^\star$~8.21e-17 (25.4\%)}\\
&{\color{red}p-H$_2$O}&{\color{red}$J_{\mathrm{K}_\mathrm{a}, \mathrm{K}_\mathrm{c}}=7_{1,7} - 6_{0,6}$}&{\color{red}71.54}&{\color{red}1.98e-16 (23.3\%)}&{\color{red}$^\star$~1.28e-16 (27.7\%)}&{\color{red}5.27e-17 (38.6\%)}&{\color{red}$^\star$~7.35e-17 (41.7\%)}&{\color{red}/}&{\color{red}$^\star$~9.65e-17 (26.2\%)}\\
&p-H$_2$O&$J_{\mathrm{K}_\mathrm{a}, \mathrm{K}_\mathrm{c}}=5_{2,4} - 4_{1,3}$&71.07&3.41e-16 (23.4\%)&1.38e-16 (24.4\%)&/&1.01e-16 (25.8\%)&/&/\\
&CO&$J=37 - 36$&70.91&4.42e-16 (21.1\%)&$^\star$~3.53e-16 (21.2\%)&9.89e-17 (26.4\%)&1.18e-16 (24.2\%)&/&1.37e-16 (24.5\%)\\
&o-H$_2$O&$J_{\mathrm{K}_\mathrm{a}, \mathrm{K}_\mathrm{c}}=8_{2,7} - 8_{1,8}$&70.70&1.37e-16 (28.9\%)&/&/&/&/&/\\
&{\color{red}CO}&{\color{red}$J=38 - 37$}&{\color{red}69.07}&{\color{red}4.49e-16 (22.0\%)}&{\color{red}$^\star$~3.65e-16 (22.0\%)}&{\color{red}/}&{\color{red}/}&{\color{red}/}&{\color{red}1.44e-16 (23.6\%)}\\
&CO&$J=39 - 38$&67.34&$^\star$~4.44e-16 (24.7\%)&$^\star$~2.89e-16 (21.8\%)&1.44e-16 (27.8\%)&/&/&1.16e-16 (27.1\%)\\
&o-H$_2$O&$J_{\mathrm{K}_\mathrm{a}, \mathrm{K}_\mathrm{c}}=3_{3,0} - 3_{0,3}$&67.27&$^\star$~2.06e-16 (35.7\%)&9.80e-17 (29.5\%)&$^\star$~7.50e-17 (46.5\%)&/&/&/\\
&p-H$_2$O&$J_{\mathrm{K}_\mathrm{a}, \mathrm{K}_\mathrm{c}}=3_{3,1} - 2_{2,0}$&67.09&$^\star$~5.01e-16 (23.0\%)&$^\star$~2.48e-16 (22.4\%)&$^\star$~1.34e-16 (34.9\%)&1.04e-16 (29.1\%)&/&6.92e-17 (35.4\%)\\
&$^{13}$CO&$J=41 - 40$&67.04&/&/&1.86e-17 (87.9\%)&/&/&1.23e-16 (27.0\%)\\
&{\color{red}o-H$_2$O}&{\color{red}$J_{\mathrm{K}_\mathrm{a}, \mathrm{K}_\mathrm{c}}=3_{3,0} - 2_{2,1}$}&{\color{red}66.44}&{\color{red}8.43e-16 (20.7\%)}&{\color{red}5.03e-16 (20.4\%)}&{\color{red}$^\star$~3.02e-16 (22.7\%)}&{\color{red}$^\star$~2.01e-16 (25.0\%)}&{\color{red}/}&{\color{red}$^\star$~2.06e-16 (22.1\%)}\\
&o-H$_2$O&$J_{\mathrm{K}_\mathrm{a}, \mathrm{K}_\mathrm{c}}=7_{1,6} - 6_{2,5}$&66.09&2.35e-16 (28.4\%)&$^\star$~1.87e-16 (24.8\%)&/&/&/&/\\
&CO&$J=40 - 39$&65.69&$^\star$~4.30e-16 (25.9\%)&$^\star$~2.98e-16 (22.2\%)&6.68e-17 (37.5\%)&/&/&1.07e-16 (25.3\%)\\
&o-H$_2$O&$J_{\mathrm{K}_\mathrm{a}, \mathrm{K}_\mathrm{c}}=6_{2,5} - 5_{1,4}$&65.17&$^\star$~5.58e-16 (22.7\%)&3.13e-16 (22.8\%)&1.38e-16 (32.0\%)&1.21e-16 (28.9\%)&/&8.93e-17 (28.8\%)\\
&CO&$J=41 - 40$&64.12&/&1.64e-16 (27.6\%)&/&/&/&/\\
&p-H$_2$O&$J_{\mathrm{K}_\mathrm{a}, \mathrm{K}_\mathrm{c}}=8_{0,8} - 7_{1,7}$&63.46&$^\star$~2.87e-16 (28.5\%)&$^\star$~1.36e-16 (29.7\%)&/&/&/&/\\
&o-H$_2$O&$J_{\mathrm{K}_\mathrm{a}, \mathrm{K}_\mathrm{c}}=8_{1,8} - 7_{0,7}$&63.32&$^\star$~3.95e-16 (38.6\%)&$^\star$~3.05e-16 (23.6\%)&/&/&/&/\\
&$^{13}$CO&$J=45 - 44$&61.21&$^\star$~2.76e-16 (30.9\%)&$^\star$~1.26e-16 (35.2\%)&/&/&/&/\\
&CO&$J=43 - 42$&61.20&Blended&Blended&/&/&/&/\\
&CO&$J=44 - 43$&59.84&$^\star$~3.34e-16 (27.6\%)&/&/&/&/&/\\
&o-H$_2$O&$J_{\mathrm{K}_\mathrm{a}, \mathrm{K}_\mathrm{c}}=4_{3,2} - 3_{2,1}$&58.70&$^\star$~6.92e-16 (23.2\%)&$^\star$~4.80e-16 (20.9\%)&1.92e-16 (26.0\%)&1.30e-16 (26.9\%)&/&1.29e-16 (27.4\%)\\
&CO&$J=45 - 44$&58.55&1.59e-16 (42.1\%)&$^\star$~1.35e-16 (39.6\%)&/&/&/&$^\star$~9.61e-17 (34.8\%)\\
&p-H$_2$O&$J_{\mathrm{K}_\mathrm{a}, \mathrm{K}_\mathrm{c}}=6_{4,2} - 7_{1,7}$&58.38&/&1.20e-16 (31.2\%)&/&/&/&/\\
&p-H$_2$O&$J_{\mathrm{K}_\mathrm{a}, \mathrm{K}_\mathrm{c}}=4_{2,2} - 3_{1,3}$&57.64&$^\star$~5.12e-16 (29.1\%)&2.32e-16 (24.6\%)&/&/&/&/\\
&o-H$_2$O&$J_{\mathrm{K}_\mathrm{a}, \mathrm{K}_\mathrm{c}}=9_{0,9} - 8_{1,8}$&56.82&$^\star$~4.35e-16 (33.0\%)&1.54e-16 (28.0\%)&/&/&/&/\\
&p-H$_2$O&$J_{\mathrm{K}_\mathrm{a}, \mathrm{K}_\mathrm{c}}=9_{1,9} - 8_{0,8}$&56.77&Blended&/&/&/&/&/\\
&$^{13}$CO&$J=49 - 48$&56.34&$^\star$~5.57e-16 (27.6\%)&$^\star$~2.91e-16 (25.1\%)&/&/&/&/\\
&p-H$_2$O&$J_{\mathrm{K}_\mathrm{a}, \mathrm{K}_\mathrm{c}}=4_{3,1} - 3_{2,2}$&56.32&Blended&Blended&/&/&/&/\\
&o-H$_2$O&$J_{\mathrm{K}_\mathrm{a}, \mathrm{K}_\mathrm{c}}=8_{2,7} - 7_{1,6}$&55.13&/&/&/&/&/&$^\star$~1.72e-16 (27.1\%)\\\hline
\end{longtable}
\end{landscape}
%\end{longtab}

\addtocounter{table}{-1}
%\begin{longtab}
\begin{landscape}
\tiny
\begin{longtable}{lllrrrrrrr}
\caption{{Integrated line strengths $I_\mathrm{int}$ (W m$^{-2}$) for a selection of lines in the PACS spectra of OT2 carbon stars observed in the old observation setting (see Sect.~\ref{sect:targetsel}) and for the additional line scan of LL~Peg. See Table~\ref{table:intintmess} for further clarification of the given information.}}\label{table:intintot2old}\\
\hline\hline
&&&&QZ~Mus&V821~Her&V1417~Aql&S~Cep&RV~Cyg&LL~Peg \\\hline
PACS&Molecule&Rotational&$\lambda_0$&\multicolumn{6}{c}{$I_\mathrm{int}$} \\
band&&transition&$\mu$m&\multicolumn{6}{c}{(W m$^{-2}$)} \\\hline
\endfirsthead
\caption{continued.}\\
\hline\hline
&&&&QZ~Mus&V821~Her&V1417~Aql&S~Cep&RV~Cyg&LL~Peg \\\hline
PACS&Molecule&Rotational&$\lambda_0$&\multicolumn{6}{c}{$I_\mathrm{int}$} \\
band&&transition&$\mu$m&\multicolumn{6}{c}{(W m$^{-2}$)} \\\hline
\endhead
\hline 
% \multicolumn{10}{c}{\tablefoot{\tablefoottext{$\star$}{\parbox{\LTcapwidth}{Line strengths flagged for potential line blends. Transitions that might cause the line blend are mentioned immediately below the flagged transition.}}
% }}
\endfoot
R1B&o-H$_2$O&$J_{\mathrm{K}_\mathrm{a}, \mathrm{K}_\mathrm{c}}=2_{2,1} - 2_{1,2}$&180.49&3.45e-17 (21.9\%)&6.34e-17 (20.9\%)&5.01e-17 (20.9\%)&4.17e-17 (21.7\%)&/&/\\
&o-H$_2$O&$J_{\mathrm{K}_\mathrm{a}, \mathrm{K}_\mathrm{c}}=2_{1,2} - 1_{0,1}$&179.53&1.50e-16 (20.2\%)&1.54e-16 (20.2\%)&1.75e-16 (20.1\%)&1.92e-16 (20.1\%)& 1.77e-17 (25.5\%) &4.26e-17 (22.4\%)\\
&CO&$J=15 - 14$&173.63&1.41e-16 (20.1\%)&2.17e-16 (20.2\%)&2.35e-16 (20.1\%)&1.43e-16 (20.3\%)&3.07e-17 (22.4\%)&2.48e-16 (20.1\%)\\\hline
R1A&CO&$J=18 - 17$&144.78&1.26e-16 (20.1\%)&2.43e-16 (20.1\%)&2.35e-16 (20.1\%)&2.08e-16 (20.1\%)&3.37e-17 (27.0\%)&2.76e-16 (20.0\%)\\
&p-H$_2$O&$J_{\mathrm{K}_\mathrm{a}, \mathrm{K}_\mathrm{c}}=4_{1,3} - 3_{2,2}$&144.52&1.67e-17 (26.0\%)&1.35e-17 (33.2\%)&2.01e-17 (31.9\%)&3.22e-17 (22.6\%)&/&/\\
&$^{13}$CO&$J=19 - 18$&143.49&6.85e-18 (31.6\%)&2.37e-17 (23.1\%)&/&/&/&6.38e-17 (20.3\%)\\
&p-H$_2$O&$J_{\mathrm{K}_\mathrm{a}, \mathrm{K}_\mathrm{c}}=8_{4,4} - 7_{5,3}$&138.64&$^\star$~8.67e-17 (20.1\%)&$^\star$~1.04e-16 (20.4\%)&$^\star$~1.27e-16 (20.2\%)&$^\star$~1.30e-16 (20.1\%)&$^\star$~6.65e-18 (32.5\%)&/\\
&p-H$_2$O&$J_{\mathrm{K}_\mathrm{a}, \mathrm{K}_\mathrm{c}}=3_{1,3} - 2_{0,2}$&138.53&Blended&Blended&Blended&Blended&Blended&/\\
&CO&$J=24 - 23$&108.76&1.34e-16 (20.5\%)&2.28e-16 (20.2\%)&$^\star$~1.91e-16 (22.4\%)&2.32e-16 (20.1\%)&$^\star$~3.60e-17 (22.9\%)&1.94e-16 (21.2\%)\\
&o-H$_2$O&$J_{\mathrm{K}_\mathrm{a}, \mathrm{K}_\mathrm{c}}=2_{2,1} - 1_{1,0}$&108.07&1.81e-16 (20.3\%)&2.43e-16 (20.1\%)&3.02e-16 (20.0\%)&2.84e-16 (20.1\%)&2.22e-17 (23.7\%)&/\\\hline
B2B&CO&$J=29 - 28$&90.16&1.21e-16 (20.4\%)&2.25e-16 (20.4\%)&2.00e-16 (20.2\%)&1.96e-16 (20.1\%)&4.90e-17 (36.1\%)&1.46e-16 (20.6\%)\\
&p-H$_2$O&$J_{\mathrm{K}_\mathrm{a}, \mathrm{K}_\mathrm{c}}=3_{2,2} - 2_{1,1}$&89.99&8.70e-17 (20.6\%)&9.63e-17 (21.8\%)&1.19e-16 (20.6\%)&1.15e-16 (20.4\%)&/&/\\
&CO&$J=30 - 29$&87.19&1.12e-16 (26.0\%)&2.08e-16 (20.8\%)&$^\star$~2.25e-16 (21.0\%)&1.49e-16 (20.9\%)&/&1.28e-16 (21.6\%)\\
&o-H$_2$O&$J_{\mathrm{K}_\mathrm{a}, \mathrm{K}_\mathrm{c}}=7_{0,7} - 6_{1,6}$&71.95&1.10e-16 (20.6\%)&1.02e-16 (20.8\%)&1.36e-16 (20.8\%)&1.45e-16 (20.5\%)&/&/\\
&p-H$_2$O&$J_{\mathrm{K}_\mathrm{a}, \mathrm{K}_\mathrm{c}}=7_{1,7} - 6_{0,6}$&71.54&7.61e-17 (21.5\%)&/&/&/&/&/\\
&CO&$J=38 - 37$&69.07&4.97e-17 (32.7\%)&/&/&$^\star$~1.19e-16 (28.1\%)&/&/\\\hline
B3A&o-H$_2$O&$J_{\mathrm{K}_\mathrm{a}, \mathrm{K}_\mathrm{c}}=3_{3,0} - 2_{2,1}$&66.44&2.48e-16 (20.1\%)&2.97e-16 (20.1\%)&3.70e-16 (20.1\%)&3.19e-16 (20.1\%)&/&/\\
\end{longtable}
\tiny
\begin{longtable}{lllrrrrrrrr}
\caption{{Integrated line strengths $I_\mathrm{int}$ (W m$^{-2}$) for a selection of lines in the PACS spectra of OT2 carbon stars observed in the new observation setting (see Sect.~\ref{sect:targetsel}). See Table~\ref{table:intintmess} for further clarification of the given information.}}\label{table:intintot2new}\\
\hline\hline
&&&&V384~Per&R~Lep&W~Ori&S~Aur&U~Hya&Y~CVn&AFGL~4202 \\\hline
PACS&Molecule&Rotational&$\lambda_0$&\multicolumn{7}{c}{$I_\mathrm{int}$} \\
band&&transition&$\mu$m&\multicolumn{7}{c}{(W m$^{-2}$)} \\\hline
\endfirsthead
\caption{continued.}\\
\hline\hline
&&&&V384~Per&R~Lep&W~Ori&S~Aur&U~Hya&Y~CVn&AFGL~4202 \\\hline
PACS&Molecule&Rotational&$\lambda_0$&\multicolumn{7}{c}{$I_\mathrm{int}$} \\
band&&transition&$\mu$m&\multicolumn{7}{c}{(W m$^{-2}$)} \\\hline
\endhead
\hline 
% \multicolumn{11}{c}{\tablefoot{\tablefoottext{$\star$}{\parbox{\LTcapwidth}{Line strengths flagged for potential line blends. Transitions that might cause the line blend are mentioned immediately below the flagged transition.}}
% }}
\endfoot
R1B&o-H$_2$O&$J_{\mathrm{K}_\mathrm{a}, \mathrm{K}_\mathrm{c}}=2_{2,1} - 2_{1,2}$&180.49&2.72e-17 (23.3\%)&6.41e-17 (20.5\%)&/&/&/&/&3.45e-17 (22.9\%)\\
&o-H$_2$O&$J_{\mathrm{K}_\mathrm{a}, \mathrm{K}_\mathrm{c}}=2_{1,2} - 1_{0,1}$&179.53&8.45e-17 (20.2\%)&2.26e-16 (20.1\%)&2.38e-17 (23.1\%)&5.87e-17 (20.5\%)&$^\star$~1.71e-17 (28.0\%)&3.43e-17 (21.9\%)&9.28e-17 (20.7\%)\\
&o-H$_2$O&$J_{\mathrm{K}_\mathrm{a}, \mathrm{K}_\mathrm{c}}=3_{0,3} - 2_{1,2}$&174.63&$^\star$~8.03e-17 (20.5\%)&$^\star$~1.86e-16 (20.1\%)&$^\star$~2.14e-17 (25.8\%)&$^\star$~5.70e-17 (20.8\%)&$^\star$~1.57e-17 (24.3\%)&$^\star$~2.29e-17 (24.0\%)&$^\star$~5.53e-17 (21.3\%)\\
&p-H$_2$O&$J_{\mathrm{K}_\mathrm{a}, \mathrm{K}_\mathrm{c}}=5_{3,3} - 6_{0,6}$&174.61&Blended&Blended&Blended&Blended&Blended&Blended&Blended\\
&CO&$J=15 - 14$&173.63&2.11e-16 (20.1\%)&1.54e-16 (20.2\%)&4.94e-17 (21.2\%)&6.36e-17 (20.6\%)&5.36e-17 (20.5\%)&4.88e-17 (21.1\%)&1.53e-16 (20.2\%)\\\hline
R1A&CO&$J=18 - 17$&144.78&2.14e-16 (20.0\%)&1.75e-16 (20.2\%)&4.45e-17 (20.4\%)&5.84e-17 (20.1\%)&4.13e-17 (20.4\%)&5.70e-17 (20.3\%)&1.65e-16 (20.1\%)\\
&p-H$_2$O&$J_{\mathrm{K}_\mathrm{a}, \mathrm{K}_\mathrm{c}}=4_{1,3} - 3_{2,2}$&144.52&9.07e-18 (31.9\%)&3.98e-17 (23.0\%)&/&7.95e-18 (25.9\%)&/&/&$^\star$~3.25e-17 (24.8\%)\\
&$^{13}$CO&$J=19 - 18$&143.49&/&/&/&/&7.74e-18 (27.5\%)&2.19e-17 (20.8\%)&1.86e-17 (24.1\%)\\
&p-H$_2$O&$J_{\mathrm{K}_\mathrm{a}, \mathrm{K}_\mathrm{c}}=8_{4,4} - 7_{5,3}$&138.64&$^\star$~7.12e-17 (20.1\%)&$^\star$~1.87e-16 (20.1\%)&$^\star$~1.49e-17 (21.9\%)&$^\star$~3.68e-17 (20.2\%)&$^\star$~1.02e-17 (25.7\%)&$^\star$~2.72e-17 (21.8\%)&$^\star$~4.88e-17 (21.1\%)\\
&p-H$_2$O&$J_{\mathrm{K}_\mathrm{a}, \mathrm{K}_\mathrm{c}}=3_{1,3} - 2_{0,2}$&138.53&Blended&Blended&Blended&Blended&Blended&Blended&Blended\\
&CO&$J=24 - 23$&108.76&1.67e-16 (20.4\%)&2.22e-16 (20.1\%)&$^\star$~6.07e-17 (22.7\%)&6.12e-17 (20.4\%)&4.83e-17 (21.7\%)&$^\star$~6.76e-17 (23.3\%)&1.02e-16 (20.6\%)\\
&o-H$_2$O&$J_{\mathrm{K}_\mathrm{a}, \mathrm{K}_\mathrm{c}}=2_{2,1} - 1_{1,0}$&108.07&1.13e-16 (20.8\%)&3.00e-16 (20.1\%)&3.34e-17 (26.3\%)&8.40e-17 (20.2\%)&1.17e-17 (25.6\%)&$^\star$~6.02e-17 (23.3\%)&9.10e-17 (20.6\%)\\\hline
B2B&CO&$J=29 - 28$&90.16&1.29e-16 (20.4\%)&2.10e-16 (20.2\%)&$^\star$~6.44e-17 (22.4\%)&$^\star$~6.11e-17 (21.9\%)&6.83e-17 (20.9\%)&5.07e-17 (24.6\%)&8.47e-17 (21.0\%)\\
&p-H$_2$O&$J_{\mathrm{K}_\mathrm{a}, \mathrm{K}_\mathrm{c}}=3_{2,2} - 2_{1,1}$&89.99&6.61e-17 (21.6\%)&1.57e-16 (20.4\%)&/&$^\star$~4.37e-17 (23.7\%)&1.00e-17 (34.1\%)&/&4.72e-17 (22.8\%)\\
&CO&$J=30 - 29$&87.19&1.26e-16 (20.6\%)&2.35e-16 (20.2\%)&4.30e-17 (21.8\%)&$^\star$~4.29e-17 (23.9\%)&5.99e-17 (21.2\%)&3.71e-17 (24.0\%)&$^\star$~9.70e-17 (21.4\%)\\
&CO&$J=36 - 35$&72.84&5.39e-17 (22.5\%)&1.70e-16 (20.8\%)&4.52e-17 (21.7\%)&3.25e-17 (21.9\%)&4.89e-17 (22.1\%)&/&/\\
&o-H$_2$O&$J_{\mathrm{K}_\mathrm{a}, \mathrm{K}_\mathrm{c}}=7_{0,7} - 6_{1,6}$&71.95&5.64e-17 (22.0\%)&2.52e-16 (20.4\%)&/&3.83e-17 (21.3\%)&$^\star$~2.45e-17 (28.9\%)&/&/\\
&p-H$_2$O&$J_{\mathrm{K}_\mathrm{a}, \mathrm{K}_\mathrm{c}}=7_{1,7} - 6_{0,6}$&71.54&/&1.16e-16 (27.1\%)&/&/&3.83e-17 (32.7\%)&/&/\\
&CO&$J=38 - 37$&69.07&/&1.31e-16 (23.0\%)&/&/&3.84e-17 (32.1\%)&/&5.36e-17 (38.5\%)\\\hline
B3A&o-H$_2$O&$J_{\mathrm{K}_\mathrm{a}, \mathrm{K}_\mathrm{c}}=3_{3,0} - 2_{2,1}$&66.44&1.61e-16 (20.3\%)&4.15e-16 (20.1\%)&1.17e-17 (42.6\%)&/&$^\star$~2.20e-17 (30.9\%)&2.67e-17 (22.1\%)&1.24e-16 (20.3\%)\\
\end{longtable}
\end{landscape}
%\end{longtab}

\addtocounter{table}{-1}
%\setlength\LTleft{2cm}
%\setlength\LTright{2cm}
%\begin{longtab}
\begin{landscape}
\tiny
\begin{longtable}{lrrrrrrr}
\caption{Integrated line strengths $I_\mathrm{int}$ (W m$^{-2}$) for additional emission lines in the PACS spectra of OT2 carbon stars observed in the old observation setting, and LL~Peg, which cannot be attributed to CO or \water and remain unidentified. The approximate central wavelength $\lambda_0$ (\mic) of the emission line is indicated. The percentages between brackets give the uncertainty on $I_\mathrm{int}$, which includes both the fitting uncertainty and the PACS absolute-flux calibration uncertainty of 20\%. Line strengths indicated with $\star$ are flagged for potential line blends.}\label{table:unidentified}\\
\hline\hline
&&QZ~Mus&V821~Her&V1417~Aql&S~Cep&RV~Cyg&LL~Peg \\\hline
PACS&$\lambda_0$&\multicolumn{6}{c}{$I_\mathrm{int}$} \\
band&$\mu$m&\multicolumn{6}{c}{(W m$^{-2}$)} \\\hline
\endfirsthead
\caption{continued.}\\
\hline\hline
&&QZ~Mus&V821~Her&V1417~Aql&S~Cep&RV~Cyg&LL~Peg \\\hline
PACS&$\lambda_0$&\multicolumn{6}{c}{$I_\mathrm{int}$} \\
band&$\mu$m&\multicolumn{6}{c}{(W m$^{-2}$)} \\\hline
\endhead
\hline 
% \multicolumn{8}{c}{\tablefoot{\tablefoottext{$\star$}{\parbox{\LTcapwidth}{Line strengths flagged for potential line blends.}}
% }}
\endfoot
R1B&180.7&$^\star$~1.90e-17 (26.55\%)&$^\star$~2.00e-17 (27.29\%)&$^\star$~2.80e-17 (23.79\%)&$^\star$~1.30e-17 (35.75\%)&/&/\\
&173.9&$^\star$~1.49e-17 (31.97\%)&/& $^\star$~2.31e-17 (29.72\%)  &/          &/&6.05e-17 (22.01\%)\\
&173.4&$^\star$~1.90e-17 (26.55\%)&/& 1.60e-17 (31.40\%)  /&/          &/&/\\
&172.8&1.47e-17 (33.62\%)&/&     /&/          &/&/\\
R1A&145.0&/&/&/&/&/&$^\star$~8.79e-17 (20.70\%)\\
&143.2&1.41e-17 (22.58\%)&2.74e-17 (22.99\%)&2.83e-17 (22.20\%)&2.70e-17 (21.70\%)  &/&/\\
&139.3&1.97e-17 (23.64\%)&/&     3.31e-17 (27.02\%)&4.87e-17 (20.94\%)  &1.36e-17 (23.48\%)&/\\
&137.9&1.45e-17 (22.25\%)&/&     2.80e-17 (23.83\%)&4.61e-17 (21.45\%)  &1.59e-17 (22.93\%)&/\\
&109.0&/&2.83e-17 (26.78\%)& 5.03e-17 (21.45\%)  &$^\star$~4.16e-17 (23.67\%)  &/ &9.49e-17 (23.80\%)\\
&108.5&/&3.09e-17 (26.27\%)& $^\star$~1.87e-17 (39.49\%)&$^\star$~5.26e-17 (23.64\%)  &/ &/\\
B2B&90.3&/&/&/           &1.45e-17 (33.86\%)    &/&/\\
&89.5&$^\star$~5.84e-17 (24.64\%)&/& $^\star$~6.78e-17 (23.04\%)&5.22e-17 (23.03\%)  &/&$^\star$~8.10e-17 (25.62\%)\\
&87.3&/&5.96e-17 (27.94\%)&      5.61e-17 (26.36\%)&3.51e-17 (30.60\%)  &/&/\\
&71.8&/&/&/&             1.57e-17 (35.75\%)  &/&/\\
&71.6&2.40e-17 (32.39\%)&/&/     &/&/&/\\

\end{longtable}
\begin{longtable}{llrrrrrrr}
\caption{{Integrated line strengths $I_\mathrm{int}$ (W m$^{-2}$) for additional emission lines in the PACS spectra of OT2 carbon stars observed in the new observation setting, which cannot be attributed to CO or \water and remain unidentified. See Table~\ref{table:unidentified} for further clarification of the given information.}}\label{table:unidentified2}\\
\hline\hline
&&V384~Per&R~Lep&W~Ori&S~Aur&U~Hya&Y~CVn&AFGL~4202 \\\hline
PACS&$\lambda_0$&\multicolumn{7}{c}{$I_\mathrm{int}$} \\
band&$\mu$m&\multicolumn{7}{c}{(W m$^{-2}$)} \\\hline
\endfirsthead
\caption{continued.}\\
\hline\hline
&&V384~Per&R~Lep&W~Ori&S~Aur&U~Hya&Y~CVn&AFGL~4202 \\\hline
PACS&$\lambda_0$&\multicolumn{7}{c}{$I_\mathrm{int}$} \\
band&$\mu$m&\multicolumn{7}{c}{(W m$^{-2}$)} \\\hline
\endhead
\hline 
% \multicolumn{9}{c}{\tablefoot{\tablefoottext{$\star$}{\parbox{\LTcapwidth}{Line strengths flagged for potential line blends.}}
% }}
\endfoot
R1B&180.7&1.95e-17 (25.73\%)&$^\star$~2.18e-17 (25.71\%)&$^\star$~5.95e-18 (35.55\%)&/  &/&/&$^\star$~2.08e-17 (27.59\%)\\
&174.4& /    &  /    &/&  /      &/&/&6.57e-18 (49.27\%)\\
&173.9&1.24e-17 (38.63\%)&1.47e-17 (37.82\%) &7.40e-18 (43.41\%)& /  &7.59e-18 (33.54\%)&3.52e-17 (22.46\%)&$^\star$~1.76e-17 (32.39\%)\\
&173.4&   /  &    /  &/&9.52e-18 (34.03\%)   &/&/&/\\
R1A&145.7&  /  &/&/&/&/&8.00e-18 (29.95\%)&/\\
&145.0&$^\star$~1.90e-17 (25.25\%)&/ &$^\star$~1.98e-17 (22.86\%)&$^\star$~1.10e-17 (24.37\%)&/&$^\star$~5.45e-17 (20.46\%)&/\\
&143.5&1.46e-17 (23.57\%)&  /  &/&      /  &/&/&/\\
&143.2&1.46e-17 (23.50\%)&3.86e-17 (22.35\%) &/&/       &6.71e-18 (26.92\%)&/&/\\
&139.3&2.06e-17 (35.10\%)&2.87e-17 (25.67\%) &1.68e-17 (21.84\%)& /  &9.31e-18 (27.22\%)&5.15e-17 (20.40\%)&/\\
&137.9&1.83e-17 (27.77\%)&2.32e-17 (22.14\%) &2.51e-17 (21.50\%)&$^\star$~1.30e-17 (22.55\%)  &/&1.84e-17 (22.59\%)&/\\
&109.0&3.60e-17 (28.44\%)&4.11e-17 (24.23\%) &/&9.03e-18 (50.00\%)   &2.19e-17 (28.87\%)&$^\star$~1.18e-16 (26.50\%)&/\\
&108.5&   /  &1.84e-17 (31.88\%) &/&1.23e-17 (26.41\%)   &/&$^\star$~7.15e-17 (25.05\%)&/\\
&107.6& /    &    /  &1.69e-17 (24.99\%)& /  &/&$^\star$~2.84e-17 (30.38\%)&/\\
B2B&89.5&5.92e-17 (22.40\%) &7.54e-17 (23.63\%) &/&  /      &3.88e-17 (28.88\%)&$^\star$~4.45e-17 (26.32\%)&/\\
&87.3&$^\star$~4.99e-17 (24.99\%)&$^\star$~8.31e-17 (21.38\%)&2.62e-17 (24.60\%)&/&$^\star$~3.55e-17 (24.79\%)&2.73e-17 (25.43\%)&2.08e-17 (29.61\%)\\
&86.9&   /  &3.69e-17 (28.39\%) &/&  /      &/&/&/\\
&72.6&2.41e-17 (28.20\%) &6.10e-17 (25.59\%) &/&1.47e-17 (29.33\%)   &1.45e-17 (34.68\%)&/&/\\
&72.5&   /  &4.67e-17 (27.97\%) &/&/       &/&/&/\\
B3A&66.4&    /  &    /  &/&      /  &/&/&$^\star$~1.66e-17 (26.53\%)\\

\end{longtable}
\end{landscape}
%\end{longtab}
\end{document}